\documentclass[fleqn,usenatbib]{mnras}

\usepackage{newtxtext}

\usepackage[T1]{fontenc}

\DeclareRobustCommand{\VAN}[3]{#2}
\let\VANthebibliography\thebibliography
\def\thebibliography{\DeclareRobustCommand{\VAN}[3]{##3}\VANthebibliography}

\usepackage{graphicx}	
\usepackage{amsmath}	
\usepackage{amssymb}	
\numberwithin{equation}{section}
\usepackage{url}

\allowdisplaybreaks[1]

\newcommand{\VEC}{\mathbfit}
\newcommand{\MAT}{\mathbfit}
\newcommand{\hMpc}{\,h^{-1}\,{\rm Mpc}}
\newcommand{\hGpc}{\,h^{-1}\,{\rm Gpc}}
\newcommand{\hk}{\,h\,{\rm Mpc^{-1}}}

\title[New constraints on cosmological modified gravity theories]{New constraints on cosmological modified gravity theories from anisotropic three-point correlation functions of BOSS DR12 galaxies}

\author[N. S. Sugiyama et al.]{
Naonori S. Sugiyama$^{1}$\thanks{E-mail: nao.s.sugiyama@gmail.com}\\
$^{1}$ National Astronomical Observatory of Japan, Mitaka, Tokyo 181-8588, Japan\\
\newauthor 
Daisuke Yamauchi$^{2}$, Tsutomu Kobayashi$^{3}$, Tomohiro Fujita$^{4,5}$, Shun Arai$^{6}$, and Shin'ichi Hirano$^{7}$\\
$^{2}$ Faculty of Engineering, Kanagawa University, Kanagawa, 221-8686, Japan\\
$^{3}$ Department of Physics, Rikkyo University, Toshima, Tokyo 171-8501, Japan\\
$^{4}$ Waseda Institute for Advanced Study, Shinjuku, Tokyo 169-8050, Japan\\
$^{5}$ Research Center for the Early Universe, The University of Tokyo, Bunkyo, Tokyo 113-0033, Japan\\
$^{6}$ Kobayashi-Maskawa Institute, Nagoya University, Nagoya 464-8602, Japan\\
$^{7}$Department of Physics, Tokyo Institute of Technology, 2-12-1 Ookayama, Meguro-ku, Tokyo 152-8551, Japan\\
\newauthor 
Shun Saito$^{8,9}$, Florian Beutler$^{10}$, and Hee-Jong Seo$^{11,12,13}$\\
$^{8}$ Institute for Multi-messenger Astrophysics and Cosmology, Department of Physics,\\
Missouri University of Science and Technology, 1315 N. Pine St., Rolla MO 65409, USA\\
$^{9}$ Kavli Institute for the Physics and Mathematics of the Universe (WPI), \\
Todai Institutes for Advanced Study, The University of Tokyo, Chiba 277-8582, Japan\\
$^{10}$ Institute for Astronomy, University of Edinburgh, Royal Observatory, Blackford Hill, Edinburgh EH9 3HJ, UK\\
$^{11}$ Department of Physics and Astronomy, Ohio University, Clippinger Labs, Athens, OH 45701, USA \\
$^{12}$ Physics Division, Lawrence Berkeley National Laboratory, 1 Cyclotron Road, Berkeley, CA 94720, USA\\
$^{13}$ Berkeley Center for Cosmological Physics, Department of Physics, University of California, Berkeley, CA 94720, USA
}

\date{}

\pubyear{}

\begin{document}
\label{firstpage}
\pagerange{\pageref{firstpage}--\pageref{lastpage}}
\maketitle

\begin{abstract}

We report a new test of modified gravity theories using the large-scale structure of the Universe. This paper is the first attempt to (1) apply a joint analysis of the anisotropic components of galaxy two- and three-point correlation functions (2 and 3PCFs) to actual galaxy data and (2) constrain the nonlinear effects of degenerate higher-order scalar-tensor (DHOST) theories on cosmological scales. Applying this analysis to the Baryon Oscillation Spectroscopic Survey (BOSS) data release 12, we obtain the lower bounds of $-1.655 < \xi_{\rm t}$ and $-0.504 < \xi_{\rm s}$ at the $95\%$ confidence level on the parameters characterising the time evolution of the tidal and shift terms of the second-order velocity field. These constraints are consistent with GR predictions of $\xi_{\rm t}=15/1144$ and $\xi_{\rm s}=0$. Moreover, they represent a $35$-fold and $20$-fold improvement, respectively, over the joint analysis with only the isotropic 3PCF. We ensure the validity of our results by investigating various quantities, including theoretical models of the 3PCF, window function corrections, cumulative ${\rm S/N}$, Fisher matrices, and statistical scattering effects of mock simulation data. We also find statistically significant discrepancies between the BOSS data and the Patchy mocks for the 3PCF measurement. Finally, we package all of our 3PCF analysis codes under the name \textsc{HITOMI} and make them publicly available so that readers can reproduce all the results of this paper and easily apply them to ongoing future galaxy surveys.

\end{abstract}




\section{Introduction}
\label{Sec:Introduction}

\subsection{Outline and summary}

This paper presents a comprehensive study of the joint analysis of galaxy two- and three-point correlation functions (2 and 3PCFs) with isotropic and anisotropic components to constrain the non-linear effects of modified gravity theories on a cosmological scale. Section~\ref{Sec:Introduction} outlines the theoretical development and the present constraints for scalar-tensor theories. We also outline the development of the measurement and analysis of the three-point correlation function of galaxies. We organize this paper such that readers unfamiliar with both or one of the two areas follow the recent developments and understand how they fit together.

Readers interested in the theoretical aspects may read Sections~\ref{Sec:ReviewOfDHOSTtheories}, \ref{Sec:CalculationsOfModels}, and \ref{Sec:FisherAnalysis}. Section~\ref{Sec:ReviewOfDHOSTtheories} reviews the non-linear evolution of the large-scale structure of the Universe in scalar-tensor theories. Section~\ref{Sec:CalculationsOfModels} presents detailed calculations of the theoretical model of the 3PCF and, in particular, investigates the dependence of the parameters that characterise the effect of scalar-tensor theories on the 3PCF model. Finally, Section~\ref{Sec:FisherAnalysis} discusses the extent to which the 3PCF contains information on the non-linear effects of scalar-tensor theories through Fisher analysis.

Readers interested in the analysis method of the 3PCF may read Sections~\ref{Sec:Measurements}, \ref{Sec:CovarianceMatrix}, \ref{Sec:AnalysisSettings}, and \ref{Sec:GoodnessOfFit}. Section~\ref{Sec:Measurements} reviews how to measure the 3PCF from galaxy data and examines the effect of the window function on the measured 3PCFs. Section~\ref{Sec:CovarianceMatrix} presents the results of the 2PCF and 3PCF covariance matrices computed from mock simulations. Section~\ref{Sec:AnalysisSettings} describes the setup for the data analysis in this paper. Finally, Section~\ref{Sec:GoodnessOfFit} discusses in detail whether the 3PCFs measured from the galaxy data in this paper can be fitted using the corresponding theoretical model in terms of $p$-values.

For readers familiar with the two areas in the literature and interested in the final results, we suggest they jump directly to Section~\ref{Sec:DataAnalysis}. The novel aspect of this paper is to focus on observationally constraining the second-order velocity field, which is a key to seeking a deviation from GR in scalar-tensor theories. We also show that the second-order velocity field imprints a unique signature in the anisotropic 3PCF on large scales. Following \citet{Yamauchi:2021arXiv210802382Y} and Section~\ref{Sec:TimeDependencyOfParameters}, we then parameterise the effects of scalar-tensor theories in the time evolution of the shift and tidal terms of the second-order velocity field using parameters $\xi_{\rm s}$ and $\xi_{\rm t}$ defined in Eq.~(\ref{Eq:TimeDependenceOfE}). Constraining these parameters using Baryon Oscillation Spectroscopic Survey Data Release 12 galaxies~\citep{Eisenstein:2011sa,Bolton:2012hz,Dawson:2012va,Alam:2015mbd}, we obtain the following lower bounds given in Eqs.~(\ref{Eq:Result_XT_95}) and (\ref{Eq:Result_XS_95}):
\begin{eqnarray}
    - 1.655 < \xi_{\rm t} \ \  \mbox{and} \ \  -0.504 < \xi_{\rm s} \quad \mbox{($95\%$CL)}. \nonumber
\end{eqnarray}
Since $\xi_{\rm t}=15/1144$ and $\xi_{\rm s}=0$ in GR, these results are consistent with GR. Finally, we summarise the final results and the various findings leading up to them in Section~\ref{Sec:Conclusions}, which concludes this paper.

We package all the code used to complete this paper under the name \textsc{HITOMI}~\footnote{\url{https://github.com/naonori/hitomi.git}} and make it publicly available. Appendix~\ref{Sec:HITOMI} summarises the structure and usage of \textsc{HITOMI}.

\subsection{General motivation}

The greatest mystery in current cosmology is the cause of the accelerated expansions that have presumably occurred twice in the cosmic expansion history: i.e., inflation and late-time acceleration.

Scalar-tensor theories, modified gravity theories that add a single scalar field degree of freedom to General Relativity (GR), have been actively studied as a promising candidate to explain these accelerated expansions~\citep[for reviews, see][]{Langlois:2018dxi,Kase:2018aps,Kobayashi:2019hrl,Amendola:2019laa,Frusciante:2019xia}\footnote{For reviews of modified gravity theories, including other theories than scalar-tensor theories, see~\citealt[][]{Nojiri:2011PhR...505...59N,Sebastiani:2016ras,Nojiri:2017PhR...692....1N,Cataneo:2018mil,Ishak:2018his,Ferreira:2019xrr,Baker:2019gxo,Arai:2022ilw}.}.

The accelerated expansion in the very early Universe, called inflation~\citep{Starobinsky:1980te,Guth:1980zm,Sato:1980yn,Linde:1981mu,Albrecht:1982wi}, is thought to be caused by a single scalar field in the simplest model, generating the seeds of the cosmic fluctuations currently observed. Furthermore, the statistical properties of these fluctuations are in excellent agreement with current observations of the cosmic microwave background~\citep[CMB;][]{Aghanim:2018eyx} and the large-scale structure~\citep[LSS;][]{Alam:2020sor}. On the other hand, the cosmological constant may explain the late-time accelerated expansion~\citep{Riess:1998cb,Perlmutter:1998np}. However, its smallness implies a serious fine-tuning problem in fundamental physics~\citep{Weinberg:1988cp,Martin:2012bt}, and in order to avoid this problem, it is preferable to adopt a scalar field that varies with time. 

In order to test scalar-tensor theories in the late-time Universe, it is crucial to follow the time evolution of the large-scale structure in detail. Examples of already completed galaxy surveys are the Baryon Oscillation Spectroscopic Survey~\citep[BOSS][]{Eisenstein:2011sa,Bolton:2012hz,Dawson:2012va,Alam:2015mbd}\footnote{\url{https://www.sdss3.org/science/boss_publications.php}} and the Extended BOSS~\citep[eBOSS;][]{Dawson:2015wdb,Alam:2020sor}\footnote{\url{https://www.sdss.org/surveys/eboss/}}. Furthermore, The next-generation galaxy surveys, such as the Dark Energy Spectroscopic Instrument~\citep[DESI;][]{Aghamousa:2016zmz}\footnote{\url{http://desi.lbl.gov/}}, Euclid~\citep{Laureijs:2011gra}\footnote{\url{www.euclid-ec.org}}, and the Subaru Prime Focus Spectrograph ~\citep[PFS;][]{Takada:2012rn}\footnote{\url{https://pfs.ipmu.jp/index.html}}, will provide unprecedented accuracy in testing scalar-tensor theories.

\subsection{DHOST theories and their constraints}

In this paper, we pay particular attention to the behaviour in the late-time Universe of Degenerate Higher-Order Scalar-Tensor (DHOST) theory~\citep[for reviews, see][]{Langlois:2018dxi,Kobayashi:2019hrl}, which are a quite general theoretical framework of scalar-tensor theories that can evade the Ostrogradsky instability~\citep{Ostrogradsky:1850fid,Woodard:2015zca,Ganz:2020skf}. Scalar-tensor theories have been developing rapidly over the last decade. In 2011, \citet{Deffayet:2011gz,Kobayashi:2011nu} rediscovered the most general theory with second-order equations of motion for metric tensor and scalar fields, Horndeski theories~\citep{Horndeski:1974wa}. To go beyond Horndeski theories, \citet{Gleyzes:2014qga,Gleyzes:2014dya} found a class of healthy theories having higher-order field equations that reduce to a second-order system by combining different components~\citep[see also][for examples beyond Horndeski]{Zumalacarregui:2013pma}. This discovery results from a degeneracy between the kinetic terms of the scalar field and the metric. This class of theories has been extended to reach DHOST theories~\citep{Langlois:2015cwa,Crisostomi:2016czh,BenAchour:2016fzp,Achour:2016rkg,Langlois:2017mdk,Langlois:2020xbc}, encompassing Horndeski and Beyond Horndeski theories\footnote{Hereafter, we do not distinguish between Beyond Horndeski theories and DHOST theories.}. So far, DHOST theories have been constrained primarily by three observations\footnote{As other probes of DHOST theories, for example, \citet{Babichev:2018rfj} shows that the scalar field in DHOST theories can significantly modify the speed of sound in the atmosphere of the Earth; \citet{Jimenez:2015bwa,Dima:2017pwp} strongly constrain DHOST models using Hulse-Taylor pulsar observations; \citet{Saltas:2019ius} proposes helioseismology as a precise way to test DHOST theories on astrophysical scales.}: gravitational waves (GW), celestial objects, and cosmological data that is the subject of this paper.

Since GW170817 was observed by LIGO and Virgo~\citep{TheLIGOScientific:2017qsa}, the situation surrounding the observational constraints of modified gravity has changed dramatically. The simultaneous observation of GRB170817~\citep{Monitor:2017mdv}, a Gamma Ray burst, confirmed that the speed of GWs matches the speed of electromagnetic waves with high accuracy, ruling out various scalar-tensor theories that change the speed of GWs at low redshifts~\citep{Lombriser:2015sxa,Lombriser:2016yzn,Creminelli:2017sry,Sakstein:2017xjx,Ezquiaga:2017ekz,Baker:2017hug,Langlois:2017dyl}. \citet{Creminelli:2018xsv,Creminelli:2019nok} pointed out that a subset of DHOST theories leads to the decay of gravitational waves, resulting in further tight constraints on DHOST theories. However, the theory of gravity considered in that paper, the class I DHOST theory~\citep{Langlois:2015cwa,Crisostomi:2016czh,Achour:2016rkg}, still survives and can modify gravity in cosmology without pathological instability~\citep{deRham:2016wji,Langlois:2017mxy,Amendola:2017orw}. Furthermore, \citet{deRham:2018red} showed that such cosmological scalar-tensor theories, which predict the speed of GWs to be different from the speed of light, break down on high energy scales ($\sim 10^2\, {\rm Hz}$) seen in neutron star mergers, indicating that the constraints from GW observations may not necessarily apply to cosmological scales. Therefore, it is essential to test modified gravity theories independently at various energy scales, such as the GW and cosmological scales.

DHOST theories generally have characteristic non-linear effects that violate the Vainshtein screening mechanism inside any gravitational source~\citep{Kobayashi:2014ida,Koyama:2015oma,Crisostomi:2017lbg,Langlois:2017dyl,Dima:2017pwp,Hirano:2019scf,Crisostomi:2019yfo}. As an alternative to the cosmological constant, scalar-tensor theories must give an ${\cal O}(1)$ modification from GR at cosmological scales, but at small scales, they must satisfy tests in weakly gravitational regions such as the solar system. The Vainshtein screening mechanism~\citep[for a review, ][]{Babichev:2013usa}, universally found in scalar-tensor theories, is a typical mechanism that satisfies these requirements, suppressing scalar interactions and restoring standard gravity through non-linear effects. While Horndeski theories allow for a natural implementation of the Vainshtein mechanism~\citep{Kimura:2011dc,Narikawa:2013pjr,Koyama:2015vza}, DHOST theories partially violate it, allowing one to test DHOST theories by examining the internal structure of objects such as Newtonian stars~\citep{Saito:2015fza,Sakstein:2015aac,Sakstein:2015zoa,Jain:2015edg,Sakstein:2016lyj,Saltas:2018mxc,Saltas:2022A&A...667A.115S}, Neutron stars~\citep{Babichev:2016jom,Sakstein:2016oel}, and galaxy clusters~\citep{Sakstein:2016ggl,Salzano:2017qac}. The Vainshtein radius, the maximum scale at which the Vainshtein mechanism works, is estimated to be ${\cal O}(100)\, \textrm{pc}$ for the Sun and ${\cal O}(1)\, \textrm{Mpc}$ for a galaxy cluster. 

DHOST theories predict a characteristic gravitational non-linear effect on even cosmological scales exceeding tens of $\textrm{Mpc}$. That is, DHOST theories violate the consistency relation for LSS~\citep{Crisostomi:2019vhj,Lewandowski:2019txi}~\citep[see also][]{Hirano:2018uar}. The LSS consistency relation~\citep{Peloso:2013zw,Kehagias:2013yd,Creminelli:2013mca} is an analogue of the consistency relation originally proposed for single-field inflation models~\citep{Maldacena:2002vr,Creminelli:2004yq}, which relates $n$-point statistics of cosmological fluctuations to $(n-1)$-point statistics in a non-perturbative matter. It is valid in the limit where the wavenumber of one of the $n$-points is hugely smaller than the others. This consistency relation is because the equations that the fluctuations obey are invariant under a Galilean transformation~\citep{Scoccimarro:1995if,Creminelli:2013mca}. In particular, in the so-called equal-time consistency relation, the Galilean transformation eliminates the large-scale flow of matter and thus cancels all non-linear contributions when calculating the $n$-point statistics. This behavior is also known as infrared (IR) cancellation~\citep{Jain:1995kx,Scoccimarro:1995if,Kehagias:2013yd,Peloso:2013zw,Sugiyama:2013pwa,Sugiyama:2013gza,Blas:2013bpa,Blas:2015qsi,Lewandowski:2017kes}. On the other hand, the LSS consistency relation breaks down when considering multiple fluids~\citep{Tseliakhovich:2010bj,Yoo:2011tq,Bernardeau:2011vy,Bernardeau:2012aq,Peloso:2013spa,Creminelli:2013poa,Lewandowski:2014rca,Slepian:2016weg} or primordial non-Gaussianities~\citep{Berezhiani:2014kga,Valageas:2016hhr,Esposito:2019jkb,Goldstein:2022PhRvD.106l3525G}, or when the equivalence principle breaks~\citep{Creminelli:2014JCAP...06..009C}. DHOST theories have a structure similar to that of multiple fluids, and on large scales, the Galilean transformation cannot make the relative velocity between the scalar field and matter vanish~\citep[for details, see][]{Crisostomi:2019vhj,Lewandowski:2019txi}. As a result, DHOST theories violate the LSS consistency relation.

Our interest in this paper is to constrain DHOST theories on cosmological scales, i.e., ${\cal O}(10-100)\, \textrm{Mpc}$ scales. However, studies using cosmological data to constrain DHOST theories are still limited~\citep{Hirano:2019nkz,Traykova:2019oyx,Peirone:2019yjs,Hiramatsu:2022arXiv220511559H}. On the other hand, many papers on Horndeski theories have used cosmological data to constrain the model~\citep{Okada:2012mn,Barreira:2014jha,Bellini:2015xja,Mueller:2016kpu,Kreisch:2017uet,Arai:2017hxj,Noller:2018eht,Noller:2018wyv,Raveri:2019mxg,Melville:2019wyy,Perenon:2019dpc,Noller:2020afd}. Therefore, exploring new cosmological methods for constraining DHOST theories is of great significance.

\subsection{Constraints on modified gravity theories using galaxy two-point statistics}

The logarithmic growth rate function $f$ of dark matter fluctuations, measured using redshift-space distortions~\citep[RSD;][]{Kaiser:1987qv}, plays an important role in constraining modified gravity theories in the late-time Universe. In the power spectrum analysis, we cannot measure the growth rate function by itself, but usually, by the combination $f\sigma_8 = d \ln \sigma_8 / d \ln a$~\citep{Song:2008qt,Percival:2008sh} using $\sigma_8$ representing the rms of matter fluctuations on the $8\hMpc$ scale. For example, the most recent observations, BOSS and eBOSS, measured $f\sigma_8$ with a precision of $\sim 5\%$ in the redshift range $0.2<z<1.0$~\citep{Alam:2020sor}. 

One concern is to test modified gravity theories directly using existing $f\sigma_8$ measurements. The standard practice is constructing a model of the non-linear galaxy power spectrum assuming GR, then using that model to measure $f\sigma_8$ from data up to the mildly non-linear region $(k\sim 0.2\hk)$~\citep[for recent studies, e.g.,][]{d'Amico2020JCAP...05..005D,Ivanov:2020JCAP...05..042I,Lange:2022MNRAS.509.1779L,Kobayashi:2022PhRvD.105h3517K,Yuan:2022MNRAS.515..871Y}. Therefore, it is worth noting that many existing analysis results using galaxy data up to the non-linear region only verify the consistency of GR. Thus, to test the gravity theory by consistently considering both linear and non-linear effects, a power spectrum model that considers non-linear effects specific to the modified gravity theory of interest is necessary. Several studies have been done on this for various modified gravity theories~\citep{Koyama:2009me,Taruya:2013quf,Taruya:2014faa,Takushima:2015iha,Bellini:2015oua,Taruya:2016jdt,Barreira:2016ovx,Bose:2016qun,Cusin:2017wjg,Cusin:2017mzw,Bose:2017myh,Bose:2018orj,Aviles:2018saf,Hernandez-Aguayo:2018oxg,Cataneo:2018cic,Valogiannis:2019nfz,Valogiannis:2019xed,Bose:2020wch}. However, only one study constrained the theory from actual galaxy data using a galaxy power spectrum model that consistently includes the non-linear effects arising from modified gravity~\citep{Song:2015oza}, where the authors focused on $f(R)$ gravity~\citep{Hu:2007nk}. 

In particular, \citet{Hirano:2020dom} pointed out that in DHOST theories, even the next-order solutions of the power spectrum in perturbation theory, the so-called one-loop solutions, are challenging to perform physically meaningful theoretical calculations due to the divergence of the wavenumber integral in the ultraviolet (UV) region. Therefore, the modelling of non-linear power spectra in DHOST theories is still highly uncertain.

\subsection{Developments in the study of galaxy three-point statistics}

A more straightforward way to investigate the non-linearity of scalar-tensor theories is to use three-point statistics of cosmological fluctuations, i.e., the 3PCF or the bispectrum. The reason is that, on large scales, the three-point statistics consist of a combination of second-order and linear-order dark matter fluctuations. The second-order fluctuations depend on two wave vectors in Fourier space and can be decomposed into three components using the angle between the two wave vectors: monopole (growth), dipole (shift), and quadrupole (tidal force)~\citep{Schmittfull:2015PhRvD..91d3530S}. For example, Horndeski theories deviate only the coefficient of the tidal term from GR while keeping the shift term among these three components~\citep{Bernardeau:2011JCAP...06..019B,Takushima:2013foa,Bartolo:2013ws,Bellini:2015wfa,Burrage:2019afs}. On the other hand, DHOST theories change both the shift and tidal terms~\citep{Hirano:2018uar,Crisostomi:2019vhj,Lewandowski:2019txi}, and this change in the shift term leads to a violation of the LSS consistency relation~\citep{Crisostomi:2019vhj,Lewandowski:2019txi}. In addition to scalar-tensor theories, there has been much researches on higher-order statistics in, for example, $f(R)$ gravity theory~\citep{Gil-Marin:2011JCAP...11..019G,Borisov:2009PhRvD..79j3506B,Hellwing:2013rxa,Bose:2018zpk,Bose:2019wuz}. Several observational proposals have been made to test modified gravity theories using cosmological three-point statistics, such as galaxy clustering~\citep{Yamauchi:2017ibz,Yamauchi:2021arXiv210802382Y}, weak lensing~\citep{Dinda:2018eyt,Munshi:2019csw,Munshi:2019jyl,Munshi:2020nfh}, and CMB lensing~\citep{Namikawa:2018erh,Namikawa:2018bju}, but none have been applied to actual observational data yet.

In the context of the galaxy three-point statistics, 3PCF resolves the degeneracy between the linear bias $b_1$ and $\sigma_8$ and allows us to directly study the evolution of dark matter density fluctuations apart from the RSD effect~\citep{Fry:1994,Frieman:1994aa,Matarrese:1997sk,Verde:1998zr,Gaztanaga:2005MNRAS.361..824G,Sefusatti:2006PhRvD..74b3522S,Greig:2013MNRAS.431.1777G,Hoffmann:2015MNRAS.447.1724H,Samushia:2021mn}. Furthermore, many previous studies have proposed to constrain primordial non-Gaussianities from the galaxy three-point statistics~\citep{Fry:1993xy,Verde:2000MNRAS.313..141V,Scoccimarro:2004PhRvD..69j3513S,Sefusatti:2007PhRvD..76h3004S,Sefusatti:2009PhRvD..80l3002S,Sefusatti:2010MNRAS.406.1014S,Liguori:2010AdAst2010E..73L,Desjacques:2010CQGra..27l4011D,Sefusatti:2012MNRAS.425.2903S,Scoccimarro:2012PhRvD..85h3002S,Alvarez:2014arXiv1412.4671A,Tellarini:2015JCAP...07..004T,Tellarini:2016JCAP...06..014T,Welling:2016JCAP...08..044W,Yamauchi:2017PhRvD..95f3530Y,Karagiannis:2018MNRAS.478.1341K,Bharadwaj:2020MNRAS.493..594B,MoradinezhadDizgah:2020arXiv201014523M,Shirasaki:2021PhRvD.103b3506S,Coulton:2022arXiv220615450C,Karagiannis:2022jc}. Recently, as in the case of the galaxy two-point statistics~\citep[e.g.,][]{Matsubara:2004ap,Taruya:2011ph}, the anisotropic component of the galaxy three-point statistics induced by the RSD effect and the Alcock-Paczy\'{n}ski (AP) effect~\citep{Alcock:1979mp} has attracted much attention, and its cosmological utility has been actively studied~\citep{Song:2015JCAP...08..007S,Gagrani:2017MNRAS.467..928G,Yankelevich:2019MNRAS.483.2078Y,Gualdi:2020JCAP...06..041G,Mazumdar:2020MNRAS.498.3975M,Sugiyama:2020uil,Agarwal:2021JCAP...03..021A,Rizzo:2022arXiv220413628R,Tsedrik:2022arXiv220713011T}. 

Based on standard perturbation theory (SPT), many theoretical studies of the galaxy three-point statistics have been conducted to calculate higher-order non-linearities, redshift-space distortions, and bias effects, and the results of these calculations have been tested for validity by comparing them with measurements from N-body simulations~\citep{Peebles:1980lssu.book.....P,Fry:1984aa,Goroff:1986ApJ...311....6G,Hivon:1995A&A...298..643H,Scoccimarro:1996jy,Scoccimarro:1997st,Jing:1997aa,Scoccimarro:1999ed,Scoccimarro:2000ApJ...544..597S,Barriga:2001wn,Barriga:2002MNRAS.333..443B,Gaztanaga:2005MNRAS.361..824G,Pan:2007MNRAS.382.1460P,Marin:2008ApJ...672..849M,Guo:2009ApJ...698..479G,Pollack:2012MNRAS.420.3469P,Lazanu:2016PhRvD..93h3517L,McCullagh:2016MNRAS.455.2945M,Lazanu:2016PhRvD..93h3517L,Lazanu:2018JCAP...04..055L,Hoffmann:2018MNRAS.476..814H,Desjacques:2018JCAP...12..035D,Child:2018arXiv181112396C,Eggemeier:2019PhRvD..99l3514E,Oddo:2020JCAP...03..056O,Eggemeier:2021arXiv210206902E,Oddo:2021JCAP...11..038O,Philcox:2022PhRvD.106d3530P}. Other approaches have been widely used in research, such as the halo models~\citep{Ma:2000ApJ...543..503M,Scoccimarro:2001ApJ...546...20S,Takada:2003MNRAS.340..580T,Fosalba:2005ApJ...632...29F,Smith:2008PhRvD..78b3523S,Yamamoto:2017PhRvD..95d3528Y,Nan:2018JCAP...07..038N} and fitting formulas~\citep{Scoccimarro:1999ApJ...520...35S,Scoccimarro:2001MNRAS.325.1312S,Gil-Marin:2012JCAP...02..047G,Gil-Marin:2014JCAP...12..029G,Takahashi:2020ApJ...895..113T}. Beyond SPT, several improved perturbation theories have been proposed. \citet{Rampf:2012JCAP...06..018R} used a resummation method based on Lagrangian perturbation theory. \citet{Baldauf:2015JCAP...05..007B,Munshi:2017JCAP...06..042M,Ivanov:2022PhRvD.105f3512I} discussed some correction terms for SPT based on the effective field theory of large-scale structure. \citet{Hashimoto:2017PhRvD..96d3526H} applied a resummation method similar to the TNS model of the power spectrum~\citep{Taruya:2010PhRvD..82f3522T}. \citet{Kuruvilla:2020JCAP...07..043K} generalised the so-called streaming model to higher-order statistics. \citet{Blas:2016sfa,Ivanov:2018gjr} developed the time-sliced perturbation theory (TSPT) to resum the IR modes of the bulk flow and describe the non-linear damping of Baryon Acoustic Oscillations~\citep[BAOs;][]{Peebles:1970ag,Sunyaev:1970eu}. \citet{Sugiyama:2020uil} constructed a new IR-resummed bispectrum model by adding a term to the model proposed by TSPT.

The measurement of three-point statistics for galaxies, galaxy clusters, and quasars has a long history. As a simple method, two-dimensional three-point angular statistics have been observed from the dawn of the study of cosmological three-point statistics to the present~\citep{Peebles:1975a,Peebles:1975b,Groth:1977gj,Fry:1980aa,Fry:1982,Sharp:1984aa,Jing:1989aa,Jing:1991aa,Toth:1989aa,Frieman:1999ApJ...521L..83F,Szapudi:2001ApJ...548..114S,deCarvalho:2020MNRAS.492.4469D}. Eventually, with the development of spectroscopic observations of galaxies, three-dimensional three-point statistics have become the primary targets observed in configuration space~\citep{Bean:1983aa,Efstathiou:1986bf,Hale-Sutton:1989,Gott:1991aa,Jing:1998qs,Jing:2004ApJ...607..140J,Kayo:2004PASJ...56..415K,Wang:2004MNRAS.353..287W,Gaztanaga:2005MNRAS.364..620G,Pan:2005MNRAS.362.1363P,Nichol:2006MNRAS.368.1507N,Kulkarni:2007MNRAS.378.1196K,Gaztanaga:2009MNRAS.399..801G,Marin:2010iv,McBride:2011ApJ...726...13M,McBride:2011ApJ...739...85M,Marin:2013bbb,Guo:2014ApJ...780..139G,Moresco:2016opr,Slepian:2017MNRAS.468.1070S,Slepian:2016kfz,Moresco:2017A&A...604A.133M,Moresco:2020arXiv201104665M} or in Fourier space~\citep{Baumgart:1991aa,Scoccimarro:2000sp,Feldman:2001PhRvL..86.1434F,Verde:2002MNRAS.335..432V,Nishimichi:2007PASJ...59...93N,Gil-Marin:2015MNRAS.451..539G,Gil-Marin:2015MNRAS.452.1914G,Gil-Marin:2017MNRAS.465.1757G,Pearson:2018MNRAS.478.4500P,Sugiyama:2018yzo,Philcox:2022PhRvD.105d3517P,Cabass:2022arXiv220107238C,DAmico2022arXiv220111518D,Cabass:2022PhRvD.106d3506C,D'Amico:2022arXiv220608327D}. As another approach, \citet{Chiang:2015JCAP...09..028C} measured the squeezed limit bispectrum by splitting the observing region and measuring the position-dependent power spectrum. Since the first measurement of the galaxy three-point statistics by~\citet{Peebles:1975a}, the three-point statistic measurement has long been limited to measuring only certain scale-dependence of the three-point statistics. However, it is now possible to perform cosmological analysis using the information on the full shape of galaxy three-point statistics at cosmological scales $(\sim 100\hMpc)$. 

In recent years, cosmological analysis of the three-point statistics of galaxies has made remarkable progress, mostly focusing on the isotropic component, i.e., \emph{monopole}, of the three-point statistics. \citet{Slepian:2016kfz} and \citet{Pearson:2018MNRAS.478.4500P} reported the detection of the BAO signal through the monopole 3PCF and the monopole bispectrum, respectively. \citet{Gil-Marin:2017MNRAS.465.1757G,d'Amico2020JCAP...05..005D,Philcox:2022PhRvD.105d3517P} performed a joint analysis of the monopole and quadrupole power spectra and the monopole bispectrum to constrain the cosmological parameters of interest. \citet{Cabass:2022arXiv220107238C,DAmico2022arXiv220111518D,Cabass:2022PhRvD.106d3506C} constrained primordial non-Gaussianities using the monopole bispectrum.

The anisotropic components, i.e., \emph{quadrupole} and \emph{hexadecapole}, of the galaxy three-point statistics have been the subject of pretty limited studies of measurements and cosmological analyses from actual galaxy data. \citet{Sugiyama:2018yzo} reported the first detection of the quadrupole bispectrum signal at the $14\sigma$ level from the BOSS DR12 galaxies. \citet{Sugiyama:2020uil} performed an anisotropic BAO analysis using the monopole and quadrupole components of the 2PCF and 3PCF for the MultiDark-Patchy mock catalogues~\citep[Patchy mocks;][]{Kitaura:2015uqa} reproducing the BOSS galaxy distribution, showing the improvement of the Hubble parameter constraint by $\sim30\%$ compared to the 2PCF-only analysis result. \citet{D'Amico:2022arXiv220608327D} performed the first joint analysis of the monopole and quadrupole components of the power and bispectra measured from the BOSS DR12 galaxy data. More recently, \citet{Ivanov:2023arXiv230204414I} presented the results of an anisotropic bispectrum analysis including quadrupole and hexadecapole components measured from the BOSS DR12 data.

\subsection{Goal of this paper}

The primary goal of this paper is to use the 3PCF of galaxies to perform a consistent cosmological analysis that constrains DHOST theories and their subclass, Horndeski theories, while accounting for linear and non-linear effects. To this end, \citet{Yamauchi:2021arXiv210802382Y} pointed out that the parameters characterising non-linear density fluctuations in DHOST theories degenerate with the non-linear bias parameter, so measuring the non-linear velocity field due to the RSD effect is essential. In addition, the authors proposed a simple parameterisation scheme that characterises the time evolution of the scale dependence of the non-linear velocity field to facilitate the combined analysis of galaxy samples at different redshifts. Specifically, the time evolution of the shift and tidal terms of the second-order velocity field is represented by $\xi_{\rm s}$ and $\xi_{\rm t}$, respectively, where $\xi_{\rm s}=0$ and $\xi_{\rm t}=15/1144$ in GR. Following the suggestion of \citet{Yamauchi:2021arXiv210802382Y}, we apply the joint analysis method of the anisotropic 2PCF and 3PCF of galaxies established by~\citet{Sugiyama:2020uil} to BOSS Data Release 12 galaxies~\citep{Eisenstein:2011sa,Bolton:2012hz,Dawson:2012va,Alam:2015mbd} to constrain these $\xi_{\rm s}$ and $\xi_{\rm t}$ parameters.

When we need to use values of fiducial cosmology parameters in our analysis, we adopt a flat $\Lambda$CDM model with the following parameters: matter density $\Omega_{\rm m0}=0.31$, Hubble constant $h\equiv H_0/(100\,{\rm km\, s^{-1}\, Mpc^{-1}})=0.676$, baryon density $\Omega_{\rm b0}h^2=0.022$, and spectral tilt $n_{\rm s}=0.97$, which are the same as those used in the final cosmological analysis in the BOSS project~\citep{Alam:2016hwk} and consistent with the best-fit values in Planck 2018~\citep{Aghanim:2018eyx}. We adopt a value for the total neutrino mass of $\sum m_{\nu}=0.06\, {\rm eV}$ close to the minimum allowed by neutrino oscillation experiments. We use these fiducial parameters to estimate the distance to galaxies from the observed redshift of each galaxy and to calculate the shape of the linear matter power spectrum at the redshifts of interest with \textsc{CLASS}~\citep{Blas:2011rf}.

\section{DHOST theories}
\label{Sec:ReviewOfDHOSTtheories}

In this section, we briefly review the analytic expressions of DHOST theories. Section~\ref{Sec:DHOSTtheories} introduces the class I DHOST theory and the perturbative solutions of the density and velocity fields of dark matter and galaxies solved up to the second-order in that theory. In Eqs.~(\ref{Eq:ActionDHOST})-(\ref{Eq:EvoOfLambda}) of this subsection, we adopt the expressions and notations given by~\citet{Hirano:2018uar}. Section~\ref{Sec:LimitationsOfOurAssumptions} discusses the limitation of the assumptions adopted to derive the perturbative solutions used in this paper. 

\subsection{Density and velocity fluctuations in DHOST theories}
\label{Sec:DHOSTtheories}

We begin by summarising the theoretical models we will investigate in this paper and the assumptions used to derive those models.

\begin{enumerate}
    \item Gravity theory is a subclass of quadratic DHOST theories, the class I DHOST theory~\citep{Crisostomi:2016czh}, which encompasses Horndeski and Beyond Horndeski theories and is free from the instabilities of a cosmological background~\citep{deRham:2016wji,Langlois:2017mxy}.
       \vspace{2mm}
    \item Matter is cold dark matter (CDM) that can be described as a pressureless perfect fluid without vorticity~\citep{Bernardeau:2001qr}.
       \vspace{2mm}
    \item Matter is minimally coupled to gravity, and the effects of the DHOST gravity appear only through the gravitational potential.
       \vspace{2mm}
    \item When solving the equations of motion of metric tensor and scalar fields in DHOST theories, the quasi-static approximation~\citep[e.g.,][]{Pace:2020qpj} is used. Then, the gravitational potential is determined by a modified Poisson equation~\citep{Hirano:2018uar,Crisostomi:2019vhj,Lewandowski:2019txi,Hirano:2020dom}.
       \vspace{2mm}
    \item Statistical properties of the CDM fluctuations are those derived in the standard theory of inflation, which satisfy the following properties: adiabaticity, negligibly weak non-Gaussianity, nearly scale-free, statistical homogeneity, statistical isotropy, and statistical parity symmetry.
       \vspace{2mm}
    \item Galaxy biases are assumed to be present only in the density field, and three biases are considered: linear bias $b_1$,  second-order local bias $b_2$, and second-order non-local bias (tidal bias) $b_{s^2}$~\citep[for a review, see e.g.,][]{Saito:2014ph,Desjacques:2016bnm}. Any bias effects related to higher-order derivatives and the velocity field of the galaxy are ignored.
\end{enumerate}

The action of quadratic DHOST theories is given by~\citep{Langlois:2015cwa,Crisostomi:2016czh}
\begin{eqnarray}
    S_{\rm DHOST} \hspace{-0.25cm}&=&\hspace{-0.25cm}  \int d^4x \sqrt{-g}
    \Big[ {\cal G}_2(\phi,X)-{\cal G}_3(\phi,X)\Box\phi +{\cal F}(\phi,X) R  \nonumber \\
    \hspace{-0.25cm}&+&\hspace{-0.25cm}  a_1\phi_{\mu\nu}\phi^{\mu\nu} + a_2(\Box \phi)^2+ a_3(\Box \phi)\phi^{\mu}\phi_{\mu\nu}\phi^{\nu}\nonumber \\
    \hspace{-0.25cm}&+&\hspace{-0.25cm}  a_4\phi^{\mu}\phi_{\mu\rho}\phi^{\rho\nu}\phi_{\nu}+a_5(\phi^{\mu}\phi_{\mu\nu}\phi^{\nu})^2\Big],
    \label{Eq:ActionDHOST}
\end{eqnarray}
where $\phi_{\mu}=\nabla_{\mu}\phi$, $\phi_{\mu\nu}=\nabla_{\mu}\nabla_{\nu}\phi$, $X=-\phi_{\mu}\phi^{\mu}/2$, and $a_i = a_i(\phi,X)$ for $i=1,\dots,5$. The functions $a_i$ ($i=1,\dots,5$) satisfy the degeneracy condition given by~\citep{Crisostomi:2016czh} to avoid the Ostrogradsky ghost~\citep{Ostrogradsky:1850fid,Woodard:2015zca}.

The density perturbation $\delta$ and velocity field $\VEC{v}$ of dark matter follow the equations of a pressureless perfect fluid without vorticity:
\begin{eqnarray}
    \dot{\delta}(\VEC{x}) + a^{-1} \partial_i \left(  (1 + \delta(\VEC{x})) v^i(\VEC{x})\right) \hspace{-0.25cm}&=&\hspace{-0.25cm} 0, \nonumber \\
    \dot{\theta}(\VEC{x}) + H \theta(\VEC{x}) + a^{-1}\partial_i\left( v^j(\VEC{x})\partial_j v^i(\VEC{x})  \right) \hspace{-0.25cm}&=&\hspace{-0.25cm}  - a^{-1}\partial^2 \Phi(\VEC{x}), 
    \label{Eq:Fluid}
\end{eqnarray}
where $a$ and $H=\dot{a}/a$ respectively denote the scale factor and the Hubble parameter, and $\theta=\partial_i v^i$ is the divergence of the velocity field. Because of no vorticity, the velocity field is represented as $v^i=(\partial_i/\partial^2)\theta$. The gravitational potential $\Phi$ is determined by the following modified Poisson equation~\citep{Hirano:2018uar}:
\begin{eqnarray}
    \frac{\partial^2 \Phi(\VEC{x})}{a^2H^2} =  \kappa \delta(\VEC{x})  + \nu\frac{\dot{\delta}(\VEC{x})}{H} + \mu \frac{\ddot{\delta}(\VEC{x})}{H^2}+ \frac{\partial^2 S_{\Phi}^{\rm NL}(\VEC{x})}{a^2H^2},
    \label{Eq:Poisson}
\end{eqnarray}
where $\kappa$, $\nu$, and $\mu$ are functions that depend only on time, and $S_{\Phi}^{\rm NL}$ is a non-linear source term obtained from the equation of motion of the scalar field.

To solve the above equations, we expand all the fluctuations as follows: $X = \sum_n X_n$, where $X = \{\delta, \theta, \Phi, S_{\Phi}^{\rm NL}\}$, and $X_n={\cal O}(\delta_1^n)$.
Then, the non-linear source $S_{\Phi}^{\rm NL}$ up to the second-order is given by
\begin{eqnarray}
    \frac{\partial^2 S_{\Phi, 1}^{\rm NL}(\VEC{x})}{a^2H^2} \hspace{-0.25cm}&=&\hspace{-0.25cm} 0, \nonumber \\
    \frac{\partial^2 S_{\Phi, 2}^{\rm NL}(\VEC{x})}{a^2H^2} 
    \hspace{-0.25cm}&=&\hspace{-0.25cm} \tau_{\alpha} W_{\alpha}(\VEC{x}) - \tau_{\gamma} W_{\gamma}(\VEC{x}),
    \label{Eq:S_NL}
\end{eqnarray}
where
\begin{eqnarray}
W_{\alpha}(\VEC{x}) \hspace{-0.25cm}&=&\hspace{-0.25cm} \left[\delta_1(\VEC{x})\right]^2 + \left[ \frac{\partial_i}{\partial^2} \delta_1(\VEC{x})  \right] \left[ \partial_i\delta_1(\VEC{x}) \right], \nonumber \\
W_{\gamma}(\VEC{x}) \hspace{-0.25cm}&=&\hspace{-0.25cm} \left[\delta_1(\VEC{x})\right]^2 - \left[ \frac{\partial_i\partial_j}{\partial^2}\delta_1(\VEC{x}) \right]^2.
\end{eqnarray}
The evolution of the density perturbation follows
\begin{eqnarray}
    \ddot{\delta}(\VEC{x}) + (2+\varsigma)H\dot{\delta}(\VEC{x}) - \frac{3}{2}\Omega_{\rm m}\Xi H^2 \delta(\VEC{x})= H^2 S_{\delta}^{\rm NL}(\VEC{x}),
    \label{Eq:delta_evo}
\end{eqnarray}
where $\varsigma=(2\mu-\nu)/(1-\mu)$, $(3/2)\Omega_{\rm m}\Xi=\kappa/(1-\mu)$, and $S_{\delta}^{\rm NL}$ is a non-linear source of the density perturbation, vanishing at linear order and given at second-order by
\begin{eqnarray}
    S_{\delta, 2}^{\rm NL}=  S_{\alpha} W_{\alpha}(\VEC{x}) - S_{\gamma} W_{\gamma}(\VEC{x})
\end{eqnarray}
with
\begin{eqnarray}
    (1-\mu) S_{\alpha} \hspace{-0.25cm}&=&\hspace{-0.25cm} 2f^2+\frac{3}{2}\Omega_{\rm m}\Xi - \varsigma f +  \tau_{\alpha}, \nonumber \\
    (1-\mu) S_{\gamma} \hspace{-0.25cm}&=&\hspace{-0.25cm} f^2 +  \tau_{\gamma}.
\end{eqnarray}
Once the solution of $\delta$ is obtained, the solution of $\theta$ is also derived from the continuity equation in Eq.~(\ref{Eq:Fluid}). In Fourier space\footnote{Our convention for the Fourier transform is 
    \begin{eqnarray}
        \widetilde{f}(\VEC{k}) = \int d^3x e^{-i\VEC{k}\cdot\VEC{x}} f(\VEC{x}). \nonumber
    \end{eqnarray}
}, Eqs.~(\ref{Eq:Fluid}) and (\ref{Eq:Poisson}) determine $\delta_n(\VEC{k})$ and $\theta_n(\VEC{k})$ in terms of the linear density fluctuations to be:
\begin{eqnarray}
    \widetilde{\delta}_n(\VEC{k}) \hspace{-0.25cm}&=&\hspace{-0.25cm}  \int \frac{d^3p_1}{(2\pi)^3} \cdots \int \frac{d^3p_n}{(2\pi)^3}
    (2\pi)^3\delta_{\rm D}(\VEC{k}-\VEC{p}_{[1n]}) \nonumber \\
    \hspace{-0.25cm}&\times&\hspace{-0.25cm}
    F^{(\rm m)}_n(\VEC{p}_1,\dots,\VEC{p}_2) \delta_1(\VEC{p}_1)\cdots \delta_1(\VEC{p}_n),\nonumber \\
    \widetilde{\theta}_n(\VEC{k}) \hspace{-0.25cm}&=&\hspace{-0.25cm}  -aHf\int \frac{d^3p_1}{(2\pi)^3} \cdots \int \frac{d^3p_n}{(2\pi)^3} 
    (2\pi)^3\delta_{\rm D}(\VEC{k}-\VEC{p}_{[1n]}) \nonumber \\
    \hspace{-0.25cm}&\times&\hspace{-0.25cm}
    G^{(\rm m)}_n(\VEC{p}_1,\dots,\VEC{p}_2) \delta_1(\VEC{p}_1)\cdots \delta_1(\VEC{p}_n),
\end{eqnarray}
where $\VEC{p}_{[1n]}=\VEC{p}_1 + \dots + \VEC{p}_n$, and $\delta_{\rm D}$ is the delta function. The functions $F^{(\rm m)}_2$ and $G^{(\rm m)}_2$ are kernel functions that characterise the gravitational non-linear effects, and the superscript $(\rm m)$ stands for ``matter''. In the second-order, $F^{(\rm m)}_2$ and $G^{(\rm m)}_2$ are given by
\begin{eqnarray}
    F^{(\rm m)}_2(\VEC{p}_1,\VEC{p}_2) = \kappa_{\delta} \alpha_{\rm s}(\VEC{p}_1,\VEC{p}_2) - \frac{2}{7}\lambda_{\delta} \gamma(\VEC{k}_1,\VEC{k}_2) \nonumber \\
    G^{(\rm m)}_2(\VEC{p}_1,\VEC{p}_2) = \kappa_{\theta} \alpha_{\rm s}(\VEC{p}_1,\VEC{p}_2) - \frac{4}{7}\lambda_{\theta} \gamma(\VEC{k}_1,\VEC{k}_2),
    \label{Eq:FG_m}
\end{eqnarray}
where
\begin{eqnarray}
    \alpha_{\rm s}(\VEC{k}_1,\VEC{k}_2) \hspace{-0.25cm}&=&\hspace{-0.25cm} 1 + (\hat{k}_1\cdot\hat{k}_2)\frac{(k_1^2+k_2^2)}{2k_1k_2},\nonumber \\
    \gamma(\VEC{k}_1,\VEC{k}_2)         \hspace{-0.25cm}&=&\hspace{-0.25cm} 1 - (\hat{k}_1\cdot \hat{k}_2)^2,
    \label{Eq:alpha_gamma}
\end{eqnarray}
and
\begin{eqnarray}
    \kappa_{\theta}  \hspace{-0.25cm}&=&\hspace{-0.25cm} 2\kappa_{\delta} \left[ 1 + \frac{1}{2f}\frac{d \ln \kappa_{\delta}}{d \ln a}  \right]- 1,  \nonumber \\
    \lambda_{\theta} \hspace{-0.25cm}&=&\hspace{-0.25cm}  \lambda_{\delta} \left[  1 + \frac{1}{2f}\frac{d \ln \lambda_{\delta}}{d \ln a}\right].
    \label{Eq:kappa_theta_and_lambda_theta}
\end{eqnarray}
The evolutions of $\kappa_{\delta}$ and $\lambda_{\delta}$ follow
\begin{eqnarray}
    && \ddot{\kappa}_{\delta} + [4f+(2+\varsigma)]H\dot{\kappa}_{\delta}+H^2\left( 2f^2+\frac{3}{2}\Omega_{\rm m}\Xi \right)\kappa_{\delta} \nonumber \\
    &&= H^2 S_{\alpha}, 
   \label{Eq:EvoOfKappa}
\end{eqnarray}
\begin{eqnarray}
    && \ddot{\lambda}_{\delta} + [4f+(2+\varsigma)]H\dot{\lambda}_{\delta}+H^2\left( 2f^2+\frac{3}{2}\Omega_{\rm m}\Xi \right)\lambda_{\delta}  \nonumber \\
   &&= \frac{7}{2}H^2 S_{\gamma}.
   \label{Eq:EvoOfLambda}
\end{eqnarray}

Since the galaxy density field is a biased quantity, we assume the linear bias $b_1$, the second-order local bias $b_2$, and the second-order tidal bias $b_{s^2}$ as the bias parameters that describe the galaxy density fluctuation up to second order~\citep[e.g.,][]{Desjacques:2016bnm}:
\begin{eqnarray}
    \delta^{(\rm g)}(\VEC{x}) = b_1\delta(\VEC{x}) + \frac{b_2}{2} \left[  \delta(\VEC{x})\right]^2 + b_{s^2} [s_{ij}]^2,
    \label{Eq:bias}
\end{eqnarray}
where the superscript $(\rm g)$ stands for ``galaxy'', and $[s_{ij}]^2$ is given by
\begin{eqnarray}
    [s_{ij}]^2 = \left[\frac{\partial_i\partial_j}{\partial^2}\delta(\VEC{x})  \right]^2 - \frac{1}{3}[\delta(\VEC{x})]^2.
    \label{Eq:s_ij}
\end{eqnarray}
Then, the second-order kernel functions for galaxies are given by
\begin{eqnarray}
    F_2^{\rm (g)} \hspace{-0.25cm}&=&\hspace{-0.25cm}  b_1F_2^{\rm (m)}(\VEC{p}_1,\VEC{p}_2) + \frac{1}{2}b_2 +b_{s^2} \left[ (\hat{p}_1\cdot\hat{p}_2)^2 - \frac{1}{3} \right], \nonumber \\
    G_2^{\rm (g)} \hspace{-0.25cm}&=&\hspace{-0.25cm}  G_2^{\rm (m)}(\VEC{p}_1,\VEC{p}_2).
    \label{Eq:BiasedFG}
\end{eqnarray}

The RSD effect shifts the observed position of galaxies $\VEC{x}_{\rm red}$ from their real-space position $\VEC{x}'$ due to the peculiar velocity of galaxies along the line-of-sight (LOS) direction:
\begin{eqnarray}
    \VEC{x}_{\rm red}(\VEC{x}') = \VEC{x}' + \frac{\VEC{v}(\VEC{x}')\cdot\hat{n}}{aH} \hat{n},
    \label{Eq:real_to_redshift}
\end{eqnarray}
where $\hat{n}$ is a unit vector pointing to the galaxy from the origin. The observed galaxy density fluctuation is then distorted along the LOS direction as follows:
\begin{eqnarray}
    \delta_{\rm s}^{(\rm g)}(\VEC{x}) = 
    \int d^3x' \left( 1 + \delta^{(\rm g)}(\VEC{x}') \right) 
    \delta_{\rm D}\left( \VEC{x} - \VEC{x}_{\rm red}(\VEC{x}') \right)-1.
    \label{Eq:real_to_red}
\end{eqnarray}
In Fourier space, the $n$-th order solution of $\delta_{\rm s}^{(\rm g)}$ is represented as
\begin{eqnarray}
    \widetilde{\delta}_{{\rm s},\, n}^{(\rm g)}(\VEC{k}) \hspace{-0.25cm}&=&\hspace{-0.25cm}  \int \frac{d^3p_1}{(2\pi)^3} \cdots \int \frac{d^3p_n}{(2\pi)^3}
    (2\pi)^3\delta_{\rm D}(\VEC{k}-\VEC{p}_{[1n]}) \nonumber \\
    \hspace{-0.25cm}&\times&\hspace{-0.25cm}
    Z_n(\VEC{p}_1,\dots,\VEC{p}_2) \delta_1(\VEC{p}_1)\cdots \delta_1(\VEC{p}_n).
\end{eqnarray}
The first and second-order kernel functions are given by~\citep{Scoccimarro:1999ed}
\begin{eqnarray}
    Z_1 \hspace{-0.25cm}&=& \hspace{-0.25cm}   b_1 + f(\hat{p}\cdot\hat{n})^2, \nonumber\\
    Z_2 \hspace{-0.25cm}&=& \hspace{-0.25cm} F_2^{(g)}(\VEC{p}_1,\VEC{p}_2) + f (\hat{k}\cdot\hat{n})^2 G^{(g)}_2(\VEC{p}_1,\VEC{p}_2) \nonumber \\
    \hspace{-0.25cm}&+& \hspace{-0.25cm} \frac{f(\VEC{k}\cdot\hat{n})}{2} \left[ \frac{(\hat{p}_1\cdot\hat{n})}{p_1}Z_1(\VEC{p}_2) +  \frac{(\hat{p}_2\cdot\hat{n})}{p_2}Z_1(\VEC{p}_1)  \right],
\end{eqnarray}
where $\VEC{k}=\VEC{p}_1+\VEC{p}_2$. In the rest of this paper, we focus only on the galaxy density fluctuation with RSDs, so for simplicity of notation, we refer to it simply as $\delta$ instead of $\delta_{\rm s}^{(\rm g)}$. We also omit the angle-dependence $\hat{n}$ of any function that includes RSDs.

At the leading-order in perturbation theory, the redshift-space power spectrum and bispectrum are represented as
\begin{eqnarray}
    P(\VEC{k}) \hspace{-0.25cm}&=&\hspace{-0.25cm} [Z_1(\VEC{k})]^2 P_{\rm lin}(k), \nonumber \\
    B(\VEC{k}_1,\VEC{k}_2,\VEC{k}_3) \hspace{-0.25cm}&=&\hspace{-0.25cm} 2 Z_2(\VEC{k}_1,\VEC{k}_2) Z_1(\VEC{k}_1)Z_1(\VEC{k}_2) P_{\rm lin}(k_1)P_{\rm lin}(k_2) \nonumber \\
   \hspace{-0.25cm} &+&\hspace{-0.25cm} \mbox{2 perms.},
   \label{Eq:Tree_PB}
\end{eqnarray}
where $\VEC{k}_1+\VEC{k}_2+\VEC{k}_3=0$, and $P_{\rm lin}$ is the linear matter power spectrum. In what follows, we omit the $\VEC{k}_3$-dependence of the bispectrum for notational simplicity: $B(\VEC{k}_1,\VEC{k}_2)=B(\VEC{k}_1,\VEC{k}_2,\VEC{k}_3=-\VEC{k}_1-\VEC{k}_2)$.

Finally, we conclude this subsection by summarising the key points about galaxy fluctuations from a theoretical point of view. First, in the case of $\nu=\mu=\tau_{\alpha}=0$ in Eqs.~(\ref{Eq:Poisson}) and (\ref{Eq:S_NL}), Horndeski theories are recovered; a $\Lambda$CDM model additionally has $\kappa=(3/2)\Omega_{\rm m}(z)$ and $\tau_{\gamma}=0$; in both Horndeski theories and $\Lambda$CDM, $\kappa_{\delta}=\kappa_{\theta}=1$ from Eq.~(\ref{Eq:EvoOfKappa}), and $\lambda_{\delta}$ and $\lambda_{\theta}$ are still time-dependent from Eq.~(\ref{Eq:EvoOfLambda}); for the approximation $f^2=\Omega_{\rm m}$ in $\Lambda$CDM, $\lambda_{\delta}=\lambda_{\theta}=1$. Second, since the linear equation of the density fluctuation (\ref{Eq:delta_evo}) omits space-dependence as in the $\Lambda$CDM case, under the assumption that the scalar field becomes to prevail during the accelerated Universe, the shape of the linear matter power spectrum can be the usual $\Lambda$CDM one determined in the matter-dominant era. In other words, the characteristic scale-dependences in $\delta$ and $v$ due to scalar-tensor theories appear only through the non-linear kernel functions $F^{(m)}_{n\geq2}$ and $G^{(m)}_{n\geq2}$. Third, the non-linear terms that appear in the fluid equation in Eq.~(\ref{Eq:Fluid}) and the Poisson equation in Eq.~(\ref{Eq:Poisson}), such as $\partial_i(\delta v^i)$, $\partial_i(v^j\partial_j v^i)$, $W_{\alpha}$ and $W_{\gamma}$, become zero when the volume average or ensemble average is calculated. Therefore, the resulting non-linear solutions satisfy $\int d^3x \delta_{n}=\langle \delta_{n}\rangle=0$ and $\int d^3x \theta_{n}=\langle \theta_{n}\rangle=0$ for $n\geq2$, and the corresponding kernel functions satisfy $F^{(\rm m)}_{n\geq2}=G^{(\rm m)}_{n\geq2}=0$ when $\VEC{p}_1+\dots+\VEC{p}_n=0$ as known in the case of $\Lambda$CDM. This condition partially breaks when the non-linear bias effect is taken into account, resulting in $F_2^{(\rm g)}(\VEC{p},-\VEC{p})\neq 0$ and $G_2^{(\rm g)}(\VEC{p},-\VEC{p})= 0$ (see Eq.~(\ref{Eq:BiasedFG})).

\subsection{Limitation of our assumptions}
\label{Sec:LimitationsOfOurAssumptions}

This subsection discusses the possible cases where the assumptions adopted in building the theoretical model in the previous subsection are violated, introducing some previous studies. The following bullet labels correspond to those in Section~\ref{Sec:DHOSTtheories}.

\begin{enumerate}
    \item Besides scalar-tensor theories, two other examples of modified gravity theories have been widely studied in cosmology: the Hu-Sawicki model~\citep{Hu:2007nk} of $f(R)$ gravity~\citep[see][for reviews]{Capozziello:2007ec,Sotiriou:2008rp} and the normal branch of the 5D brane-world Dvali-Gabadadze-Porrati model~\citep[nDGP;][]{Dvali:2000hr}. These two models have been investigated in detail by~\citet{Alam:2020jdv} as representative targets in DESI. Focusing on the non-linear effects, the nDGP model generates a scale dependence of the same form as Horndeski theories, characterised by the function $\gamma(\VEC{p}_1,\VEC{p}_2)$ (\ref{Eq:alpha_gamma}). On the other hand, the Hu-Sawicki $f(R)$ model produces a kernel function different from the one predicted by scalar-tensor theories. Specifically, in the modified Poisson equation of Eq.~(\ref{Eq:Poisson}), $\kappa$ is scale-dependent, resulting in the linear growth function that depends on the wavenumber. In addition, the non-linear source $S_{\Phi}^{\rm NL}$ for the Hu-Sawicki model also appears as a form that cannot be described by $W_{\alpha}$ and $W_{\gamma}$, unlike Eq.~(\ref{Eq:S_NL}). Such non-linearities in the density field specific to the Hu-Sawicki model have been studied by~\citep{Koyama:2009me,Taruya:2016jdt} in the context of perturbation theory, and the model has been tested by applying the theory to BOSS galaxy data~\citep{Song:2015oza}.     \vspace{2mm}
    \vspace{2mm}
    \item The effect of the relative velocity of baryons and CDM enters the galaxy density fluctuation quadratically together with the corresponding bias parameter~\citep{Dalal:2010yt}, thus modifying the shape of the measured bispectrum. In particular, as in the case of the $\kappa_{\delta}$ parameter in DHOST theories, it corrects the term in $F_2^{(\rm g)}(\VEC{p}_1,\VEC{p}_2)$ that depends on $(\hat{p}_1\cdot\hat{p}_2)$ called the shift term~\citep{Yoo:2011tq}. The relative velocity effect on galaxy clustering has been measured using the galaxy power spectrum~\citep{Yoo:2013qla,Beutler:2015tla} and 3PCF~\citep{Slepian:2016nfb}, but any signature has not yet been detected.
        
        Although massive neutrinos can also change the shape of the bispectrum, the results of simulations performed by~\citet{Ruggeri:2018JCAP...03..003R} confirm that the CDM component in the bispectrum is dominant; Interestingly, \citet{Kamalinejad:2020arXiv201100899K} has shown that the effect of neutrino corrections appears in the shift term as well as the growth and tidal terms in the second-order velocity field~(\ref{Eq:Param_FG}). Hence, the anisotropic 3PCF (or bispectrum) may help to constrain the neutrino masses~\citep[see e.g.,][]{Saito:2009ph,Levi:2016ar,Yoshikawa:2020ap}.
    \vspace{2mm}
    \item The case of non-minimally coupled scalar fields with CDM has already been the subject of several studies in the context of cosmology~\citep{Kimura:2017fnq,Chibana:2019jrf,Kase:2019veo,Chiba:2020mte,Kase:2020hst}. For example, \citet{Kimura:2017fnq,Chibana:2019jrf} have shown that in this case, the continuity equation~(\ref{Eq:Fluid}) is modified, and thus the relation between the density fluctuations in real and redshift spaces, i.e. the Kaiser formula in linear theory~\citep{Kaiser:1987qv}, is also modified. 
    \vspace{2mm}
    \item The quasi-static approximation breaks when the scale of interest is close to the sound horizon scale. Even in GR, it is known that there are relativistic corrections to $F_2^{(\rm g)}$ when approaching the horizon size~\citep{Tram:2016cpy,Jolicoeur:2017JCAP...09..040J,Jolicoeur:2018JCAP...03..036J,Koyama:2018ttg,Castiblanco:2018qsd,Umeh:2019qyd,Calles:2019prs,deWeerd:2020JCAP...05..018D}. 
    \vspace{2mm}
    \item Various possibilities have been proposed for how the initial conditions of cosmic fluctuations predicted by inflation theory could affect observables. One of the most critical examples relevant to this paper is the existence of primordial non-Gaussianity, which breaks the LSS consistency relation~\citep{Berezhiani:2014kga,Valageas:2016hhr,Esposito:2019jkb}.
    \vspace{2mm}
\item \citet{Fujita:2020JCAP...10..059F} proposed a bias expansion formalism dubbed ``Monkey bias'' based on the LSS consistency relation and showed that it is equivalent to the existing bias expansion framework. In other words, in DHOST theories, which violate the LSS consistency relation, the existing bias expansion we adopted~(\ref{Eq:bias}) may not be valid, and a new bias in the shift term of non-linear galaxy density fluctuations, i.e., the shift bias parameter, may appear. Moreover, the shift bias may also induce velocity bias effects.
        
    In Section~\ref{Sec:CommentsOnShiftBias}, we will discuss and clarify which parts of theories can be tested with the anisotropic 3PCF, even in the presence of the shift and velocity biases.

\end{enumerate}

\section{Theoretical models}
\label{Sec:CalculationsOfModels}

This section describes how to calculate the theoretical models of multipole 2PCFs and 3PCFs. Section~\ref{Sec:3PCFs} summarises the decomposition formalism for the anisotropic three-point statistics (bispectra and 3PCFs). Section~\ref{Sec:3PCFsBAO} introduces the power and bispectrum models used to compute the 2PCF and 3PCF. Section~\ref{Sec:Parameters} discusses what parameters should be varied to perform the cosmological analysis and shows the specific parameter dependence of the bispectrum model we use. Section~\ref{Sec:TimeDependencyOfParameters} reviews new parameters helpful in testing DHOST theories proposed by~\citet{Yamauchi:2021arXiv210802382Y} and their time evolution. Section~\ref{Sec:LimitationsOf2PCFand3PCF} discusses the limits of applying our theoretical models of the 2PCF and 3PCF to the data analysis.

\subsection{Decomposition formalisms of the 2PCF and 3PCF}
\label{Sec:3PCFs}

We follow the decomposition formalism of redshift-space bispectra proposed by~\citet{Sugiyama:2018yzo} using the tri-polar spherical harmonics (TripoSH) as a basis function. In that formalism, under statistical homogeneity, isotropy, and parity-symmetry assumptions, we define the base function to expand the bispectrum using three spherical harmonics $Y_{\ell m}$ as
\begin{eqnarray}
    {\cal S}_{\ell_1\ell_2\ell}(\hat{k}_1,\hat{k}_2,\hat{n}) 
	\hspace{-0.25cm}&=&\hspace{-0.25cm}
   \frac{4\pi}{h_{\ell_1\ell_2\ell}} 
   \sum_{m_1m_2m} 
   \left( \begin{smallmatrix} \ell_1 & \ell_2 & \ell \\ m_1 & m_2 & m \end{smallmatrix}  \right)\nonumber \\ 
    \hspace{-0.25cm}&\times&\hspace{-0.25cm}
    Y_{\ell_1 m_1}(\hat{k}_1) Y_{\ell_2 m_2}(\hat{k}_2) Y_{\ell m}(\hat{n}),
    \label{Eq:Slll}
\end{eqnarray} 
where
\begin{eqnarray}
    h_{\ell_1\ell_2\ell}= \sqrt{ \frac{(2\ell_1+1)(2\ell_2+1)(2\ell+1)}{4\pi}}
   \left( \begin{smallmatrix} \ell_1 & \ell_2 & \ell \\ 0 & 0 & 0 \end{smallmatrix}  \right),
\end{eqnarray}
and the circle bracket with $6$ multipole indices, $(\dots)$, denotes the Wigner-3j symbol. The bispectrum is then expanded as
\begin{eqnarray}
    B(\VEC{k}_1,\VEC{k}_2,\hat{n})=\hspace{-0.25cm} \sum_{\ell_1+\ell_2+\ell={\rm even}}\hspace{-0.25cm}  B_{\ell_1\ell_2\ell}(k_1,k_2){\cal S}_{\ell_1\ell_2\ell}(\hat{k}_1,\hat{k}_2,\hat{n}),
\end{eqnarray}
and the corresponding multipole components are given by
\begin{eqnarray}
    B_{\ell_1\ell_2\ell}(k_1,k_2) \hspace{-0.25cm}&=&\hspace{-0.25cm} 4\pi h_{\ell_1\ell_2\ell}^2 \int \frac{d^2\hat{k}_1}{4\pi}\int \frac{d^2\hat{k}_2}{4\pi}\int \frac{d^2\hat{n}}{4\pi} \nonumber \\
    \hspace{-0.25cm}&\times&\hspace{-0.25cm} {\cal S}^*_{\ell_1\ell_2\ell}(\hat{k}_1,\hat{k}_2) B(\VEC{k}_1,\VEC{k}_2).
\end{eqnarray}
Since the bispectrum multipoles defined here are independent of the coordinate system in which they are calculated, it is possible to compare theoretical calculations with observations in different coordinate systems. Specifically, we use the following coordinate system with $\hat{k}_1$ as the $z$-axis for theoretical calculations:
\begin{eqnarray}
    \hat{k}_1 \hspace{-0.25cm}&=&\hspace{-0.25cm} \{0,0,1\} \nonumber \\
    \hat{k}_2 \hspace{-0.25cm}&=&\hspace{-0.25cm} \{\sin\theta_{k_2},0,\cos\theta_{k_2}\} \nonumber \\
    \hat{n}   \hspace{-0.25cm}&=&\hspace{-0.25cm} \{\sin \theta \cos\varphi, \sin \theta \sin \varphi, \cos\theta\}.
    \label{Eq:coordinates}
\end{eqnarray}
On the other hand, when measuring the bispectrum from galaxy data, we use the Cartesian coordinate and take the north pole as our $z$-axis (see Section~\ref{Sec:Estimator}). 

We perform the expansion of the 3PCF in the same way as for the bispectrum. The resulting 3PCF multipoles are related to $B_{\ell_1\ell_2\ell}$ through a two-dimensional Hankel transform:
\begin{eqnarray}
	\zeta_{\ell_1\ell_2\ell}(r_1,r_2) 
    \hspace{-0.25cm}&=&\hspace{-0.25cm}
    i^{\ell_1+\ell_2} \int \frac{dk_1k_1^2}{2\pi^2} \int \frac{dk_2k_2^2}{2\pi^2} \nonumber \\
	\hspace{-0.25cm}&\times&\hspace{-0.25cm}
    j_{\ell_1}(r_1k_1) j_{\ell_2}(r_2k_2) B_{\ell_1\ell_2\ell}(k_1,k_2),
    \label{Eq:B_to_zeta}
\end{eqnarray}
where $j_{\ell}$ is the spherical Bessel function at the $\ell$-th order. This relation means that $\zeta_{\ell_1\ell_2\ell}$ have in principle the same information as $B_{\ell_1\ell_2\ell}$, facilitating the comparison of the configuration-space and Fourier-space analyses. 

Note that $B_{\ell_1\ell_2\ell}(k_1,k_2)=B_{\ell_2\ell_1\ell}(k_2,k_1)$ and $\zeta_{\ell_1\ell_2\ell}(r_1,r_2)=\zeta_{\ell_2\ell_1\ell}(r_2,r_1)$. From this relation, when $\ell_1=\ell_2$, only $k_1\geq k_2$ and $r_1\geq r_2$ need to be computed for the bispectrum and 3PCF, respectively. Also, when $\ell>0$, only $\ell_1\geq\ell_2$ should be considered.

In the case of the power spectrum, it is common to expand the power spectrum using Legendre polynomial functions ${\cal L}_{\ell}$~\citep[e.g.,][]{Hamilton:1997zq}:
\begin{eqnarray}
    P(\VEC{k}) = \sum_{\ell} P_{\ell}(k) {\cal L}_{\ell}(\hat{k}\cdot\hat{n}),
\end{eqnarray}
and the corresponding multipole components of the 2PCF are given by
\begin{eqnarray}
    \xi_{\ell}(r) 
    \hspace{-0.25cm}&=&\hspace{-0.25cm}
    i^{\ell} \int \frac{dkk^2}{2\pi^2} j_{\ell}(rk)P_{\ell}(k).
\end{eqnarray}

This paper tests DHOST theories by measuring $\xi_{\ell}$ and $\zeta_{\ell_1\ell_2\ell}$ from the BOSS galaxy data and comparing them with the corresponding theoretical models. The index $\ell$ that is common for both $\xi_{\ell}$ and $\zeta_{\ell_1\ell_2\ell}$ represents the decomposition related to the RSD or AP effect, where $\ell=0$ means monopole, $\ell=2$ quadrupole, and $\ell=4$ hexadecapole. Relativistic effects can generate $\ell={\rm odd}$ components~\citep[e.g.,][]{McDonald:2009ud,Desjacques:2018pfv,Clarkson:2019MNRAS.486L.101C}, but we ignore them here. Furthermore, we also ignore the $\ell=4$ modes; although the signal of the $\ell=4$ modes is too small to be detected in the BOSS data, it should be taken into account in the future as it helps to improve the constraints on the cosmological parameters~\citep{Beutler:2016arn,Sugiyama:2018yzo}. Therefore, in this paper, we focus on only two modes, $\ell=0$ and $\ell=2$. In particular, for the 3PCF, we consider the first two terms of the monopole ($\zeta_{000}$ and $\zeta_{110}$) and the first two terms of the quadrupole ($\zeta_{202}$ and $\zeta_{112}$).

Finally, we discuss the relation with the widely used decomposition formalism of the bispectrum proposed by~\citet{Scoccimarro:1999ed}. As in Eq.~(\ref{Eq:coordinates}), this formalism decomposes the bispectrum by choosing the coordinate system with $k_1$ as the $z$-axis and using the spherical harmonic function for the LOS direction: $B(\VEC{k}_1,\VEC{k}_2,\hat{n}) = \sum_{LM} B_{LM}(\VEC{k}_1,\VEC{k}_2) Y_{LM}(\hat{n})$. The relation between \citet{Scoccimarro:1999ed}'s decomposition formalism and our TripoSH decomposition has already been shown in Eq.~(25) of \citet{Sugiyama:2018yzo}. According to the relation, $\zeta_{202}$ contains only $M=0$ mode in \citet{Scoccimarro:1999ed}'s formalism, while $\zeta_{112}$ further contains the $M\neq0$ modes in addition to the $M=0$ mode. The ability to handle the $M\neq 0$ modes, including window function corrections (see Section~\ref{Sec:WindowCorrections}), is one advantage of our TripoSH decomposition formalism. For example, studies of the quadrupole bispectrum using \citet{Scoccimarro:1999ed}'s method have mainly dealt only with the $M=0$ mode~\citep{D'Amico:2022arXiv220608327D}. One reason is that the correction formula for the window function effect is only given for the $M=0$ case~\citep{Pardede:2022JCAP...10..066P}. Moreover, we show in Section~\ref{Sec:FisherAnalysis} that $\zeta_{112}$ gives additional cosmological information to $\zeta_{202}$, pointing out the importance of the $M\neq0$ modes.

\subsection{IR-resummed power spectrum and bispectrum models}
\label{Sec:3PCFsBAO}

In this paper, we focus on the 2PCF and 3PCF at scales above $80\hMpc$ (Section~\ref{Sec:DataAnalysis}), where we can ignore loop corrections arising from higher-order non-linear effects. The power spectrum and bispectrum shapes can be described at those scales by their leading solutions, the so-called tree-level solutions (\ref{Eq:Tree_PB}). However, we need to consider the non-linear damping effect of BAOs due to the linear gravity that shifts the position of galaxies.

The non-linear damping of BAO can be described by a large-scale bulk flow that is position-independent in a given observed region~\citep{Eisenstein:2006nj,Crocce:2007dt,Matsubara:2007wj,Sugiyama:2013gza,Baldauf:2015xfa}, called the infra-red (IR) flow. In the limit where the IR flow does not correlate with small-scale density fluctuations, based on the Galilean invariance of the system of equations in the IR limit, all the effects of the IR flow are cancelled out in equal-time $n$-point statistics~\citep{Jain:1995kx,Scoccimarro:1995if,Kehagias:2013yd,Peloso:2013zw,Sugiyama:2013pwa,Sugiyama:2013gza,Blas:2013bpa,Blas:2015qsi,Lewandowski:2017kes}. However, when we deviate from such an extreme situation, we find a correlation between the IR flow and the small-scale density field. By extracting this correlation in the full perturbative order only for the BAO signal, it becomes possible to describe the non-linear effects of BAOs. This kind of construction of $n$-point statistics models is called the IR resummation method~\citep{Crocce:2007dt,Matsubara:2007wj,Sugiyama:2013gza,Senatore:2014via,Baldauf:2015xfa,Blas:2016sfa,Senatore:2017pbn,Ivanov:2018gjr,Lewandowski:2018ywf,Sugiyama:2020uil}. In this paper, we will use the IR resummed power and bispectrum models given in Eqs.~(\ref{Eq:power_IR}) and (\ref{Eq:bispec_IR}), even in DHOST theories that break the IR cancellation, but we will mention the issues that may arise in this case in Section~\ref{Sec:LimitationsOf2PCFand3PCF}.

For the power spectrum, we adopt the following IR-resummed model:
\begin{eqnarray}
    P(\VEC{k}) = \left[ Z_1(\VEC{k}) \right]^2 \left[{\cal D}^2(\VEC{k})P_{\rm w}(k) + P_{\rm nw}(k) \right],
    \label{Eq:power_IR}
\end{eqnarray}
where $P_{\rm lin}$ is decomposed into two parts: the "no-wiggle (nw)" part $P_{\rm nw}$ that is a smooth version of $P_{\rm lin}$ with the baryon oscillations removed~\citep{Eisenstein:1997ik}, and the "wiggle (w)" part defined as $P_{\rm w}=P_{\rm lin}-P_{\rm nw}$. The non-linear BAO degradation is represented by the two-dimensional Gaussian damping factor derived from a differential motions of Lagrangian displacements~\citep{Eisenstein:2006nj,Crocce:2007dt,Matsubara:2007wj}:
\begin{eqnarray}
    {\cal D}(\VEC{k}) = \exp\left( -\frac{ k^2(1-\mu^2)\sigma_{\perp}^2 + k^2\mu^2 \sigma^2_{\parallel} }{2} \right),
    \label{Eq:Damping}
\end{eqnarray}
where $\mu=\hat{k}\cdot\hat{n}$. We compute the radial and transverse components of smoothing parameters, $\sigma_{\perp}$ and $\sigma_{\parallel}$, using the Zel'dovich approximation~\citep{Zeldovich:1970,Crocce:2007dt,Matsubara:2007wj}:
\begin{eqnarray}
    \sigma^2_{\perp} \hspace{-0.25cm}&=&\hspace{-0.25cm} 
    \frac{1}{3}\int \frac{dp}{2\pi^2} P_{\rm lin}(p), \nonumber \\
    \sigma^2_{\parallel} \hspace{-0.25cm}&=&\hspace{-0.25cm} (1+f)^2\, \sigma^2_{\perp}.
    \label{Eq:sigma_sigma}
\end{eqnarray}
The power spectrum model in Eq.~(\ref{Eq:power_IR}) was first proposed empirically by~\citet{Eisenstein:2006nj}. Subsequently, the damping factor ${\cal D}^2$ in front of $P_{\rm lin}$ was derived in the context of perturbation theory by~\citet{Crocce:2007dt,Matsubara:2007wj}; an additional term to recover a smooth linear power spectrum without BAOs, $(1-{\cal D}^2)P_{\rm nw}$, was derived using the IR resummation method~\citep{Sugiyama:2013gza,Baldauf:2015xfa,Blas:2016sfa,Ivanov:2018gjr,Sugiyama:2020uil}.

For the bispectrum, we adopt the following IR-resummed model~\citep{Sugiyama:2020uil}:
\begin{eqnarray}
    B(\VEC{k}_1,\VEC{k}_2) 
    \hspace{-0.25cm}&=&\hspace{-0.25cm}  2\, Z_2(\VEC{k}_1,\VEC{k}_2)Z_1(\VEC{k}_1)Z_1(\VEC{k}_2) \nonumber \\
    \hspace{-0.25cm}&\times&\hspace{-0.25cm} \Big\{ {\cal D}(\VEC{k}_1){\cal D}(\VEC{k}_2){\cal D}(\VEC{k}_3) P_{\rm w}(k_1)P_{\rm w}(k_2) \nonumber \\
    \hspace{-0.25cm}&+&\hspace{-0.25cm} {\cal D}^2(\VEC{k}_1) P_{\rm w}(k_1)P_{\rm nw}(k_2) + {\cal D}^2(\VEC{k}_2) P_{\rm nw}(k_1)P_{\rm w}(k_2) \nonumber \\
    \hspace{-0.25cm}&+&\hspace{-0.25cm} P_{\rm nw}(k_1)P_{\rm nw}(k_2)\Big\} + \mbox{2 perms.},
    \label{Eq:bispec_IR}
\end{eqnarray}
where $\VEC{k}_1+\VEC{k}_2+\VEC{k}_3=0$. As in the case of the power spectrum, this bispectrum model restores the tree-level solution~(\ref{Eq:Tree_PB}) consisting of a smooth version (without BAOs) of the linear power spectrum after degrading the BAO signature~\footnote{\citet{Blas:2016sfa,Ivanov:2018gjr} proposed a bispectrum model similar to Eq.~(\ref{Eq:bispec_IR}). However, the authors ignore the ${\cal O}(P^2_{\rm w}/P^2_{\rm nw})$ term, so their model does not include the second line term, $D(\VEC{k}_1)D(\VEC{k}_2)D(\VEC{k}_3)P_{\rm w}(k_1)P_{\rm w}(k_2)$, in Eq.~(\ref{Eq:bispec_IR}). This term added by~\citet{Sugiyama:2020uil} contains the full tree-level solution.}.

\subsection{Parameterization method for the bispectrum}
\label{Sec:Parameters}

The non-linear kernel functions $F^{(\rm m)}_2$ and $G^{(\rm m)}_2$ can be decomposed into three terms using Legendre polynomial functions ${\cal L}_{\ell}(\hat{p}_1\cdot\hat{p}_2)$: i.e., monopole, dipole, and quadrupole components~\citep{Schmittfull:2015PhRvD..91d3530S}. They are called the growth, shift, and tidal terms, and are understood in $\Lambda$CDM as follows: the growth term represents the spherical collapse of density fluctuations~\citep{Fosalba:1997tn}; the shift term appears in the form $\Psi_1^i\partial_i \delta_1$ or $\Psi_1^i\partial_i \theta_1$ as a coordinate transformation of $\delta$ or $\theta$ by the displacement vector $\mathbf{\Psi}$; the last term represents the tidal force (\ref{Eq:s_ij}). Then, $F^{(\rm m)}_2$ and $G^{(\rm m)}_2$ (\ref{Eq:FG_m}) are rewritten as~\citep[e.g.,][]{Bouchet:1992xz,Sherwin:2012PhRvD..85j3523S,Baldauf:2012PhRvD..86h3540B,Schmittfull:2015PhRvD..91d3530S}
\begin{eqnarray}
    F^{(\rm m)}_2 \hspace{-0.25cm}&=&\hspace{-0.25cm}  \left( \kappa_{\delta} - \frac{4}{21}\lambda_{\delta} \right) 
    + \kappa_{\delta} S(\VEC{k}_1,\VEC{k}_2) + \frac{2}{7} \lambda_{\delta} T(\VEC{k}_1,\VEC{k}_2), \nonumber \\
    G^{(\rm m)}_2 \hspace{-0.25cm}&=&\hspace{-0.25cm}  \left( \kappa_{\theta} - \frac{8}{21}\lambda_{\theta} \right) 
    + \kappa_{\theta} S(\VEC{k}_1,\VEC{k}_2) + \frac{4}{7} \lambda_{\theta} T(\VEC{k}_1,\VEC{k}_2),
\end{eqnarray}
where $S$ and $T$ are the scale-dependent functions characterising the shift and tidal terms:
\begin{eqnarray}
    S(\VEC{k}_1,\VEC{k}_2) \hspace{-0.25cm}&=& \hspace{-0.25cm} \frac{1}{2} (\hat{k}_1\cdot\hat{k}_2) \left( \frac{k_1}{k_2} + \frac{k_2}{k_1} \right), \nonumber \\
    T(\VEC{k}_1,\VEC{k}_2) \hspace{-0.25cm}&=& \hspace{-0.25cm}  (\hat{k}_1\cdot\hat{k}_2)^2 - \frac{1}{3}.
    \label{Eq:ST}
\end{eqnarray}
As mentioned in Section~\ref{Sec:DHOSTtheories}, the coefficients of the growth, shift, and tidal terms are not independent of each other but are related to under the condition that $F_2^{(\rm m)}(\VEC{p},-\VEC{p})=G_2^{(\rm m)}(\VEC{p},-\VEC{p})=0$. Therefore, the coefficient of the growth term is determined from the coefficients of the shift and tidal terms. 

Considering the linear and non-linear bias effects, that the second-order fluctuations are proportional to $\sigma_8^2$, and that $G_2^{(\rm m)}$ always appears with $f$, we introduce the following parameterisation,
\begin{eqnarray}
    F^{\rm (g)}_2\sigma_8^2 \hspace{-0.25cm}&=&\hspace{-0.25cm} (b_1\sigma_8) \left[(F_{\rm g}\sigma_8) + (F_{\rm s}\sigma_8) S + (F_{\rm t}\sigma_8) T  \right], \nonumber \\
    f G^{\rm (g)}_2\sigma_8^2 \hspace{-0.25cm}&=&\hspace{-0.25cm} (f\sigma_8) \left[(G_{\rm g}\sigma_8) + (G_{\rm s}\sigma_8) S + (G_{\rm t}\sigma_8) T  \right].
    \label{Eq:F_2G_2}
\end{eqnarray}
DHOST theories have $G_{\rm g} = G_{\rm s}- (2/3)G_{\rm t}$ from the condition $G_2^{(\rm g)}(\VEC{p},-\VEC{p})=0$; Horndeski theories further have $F_{\rm s}=G_{\rm s}=1$. The specific form of each coefficient in DHOST theories is given by
\begin{eqnarray}
    F_{\rm g} \hspace{-0.25cm}&=&\hspace{-0.25cm} \kappa_{\delta} - \frac{4}{21}\lambda_{\delta} + \frac{1}{2} \frac{b_2}{b_1}, \nonumber \\
    F_{\rm s} \hspace{-0.25cm}&=&\hspace{-0.25cm} \kappa_{\delta}, \nonumber \\
    F_{\rm t} \hspace{-0.25cm}&=&\hspace{-0.25cm} \frac{2}{7}\lambda_{\delta} + \frac{b_{s^2}}{b_1}, \nonumber \\
    G_{\rm g} \hspace{-0.25cm}&=&\hspace{-0.25cm} \kappa_{\theta} - \frac{8}{21}\lambda_{\theta}, \nonumber \\
    G_{\rm s} \hspace{-0.25cm}&=&\hspace{-0.25cm} \kappa_{\theta}, \nonumber \\
    G_{\rm t} \hspace{-0.25cm}&=&\hspace{-0.25cm} \frac{4}{7}\lambda_{\theta}.
    \label{Eq:Param_FG}
\end{eqnarray}
In Eq.~(\ref{Eq:Param_FG}), $F_{\rm g}$ and $F_{\rm t}$ do not contain any cosmological information because they are degenerate with the non-linear bias parameters, and $G_{\rm g}$ is determined from $G_{\rm t}$ and $G_{\rm s}$. Thus, cosmologically meaningful parameters are $F_{\rm s}$, $G_{\rm s}$, and $G_{\rm t}$. 

Following the method proposed by \citet{Sugiyama:2020uil}, we decompose the IR-resummed bispectrum model into
\begin{eqnarray}
    B(\VEC{k}_1,\VEC{k}_2) = \sum_{p=1}^{22} X^{(p)}B^{(p)}(\VEC{k}_1,\VEC{k}_2),
    \label{Eq:Decomposed_B}
\end{eqnarray}
with
\begin{eqnarray}
    \hspace{-0.5cm}   B^{(p)}(\VEC{k}_1,\VEC{k}_2) \hspace{-0.25cm}&=&\hspace{-0.25cm}  2\, H^{(p)}(\VEC{k}_1,\VEC{k}_2)  \nonumber \\
    \hspace{-0.25cm}&\times&\hspace{-0.25cm} \Big\{ {\cal D}(\VEC{k}_1){\cal D}(\VEC{k}_2){\cal D}(\VEC{k}_3) P^{\rm (n)}_{\rm w}(k_1)P^{\rm (n)}_{\rm w}(k_2) 
        \nonumber \\
    \hspace{-0.25cm}&+&\hspace{-0.25cm} {\cal D}^2(\VEC{k}_1) P^{\rm (n)}_{\rm w}(k_1)P^{\rm (n)}_{\rm nw}(k_2) \nonumber \\
    \hspace{-0.25cm}&+&\hspace{-0.25cm} {\cal D}^2(\VEC{k}_2) P^{\rm (n)}_{\rm nw}(k_1)P^{\rm (n)}_{\rm w}(k_2) \nonumber \\
    \hspace{-0.25cm}&+&\hspace{-0.25cm} P^{\rm (n)}_{\rm nw}(k_1)P^{\rm (n)}_{\rm nw}(k_2)\Big\}+ \mbox{2 perms.},
    \label{Eq:BP}
\end{eqnarray}
where $P_{\rm w}^{(\rm n)}$ and $P_{\rm nw}^{(\rm n)}$ are respectively the wiggle and no-wiggle linear matter power spectra normalized by $\sigma^2_8$: $P_{\rm w}^{(\rm n)}=P_{\rm w}/\sigma_8^2$ and $P_{\rm nw}^{(\rm n)}=P_{\rm nw}/\sigma_8^2$. The functions $X^{(p)}$ ($p=1-22$) represent the combinations of the parameters of interest and are given by
\begin{eqnarray}
    X^{(1)}  \hspace{-0.25cm}&=&\hspace{-0.25cm} (F_{\rm g}\sigma_8)(b_1\sigma_8)^3             , \nonumber \\
    X^{(2)}  \hspace{-0.25cm}&=&\hspace{-0.25cm} (F_{\rm s}\sigma_8)(b_1\sigma_8)^3             , \nonumber \\
    X^{(3)}  \hspace{-0.25cm}&=&\hspace{-0.25cm} (F_{\rm t}\sigma_8)(b_1\sigma_8)^3             , \nonumber \\
    X^{(4)}  \hspace{-0.25cm}&=&\hspace{-0.25cm} (F_{\rm g}\sigma_8)(b_1\sigma_8)^2(f\sigma_8)  , \nonumber \\
    X^{(5)}  \hspace{-0.25cm}&=&\hspace{-0.25cm} (F_{\rm s}\sigma_8)(b_1\sigma_8)^2(f\sigma_8)  , \nonumber \\
    X^{(6)}  \hspace{-0.25cm}&=&\hspace{-0.25cm} (F_{\rm t}\sigma_8)(b_1\sigma_8)^2(f\sigma_8)  , \nonumber \\
    X^{(7)}  \hspace{-0.25cm}&=&\hspace{-0.25cm} (F_{\rm g}\sigma_8)(b_1\sigma_8)(f\sigma_8)^2  , \nonumber \\
    X^{(8)}  \hspace{-0.25cm}&=&\hspace{-0.25cm} (F_{\rm s}\sigma_8)(b_1\sigma_8)(f\sigma_8)^2  , \nonumber \\
    X^{(9)}  \hspace{-0.25cm}&=&\hspace{-0.25cm} (F_{\rm t}\sigma_8)(b_1\sigma_8)(f\sigma_8)^2  , \nonumber \\
    X^{(10)} \hspace{-0.25cm}&=&\hspace{-0.25cm} (G_{\rm g}\sigma_8)(b_1\sigma_8)^2(f\sigma_8)  , \nonumber \\
    X^{(11)} \hspace{-0.25cm}&=&\hspace{-0.25cm} (G_{\rm s}\sigma_8)(b_1\sigma_8)^2(f\sigma_8)  , \nonumber \\
    X^{(12)} \hspace{-0.25cm}&=&\hspace{-0.25cm} (G_{\rm t}\sigma_8)(b_1\sigma_8)^2(f\sigma_8)  , \nonumber \\
    X^{(13)} \hspace{-0.25cm}&=&\hspace{-0.25cm} (G_{\rm g}\sigma_8)(b_1\sigma_8)(f\sigma_8)^2  , \nonumber \\
    X^{(14)} \hspace{-0.25cm}&=&\hspace{-0.25cm} (G_{\rm s}\sigma_8)(b_1\sigma_8)(f\sigma_8)^2  , \nonumber \\
    X^{(15)} \hspace{-0.25cm}&=&\hspace{-0.25cm} (G_{\rm t}\sigma_8)(b_1\sigma_8)(f\sigma_8)^2  , \nonumber \\
    X^{(16)} \hspace{-0.25cm}&=&\hspace{-0.25cm} (G_{\rm g}\sigma_8)(f\sigma_8)^3  , \nonumber \\
    X^{(17)} \hspace{-0.25cm}&=&\hspace{-0.25cm} (G_{\rm s}\sigma_8)(f\sigma_8)^3  , \nonumber \\
    X^{(18)} \hspace{-0.25cm}&=&\hspace{-0.25cm} (G_{\rm t}\sigma_8)(f\sigma_8)^3  ,  \nonumber \\
    X^{(19)} \hspace{-0.25cm}&=&\hspace{-0.25cm} (b_1\sigma_8)^3(f\sigma_8)  , \nonumber \\
    X^{(20)} \hspace{-0.25cm}&=&\hspace{-0.25cm} (b_1\sigma_8)^2(f\sigma_8)^2   , \nonumber \\
    X^{(21)} \hspace{-0.25cm}&=&\hspace{-0.25cm} (b_1\sigma_8)(f\sigma_8)^3  , \nonumber \\
    X^{(22)} \hspace{-0.25cm}&=&\hspace{-0.25cm} (f\sigma_8)^4.
    \label{Eq:X}
\end{eqnarray}
The scale-dependent functions $H^{(p)}$ ($p=1-22$) are derived by decomposing the non-linear kernel functions $Z_1Z_1Z_2$ in terms of the parameters, given by
\begin{eqnarray}
    H^{(1)}  \hspace{-0.25cm}&=&\hspace{-0.25cm} 1  , \nonumber \\
    H^{(2)}  \hspace{-0.25cm}&=&\hspace{-0.25cm} S(\VEC{k}_1,\VEC{k}_2), \nonumber \\
    H^{(3)}  \hspace{-0.25cm}&=&\hspace{-0.25cm} T(\VEC{k}_1,\VEC{k}_2), \nonumber \\
    H^{(4)}  \hspace{-0.25cm}&=&\hspace{-0.25cm} (\mu_1^2+\mu_2^2)                                 , \nonumber \\
    H^{(5)}  \hspace{-0.25cm}&=&\hspace{-0.25cm} S(\VEC{k}_1,\VEC{k}_2)(\mu_1^2+\mu_2^2)           , \nonumber \\
    H^{(6)}  \hspace{-0.25cm}&=&\hspace{-0.25cm} T(\VEC{k}_1,\VEC{k}_2)(\mu_1^2+\mu_2^2)           , \nonumber \\
    H^{(7)}  \hspace{-0.25cm}&=&\hspace{-0.25cm} (\mu_1^2\mu_2^2)                                  , \nonumber \\
    H^{(8)}  \hspace{-0.25cm}&=&\hspace{-0.25cm} S(\VEC{k}_1,\VEC{k}_2)(\mu_1^2\mu_2^2)            , \nonumber \\
    H^{(9)}  \hspace{-0.25cm}&=&\hspace{-0.25cm} T(\VEC{k}_1,\VEC{k}_2)(\mu_1^2\mu_2^2)            , \nonumber \\
    H^{(10)} \hspace{-0.25cm}&=&\hspace{-0.25cm} (\mu^2)                                           , \nonumber \\
    H^{(11)} \hspace{-0.25cm}&=&\hspace{-0.25cm} S(\VEC{k}_1,\VEC{k}_2)(\mu^2)                     , \nonumber \\
    H^{(12)} \hspace{-0.25cm}&=&\hspace{-0.25cm} T(\VEC{k}_1,\VEC{k}_2)(\mu^2)                     , \nonumber \\
    H^{(13)} \hspace{-0.25cm}&=&\hspace{-0.25cm} (\mu^2)(\mu_1^2+\mu_2^2)                          , \nonumber \\
    H^{(14)} \hspace{-0.25cm}&=&\hspace{-0.25cm} S(\VEC{k}_1,\VEC{k}_2)(\mu^2)(\mu_1^2+\mu_2^2)    , \nonumber \\
    H^{(15)} \hspace{-0.25cm}&=&\hspace{-0.25cm} T(\VEC{k}_1,\VEC{k}_2)(\mu^2)(\mu_1^2+\mu_2^2)    , \nonumber \\
    H^{(16)} \hspace{-0.25cm}&=&\hspace{-0.25cm} (\mu^2)(\mu_1^2\mu_2^2)                           , \nonumber \\
    H^{(17)} \hspace{-0.25cm}&=&\hspace{-0.25cm} S(\VEC{k}_1,\VEC{k}_2)(\mu^2)(\mu_1^2\mu_2^2)    , \nonumber \\
    H^{(18)} \hspace{-0.25cm}&=&\hspace{-0.25cm} T(\VEC{k}_1,\VEC{k}_2)(\mu^2)(\mu_1^2\mu_2^2)    , \nonumber \\
    H^{(19)} \hspace{-0.25cm}&=&\hspace{-0.25cm} DV(\VEC{k}_1,\VEC{k}_2), \nonumber \\
    H^{(20)} \hspace{-0.25cm}&=&\hspace{-0.25cm} DV(\VEC{k}_1,\VEC{k}_2)(\mu_1^2+\mu_2^2) + V(\VEC{k}_1,\VEC{k}_2)   , \nonumber \\
    H^{(21)} \hspace{-0.25cm}&=&\hspace{-0.25cm} DV(\VEC{k}_1,\VEC{k}_2)(\mu_1^2\mu_2^2) + V(\VEC{k}_1,\VEC{k}_2)(\mu_1^2+\mu_2^2)   , \nonumber \\
    H^{(22)} \hspace{-0.25cm}&=&\hspace{-0.25cm}  V(\VEC{k}_1,\VEC{k}_2)(\mu_1^2\mu_2^2),
    \label{Eq:Bp}
\end{eqnarray}
where $\VEC{k}=\VEC{k}_1+\VEC{k}_2$, $\mu = \hat{k}\cdot\hat{n}$, $\mu_1 = \hat{k}_1\cdot\hat{n}$, $\mu_2 = \hat{k}_2\cdot\hat{n}$, and
\begin{eqnarray}
    V(\VEC{k}_1,\VEC{k}_2) \hspace{-0.25cm} &=&\hspace{-0.25cm} \frac{1}{2} \frac{k^2}{k_1k_2}\mu^2 \mu_1  \mu_2, \nonumber \\
    DV(\VEC{k}_1,\VEC{k}_2)\hspace{-0.25cm} &=&\hspace{-0.25cm} \frac{1}{2} k \mu \left[ \frac{\mu_1}{k_1} + \frac{\mu_2}{k_2}\right].
\end{eqnarray}
We pre-compute $B^{(p)}(\VEC{k}_1,\VEC{k}_2)$ using the fiducial cosmology introduced in Section~\ref{Sec:Introduction} and save the resulting data in a file. In this way, when constraining $X^{(p)}$ from the BOSS data, we can quickly calculate the bispectrum by loading the data file containing $B^{(p)}$ and substituting them into Eq.~(\ref{Eq:Decomposed_B}) along with $X^{(p)}$.

\begin{figure*}
    \scalebox{0.9}{\includegraphics[width=\textwidth]{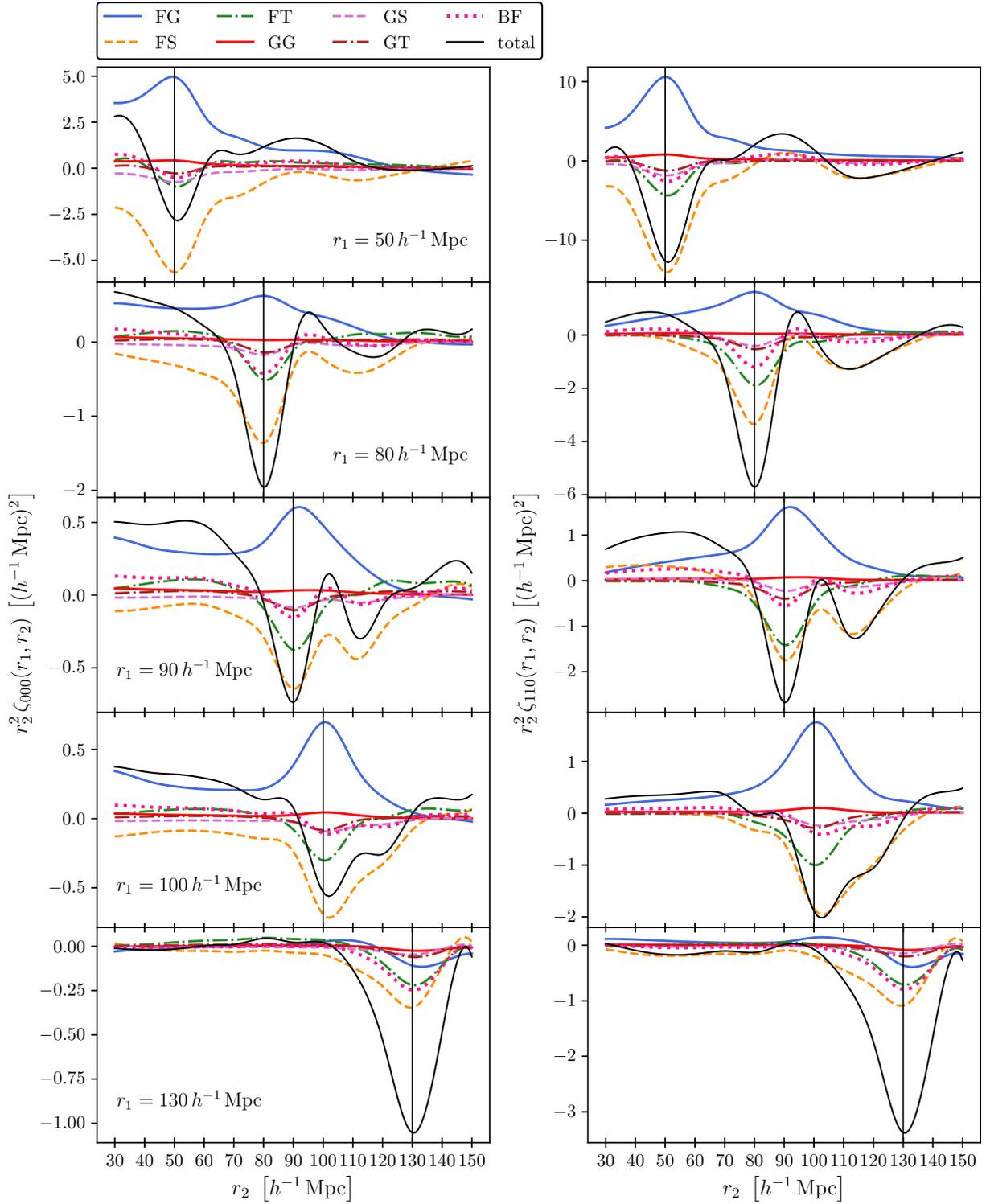}}
    \caption{The monopole 3PCFs, $\zeta_{000}$ (left) and $\zeta_{110}$ (right), calculated from the decomposed bispectra (\ref{Eq:Decom_B}) according to the parameter dependence, are shown as a function of $r_2$ after fixing $r_1$ to $50$, $80$, $90$, $100$, and $130\hMpc$. The ``FG'', ``FS'', and ``FT'' terms arise from the growth, shift and tidal effects of the non-linear density fluctuation; the ``GG'', ``GS'', and ``GT'' terms arise from those of the non-linear velocity field; the ``BF'' term consists only of linear density and linear velocity fields; the ``total'' term is the sum of all the decomposed components. For these calculations, the $\Lambda$CDM model at $z=0.61$, the linear bias $b_1=2.0$, and no non-linear bias are assumed.}
	\label{fig:decom_3PCF_mono}
\end{figure*}

\begin{figure*}
    \scalebox{0.9}{\includegraphics[width=\textwidth]{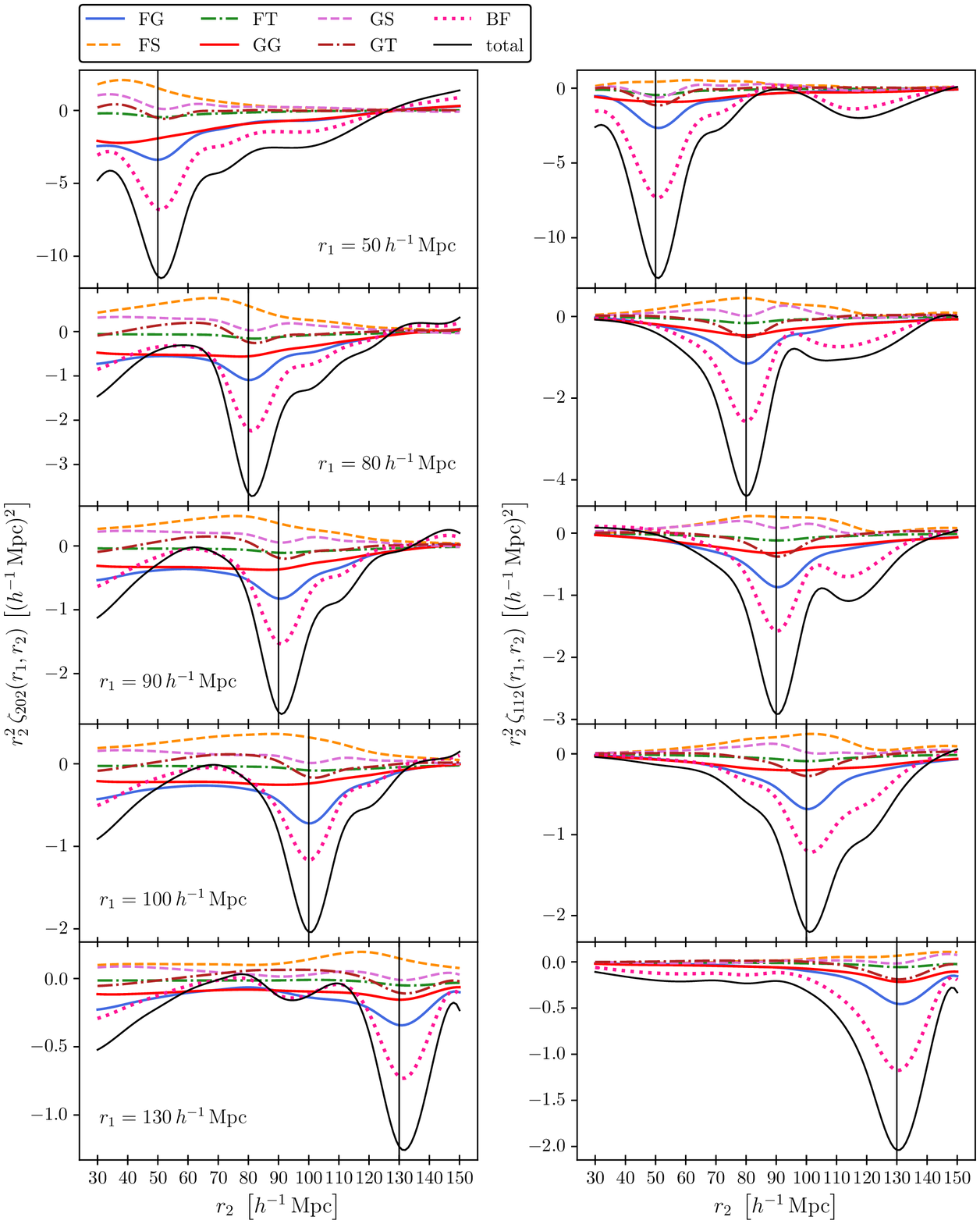}}
    \caption{Same as Figure~\ref{fig:decom_3PCF_mono}, except that the quadrupole 3PCFs, $\zeta_{202}$ and $\zeta_{112}$, are shown.}
	\label{fig:decom_3PCF_quad}
    \vspace{1cm}
\end{figure*}

Here we demonstrate how the growth, shift, and tidal terms of the second-order density and velocity fields affect the multipole components of the 3PCF. To do so, we consider the following seven bispectra:
\begin{eqnarray}
    B_{\rm FG}(\VEC{k}_1,\VEC{k}_2) \hspace{-0.25cm}&=&\hspace{-0.25cm} \sum_{p=1,4,7} X^{(p)}B^{(p)}(\VEC{k}_1,\VEC{k}_2), \nonumber \\
    B_{\rm FS}(\VEC{k}_1,\VEC{k}_2) \hspace{-0.25cm}&=&\hspace{-0.25cm} \sum_{p=2,5,8} X^{(p)}B^{(p)}(\VEC{k}_1,\VEC{k}_2), \nonumber \\
    B_{\rm FT}(\VEC{k}_1,\VEC{k}_2) \hspace{-0.25cm}&=&\hspace{-0.25cm} \sum_{p=3,6,9} X^{(p)}B^{(p)}(\VEC{k}_1,\VEC{k}_2), \nonumber \\
    B_{\rm GG}(\VEC{k}_1,\VEC{k}_2) \hspace{-0.25cm}&=&\hspace{-0.25cm} \sum_{p=10,13,16} X^{(p)}B^{(p)}(\VEC{k}_1,\VEC{k}_2), \nonumber \\
    B_{\rm GS}(\VEC{k}_1,\VEC{k}_2) \hspace{-0.25cm}&=&\hspace{-0.25cm} \sum_{p=11,14,17} X^{(p)}B^{(p)}(\VEC{k}_1,\VEC{k}_2), \nonumber \\
    B_{\rm GT}(\VEC{k}_1,\VEC{k}_2) \hspace{-0.25cm}&=&\hspace{-0.25cm} \sum_{p=12,15,18} X^{(p)}B^{(p)}(\VEC{k}_1,\VEC{k}_2), \nonumber \\
    B_{\rm BF}(\VEC{k}_1,\VEC{k}_2) \hspace{-0.25cm}&=&\hspace{-0.25cm} \sum_{p=19,20,21,22} X^{(p)}B^{(p)}(\VEC{k}_1,\VEC{k}_2).
    \label{Eq:Decom_B}
\end{eqnarray}
where $B_{\rm FG}$, $B_{\rm FS}$, $B_{\rm FT}$, $B_{\rm GG}$, $B_{\rm GS}$, and $B_{\rm GT}$ are proportional to $(F_{\rm g}\sigma_8)$, $(F_{\rm s}\sigma_8)$, $(F_{\rm t}\sigma_8)$, $(G_{\rm g}\sigma_8)$, $(G_{\rm s}\sigma_8)$, and $(G_{\rm t}\sigma_8)$, respectively, and $B_{\rm BF}$ depends only on $(b_1\sigma_8)$ and $(f\sigma_8)$. When computing the above seven bispectra, we assume the cosmological parameters in $\Lambda$CDM given in Section~\ref{Sec:Introduction}, the linear bias parameter $b_1=2$, no non-linear bias, i.e. $b_2=b_{s^2}=0$, and the redshift $z=0.61$. Next, we decompose the seven bispectra using TripoSHs according to Section~\ref{Sec:3PCFs} and compute the 3PCF multipoles via the 2D Hankel transform (\ref{Eq:B_to_zeta}). We plot the resulting 3PCF multipoles in Figures~\ref{fig:decom_3PCF_mono} and~\ref{fig:decom_3PCF_quad} as a function of $r_2$ after fixing $r_1$ to $50$, $80$, $90$, $100$, and $130\hMpc$.

As shown in~\citet{Sugiyama:2020uil}, in the monopole component (Figure~\ref{fig:decom_3PCF_mono}), the growth term (``FG'') is positive for scales smaller than $\sim130\hMpc$ and has a peak at $r_1=r_2$, while it goes from positive to negative and behaves like a trough for scales above $\sim130\hMpc$. On the other hand, the shift (``FS'') and tidal (``FT'') terms have troughs for any scale. Depending on the scale of interest, the shift term dominates for scales above $\sim30\hMpc$, and the total 3PCF (``total''), which is the sum of all components, is found to have a trough. To illustrate the trough-like behavior of the 3PCF at $r_1 = r_2$, we have drawn vertical black lines representing $r_1=r_2$ in Figures~\ref{fig:decom_3PCF_mono} and~\ref{fig:decom_3PCF_quad}. It can be seen that the bottom of the trough of the black curve representing the total 3PCF is always on the line $r_1=r_2$. Around $r_1\sim100\hMpc$, the BAO peak appears and has a wavy shape as it cancels out the trough due to non-linear gravity effects (e.g., see the middle panels). At $r_1=130\hMpc$ (the bottom panels), almost all the components have troughs, so the 3PCF has a more significant trough at $r_1=r_2$.

The quadrupole component (Figure~\ref{fig:decom_3PCF_quad}) of the 3PCF only shows an overall trough behaviour because the BAO signal is sufficiently non-linearly damped. The most dominant term in the quadrupole 3PCF is the ``BF'' term, which does not depend on any non-linear coefficients such as $F_{\rm g}$ or $G_{\rm g}$. This ``BF'' term consists of two effects: first, a term expressed as the product of a linear density field and a linear velocity field, and second, a term expressed as the square of the linear velocity field. In particular, the former can be interpreted as a new shift term resulting from the coordinate transformation from real to redshift space~(\ref{Eq:real_to_redshift}), and it dominates the ``BF'' term. Therefore, it behaves similarly to the shift term in the monopole 3PCF and explains most of the trough structure in the quadrupole 3PCF. The growth (``GG''), shift (``GS''), and tidal (``GT'') terms in the non-linear velocity field contribute to the quadrupole 3PCF comparably to those in the non-linear density field, and thus we can use the quadrupole 3PCF to determine the ``GG'', ``GS'', and ``GT'' terms. In contrast to the monopole case, the growth terms (``FG'' and ``GG'') are negative and behave as troughs, while the shift terms (``FS'' and ``GS'') are positive.


\subsection{Time-dependences of parameters}
\label{Sec:TimeDependencyOfParameters}

We review the discussion by~\citet{Yamauchi:2021arXiv210802382Y} on introducing new parameters to test DHOST theories and their time-dependences.

Note that some previous works predict that constraining $\sigma_8$ alone from the 3PCF can break the degeneracy between $f\sigma_8$ and $\sigma_8$, but this no longer happens in the framework of DHOST theories. To illustrate this fact in the context of our parameterisation, we can see from Eq.~(\ref{Eq:F_2G_2}) that the coefficient of the shift term in the second-order density fluctuation in $\Lambda$CDM ($F_{\rm s}=1$) determines $\sigma_8$ because both the growth and tidal terms are degenerate with the non-linear bias parameters~\citep{Schmittfull:2015PhRvD..91d3530S}. However, in the case of DHOST theories, there appears the parameter $\kappa_{\delta}$ in the coefficient of the shift term, which makes it impossible to measure $\sigma_8$ alone. Therefore, we introduce three new parameters that are not degenerate with $\sigma_8$ following~\citet{Yamauchi:2021arXiv210802382Y}: 
\begin{eqnarray}
    E_{f} \hspace{-0.25cm}&=&\hspace{-0.25cm}  \frac{f}{\kappa_{\delta}}= \frac{f\sigma_8}{F_{\rm s}\sigma_8} , \nonumber\\
    E_{\rm s} \hspace{-0.25cm}&=&\hspace{-0.25cm}  \frac{\kappa_{\theta}}{\kappa_{\delta}}= \frac{G_{\rm s}\sigma_8}{F_{\rm s}\sigma_8} , \nonumber\\
    E_{\rm t} \hspace{-0.25cm}&=&\hspace{-0.25cm} \frac{\lambda_{\theta}}{\kappa_{\delta}} = \frac{7}{4} \frac{G_{\rm t}\sigma_8}{F_{\rm s}\sigma_8}.
    \label{Eq:Est}
\end{eqnarray}
In GR or Horndeski theories, $E_f = f$, $E_{\rm s} = 1$ and $E_{\rm t}=\lambda_{\theta}$, because $\kappa_{\delta}=\kappa_{\theta}=1$. Horndeski theories differ from $\Lambda$CDM only in $f$ and $E_{\rm t}$ while keeping $E_{\rm s}=1$. If $E_{\rm s}\neq 1$, then the signal is specific to DHOST theories; $E_{\rm s}\neq1$ is a sufficient condition for detecting DHOST theories because there can be DHOST theories satisfying $E_{\rm s}=1$.

It has been known for a long time that the coefficient of the tidal term in the non-linear density field, $\lambda_{\delta}$, is time-dependent in GR~\citep[e.g.,][]{Bouchet:1992xz}, and in the case of $\Lambda$CDM, the following approximation holds well with an precision better than $0.6\%$~\citep{Bouchet:1995A&A...296..575B,Yamauchi:2017ibz}\footnote{The original derivation of the equation was calculated in the Lagrangian picture and is given in the form~\citep{Bouchet:1995A&A...296..575B}
\begin{eqnarray}
    \frac{2}{7}\lambda_{\delta}  = \frac{1}{2}\left[ 1 - \frac{3}{7} \Omega_{\rm m}^{-1/143} \right].
\end{eqnarray}
This equation can be rewritten to Eq.~(\ref{Eq:lambda_delta}) under the condition $(1-\Omega_{\rm m})\ll 1$.
}:
\begin{eqnarray}
    \lambda_{\delta} \sim \Omega_{\rm m}^{3/572}.
    \label{Eq:lambda_delta}
\end{eqnarray}
Through Eq.~(\ref{Eq:kappa_theta_and_lambda_theta}), the coefficient of the tidal term in the non-linear velocity field, $\lambda_{\theta}$, is also given by~\citep{Yamauchi:2021arXiv210802382Y}
\begin{eqnarray}
    \lambda_{\theta} \sim \Omega_{\rm m}^{15/1144}.
    \label{Eq:lambda_theta}
\end{eqnarray}

\citet{Yamauchi:2017ibz} extended the above discussion to Horndeski theories and showed that $\lambda_{\delta}$ is parameterised as a power of $\Omega_{\rm m}$ in Horndeski theories. In addition, \citet{Yamauchi:2021arXiv210802382Y} performed similar calculations for DHOST theories and showed that the coefficient of the shift term, $\kappa_{\delta}$, is also described by a power of $\Omega_{\rm m}$. The coefficients of the shift and tidal terms in the non-linear velocity field can be calculated through Eq.~(\ref{Eq:kappa_theta_and_lambda_theta}), and they also follow the powers of $\Omega_{\rm m}$. Therefore, we can parameterise the time-dependences of $E_{\rm f}$, $E_{\rm s}$, and $E_{\rm t}$ as follows:
\begin{eqnarray}
    E_f  \sim \Omega_{\rm m}^{\xi_f}, \quad
    E_{\rm s} \sim \Omega_{\rm m}^{\xi_{\rm s}}, \quad
    E_{\rm t} \sim \Omega_{\rm m}^{\xi_{\rm t}}.
    \label{Eq:TimeDependenceOfE}
\end{eqnarray}
In GR, we have
\begin{eqnarray}
    \xi_f = \frac{6}{11}, \quad
    \xi_{\rm s}= 0, \quad
    \xi_{\rm t}= \frac{15}{1144}.
    \label{Eq:Xi_GT}
\end{eqnarray}
In summary, we parameterise the second-order kernel function of the velocity field (\ref{Eq:F_2G_2}) as
\begin{eqnarray}
    f G_2^{\rm (g)}\sigma_8^2 
    = \Omega_{\rm m}^{\xi_{f}}\, (F_{\rm s}\sigma_8)^2 
    \Bigg[ \left( G_{\rm g}\right) + \Omega_{\rm m}^{\xi_{\rm s}}\, S + \frac{4}{7}\Omega_{\rm m}^{\xi_{\rm t}}\, T\Bigg],
\end{eqnarray}
where $(G_{\rm g}) = \Omega_{\rm m}^{\xi_{\rm s}} - (8/21) \Omega_{\rm m}^{\xi_{\rm t}}$, and the functions $S$ and $T$ are given in Eq.~(\ref{Eq:ST}). We will test the theory of gravity by measuring the above three parameters, $\xi_f$, $\xi_{\rm s}$, and $\xi_{\rm t}$, from the BOSS data in Section~\ref{Sec:DataAnalysis}.

    In DHOST theories, the Planck mass is time-varying, and the time variation of the Hubble parameter is different from GR. Therefore, one may be concerned that the time dependence of $\Omega_{\rm m}$ is different from $\Omega_{\rm m}^{\rm GR}$ that is calculated assuming GR. However, Appendix $C$ in \citet{Yamauchi:2021arXiv210802382Y} showed that the difference between DHOST theories and GR is suppressed by $(1-\Omega_{\rm m}^{\rm GR})$. Hence, we can replace $\Omega_{\rm m}$ in Eq.~(\ref{Eq:TimeDependenceOfE}) with $\Omega_{\rm m}^{\rm GR}$ as an approximation and perform the analysis to constrain $\xi_f$, $\xi_{\rm s}$, and $\xi_{\rm t}$. 

\subsection{Limitations of our theoretical approach to the 2PCF and 3PCF}
\label{Sec:LimitationsOf2PCFand3PCF}

In this subsection, we discuss the validity of the calculation methods of the 2PCF and 3PCF models described so far and the limitations of their application.

First, we can use the TripoSH decomposed 3PCF (\ref{Eq:B_to_zeta}) to constrain all the scale dependencies in the 3PCF, such as the shift and tidal terms, as shown Figures~\ref{fig:decom_3PCF_mono} and~\ref{fig:decom_3PCF_quad}, because it does not focus only on specific scale dependencies such as the squeezed limit. However, our analysis that uses only some multipoles of the TripoSH decomposition does not fully utilize the information on the scale dependence of the 3PCF. The reason for restricting the multipole components used in this work is to keep the number of data bins much smaller than the number of mock simulations used to compute the covariance matrix (Section~\ref{Sec:CovarianceMatrix}). Therefore, increasing the number of multipoles in the 3PCF to be considered will improve the results of this work when more mock catalogues are created in the future.

Second, note that the power spectrum and bispectrum models in Eqs.~(\ref{Eq:power_IR}) and (\ref{Eq:bispec_IR}) are valid for any theory in which the IR cancellation occurs based on the Galilean invariance of the system of equations in the IR limit: i.e., these models hold not only for $\Lambda$CDM but also for Horndeski theories~\citep{Crisostomi:2019vhj}. On the other hand, as \citet{Lewandowski:2019txi} pointed out in the power spectrum case, additional terms arise when performing the IR resummation in DHOST theories because of the violation of the IR cancellation. Specifically, when one applies the IR limit to the one-loop solution of the power spectrum in DHOST theories, a term proportional to $k^2 P_{\rm lin}(k)$ appears, changing the shape of the power spectrum~\citep{Crisostomi:2019vhj,Lewandowski:2019txi,Hirano:2020dom}. Note also that this additional term is degenerate with the higher-order derivative bias, which is ignored in this paper. However, since this additional term in the IR limit is proportional to $k^2 P_{\rm lin}(k)$, it is considered to be negligible at the large scales of interest in this paper ($\geq 80\hMpc$). Assuming that the same should happen in the bispectrum, we directly use the power and bispectrum models in Eqs.~(\ref{Eq:power_IR}) and (\ref{Eq:bispec_IR}) in the present analysis. In addition, it should be noted that \citet{Hirano:2020dom} showed that in DHOST theories, a term consisting of the product of first- and third-order fluctuations in the one-loop power spectrum causes UV divergence. Further model development is thus needed to take advantage of smaller-scale information by solving these problems.

Third, since the linear equation for density fluctuations is scale-independent (\ref{Eq:delta_evo}), we assume that we can use the shape of the linear matter power spectrum determined in the high-$z$ region, where the scalar field is expected to be sub-dominant. Thus, we can pre-compute the $\sigma_8^2$-normalized wiggle and no-wiggle power spectra, $P_{\rm w}^{(\rm n)}$ and $P_{\rm nw}^{(\rm n)}$, appearing in the $B^{(p)}$ terms (\ref{Eq:BP}), using a $\Lambda$CDM model. 

Fourth, there is a concern about the pre-computation of ${\cal D}(\VEC{k})$ (\ref{Eq:Damping}) appearing in the $B^{(p)}$ terms (\ref{Eq:BP}). It is known that $\sigma_{\perp}$ and $\sigma_{\parallel}$, which characterise ${\cal D}(\VEC{k})$, can be calculated successfully using linear displacement vectors~\citep[e.g.,][]{Matsubara:2007wj}, and we adopt the same calculation in this paper (\ref{Eq:sigma_sigma}). Since $\sigma_{\perp}$ and $\sigma_{\parallel}$ in the linear theory depend on $f$ and $\sigma_8$, their values should differ for different gravity theories. For this reason, it is desirable to vary $\sigma_{\perp}$ and $\sigma_{\parallel}$ as free parameters in the data analysis. However, to do so, the bispectrum decomposition method in Eq.~(\ref{Eq:Decomposed_B}) cannot be applied, and the computation time of the bispectrum model increases significantly, making it challenging to perform cosmological analysis. Fortunately, the BAO signal does not significantly impact the shape of the 3PCF. The reason is that the BAO signal is maximized when $r_1\sim r_2 \sim 100\hMpc$, while $r_1$ and $r_2$ can take various combinations in the 3PCF~\citep{Sugiyama:2020uil}. Therefore, in this paper, we ignore the concern about ${\cal D}(\VEC{k})$ and pre-compute $\sigma_{\perp}$ and $\sigma_{\parallel}$ using the linear theory in $\Lambda$CDM. Furthermore, to keep consistency with the 3PCF calculation, we fix $\sigma_{\perp}$ and $\sigma_{\parallel}$ to those calculated using the $\Lambda$CDM model in the 2PCF calculation as well.

Finally, to simplify the analysis, we ignore the AP effect~\citep{Alcock:1979mp}, which can directly measure the Hubble parameter and angular radial distance at the redshift of the galaxy distribution of interest. Ignoring the AP effect means that the values of the angular diameter distance and the Hubble parameter, which should be constrained by the AP effect, are given by the fiducial $\Lambda$CDM. In this sense, this paper assumes the fiducial $\Lambda$CDM for the expansion of the Universe in the background spacetime. However, the AP effect can be determined by the 2PCF by a few percent and is not expected to significantly affect the constraint results for the parameters that characterize the nonlinear fluctuations of interest in this paper, such as $\xi_{\rm t}$ and $\xi_{\rm s}$. Since DHOST theories vary these parameter values, the AP effect is expected to provide further information into the constraint on DHOST theories. \citet{Sugiyama:2020uil} have performed a joint analysis of the anisotropic 2PCF and 3PCF to constrain the AP parameters under the GR assumption. Combining that method with the analysis method developed in this paper allows for consistent DHOST theory constraints that simultaneously account for the AP and nonlinear gravity effects, which is left as future work.

\section{Measurements}
\label{Sec:Measurements}

This section summarises how to measure multipole 2PCFs and 3PCFs from BOSS galaxy data according to the method proposed by \citet{Sugiyama:2020uil}. First, Section~\ref{Sec:Data} introduces the BOSS galaxy data used in this paper and the mock simulation data designed to reproduce it. Then, Section~\ref{Sec:Estimator} describes the measurements of the multipole 2PCFs and 3PCFs. Finally, Section~\ref{Sec:WindowCorrections} explains how to correct for the window function effects on the measured 2PCF and 3PCF.

\subsection{Data}
\label{Sec:Data}

We use the final galaxy clustering data set, Data Release 12~\citep[DR12;][]{Alam:2015mbd}, from the Baryon Oscillation Spectroscopic Survey~\citep[BOSS;][]{Dawson:2012va}. The BOSS survey is part of the Sloan Digital Sky Survey III~\citep[SDSS III][]{Eisenstein:2011sa}, selected galaxies from multicolour SDSS imaging~\citep{Fukugita:1996qt,Gunn:1998vh,Smith:2002pca,Gunn:2006tw,Doi:2010rf} and used the SDSS multi-fibre spectrograph~\citep{Bolton:2012hz,Smee:2013AJ....146...32S} to measure spectroscopic redshifts of the galaxies. As detailed in~\citet{Reid:2015gra}, the BOSS survey has four samples, CMASS, LOWZ, LOWZ2 and LOWZ3, and those four samples are combined into one sample. In brief, the survey footprint, veto masks and survey-related systematics (such as fibre collisions and redshift failures) are considered to construct data and random catalogues for the DR12 BOSS galaxies. This DR12 combined sample comprises $1.2$ million massive galaxies over an effective area of $9329\, {\rm deg}^2$ and covers a redshift range of $0.2-0.75$. In our analysis, we split this redshift range into two redshift bins defined by $0.2 < z < 0.5$ and $0.5 < z < 0.75$ with the effective redshifts $z_{\rm eff}=0.38$ and $0.61$, respectively, where the effective redshifts are calculated as the weighted average over all galaxies~\citep[see e.g. Eq.~(67) in][]{Beutler:2014MNRAS.443.1065B}. The DR12 combined sample is observed across the two Galactic hemispheres, referred to as the Northern and Southern galactic caps (NGC and SGC, respectively), and the NGC and SGC samples probe slightly different galaxy populations in the low-redshift part of the combined sample~\citep[see Appendix~A in][]{Alam:2015mbd}.

To derive the covariance matrices of the 2PCF and 3PCF and test the validity of the 2PCF and 3PCF models given in Eqs.~(\ref{Eq:power_IR}) and (\ref{Eq:bispec_IR}), we use the MultiDark-Patchy mock catalogues~\citep[Patchy mocks;][]{Kitaura:2015uqa}. The Patchy mocks have been calibrated to an $N$-body simulation-based reference sample using approximate galaxy solvers and analytical-statistical biasing models and incorporate observational effects including the survey geometry, veto mask and fiber collisions. The reference catalogue is extracted from one of the BigMultiDark simulations~\citep{Klypin:2014kpa}, which was performed using \textsc{GADGET-2}~\citep{Springel:2005MNRAS.364.1105S} with $3840^3$ particles on a volume of $(2.5\hMpc)^3$. Halo abundance matching is used to reproduce the observed BOSS two and three-point clustering measurements~\citep{Rodriguez-Torres:2016MNRAS.460.1173R}. There are $2048$ catalogues available for each the NGC and SGC over the redshift range $z=0.2-0.75$. The fiducial cosmology for these mocks assumes a $\Lambda$CDM cosmology with $(\Omega_{\rm \Lambda}, \Omega_{\rm m}, \Omega_{\rm b}, \sigma_8, h)$ $=(0.692885, 0.307115, 0.048206, 0.8288, 0.6777)$. These fiducial parameters are slightly different from those used in our analysis of the BOSS galaxy data introduced in the introduction~(Section~\ref{Sec:Introduction}), but we expect that such differences do not significantly affect the covariance matrix estimations of the 2PCF and 3PCF.

We include three different incompleteness weights to account for shortcomings of the BOSS dataset: a fiber collision weight, $w_{\rm cp}$, a redshift failure weight, $w_{\rm noz}$, and a systematics weight, $w_{\rm sys}$, which is a combination of a stellar density weight and a seeing condition weight. Each galaxy observed at position $\VEC{x}$ is counted with the following weight~\citep{Ross:2012qm,Anderson:2013zyy,Reid:2015gra}:
\begin{eqnarray}
    w_{\rm c}(\VEC{x}) = w_{\rm sys}(\VEC{x})\left( w_{\rm cp}(\VEC{x}) + w_{\rm noz}(\VEC{x}) - 1\right).
    \label{Eq:w_c}
\end{eqnarray}
In addition,  we use a signal-to-noise weight, the so-called FKP weight, proposed by~\citet{Feldman:1994ApJ...426...23F}, $w_{\rm FKP}(\VEC{x})=1/[1 + \bar{n}_0(\VEC{x})\bar{P}]$, where $\bar{P}=10^4\, (h^{-1}\, {\rm Mpc})^3$. The FKP weight function is effective not only for the power spectrum but also for the bispectrum when assuming Gaussian errors~\citep{Scoccimarro:2000ApJ...544..597S}, and bispectrum measurements from the Patchy mock catalogue confirm that the FKP weight improves the bispectrum signal-to-noise ratio even when including non-Gaussian errors~\citep[see Appendix D in][]{Sugiyama:2018yzo}. We expect the validity of the FKP weight to hold for the 2PCF and 3PCF in configuration space because we measure the 2PCF and 3PCF as Fourier transforms of the power spectrum and bispectrum, respectively (Section~\ref{Sec:Estimator}). For the galaxy data, multiplying the completeness weights by the FKP weights yields the local weight function that is used in our analysis, while the random catalogues have only the FKP weights:
\begin{eqnarray}
    w^{(\rm gal)}(\VEC{x}) \hspace{-0.25cm} &=&\hspace{-0.25cm}  w_{\rm c}(\VEC{x})\, w_{\rm FKP}(\VEC{x}), \nonumber \\
    w^{(\rm ran)}(\VEC{x}) \hspace{-0.25cm} &=&\hspace{-0.25cm}  w_{\rm FKP}(\VEC{x}),
\end{eqnarray}
where the superscripts, ``(gal)'' and ``(ran)'', stand for ``galaxy'' and ``random''.

\subsection{Estimators of the 2PCF and 3PCF}
\label{Sec:Estimator}

We measure the number densities of both real and random galaxies weighted by the spherical harmonic function $Y_{\ell m}$:
\begin{eqnarray}
    \hspace{-0.5cm} D_{\ell m}(\VEC{x}) \hspace{-0.25cm}&=&\hspace{-0.25cm} \sum_i^{N_{\rm gal}} w^{(\rm gal)}(\VEC{x}_i)Y_{\ell m}^{*}\left(\hat{x}^{(\rm gal)}_i  \right) \delta_{\rm D}\left( \VEC{x} - \VEC{x}^{(\rm gal)}_i \right), \nonumber \\
    \hspace{-0.5cm}R_{\ell m}(\VEC{x}) \hspace{-0.25cm}&=&\hspace{-0.25cm} \sum_j^{N_{\rm ran}} w^{(\rm ran)}(\VEC{x}_j)Y_{\ell m}^{*}\left(\hat{x}^{(\rm ran)}_j  \right) \delta_{\rm D}\left( \VEC{x} - \VEC{x}^{(\rm ran)}_j \right),
\end{eqnarray}
where $N_{\rm gal}$ and $N_{\rm ran}$ are the total number of real and random galaxies, respectively, and the normal number densities are given by $D(\VEC{x})=\sqrt{4\pi}D_{00}(\VEC{x})$ and $R(\VEC{x})=\sqrt{4\pi}R_{00}(\VEC{x})$. Defining $N'_{\rm gal}\equiv \int d^3x D(\VEC{x})$ and $N'_{\rm ran}\equiv \int d^3x R(\VEC{x})$, we can estimate the survey volume as
\begin{eqnarray}
    V = \frac{N_{\rm ran}^{'2}}{\int d^3x\, [R(\VEC{x})]^2}.
    \label{Eq:volume}
\end{eqnarray}
Then, the observed density fluctuation weighted by $Y_{\ell m}$ is 
\begin{eqnarray}
    \delta_{\rm obs, \ell m}(\VEC{x}) = V\left[ D_{\ell m}(\VEC{x})/N'_{\rm gal} - R_{\ell m}(\VEC{x})/N'_{\rm ran} \right],
    \label{Eq:delta_obs_lm}
\end{eqnarray}
and
\begin{eqnarray}
    \delta_{\rm obs}(\VEC{x}) = \sqrt{4\pi} \delta_{\rm obs,\, 00}(\VEC{x}).
\end{eqnarray}
We use the fast Fourier transform (FFT) algorithm~\footnote{\url{http://fftw.org/}} to calculate
\begin{eqnarray}
    \widetilde{\delta}_{\rm obs, \ell m}(\VEC{k}) = \frac{1}{W_{\rm mass}(\VEC{k})}\int d^3x e^{-i\VEC{k}\cdot\VEC{x}} \delta_{\rm obs, \ell m}(\VEC{x}),
    \label{Eq:FFT}
\end{eqnarray}
where the Fourier transform of the normal density fluctuation is given by $\widetilde{\delta}_{\rm obs}(\VEC{k}) = \sqrt{4\pi}\,\widetilde{\delta}_{\rm obs,00}(\VEC{k})$, and $W_{\rm mass}(\VEC{k})$ is the mass assignment function that corrects for the effect when arising assign particles on a regular grid in position space~\citep{Jing:2005ApJ...620..559J}. The most popular mass assignment function is given by~\citep{Hockney:1981csup.book.....H}
\begin{eqnarray}
    W_{\rm mass}(\VEC{k}) = \prod_{i=x,y,z} \left[ {\rm sinc}\left( \frac{\pi k_i}{2k_{\rm N, i}} \right) \right]^p,
\end{eqnarray}
where $k_{\rm N, i}$ is the Nyquist frequency of $i$-axis with the grid spacing $H_i$ on the axis. The indexes $p = 1$, $p = 2$, and $p = 3$ correspond to the nearest grid point (NGP), cloud-in-cell (CIC), and triangular-shaped cloud (TSC) assignment functions, respectively.

The FFT-based estimator of the multipole 2PCFs is given by~\citep{Hand:2017JCAP...07..002H,Sugiyama:2018MNRAS.473.2737S}~\citep[see also][]{Bianchi:2015MNRAS.453L..11B,Scoccimarro:2015PhRvD..92h3532S}
\begin{eqnarray}
    \widehat{\xi}_{\ell}(r) \hspace{-0.25cm}&=&\hspace{-0.25cm}
    \frac{(4\pi)}{V}\sum_m\int \frac{d^2\hat{r}}{4\pi} Y_{\ell m}(\hat{r})\int \frac{d^3k}{(2\pi)^3} e^{i\VEC{k}\cdot\VEC{r}}\nonumber \\
    \hspace{-0.25cm}&\times&\hspace{-0.25cm}
    \bigg[  \widetilde{\delta}_{\rm obs, \ell m}(\VEC{k})\widetilde{\delta}^*_{\rm obs}(\VEC{k}) - S_{\ell m}(\VEC{k}) \bigg].
    \label{Eq:2PCF_estimator}
\end{eqnarray}
The shot-noise term $S_{\ell m}(\VEC{k})$ is given by
\begin{eqnarray}
    \hspace{-0.25cm}  S_{\ell m}(\VEC{k}) \hspace{-0.25cm}&=&\hspace{-0.25cm} \frac{C_{\rm shot}(\VEC{k})}{W_{\rm mass}^2(\VEC{k})}  \left(\frac{V}{N'_{\rm gal}}  \right)^2 \nonumber \\
    \hspace{-0.25cm}&\times& \hspace{-0.25cm}
    \Bigg[ \sum_i^{N_{\rm gal}}\left[ w^{(\rm gal)}(\VEC{x}_i) \right]^2 Y_{\ell m}^{*}\left(\hat{x}^{(\rm gal)}_i  \right) \nonumber \\
    \hspace{-0.25cm}\hspace{-0.25cm}&+&\hspace{-0.25cm}\left(\frac{N'_{\rm gal}}{N'_{\rm ran}}  \right)^2
    \sum_j^{N_{\rm ran}}\left[ w^{(\rm ran)}(\VEC{x}_j) \right]^2Y_{\ell m}^{*}\left(  \hat{x}^{(\rm ran)}_j\right) \Bigg].
    \label{Eq:2PCF_shotnoise}
\end{eqnarray}
where $C_{\rm shot}(\VEC{k})$ represents the correction for the assignment effect to the shot-noise term, given by~\cite[Eq.~(20) in][]{Jing:2005ApJ...620..559J}
\begin{eqnarray}
    \hspace{-1.2cm}&& C_{\rm shot}(\VEC{k})   \nonumber \\
    \hspace{-1.2cm}&=& \hspace{-0.25cm}
    \begin{cases} 1, \hspace{5.35cm} \mbox{NGP};  \\ 
        \prod_i\left[ 1 - \frac{2}{3}\sin^2\left( \frac{\pi k_i}{2 k_{\rm N, i}} \right) \right], \hspace{2.2cm} \mbox{CIC}; \\ 
        \prod_i\left[ 1 - \sin^2\left( \frac{\pi k_i}{2 k_{\rm N, i}} \right) + \frac{2}{15}\sin^4\left( \frac{\pi k_i}{2 k_{\rm N, i}} \right)\right], \mbox{TSC}.
    \end{cases}
\end{eqnarray}
The angle integral $\int d^2\hat{r}/(4\pi)$ in Eq.~(\ref{Eq:2PCF_estimator}) can be rewritten as 
\begin{eqnarray}
    \int \frac{d^2\hat{r}}{4\pi} = \frac{1}{N_{r}(r)} \sum_{r- \Delta r /2 < r < r + \Delta r / 2},
\end{eqnarray}
where $\Delta r$ is the width of the $r$-bins, and $N_r(r)$ is the number of three-dimensional data contianed in each $r$-bin width. From the expression of the shot noise term in the 2PCF given in Eq.~(\ref{Eq:2PCF_shotnoise}), we compute the weighted mean number density as
\begin{eqnarray}
    \bar{n} = \left\{  \frac{V}{N^{'2}_{\rm gal}}\sum_i^{N_{\rm gal}}
    \left[ w^{(\rm gal)}(\VEC{x}_i) \right]^2 \right\}^{-1}.
    \label{Eq:mean_n}
\end{eqnarray}

The FFT-based estimator of the multipole 3PCFs is given by~\citep{Sugiyama:2018yzo}~\citep[see also][]{Scoccimarro:2015PhRvD..92h3532S,Slepian:2016MNRAS.455L..31S}
\begin{eqnarray}
    \hspace{-0.4cm}\widehat{\zeta}_{\ell_1\ell_2\ell}(r_1,r_2)
	\hspace{-0.25cm}&=&\hspace{-0.25cm} 
    \frac{(4\pi)^2 h_{\ell_1\ell_2\ell}}{V} \sum_{m_1m_2m}	\left( \begin{smallmatrix} \ell_1 & \ell_2 & \ell \\ m_1 & m_2 & m \end{smallmatrix}  \right)  \nonumber \\
	\hspace{-0.25cm} &\times&\hspace{-0.25cm}
    \Bigg[
    \int d^3x F_{\ell_1 m_1}(\VEC{x};r_1) F_{\ell_2 m_2}(\VEC{x};r_2) G_{\ell m}(\VEC{x}) \nonumber \\
&& - \delta_{r_1r_2}^{\rm (K)}S_{\ell_1m_1;\ell_2m_2;\ell m}(r_1) \Bigg],
	\label{Eq:3PCF_estimator}
\end{eqnarray}
where
\begin{eqnarray}
	F_{\ell m}(\VEC{x};r)
    \hspace{-0.25cm}&=& \hspace{-0.25cm}
	i^{\ell}\int \frac{d^3k}{(2\pi)^3} 
    e^{i\VEC{k}\cdot\VEC{x}} j_{\ell}(rk) Y_{\ell m}^{*}(\hat{k}) \widetilde{\delta}_{ {\rm obs}}(\VEC{k}), \nonumber \\
    G_{\ell m}(\VEC{x}) 
    \hspace{-0.25cm}&=& \hspace{-0.25cm}
    \int \frac{d^3k}{(2\pi)^3} \,
    e^{i\VEC{k}\cdot\VEC{x}} \, \widetilde{\delta}_{ {\rm obs}, \ell m}(\VEC{k}).
\end{eqnarray}
Note that the shot-noise term only contributes to the 3PCF measurement for the $r_1=r_2$ bins, represented by the Kronecker delta $\delta^{(\rm K)}_{r_1r_2}$ in Eq.~(\ref{Eq:3PCF_estimator}). To specifically calculate the shot-noise term in the 3PCF, we first measure the following density field
\begin{eqnarray}
    \hspace{-0.5cm}N(\VEC{x}) 
    \hspace{-0.25cm}&=&\hspace{-0.25cm}     \sum_i^{N_{\rm gal}} \left[ w_i^{\rm (gal)}(\VEC{x}_i) \right]^2 \delta_{\rm D}\left( \VEC{x} - \VEC{x}_i^{\rm (gal)} \right) \nonumber \\
    \hspace{-0.25cm}&+&\hspace{-0.25cm} \left( \frac{N'_{\rm gal}}{N'_{\rm ran}} \right)^2 \sum_i^{N_{\rm ran}} \left[ w_i^{\rm (ran)}(\VEC{x}_i) \right]^2 \delta_{\rm D}\left( \VEC{x} - \VEC{x}_i^{\rm (ran)} \right), 
\end{eqnarray}
and divide it by $(N'_{\rm gal}/V)$ to have 
\begin{eqnarray}
    \delta_{\rm N}(\VEC{x})=(V/N'_{\rm gal})\, N(\VEC{x}).
\end{eqnarray}
Then, we calculate the Fourier transform of $\delta_{\rm N}(\VEC{x})$ in the same manner as in Eq.~(\ref{Eq:FFT}) and denote it as $\widetilde{\delta}_{\rm N}(\VEC{k})$. Finally, we derive $S_{\ell_1m_1;\ell_2m_2;\ell m}(r)$ by substituting $\widetilde{\delta}_{\rm N}(\VEC{k})$ into the following equation
\begin{eqnarray}
    S_{\ell_1m_1;\ell_2m_2;\ell m}(r)
    \hspace{-0.25cm}&=& \hspace{-0.25cm}
    \left( \frac{1}{4\pi r^2 \Delta r} \right)
    \left( \frac{V}{N'_{\rm gal}} \right) (-1)^{\ell_1+\ell_2} \nonumber \\
    \hspace{-0.25cm}&\times&\hspace{-0.25cm} \int \frac{d^2\hat{r}}{4\pi} Y^*_{\ell_1m_1}(\hat{r}) Y^*_{\ell_2m_2}(\hat{r})
    \int \frac{d^3k}{(2\pi)^3}e^{i\VEC{k}\cdot\VEC{r}} \nonumber \\
    \hspace{-0.25cm}&\times&\hspace{-0.25cm} \left[   \widetilde{\delta}_{\ell m}(\VEC{k})\widetilde{\delta}_{\rm N}^*(\VEC{k}) - S^{(\rm 3PCF)}_{\ell m}(\VEC{k})\right],
\end{eqnarray}
where
\begin{eqnarray}
    \hspace{-0.45cm}  S^{\rm (3PCF)}_{\ell m}(\VEC{k}) \hspace{-0.25cm}&=&\hspace{-0.25cm} \frac{C_{\rm shot}(\VEC{k})}{W_{\rm mass}^2(\VEC{k})} \left( \frac{V}{N'_{\rm gal}}  \right)^2 \nonumber \\
    \hspace{-0.25cm}&\times& \hspace{-0.25cm}
    \Bigg[ \sum_i^{N_{\rm gal}}\left[ w^{(\rm gal)}(\VEC{x}_i) \right]^3 Y_{\ell m}^{*}\left(\hat{x}^{(\rm gal)}_i  \right) \nonumber \\
    \hspace{-0.25cm}\hspace{-0.25cm}&-&\hspace{-0.25cm}\left(  \frac{N'_{\rm gal}}{N'_{\rm ran}}\right)^3 \sum_j^{N_{\rm ran}}\left[ w^{(\rm ran)}(\VEC{x}_j) \right]^3Y_{\ell m}^{*}\left(  \hat{x}^{(\rm ran)}_j\right) \Bigg].
\end{eqnarray}
The factor $(1/(4\pi r^2 \Delta r))$ can be rewritten as
\begin{eqnarray}
    \frac{1}{4\pi r^2 \Delta r} = \frac{1}{N_r(r)}\frac{N_{\rm grid}}{V_{\rm FFT}},
\end{eqnarray}
where $V_{\rm FFT}$ is the volume of the Cartesian box in which the galaxies are placed before the FFT is performed, and $N_{\rm grid}$ is the number of FFT grid cells.

In the scale range of $80\leq r \leq 150\hMpc$, we choose $\Delta r = 5\hMpc$ for the 2PCF and $\Delta r = 10\hMpc$ for the 3PCF. Considering $\zeta_{\ell_1\ell_2\ell}(r_1,r_2)=\zeta_{\ell_2\ell_1\ell}(r_2,r_1)$, the numbers of data bins for the 2PCF and 3PCF multipoles are $15$, $15$, $36$, $36$, $64$, and $36$ for $\xi_0$, $\xi_2$, $\zeta_{000}$, $\zeta_{110}$, $\zeta_{202}$, and $\zeta_{112}$, respectively. 

We use the Cartesian coordinates $\VEC{x}=\{x,y,z\}$ with the $z$-axis pointing to the north pole to define a cuboid of dimension $\VEC{L}[\hMpc]=(L_x, L_y, L_z)$ containing the galaxy sample; to perform the FFT, each axis of this cuboid is delimited into $\VEC{N}=(N_x,N_y,N_z)$ grids. We then distribute the galaxies on the FFT grid using the TSC assignment function. We adopt the same values for $\VEC{L}$ and $\VEC{N}$ that were used by the Fourier space analysis of the two-point statistics performed by~\citet{Beutler:2016arn}. They are chosen so that the width of each grid is $\sim5\hMpc$, which is well below the scales $r\geq 80\hMpc$ that we are interested in. We summarise the specific values of $\VEC{L}$ and $\VEC{N}$, as well as the survey volume (\ref{Eq:volume}) and the weighted mean number density (\ref{Eq:mean_n}) computed using these values of $\VEC{L}$ and $\VEC{N}$ in Table~\ref{Table:Grid}.

\begin{table*}

\centering
\begin{tabular}{lcccc}
\hline\hline
& $(L_x,L_y,L_z)\, [\hMpc]$ & $(N_x,N_y,N_z)$ & $V\, [(\hGpc)^3]$ & $\bar{n}/10^{-4}\, [(\hMpc)^{-3}]$\\
\hline
    NGC at $z_{\rm eff}=0.38$ ($0.2<z<0.5$)  & (1350, 2450, 1400) & (250, 460, 260) & 1.51 & 2.65 \\
    SGC at $z_{\rm eff}=0.38$ ($0.2<z<0.5$)  & (1000, 1900, 1100) & (190, 360, 210) & 0.56 & 2.88 \\
\hline
    NGC at $z_{\rm eff}=0.61$ ($0.5<z<0.75$) & (1800, 3400, 1900) & (340, 650, 360) & 2.35 & 1.37 \\
    SGC at $z_{\rm eff}=0.61$ ($0.5<z<0.75$) & (1000, 2600, 1500) & (190, 500, 280) & 0.87 & 1.29 \\
\hline  
\end{tabular}
\caption{
    The length of each side of the cube containing the observed galaxies, defined for performing FFTs, and the number of grids on which the cube is delimited are shown for the four BOSS samples (Section~\ref{Sec:Data}). Also shown are the survey volume (\ref{Eq:volume}) and the mean galaxy number density (\ref{Eq:mean_n}), calculated using the values of these parameters.
}
\label{Table:Grid}
\end{table*}
 
\subsection{Window function corrections}
\label{Sec:WindowCorrections}

When measuring 2PCFs and 3PCFs in configuration space from galaxy data, if we directly measure their angle-averaged multipole components, we can not eliminate the effect of the window function~\citep[Appendix A in ][]{Sugiyama:2020uil}. The FFT-based estimators introduced in Section~(\ref{Sec:Estimator}) are a typical example of this, but even when measuring multipole 3PCFs without using the FFT, we need to be aware of the window function effect~\citep{Slepian:2015MNRAS.454.4142S,Slepian:2018MNRAS.478.1468S}. Since the window function $W(\VEC{x})$ characterising the geometry of the observed region can be estimated as $W(\VEC{x})=(V/N'_{\rm ran})\,R(\VEC{x})$, we can quantitatively estimate the corrections due to the window function by measuring the multipole 2PCFs and 3PCFs from the random catalogue.

For the 2PCF, we compute 
\begin{eqnarray}
    Q_{\ell}(r) \hspace{-0.25cm}&=&\hspace{-0.25cm}
    \frac{(4\pi)}{V}\sum_m\int \frac{d^2\hat{r}}{4\pi} Y_{\ell m}(\hat{r})\int \frac{d^3k}{(2\pi)^3} e^{i\VEC{k}\cdot\VEC{r}}\nonumber \\
    \hspace{-0.25cm}&\times&\hspace{-0.25cm}
    \bigg[  \widetilde{W}_{\ell m}(\VEC{k})\widetilde{W}^*(\VEC{k}) - S^{\rm (w)}_{\ell m}(\VEC{k}) \bigg].
    \label{Eq:2PCF_window}
\end{eqnarray}
where $\widetilde{W}_{\ell m}(\VEC{k})$ is the Fourier transform of $(V/N'_{\rm ran})R_{\ell m}(\VEC{x})$ computed in the same manner as in Eq.~(\ref{Eq:FFT}), $\widetilde{W}(\VEC{k})=\sqrt{4\pi}\,\widetilde{W}_{00}(\VEC{k})$, and the shot-noise term is given by
\begin{eqnarray}
    \hspace{-0.35cm}  S^{\rm (w)}_{\ell m}(\VEC{k}) \hspace{-0.25cm}&=&\hspace{-0.25cm} \frac{C_{\rm shot}(\VEC{k})}{W_{\rm mass}^2(\VEC{k})}  \left( \frac{V}{N'_{\rm ran}} \right)^2 \nonumber \\
     \hspace{-0.25cm}&\times&  \hspace{-0.25cm}
    \Bigg[ \sum_i^{N_{\rm ran}}\left[ w^{(\rm ran)}(\VEC{x}_i) \right]^2 Y_{\ell m}^{*}\left(\hat{x}^{(\rm ran)}_i  \right) \Bigg].
\end{eqnarray}
Then, we have the theoretical model of $\xi_{\ell}(r)$ taking the survey window effect into account as follows~\citep{Wilson:2017MNRAS.464.3121W,Beutler:2016arn}:
\begin{eqnarray}
    \xi^{(\rm w)}_{\ell}(r) = (2\ell+1)\sum_{\ell_1\ell_2} 
     \left( \begin{smallmatrix} \ell_1 & \ell_2 & \ell \\ 0 & 0 & 0 \end{smallmatrix}  \right)^2 Q_{\ell_1}(r)\, \xi_{\ell_2}(r).
     \label{Eq:xi_window}
\end{eqnarray}

For the 3PCF, we compute 
\begin{eqnarray}
    \hspace{-0.4cm}  Q_{\ell_1\ell_2\ell}(r_1,r_2)
	\hspace{-0.25cm}&=&\hspace{-0.25cm} 
    \frac{(4\pi)^2 h_{\ell_1\ell_2\ell}}{V} \sum_{m_1m_2m}	\left( \begin{smallmatrix} \ell_1 & \ell_2 & \ell \\ m_1 & m_2 & m \end{smallmatrix}  \right)  \nonumber \\
	\hspace{-0.25cm} &\times&\hspace{-0.25cm}
    \Bigg[
        \int d^3x F_{\ell_1m_1}^{({\rm w})}(\VEC{x};r_1) F_{\ell_2m_2}^{({\rm w})}(\VEC{x};r_2) G_{\ell m}^{({\rm w})}(\VEC{x}) \nonumber \\
&& - \delta_{r_1r_2}^{\rm (K)}S^{({\rm w})}_{\ell_1m_1;\ell_2m_2;\ell m}(r_1) \Bigg],
	\label{Eq:3PCF_window}
\end{eqnarray}
where
\begin{eqnarray}
    F_{\ell m}^{({\rm w})}(\VEC{x};r)
    \hspace{-0.25cm}&=& \hspace{-0.25cm}
	i^{\ell}\int \frac{d^3k}{(2\pi)^3} 
    e^{i\VEC{k}\cdot\VEC{x}} j_{\ell}(rk) Y_{\ell m}^{*}(\hat{k}) \widetilde{W}(\VEC{k}), \nonumber \\
    G_{\ell m}^{({\rm w})}(\VEC{x}) 
    \hspace{-0.25cm}&=& \hspace{-0.25cm}
    \int \frac{d^3k}{(2\pi)^3} \,
    e^{i\VEC{k}\cdot\VEC{x}} \, \widetilde{W}_{\ell m}(\VEC{k}).
\end{eqnarray}
The shot-noise term is given by
\begin{eqnarray}
    S^{{\rm (w)}}_{\ell_1m_1;\ell_2m_2;\ell m}(r)
    \hspace{-0.25cm}&=& \hspace{-0.25cm}
    \left( \frac{1}{4\pi r^2 \Delta r} \right) \left( \frac{V}{N'_{\rm ran}} \right)
     (-1)^{\ell_1+\ell_2} \nonumber \\
    \hspace{-0.25cm}&\times&\hspace{-0.25cm} \int \frac{d^2\hat{r}}{4\pi} Y^*_{\ell_1m_1}(\hat{r}) Y^*_{\ell_2m_2}(\hat{r})
    \int \frac{d^3k}{(2\pi)^3}e^{i\VEC{k}\cdot\VEC{r}} \nonumber \\
    \hspace{-0.25cm}&\times&\hspace{-0.25cm} \left[   \widetilde{W}_{\ell m}(\VEC{k})\widetilde{\delta}_{\rm N}^{{(\rm w)*}}(\VEC{k}) - S^{(\rm 3PCF,w)}_{\ell m}(\VEC{k})\right],
\end{eqnarray}
where
\begin{eqnarray}
    \hspace{-0.45cm}  S^{\rm (3PCF,w)}_{\ell m}(\VEC{k}) \hspace{-0.25cm}&=&\hspace{-0.25cm} \frac{C_{\rm shot}(\VEC{k})}{W_{\rm mass}^2(\VEC{k})}\left( \frac{V}{N'_{\rm ran}} \right)^2  \nonumber \\
    \hspace{-0.25cm}&\times& \hspace{-0.25cm}
    \Bigg[ \sum_i^{N_{\rm ran}}\left[ w^{(\rm ran)}(\VEC{x}_i) \right]^3 Y_{\ell m}^{*}\left(\hat{x}^{(\rm ran)}_i  \right) \Bigg],
\end{eqnarray}
and $\widetilde{\delta}_{\rm N}^{\rm (w)}(\VEC{k})$ is the Fourier transform of 
\begin{eqnarray}
    \delta_{\rm N}^{(\rm w)}(\VEC{x}) = \left( \frac{V}{N'_{\rm ran}} \right)\sum_i^{N_{\rm ran}} \left[ w_i^{\rm (ran)} \right]^2 \delta_{\rm D}\left( \VEC{x} - \VEC{x}_i^{(\rm ran)} \right).
\end{eqnarray}
Then, we have the theoretical model of $\zeta_{\ell_1\ell_2\ell}(r_1,r_2)$ taking the survey window effect into account as follows~\citep{Sugiyama:2018yzo,Sugiyama:2020uil}:
\begin{eqnarray}
     \zeta^{(\rm w)}_{\ell_1\ell_2\ell}(r_1,r_2)
	\hspace{-0.25cm}&=&\hspace{-0.25cm} 
    (4\pi) \sum_{\ell'_1+\ell'_2+\ell'={\rm even}}\ \ \sum_{\ell''_1+\ell''_2+\ell''={\rm even} } \nonumber \\
	\hspace{-0.25cm}&\times&\hspace{-0.25cm} 
	\left\{ \begin{smallmatrix} \ell''_1 & \ell''_2 & \ell'' \\   \ell'_1 & \ell'_2 & \ell' \\   \ell_1 & \ell_2 & \ell \end{smallmatrix}  \right\}
	\left[\frac{h_{\ell_1\ell_2\ell}h_{\ell_1\ell'_1\ell''_1}h_{\ell_2\ell'_2\ell''_2}h_{\ell \ell'\ell''}}{h_{\ell'_1\ell'_2\ell'}h_{\ell''_1\ell''_2\ell''}} \right]
	\nonumber \\
	\hspace{-0.25cm}&\times&\hspace{-0.25cm} 
			Q_{\ell''_1\ell''_2\ell''}(r_1,r_2)\, \zeta_{\ell'_1\ell'_2\ell'}(r_1,r_2),
	\label{Eq:zeta_window}
\end{eqnarray}
where the bracket with $9$ multipole indices, $\{\dots\}$, denotes the Wigner-$9$j symbol. In the likelihood fitting performed in Section~\ref{Sec:DataAnalysis}, we use $\xi_{\ell}^{(\rm w)}$ and $\zeta_{\ell_1\ell_2\ell}^{(\rm w)}$ to compare the measured multipole 2PCF and 3PCF estimators with the theoretical models given in Eqs. (\ref{Eq:power_IR}) and (\ref{Eq:bispec_IR}). In this paper, we ignore the contribution from the integral constraint~\citep{Peacock1991} for both the 2PCF and the 3PCF.

In the 2PCF case, the correction equation for the window function effect shown in Eq.~(\ref{Eq:xi_window}) calculates only the three multipole components for both $Q_{\ell_1}$ and $\xi_{\ell_2}$, i.e., $\ell_1,\ell_2=0,2,4$. The reason is that our analysis focuses only on large scales above $80\hMpc$, where the linear theory is dominant, and the linear Kaiser effect gives only up to the hexadecapole $\ell=4$. For the window correction formula of the 3PCF~(\ref{Eq:zeta_window}), \citet{Sugiyama:2020uil} examined in detail which multipole components contribute to the observed estimator~(\ref{Eq:3PCF_estimator}) and to what extent, for the NGC sample at $0.4<z<0.6$, and showed that a finite number of multipole components can correct for the window effect on the 3PCF with sufficiently good accuracy. Assuming that this result is not significantly different for the other BOSS samples, we calculate a total of $14$ multipole components for both $Q_{\ell_1''\ell_2''\ell''}$ and $\zeta_{\ell_1'\ell_2'\ell'}$ as follows: $(\ell_1,\ell_2,\ell)=(0,0,0)$, $(1,1,0)$, $(2,2,0)$, $(3,3,0)$ and $(4,4,0)$ for the monopole 3PCF ($\ell=0$), and $(\ell_1,\ell_2,\ell)=(0,2,2)$, $(1,1,2)$, $(2,0,2)$, $(1,3,2)$, $(2,2,2)$, $(3,1,2)$, $(2,4,2)$, $(3,3,2)$ and $(4,2,2)$ for the quadrupole 3PCF ($\ell=2$).

\begin{figure*}
    \scalebox{0.95}{\includegraphics[width=\textwidth]{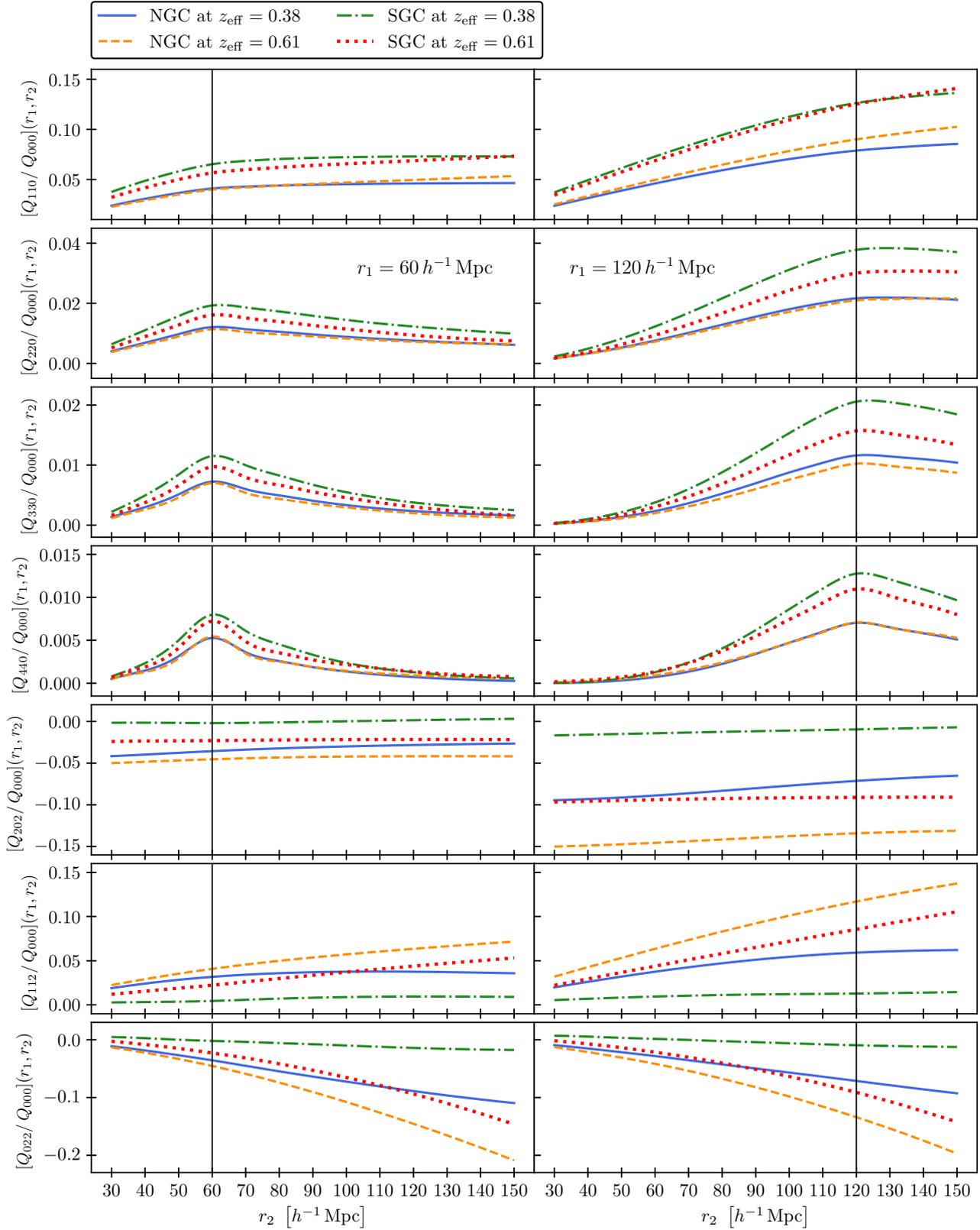}}
    \caption{The monopole and quadrupole components of the window 3PCF (\ref{Eq:3PCF_window}), $Q_{000}$, $Q_{110}$, $Q_{220}$, $Q_{330}$, $Q_{440}$, $Q_{202}$, $Q_{112}$, and $Q_{022}$, measured from the four BOSS samples are shown as a function of $r_2$ after fixing $r_1$ to $60\hMpc$ (left) and $120\hMpc$ (right).}
	\label{fig:window1}
\end{figure*}

\begin{figure*}
	\includegraphics[width=\textwidth]{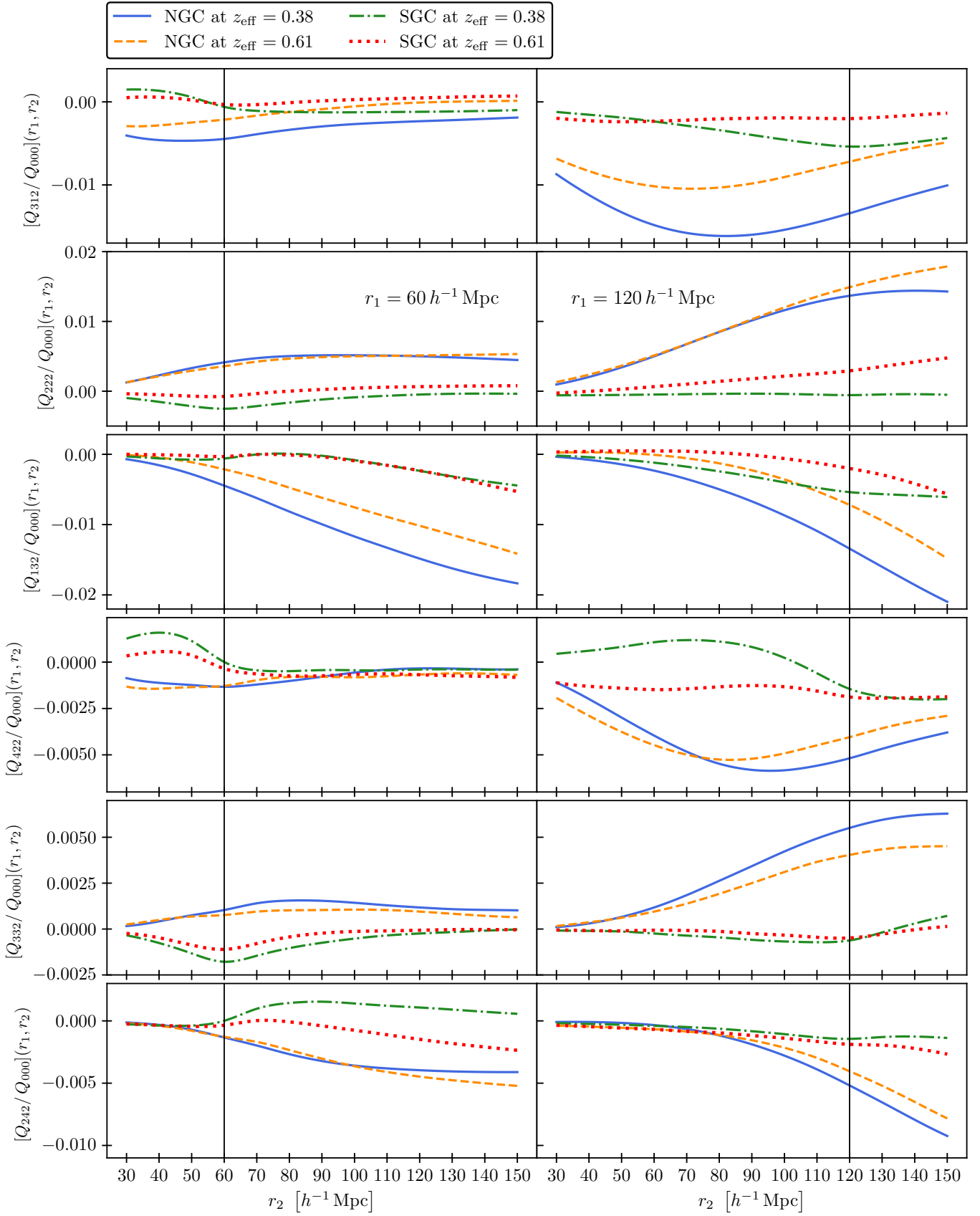}
    \caption{Same as Figure~\ref{fig:window1}, except that the higher-order quadrupole components of the window 3PCF, $Q_{312}$, $Q_{222}$, $Q_{132}$, $Q_{422}$, $Q_{332}$, and $Q_{242}$, are shown.}
    \label{fig:window2}
\end{figure*}

Figures~\ref{fig:window1} and~\ref{fig:window2} plot the $13$ window 3PCF multipoles normalized by $Q_{000}$ as a function of $r_2$ after fixing $r_1$ to $60$ and $120\hMpc$. For the monopole components ($Q_{110}$, $Q_{220}$, $Q_{330}$, and $Q_{440}$), we find that the window 3PCF multipoles measured at different redshift bins 
in each sky region (NGC or SGC) behave similarly (see, for example, the solid blue and dashed orange lines). On the other hand, for the quadrupole component, we see that the four BOSS samples may behave differently. The first few terms of the monopole and quadrupole components, such as $Q_{110}$, $Q_{220}$, $Q_{202}$, $Q_{112}$, and $Q_{022}$, have values of ${\cal O}(0.01)-{\cal O}(0.1)$, while the higher-order terms have values of ${\cal O}(0.01)$ or less. Therefore, we can conclude that the higher-order window 3PCF multipoles have no significant effect on the final $\zeta^{(\rm w)}_{\ell_1\ell_2\ell}(r_1,r_2)$, as long as we measure the first few terms of the monopole and quadrupole components, i.e., $\zeta^{(\rm w)}_{000}(r_1,r_2)$, $\zeta^{(\rm w)}_{110}(r_1,r_2)$, $\zeta^{(\rm w)}_{202}(r_1,r_2)$, and $\zeta^{(\rm w)}_{112}(r_1,r_2)$.

\begin{figure*}
    \scalebox{0.9}{\includegraphics[width=\textwidth]{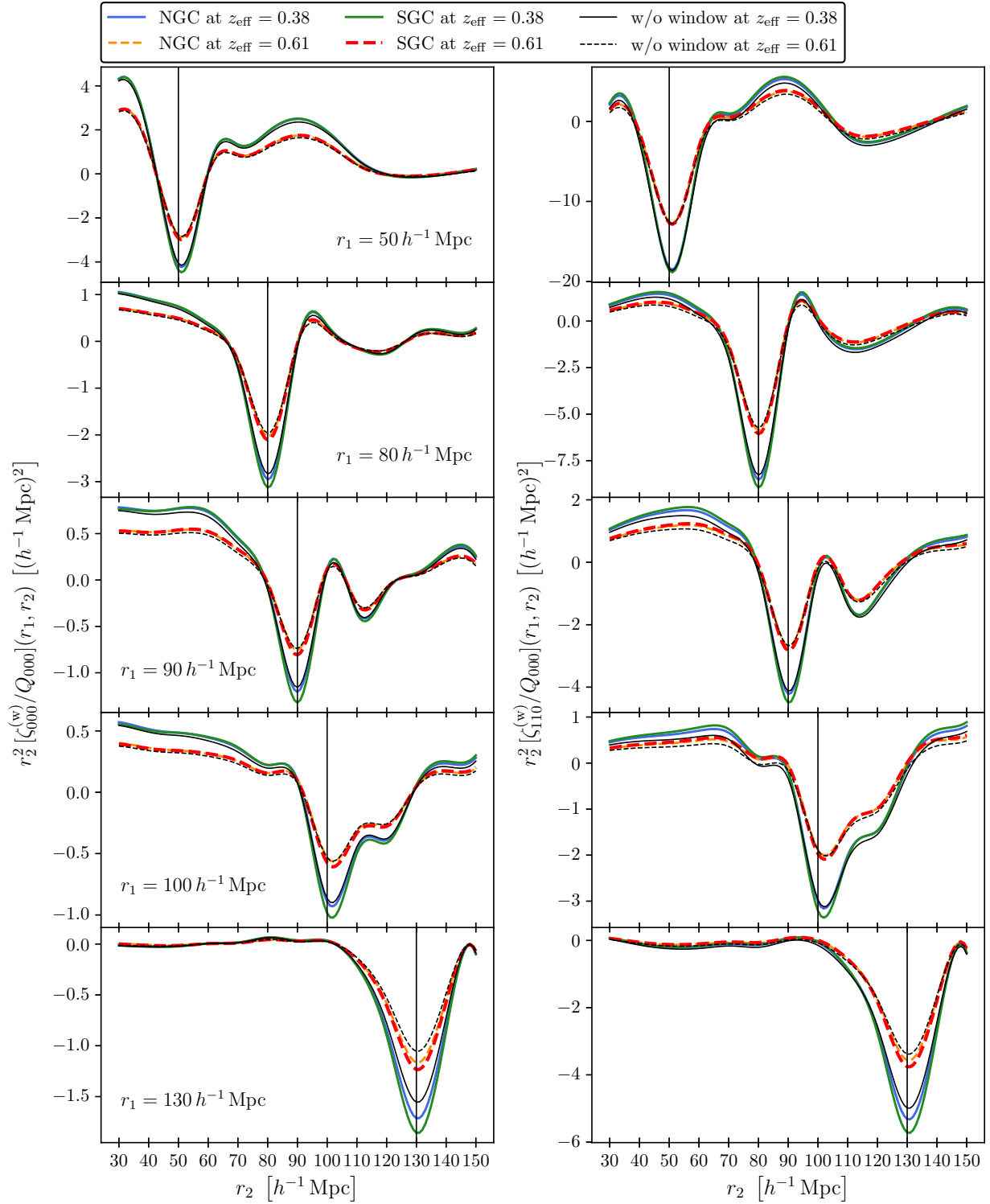}}
    \caption{The monopole 3PCFs, $\zeta_{000}$ (left) and $\zeta_{110}$ (right), that include the window function effect (\ref{Eq:zeta_window}) are shown for the four BOSS samples. The results are plotted as a function of $r_2$ after fixing $r_1$ to $50$, $80$, $90$, $100$, $130\hMpc$ from top to bottom panels. For these calculations, the $\Lambda$CDM model at $z=0.61$, the linear bias $b_1=2.0$, and no non-linear bias are assumed.}
	\label{fig:window_3PCF_mono}
\end{figure*}

\begin{figure*}
    \scalebox{0.9}{\includegraphics[width=\textwidth]{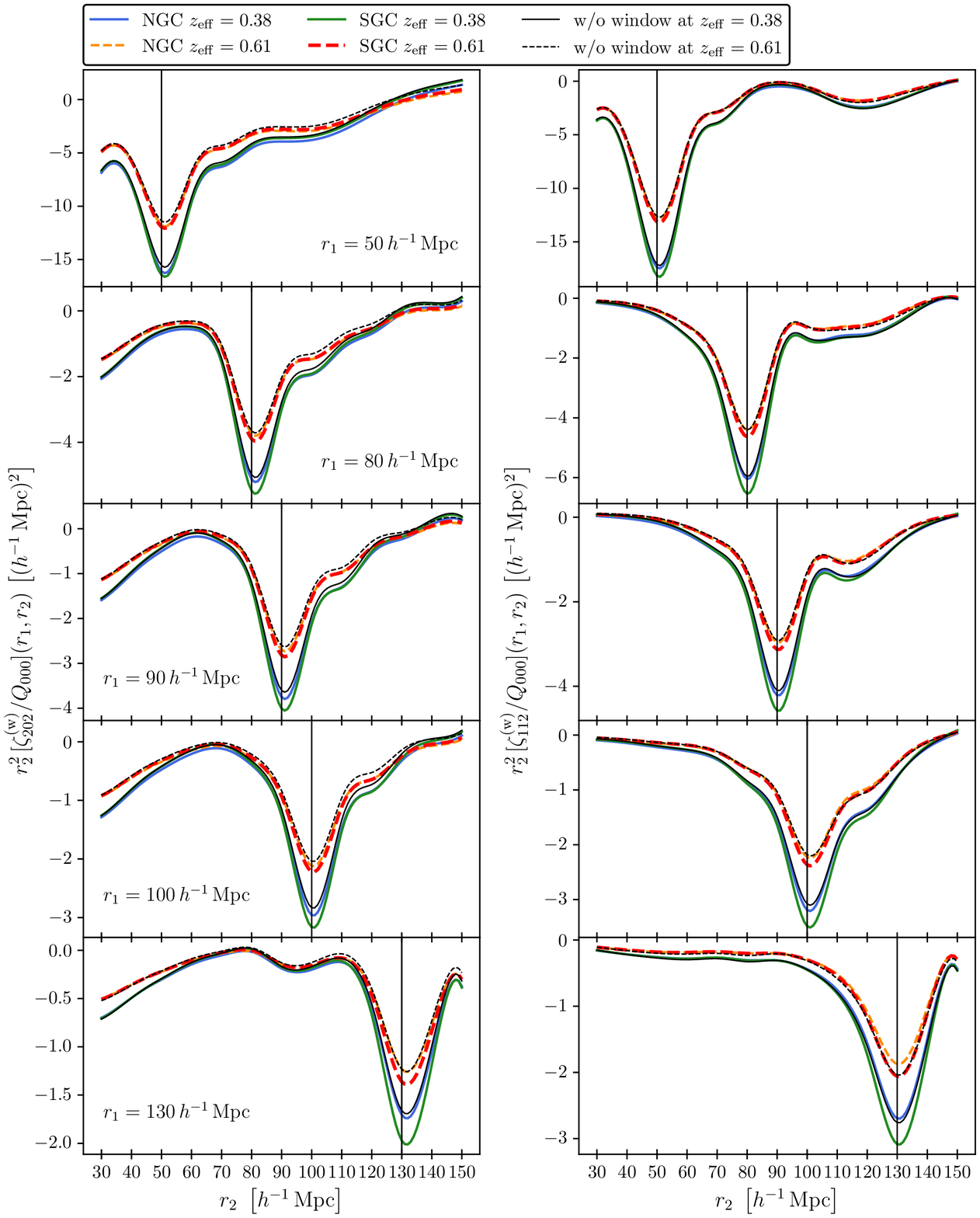}}
    \caption{Same as Figure~\ref{fig:window_3PCF_mono}, except that the quadrupole 3PCFs, $\zeta_{202}$ (left) and $\zeta_{112}$ (right), are shown.}
    \label{fig:window_3PCF_quad}
    \vspace{1cm}
\end{figure*}

Figures~\ref{fig:window_3PCF_mono} and~\ref{fig:window_3PCF_quad} plot the theoretical predictions for the 3PCF multipoles, including window function effects, corresponding to the four BOSS samples. These calculations assume the $\Lambda$CDM and linear bias as in Figures~\ref{fig:decom_3PCF_mono} and~\ref{fig:decom_3PCF_quad}, with redshifts of $0.38$ and $0.61$. As the value of $r_1$ increases, the difference between NGC and SGC due to the window function effect becomes more considerable. 

To quantitatively estimate the extent to which the multipole component of interest, $\zeta_{\ell_1\ell_2\ell}^{(\rm w)}$, is affected by the other multipole components, $\zeta_{\ell_1'\ell_2'\ell'}$, through window function effects, we compute the following quantities~\citep{Sugiyama:2020uil}:
\begin{eqnarray}
     \Delta \bar{\zeta}_{\ell'_1\ell'_2\ell'}
    \hspace{-0.25cm}&=& \hspace{-0.25cm}
    \frac{\mbox{Sum}\Big[\Delta \zeta_{\ell'_1\ell'_2\ell'}^{\ell_1\ell_2\ell}/Q_{000} \Big]}{\mbox{Sum}\left[ \zeta_{\ell_1\ell_2\ell}^{(\rm w)}/Q_{000} \right]}
    \nonumber \\
    \label{Eq:Delta_zeta}
\end{eqnarray}
with
\begin{eqnarray}
    \Delta \zeta_{\ell'_1\ell'_2\ell'}^{\ell_1\ell_2\ell}(r_1,r_2)
    \hspace{-0.25cm}&=&\hspace{-0.25cm}
    (4\pi) \ \ \sum_{\ell''_1+\ell''_2+\ell''={\rm even} }
    \left\{ \begin{smallmatrix} \ell''_1 & \ell''_2 & \ell'' \\   \ell'_1 & \ell'_2 & \ell' \\   \ell_1 & \ell_2 & \ell \end{smallmatrix}  \right\}
    \nonumber \\
	\hspace{-0.25cm}&\times&\hspace{-0.25cm} 
		\left[\frac{h_{\ell_1\ell_2\ell}h_{\ell_1\ell'_1\ell''_1}h_{\ell_2\ell'_2\ell''_2}h_{\ell \ell'\ell''}}{h_{\ell'_1\ell'_2\ell'}h_{\ell''_1\ell''_2\ell''}} \right]
	\nonumber \\
	\hspace{-0.25cm}&\times&\hspace{-0.25cm} 
			Q_{\ell''_1\ell''_2\ell''}(r_1,r_2)\, \zeta_{\ell'_1\ell'_2\ell'}(r_1,r_2)
    \label{Eq:Delta_zeta_mean}
\end{eqnarray}
and 
\begin{eqnarray}
    \hspace{-0.7cm}
    \mbox{Sum}\left[ \zeta_{\ell_1\ell_2\ell} \right] = 
    \begin{cases}
    \sum_{r_1\geq r_2} \zeta_{\ell_1\ell_2\ell}(r_1,r_2) \quad \mbox{for $\ell_1=\ell_2$};\\
    \sum_{r_1, r_2}\  \zeta_{\ell_1\ell_2\ell}(r_1,r_2) \quad \mbox{for $\ell_1\neq \ell_2$},
    \end{cases}
    \hspace{-0.5cm}
    \label{Eq:zeta_mean}
\end{eqnarray}
where $\Delta \bar{\zeta}_{\ell'_1\ell'_2\ell'}$ satisfies $\sum_{\ell_1'\ell_2'\ell'}\Delta \bar{\zeta}_{\ell'_1\ell'_2\ell'} = 1$, and the summation is performed in the range of $80\leq r\leq150\hMpc$ which we use for our data analysis.

Table~\ref{Table:zeta_window} summarises the $\Delta \bar{\zeta}_{\ell'_1\ell'_2\ell'}$ results calculated from Eq.~(\ref{Eq:Delta_zeta}) for the four BOSS samples. Naturally, the multipole component that is the same as the target one has the largest contribution. For example, for $\zeta_{000}^{(\rm w)}$ at $z_{\rm eff}=0.38$ in NGC, $95.34\%$ of the contribution comes from $\zeta_{000}$. For all four samples, multipole components other than the measured one have positive or negative values, and their overall contribution is about $5-10\%$. As expected, the contributions of higher-order components such as $\zeta_{330}$, $\zeta_{440}$, and $\zeta_{332}$ are mostly below $0.5\%$. Therefore, we conclude that the window function correction equation in Eq.~(\ref{Eq:zeta_window}) can account for the window function effect on the 3PCF in BOSS with sufficient accuracy, even if it is truncated at a finite number of 14 multipole components used in this work.

We note here the importance of $\Delta \bar{\zeta}_{112}$, which includes the $M\neq 0$ modes of \citet{Scoccimarro:1999ed}' decomposition method in the correction for window function effects: it gives a contribution comparable to $\Delta \bar{\zeta}_{202}$ and $\Delta \bar{\zeta}_{022}$, which include only the $M=0$ mode, and tends to have the opposite sign to that of $\Delta \bar{\zeta}_{202}$ and $\Delta \bar{\zeta}_{022}$. Therefore, failure to properly account for effects such as $\Delta \bar{\zeta}_{112}$ that include the $M\neq 0$ modes may result in an error of $\sim5\%$ in the correction for the window function effect.

\begin{table*}
\centering
\begin{tabular}{lcrrrrcrrrr}
    \vspace{1.0cm}\\
\hline\hline
\multicolumn{11}{c}{$z_{\rm eff} = 0.38$ ($0.2 < z< 0.5$)} \\
\hline
\multicolumn{2}{c}{} & \multicolumn{4}{c}{NGC} & \multicolumn{1}{c}{} & \multicolumn{4}{c}{SGC}\\
\hline
& $\Delta \bar{\zeta}_{\ell'_1\ell'_2\ell'}\, [\%]$ & $\zeta^{(\rm w)}_{000}$ & $\zeta^{(\rm w)}_{110}$ & $\zeta^{(\rm w)}_{202}$ & $\zeta^{(\rm w)}_{112}$ 
& & $\zeta^{(\rm w)}_{000}$ & $\zeta^{(\rm w)}_{110}$ & $\zeta^{(\rm w)}_{202}$ & $\zeta^{(\rm w)}_{112}$ \\
\hline\vspace{0.07cm}
 monopole ($\ell=0$)
   & $\Delta \bar{\zeta}_{000}$ & $\mathbf{95.34}$ & $\mathbf{1.59}$ & $\mathbf{-1.18}$ & $\mathbf{0.97}$  & & 
                                  $\mathbf{86.41}$ & $\mathbf{2.43}$ & $-0.15$ & $0.19$\\ \vspace{0.07cm}
   & $\Delta \bar{\zeta}_{110}$ & $\mathbf{9.39}$ & $\mathbf{102.60}$ & $\mathbf{1.41}$ & $\mathbf{-3.77}$ & & 
                                  $\mathbf{13.66}$ & $\mathbf{98.60}$ & $0.25$ & $\mathbf{-0.50}$   \\ \vspace{0.07cm}
   & $\Delta \bar{\zeta}_{220}$ & $-0.09$ & $\mathbf{-0.53}$ & $\mathbf{1.14}$ & $0.09$ & & 
                                  $-0.12$ & $\mathbf{-0.79}$ & $0.13$ & $0.01$  \\ \vspace{0.07cm}
   & $\Delta \bar{\zeta}_{330}$ & $0.26$ & $0.21$ & $0.08$ & $-0.01$ & & 
                                  $0.41$ & $0.35$ & $0.02$ & $-0.02$   \\ \vspace{0.07cm}
   & $\Delta \bar{\zeta}_{440}$ & $0.06$ & $0.00$ & $0.06$ & $-0.00$ & &   
                                  $0.09$ & $0.01$ & $-0.00$ & $-0.00$   \\ 
\hline\vspace{0.07cm}
 quadrupole ($\ell=2$)
  & $\Delta \bar{\zeta}_{202}$ & $\mathbf{-3.75}$ & $\mathbf{0.59}$ & $\mathbf{90.14}$ & $\mathbf{2.07}$ & &
$-0.43$ & $0.10$ & $\mathbf{89.28}$ & $\mathbf{2.38}$  \\ \vspace{0.07cm}
  & $\Delta \bar{\zeta}_{112}$ & $\mathbf{4.36}$ & $\mathbf{-2.89}$ & $\mathbf{4.21}$ & $\mathbf{96.78}$ & &
$\mathbf{0.86}$ & $-0.38$ & $\mathbf{5.06}$ & $\mathbf{92.64}$  \\ \vspace{0.07cm}
  & $\Delta \bar{\zeta}_{022}$ & $\mathbf{-4.57}$ & $\mathbf{0.81}$ & $\mathbf{0.50}$ & $\mathbf{2.75}$ & & 
$-0.48$ & $0.14$ & $\mathbf{0.51}$ & $\mathbf{3.16}$   \\ \vspace{0.07cm}
  & $\Delta \bar{\zeta}_{312}$ & $-0.10$ & $-0.38$ & $\mathbf{1.95}$ & $0.13$ & & 
$-0.05$ & $-0.05$ & $\mathbf{2.87}$ & $0.26$  \\ \vspace{0.07cm}
  & $\Delta \bar{\zeta}_{222}$ & $0.04$ & $0.04$ & $0.38$ & $0.07$  & &
 $-0.02$ & $-0.01$ & $-0.04$ & $0.12$ \\ \vspace{0.07cm}
  & $\Delta \bar{\zeta}_{132}$ & $\mathbf{-0.56}$ & $\mathbf{-1.54}$ & $0.14$ & $0.23$ & &
$-0.19$ & $-0.19$ & $0.18$ & $\mathbf{0.66}$  \\ \vspace{0.07cm}
  & $\Delta \bar{\zeta}_{422}$ & $-0.21$ & $-0.23$ & $\mathbf{1.02}$ & $0.21$ & &
 $-0.02$ & $-0.08$ & $\mathbf{1.53}$ & $0.31$  \\ \vspace{0.07cm}
  & $\Delta \bar{\zeta}_{332}$ & $0.10$ & $0.08$ & $0.02$ & $0.10$  & &
 $-0.03$ & $-0.02$ & $0.10$ & $0.24$  \\ \vspace{0.07cm}
  & $\Delta \bar{\zeta}_{242}$ & $-0.29$ & $-0.34$ & $0.12$ & $0.38$ & &
 $-0.08$ & $-0.14$ & $0.26$ & $\mathbf{0.55}$  \\ 
\hline
\vspace{1.0cm}\\
\hline\hline
\multicolumn{11}{c}{$z_{\rm eff} = 0.61$ ($0.5 < z< 0.75$)} \\
\hline
\multicolumn{2}{c}{} & \multicolumn{4}{c}{NGC} & \multicolumn{1}{c}{} & \multicolumn{4}{c}{SGC}\\
\hline
\hline
& $\Delta \bar{\zeta}_{\ell'_1\ell'_2\ell'}\, [\%]$ & $\zeta^{(\rm w)}_{000}$ & $\zeta^{(\rm w)}_{110}$ & $\zeta^{(\rm w)}_{202}$ & $\zeta^{(\rm w)}_{112}$ & &
$\zeta^{(\rm w)}_{000}$ & $\zeta^{(\rm w)}_{110}$ & $\zeta^{(\rm w)}_{202}$ & $\zeta^{(\rm w)}_{112}$ \\
\hline\vspace{0.07cm}
 monopole ($\ell=0$)
   & $\Delta \bar{\zeta}_{000}$ & $\mathbf{96.00}$ & $\mathbf{1.77}$ & $\mathbf{-2.00}$ & $\mathbf{1.78}$ & &
$\mathbf{89.49}$ & $\mathbf{2.40}$ & $\mathbf{-1.29}$ & $\mathbf{1.19}$   \\ \vspace{0.07cm}
   & $\Delta \bar{\zeta}_{110}$ & $\mathbf{10.43}$ & $\mathbf{104.18}$ & $\mathbf{2.92}$ & $\mathbf{-6.81}$ & &
$\mathbf{13.59}$ & $\mathbf{101.41}$ & $\mathbf{2.10}$ & $\mathbf{-4.36}$ \\ \vspace{0.07cm}
   & $\Delta \bar{\zeta}_{220}$ & $-0.08$ & $\mathbf{-0.60}$ & $\mathbf{1.99}$ & $0.01$   & &
$-0.06$ & $\mathbf{-0.80}$ & $\mathbf{1.33}$ & $-0.04$  \\ \vspace{0.07cm}
   & $\Delta \bar{\zeta}_{330}$ & $0.25$ & $0.21$ & $0.04$ & $0.01$     & &
$0.35$ & $0.31$ & $0.00$ & $-0.01$    \\ \vspace{0.07cm}
   & $\Delta \bar{\zeta}_{440}$ & $0.06$ & $0.01$ & $0.05$ & $-0.00$    & &
$0.08$ & $0.01$ & $0.02$ & $-0.00$    \\ 
\hline\vspace{0.07cm}
 quadrupole ($\ell=2$)
   & $\Delta \bar{\zeta}_{202}$ & $\mathbf{-6.26}$ & $\mathbf{1.21}$ & $\mathbf{86.94}$ & $\mathbf{2.72}$ & & 
$\mathbf{-3.73}$ & $\mathbf{0.84}$ & $\mathbf{86.45}$ & $\mathbf{2.94}$ \\ \vspace{0.07cm}
   & $\Delta \bar{\zeta}_{112}$ & $\mathbf{7.84}$ & $\mathbf{-5.12}$ & $\mathbf{5.36}$ & $\mathbf{97.84}$ & & 
$\mathbf{5.09}$ & $\mathbf{-3.31}$ & $\mathbf{5.98}$ & $\mathbf{94.88}$ \\ \vspace{0.07cm}
   & $\Delta \bar{\zeta}_{022}$ & $\mathbf{-7.65}$ & $\mathbf{1.65}$ & $0.48$ & $\mathbf{3.64}$  & &   
$\mathbf{-4.54}$ & $\mathbf{1.14}$ & $0.44$ & $\mathbf{3.93}$  \\ \vspace{0.07cm}
   & $\Delta \bar{\zeta}_{312}$ & $-0.01$ & $\mathbf{-0.54}$ & $\mathbf{2.25}$ & $0.09$   & & 
$0.01$ & $-0.30$ & $\mathbf{2.89}$ & $0.17$  \\ \vspace{0.07cm}
   & $\Delta \bar{\zeta}_{222}$ & $0.04$ & $0.04$ & $\mathbf{0.69}$ & $0.07$     & & 
$-0.00$ & $0.01$ & $0.42$ & $0.11$  \\ \vspace{0.07cm}
   & $\Delta \bar{\zeta}_{132}$ & $-0.28$ & $\mathbf{-2.63}$ & $0.13$ & $-0.02$  & & 
$-0.06$ & $\mathbf{-1.66}$ & $0.15$ & $0.25$ \\ \vspace{0.07cm}
   & $\Delta \bar{\zeta}_{422}$ & $-0.18$ & $-0.13$ & $\mathbf{0.98}$ & $0.20$   & & 
$-0.07$ & $-0.03$ & $\mathbf{1.21}$ & $0.27$ \\ \vspace{0.07cm}
   & $\Delta \bar{\zeta}_{332}$ & $0.08$ & $0.09$ & $0.03$ & $0.10$     & & 
$-0.02$ & $0.01$ & $0.09$ & $0.19$  \\ \vspace{0.07cm}
   & $\Delta \bar{\zeta}_{242}$ & $-0.24$ & $-0.15$ & $0.13$ & $0.37$   & & 
$-0.12$ & $-0.03$ & $0.21$ & $0.48$ \\ 
\hline

\end{tabular}
\caption{
    Contributions of other 3PCF multipole components to the observed 3PCF multipole components, as manifested through the effect of the window function, are shown for the four BOSS samples. When the contribution to the final result exceeds $0.5\%$, it is written in bold. The value of the same multipole component $\Delta \bar{\zeta}_{\ell_1\ell_2\ell}$ (\ref{Eq:Delta_zeta}) as the measured $\zeta_{\ell_1\ell_2\ell}^{(\rm w)}$ (\ref{Eq:zeta_window}) is larger (smaller) than $100\%$, when the total contribution from all the other multipole components is negative (positive).
}
\label{Table:zeta_window}
\end{table*}



\section{Covariance matrix}
\label{Sec:CovarianceMatrix}

We estimate the covariance matrix from the $2048$ Patchy mock catalogues described in Section~\ref{Sec:Data}. Let $\VEC{d}^{(r)}$ be the data vector measured from the $r$-th catalogue, and $\overline{\VEC{d}} = (1/N_{\rm s})\sum_{r=1}^{N_{\rm s}} \VEC{d}^{(r)}$ be its mean value, then the covariance matrix of the data vector is given by
\begin{eqnarray}
    \MAT{C} = \frac{1}{N_{\rm s}-1}\sum_{r = 1}^{N_{\rm s}}
    \left( \VEC{d}^{(r)} - \overline{\VEC{d}} \right)\left( \VEC{d}^{(r)} - \overline{\VEC{d}} \right)^T.
\end{eqnarray} 
where $N_{\rm s}=2048$ is the number of the Patchy mock catalogues. 

\subsection{Effects of a finite number of mocks}

The covariance matrix $\MAT{C}$ inferred from the mock catalogues suffers from noise due to the finite number of mocks, which directly leads to an increase in the uncertainty of the cosmological parameters~\citep{Hartlap:2006kj,Taylor:2012kz,Dodelson:2013uaa,Percival:2013sga,Taylor:2014ota}. This effect is decomposed into two factors. First, the inverse covariance matrix, $\MAT{C}^{-1}$, provides a biased estimate of the true inverse covariance matrix. To correct this bias, we rescale the inverse covariance matrix as~\citep{Hartlap:2006kj}
\begin{eqnarray}
    \MAT{C}_{\rm Hartlap}^{-1} = \left( \frac{N_{\rm s}-N_{\rm b}-2}{N_{\rm s}-1} \right) \MAT{C}^{-1},
    \label{Eq:Hartlap}
\end{eqnarray}
where the pre-factor on the right-hand side, $(N_{\rm s}-N_{\rm b}-2)/(N_{\rm s}-1)$, is the so-called ``Hartlap'' factor, and $N_{\rm b}$ is the number of data bins. Second, we need to consider the propagation of the error in the covariance matrix to the error on the estimated parameters. This effect is corrected by multiplying the final result of the parameter errors by the following factor~\citep{Percival:2013sga}
\begin{eqnarray}
    M_1 = \sqrt{\frac{1 + B(N_{\rm b}-N_{\rm p})}{1 + A + B(N_{\rm p}+1)}} 
    \label{Eq:M1}
\end{eqnarray}
with
\begin{eqnarray}
    A \hspace{-0.25cm} &=& \hspace{-0.25cm} \frac{2}{(N_{\rm s}-N_{\rm b}-1)(N_{\rm s}-N_{\rm b}-4)} \nonumber \\
    B \hspace{-0.25cm} &=& \hspace{-0.25cm} \frac{N_{\rm s}-N_{\rm b}-2}{(N_{\rm s}-N_{\rm b}-1)(N_{\rm s}-N_{\rm b}-4)},
\end{eqnarray}
where $N_{\rm p}$ is the number of parameters.

The derivation of the Hartlap factor (\ref{Eq:Hartlap}) assumes that the data vector follows a Gaussian distribution. On the other hand, \citet{Sellentin:2016MNRAS.456L.132S} shows that in covariance matrix estimates from simulations, the data vector follows a multivariate $t$-distribution. When the number of simulations is sufficiently larger than the number of data bins, this $t$-distribution approaches a Gaussian distribution~\citep{Heavens:2017MNRAS.472.4244H}, and the present analysis satisfies this condition. The reason is that the number of the Patchy mocks we use to estimate the covariance matrix is $2048$, while the maximum number of data in our analysis is $202$ (Section~\ref{Sec:NumberOfBins}). In addition, the derivation of the $M_1$ factor (\ref{Eq:M1}) also assumes the Gaussian distribution of the data vector, but there is no known value for the correction factor that corresponds to $M_1$ in the \citet{Sellentin:2016MNRAS.456L.132S}'s method. Therefore, in this paper, we have decided to use Eqs.~(\ref{Eq:Hartlap}) and (\ref{Eq:M1}) to correct the uncertainty in parameter estimation due to a finite number of simulations \citep[see also e.g.,][]{Percival:2022mn}.

We can therefore evaluate the effect of a finite number of mocks on the final error estimation using the square root of the Hartlap factor multiplied by the $M_1$ factor~\citep{Percival:2013sga},
\begin{eqnarray}
    M_2 = \sqrt{\frac{N_{\rm s}-1}{N_{\rm s}-N_{\rm b}-2}} M_1.
    \label{Eq:M2}
\end{eqnarray}
Note that this $M_2$ factor is not used in the actual analysis. It is essential to increase the number of simulations and reduce the number of data bins to keep the value of $M_2$ as close to $1$ as possible for a conservative analysis. The reason is that the Hartlap and $M_1$ factors cannot be always accurately correct for parameter errors for any number of simulations. For example, using both monopole and quadrupole components of the 2PCF and 3PCF, as in this paper, \citet{Sugiyama:2020uil} performed an anisotropic BAO analysis with the AP effect. The result showed that the error in the angular diameter distance for $M_2=1.32$ is underestimated by about $10\%$ compared to the case for $M_2=1.06$ by changing the number of simulations. We will calculate the $M_2$ factor in Section~\ref{Sec:M1M2} and summarise the results in Table~\ref{Table:M1M2}, where $M_2\sim1.1$, indicating that our analysis achieves $M_2$ values sufficiently close to $1$.

\subsection{Correlation matrix}

\begin{figure*}
    \vspace{2cm}
    \scalebox{0.95}{\includegraphics[width=\textwidth]{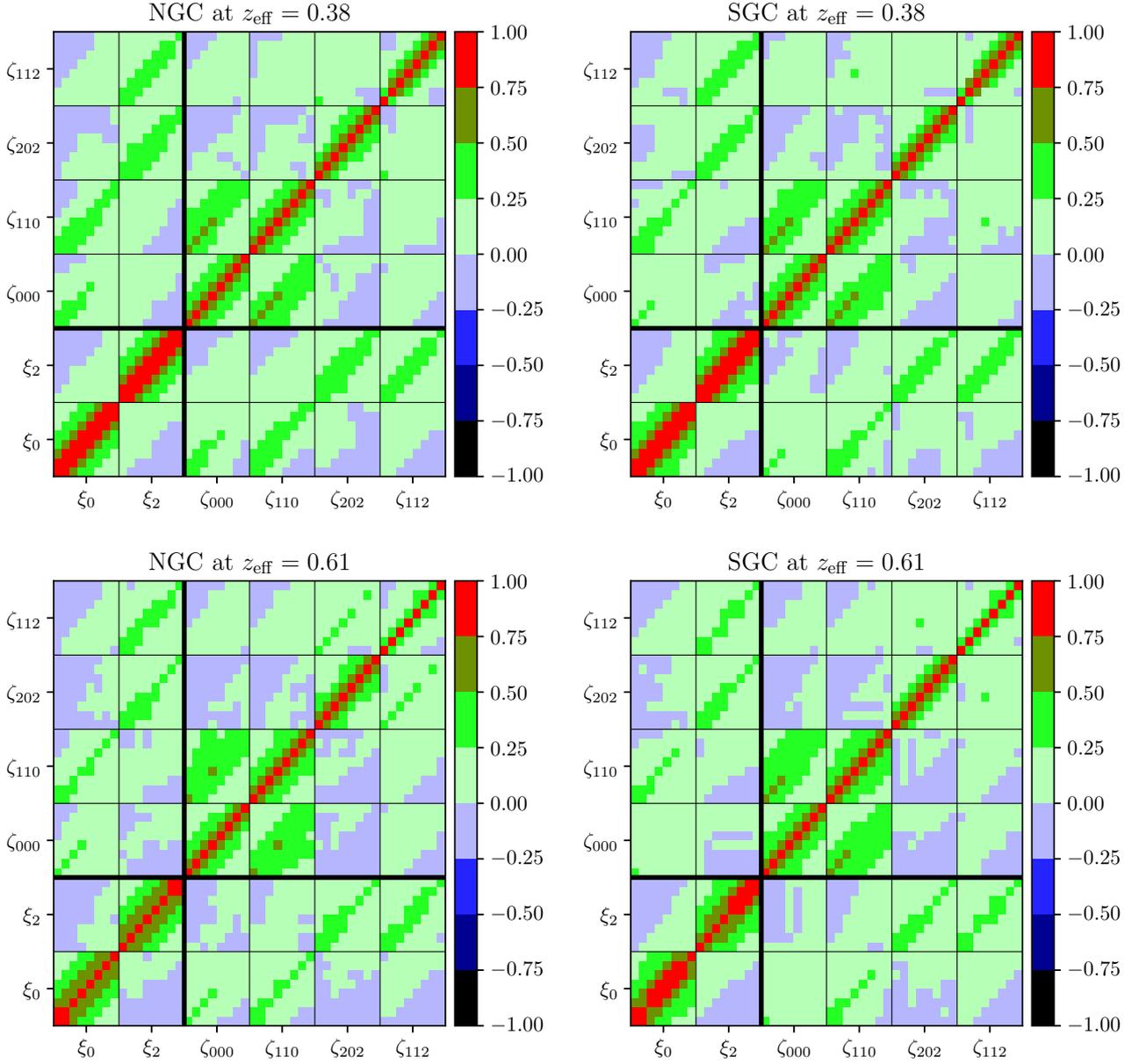}}
    \caption{The correlation matrices of the monopole and quadrupole 2PCFs ($\xi_0$ and $\xi_2$), the first two monopole 3PCFs ($\zeta_{000}$ and $\zeta_{110}$), and the first two quadrupole 3PCFs ($\zeta_{202}$ and $\zeta_{112}$) are shown for the four BOSS samples. For simplicity of the figure, only the results for the $r_1=r_2$ case multipole 3PCFs, i.e. $\zeta(r_1,r_2=r_1)$, are plotted, but in the actual analysis (Section~\ref{Sec:DataAnalysis}), the data bins for the $r_1\neq r_2$ case are also used. The plotted scale range is $80\leq r\leq 150\hMpc$, and the $r$-bin width is $\Delta r=10\hMpc$.
    }
    \vspace{2cm}
    \label{fig:corr}
\end{figure*}

The $(i,j)$ elements of the correlation matrix is computed from the covariance matrix as
\begin{eqnarray}
    r_{ij} = \frac{C_{ij}}{\sqrt{C_{ii}C_{jj}}}.
    \label{Eq:rij}
\end{eqnarray}
Considering the data vector $\VEC{d} = \{\xi_0,\xi_2,\zeta_{000},\zeta_{110},\zeta_{202},\zeta_{112}\}$, we show the results of the correlation matrix for the four BOSS samples in Figure~\ref{fig:corr}. To simplify the figure, we only plot the results for the diagonal component of the 3PCF multipoles, i.e., $\zeta_{\ell_1\ell_2\ell}(r_1,r_2=r_1)$. The range of scales shown in the figure is $80\leq r \leq 150\hMpc$, and the width of the $r$-bin is $\Delta r = 10\hMpc$. The four samples show similar results, and we summarise the overall features below. First, the monopole 2PCF and the monopole 3PCFs have a moderate correlation ($0.25<r_{ij}<0.5$); the same is true for the quadrupole 2PCF and the quadrupole 3PCFs. Next, the first two terms of the monopole 3PCFs ($\zeta_{000}$ and $\zeta_{110}$) are strongly correlated with each other ($0.5<r_{ij}<0.75$); on the other hand, the first two quadrupole 3PCFs ($\zeta_{202}$ and $\zeta_{112}$) are weakly correlated ($0.0<r_{ij}<0.25$). This result indicates that $\zeta_{202}$ and $\zeta_{112}$ have independent information from each other. These results are consistent with the results in the bispectrum presented by~\citet{Sugiyama:2018yzo}.

\subsection{Standard deviation}

\begin{figure*}
    \scalebox{0.95}{\includegraphics[width=\textwidth]{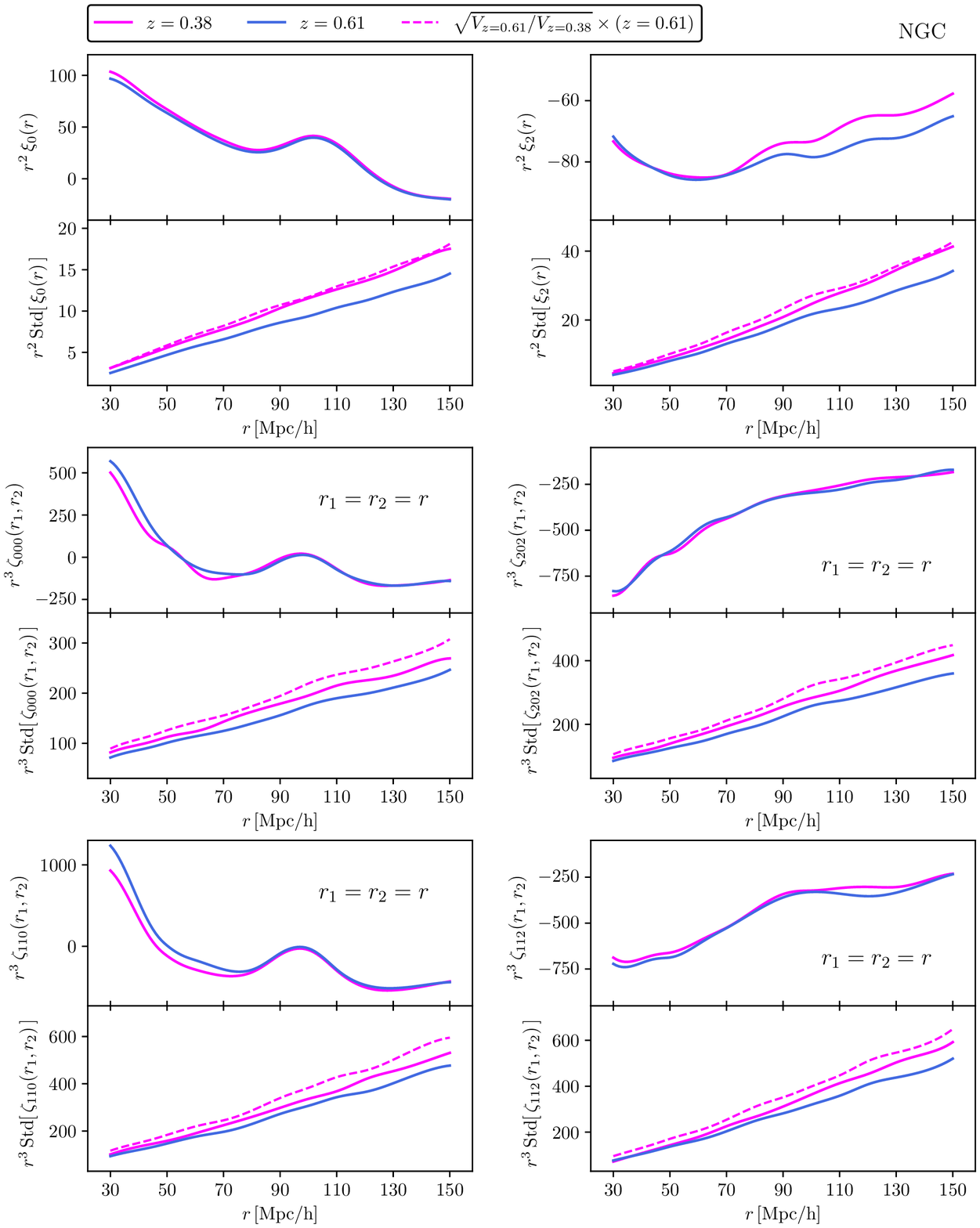}}
    \caption{The mean values and standard deviations of $\xi_{0}$, $\xi_2$, $\zeta_{000}$, $\zeta_{110}$, $\zeta_{202}$, and $\zeta_{112}$ calculated from the $2048$ Patchy mock catalogues. The results are plotted at the two redshifts, $z=0.38$ (magenta) and $0.61$ (blue), for the NGC sample. The magenta dashed lines are the standard deviation at $z=0.61$ multiplied by $\sqrt{V_{z=0.61}/V_{z=0.38}}$ and normalized to have the same survey volume as the sample at $z=0.38$, where the survey volumes at $z=0.38$ and $z=0.61$, $V_{z=0.38}$ and $V_{z=0.61}$, respectively, are given in Table~\ref{Table:Grid}. For simplicity of the figure, only the results in the $r_1=r_2$ case for the 3PCF are plotted.
    }
    \label{fig:3PCF_std}
\end{figure*}

The standard deviation is given by the square root of the diagonal components of the covariance matrix: i.e., $\sqrt{C_{ii}}$. Figure~\ref{fig:3PCF_std} shows the mean and standard deviation of $\xi_{\ell}(r)$ and $\zeta_{\ell_1\ell_2\ell}(r_1,r_2)$ calculated from the $2048$ Patchy mock simulations. The mock data used are the NGC samples at $z=0.38$ and $0.61$. For $\zeta_{\ell_1\ell_2\ell}(r_1,r_2)$, the measured values and the standard deviations are plotted as a function of the scale variable $r_1=r_2=r$ to simplify the figure.

From this figure, it can be seen that the mean values of $\xi_{\ell}$ and $\zeta_{\ell_1\ell_2\ell}$ do not differ much for the different redshifts, i.e., $z=0.38$ and $0.61$ (compare the magenta and blue lines). One may expect the amplitudes of $\xi_{\ell}$ and $\zeta_{\ell_1\ell_2\ell}$ to be larger at lower redshifts because the tree-level solutions (\ref{Eq:Tree_PB}) of $\xi_{\ell}$ and $\zeta_{\ell_1\ell_2\ell}$ are proportional to $D^2$ and $D^4$, respectively, with $D$ being the linear growth function. However, this is not the case in Figure~\ref{fig:3PCF_std}. There are two possible reasons for this. The first is the bias effect. For halos with similar mass, the lower the redshift, the smaller the value of the linear bias $b_1$ tends to be. Therefore, $b_1 D(z)$ is less time-dependent and does not show significant differences at the different redshifts, especially for the monopole components of $\xi_{\ell}$ and $\zeta_{\ell_1\ell_2\ell}$. A similar effect to the linear bias is likely to occur for the non-linear bias included in the 3PCF. Second, the product of the linear growth rate $f$ and the linear growth function $D$ is also a less time-dependent function. Therefore, the redshift dependence of $\xi_{\ell}$ and $\zeta_{\ell_1\ell_2\ell}$ is not pronounced, even for the quadrupole component.

On the other hand, the standard deviations of $\xi_{\ell}$ and $\zeta_{\ell_1\ell_2\ell}$ are significantly different for the different redshifts. In general, the so-called Gaussian terms in the covariance matrix depend only on the two-point statistic, while higher-order statistics such as the three-point statistic appear in the non-Gaussian terms. It is also known that the covariance matrix is inversely proportional to the survey volume, and that the higher the number density of observed galaxies, the smaller the covariance matrix. Therefore, the fact that the $\xi_{\ell}$ and $\zeta_{\ell_1\ell_2\ell}$ signals measured from the Patchy mock do not differ significantly at the different redshifts suggests that the redshift dependence in the standard deviation may be due to the survey volume and galaxy number density.

In Figure~\ref{fig:3PCF_std}, the standard deviation at $z=0.61$ (blue) multiplied by $\sqrt{V_{z=0.61}/V_{z=0.38}}$ is plotted as magenta dashed lines, with the survey volumes at $z=0.38$ and $z=0.61$ denoted as $V_{z=0.38}$ and $V_{z=0.61}$, respectively. In the case of the 2PCF, the magenta dashed line is similar to the result at $z=0.38$ (magenta), indicating that the difference in the standard deviation of the 2PCF due to differences in redshift can be explained mainly by differences in the survey volume. However, this is not the case for the 3PCF, where the standard deviation of the 3PCF at $z=0.38$ is smaller than the magenta dashed line. This fact suggests that the effect of the galaxy number density on the covariance matrix is more significant for the 3PCF than for the 2PCF. In other words, as can be seen from Table~\ref{Table:Grid}, the sample at $z=0.38$ has a higher galaxy number density than the sample at $z=0.61$, even though the survey volume is smaller. Therefore, the standard deviation at $z=0.38$ is smaller than the standard deviation at $z=0.61$ normalised to the survey volume at $z=0.38$. This result is consistent with the finding of \citet{Sugiyama:2020MNRAS.497.1684S} that the galaxy number density plays an essential role in the covariance matrix of the bispectrum, even on large scales. 

\subsection{Cumulative signal-to-noise ratio}
\label{Sec:CumulativeSignalToNoiseRatio}

\begin{figure*}
    \scalebox{0.95}{\includegraphics[width=\textwidth]{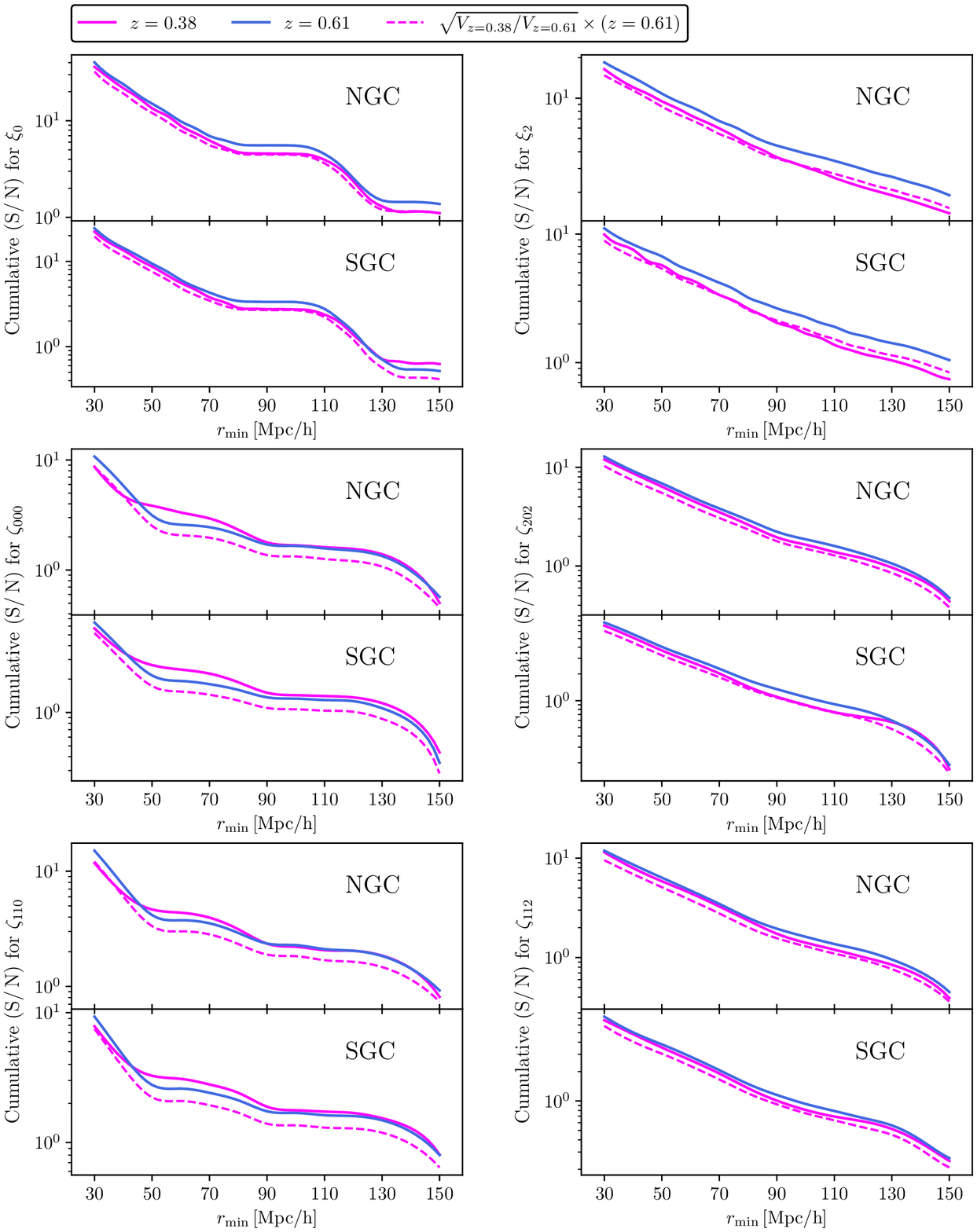}}
    \caption{Cumulative signal-to-noise ratios $({\rm S/N}s)$ for the multipole components of the 2PCF and 3PCF, where both signal and noise (covariance matrix) are computed from the $2048$ Patchy mock catalogues. The maximum scale used for the ${\rm S/N}$ calculation is fixed at $r_{\rm max}=150\hMpc$ and the ${\rm S/N}s$ are plotted as a function of the minimum scale $r_{\rm min}$. The blue and magenta solid lines show the results for the samples at $z=0.61$ and $z=0.38$, respectively. The magenta dashed lines are the ${\rm S/N}$ values in the sample at $z=0.61$ multiplied by $\sqrt{V_{z=0.38}/V_{z=0.61}}$.
    }
    \label{fig:sn}
\end{figure*}

The covariance matrix is a two-dimensional quantity in the 2PCF case and a four-dimensional quantity in the 3PCF case. Therefore, a useful way compressing and quantifying this multi-dimensional information in the covariance matrix is to estimate the cumulative signal-to-noise $({\rm S/N})$ ratios, given by
\begin{eqnarray}
    \left( \rm {\frac{S}{N}} \right) = \left(  \overline{\VEC{d}}^{T}\cdot\MAT{C}_{\rm Hartlap}^{-1}\cdot \overline{\VEC{d}}\right)^{1/2}.
\end{eqnarray}
We calculate the cumulative ${\rm S/N}$ for each multipole component of the 2PCF and 3PCF: i.e., $\overline{d} = \xi_0$, $\xi_2$, $\zeta_{000}$, $\zeta_{110}$, $\zeta_{202}$, or $\zeta_{112}$. We also fix the maximum scale $r_{\rm max}=150\hMpc$, vary the minimum scale $r_{\rm min}$ from $150\hMpc$ to $30\hMpc$, and calculate the ${\rm S/N}$ as a function of $r_{\rm min}$. In Figure~\ref{fig:sn}, we plot the ${\rm S/N}$ for the four BOSS samples, NGC and SGC at $z=0.38$ and $0.61$. Note that we do not consider cross-covariance matrices between different multipole components, e.g., between $\xi_0$ and $\zeta_{000}$. How the information in the covariance matrix, including all cross-covariance matrices, ultimately propagates to the errors in the cosmological parameters of interest will be discussed through the Fisher analysis in Section~\ref{Sec:FisherAnalysis}.

The top two panels of Figure~\ref{fig:sn} show the ${\rm S/N}$ of $\xi_0$ and $\xi_2$. In all cases shown in the panels, the ${\rm S/N}$ at $z = 0.61$ (blue line) is larger than the ${\rm S/N}$ at $z = 0.38$ (magenta line). The difference is because the ${\rm S/N}$ of the 2PCF is proportional to the square root of the survey volume $V$, and the survey volume at $z = 0.61$, denoted $V_{\rm z=0.61}$, is larger than the survey volume at $z = 0.38$, $V_{\rm z=0.38}$. Therefore, multiplying the ${\rm S/N}$ at $z=0.61$ by $\sqrt{V_{\rm z=0.38}/V_{\rm z=0.61}}$ approximately reproduces the ${\rm S/N}$ at $z=0.38$ (see magenta dashed lines). This result is consistent with the findings in the signal and standard deviation of the 2PCF in Figure~\ref{fig:3PCF_std}.

The middle and bottom results are for $\zeta_{000}$, $\zeta_{110}$, $\zeta_{202}$, and $\zeta_{112}$. These results show that, in contrast to the 2PCF case, the ${\rm S/N}$ at $z=0.38$ is comparable to the ${\rm S/N}$ at $z=0.61$. The difference in the ${\rm S/N}$ at $z=0.38$ and $z=0.61$ in the 3PCF case cannot be explained by the difference in the survey volumes (see magenta dashed lines). This behaviour of the ${\rm S/N}$ of the 3PCF can be explained by the finding shown in Figure~\ref{fig:3PCF_std} that the galaxy number density strongly influences the standard deviation of the 3PCF. In particular, in the present case, the effect of the galaxy number density is more pronounced when considering correlations between different scales, resulting in the ${\rm S/N}$ at $z=0.38$ that is comparable to the ${\rm S/N}$ at $z=0.61$. This result shows that a higher galaxy number density is as important for obtaining cosmological information from the 3PCF as increasing the survey volume.

\section{Analysis settings}
\label{Sec:AnalysisSettings}

\subsection{Likelihoods}
\label{Sec:Likelihoods}

We assume that the likelihood of the data compared to the model predictions follows a multivariate Gaussian distribution:
\begin{eqnarray}
    \ln {\cal L}(\VEC{d}|\boldsymbol{\theta}) 
    = -\frac{1}{2} \left[ \VEC{d} - \VEC{t}(\boldsymbol{\theta}) \right] \MAT{C}^{-1}_{\rm Hartlap} \left[ \VEC{d} - \VEC{t}(\boldsymbol{\theta}) \right]^T, 
    \label{Eq:likelihood}
\end{eqnarray}
where $\VEC{d}$ is the data vector, \VEC{t} is the model prediction of the data vector given the model parameters $\boldsymbol{\theta}$, and $\MAT{C}^{-1}_{\rm Hartlap}$ is the inverse of the covariance matrix after correction by the Hartlap factor (\ref{Eq:Hartlap}). We can then obtain the posterior distribution of the model parameters given the data by performing Bayesian inference:
\begin{eqnarray}
    {\cal P}(\boldsymbol{\theta}|\VEC{d}) \propto {\cal L}(\VEC{d}|\boldsymbol{\theta}) \Pi(\boldsymbol{\theta})
\end{eqnarray}
where ${\cal P}(\boldsymbol{\theta}|\VEC{d})$ is the posterior distribution of $\boldsymbol{\theta}$ given the data vector, $\VEC{d}$, and $\Pi(\boldsymbol{\theta})$ is the prior distribution.

We assume that the four BOSS galaxy samples (Table~\ref{Table:Grid}) are far enough apart that they each have independent cosmological information. Then, when constraining the model parameters common to each galaxy sample, we add up the likelihood functions of each galaxy sample. For example, when using all four galaxy samples, the total likelihood function is given by
\begin{eqnarray}
    \ln {\cal L}_{\rm total} \hspace{-0.25cm} &=& \hspace{-0.25cm}  \ln {\cal L}_{{\rm NGC\, at}\, z=0.38} + \ln {\cal L}_{{\rm SGC\, at}\, z=0.38} \nonumber \\
                         \hspace{-0.25cm} &+&  \hspace{-0.25cm}\ln {\cal L}_{{\rm NGC\, at}\, z=0.61} + \ln {\cal L}_{{\rm SGC\, at}\, z=0.61}.
\end{eqnarray}

\subsection{Multipoles used, scale range, and number of bins}
\label{Sec:NumberOfBins}

To repeat what was explained in Section~\ref{Sec:Estimator}, the scale range used for parameter estimation in Section~\ref{Sec:DataAnalysis} is $80\hMpc\leq r \leq 150\hMpc$, and we choose $\Delta r = 5\hMpc$ and $10\hMpc$ for the 2PCF and 3PCF bin widths, respectively. The 2PCF and 3PCF multipoles used are the monopole and quadrupole 2PCFs ($\xi_0$ and $\xi_2$), the two monopole 3PCFs ($\zeta_{000}$ and $\zeta_{202}$), and the two quadrupole 3PCFs ($\zeta_{202}$ and $\zeta_{112}$). Considering $\zeta_{\ell_1\ell_2\ell}(r_1,r_2)=\zeta_{\ell_2\ell_1\ell}(r_2,r_1)$, the numbers of data bins for the 2PCF and 3PCF multipoles are $15$, $15$, $36$, $36$, $64$, and $36$ for $\xi_0$, $\xi_2$, $\zeta_{000}$, $\zeta_{110}$, $\zeta_{202}$, and $\zeta_{112}$, respectively. The reason why the bin width for the 3PCF is wider than for the 2PCF is to reduce the number of data bins and to conservatively estimate the inverse covariance matrix for the 2PCF and 3PCF. The total number of data bins is then $202$, which is small enough compared to the $2048$ Patchy mock simulations (Section~\ref{Sec:CovarianceMatrix}).

\subsection{Parameter setting}
\label{Sec:ParameterSetting}

The parameters we constrain are as follows:
\begin{eqnarray}
    \boldsymbol{\theta} = \boldsymbol{\theta}_{\rm phys} + \boldsymbol{\theta}_{\rm bias},
\end{eqnarray}
where
\begin{eqnarray}
    \boldsymbol{\theta}_{\rm bias} = \{(b_1\sigma_8), (F_{\rm g}\sigma_8), (F_{\rm t}\sigma_8) \},
\end{eqnarray}
and
\begin{eqnarray}
    \boldsymbol{\theta}_{\rm phys}  =
    \begin{cases}
        \{f\sigma_8, \sigma_8\}, \quad \mbox{GR};\\
        \{\sigma_8, \xi_{f}, \xi_{\rm t}\}, \quad \mbox{Horndeski};\\
        \{F_{\rm s}\sigma_8, \xi_{f}, \xi_{\rm s}, \xi_{\rm t}\}, \quad \mbox{DHOST}.\\
    \end{cases}
    \label{Eq:param_phys}
\end{eqnarray}
The parameters $F_{\rm g,s,t}$ and $\xi_{f, \rm s, t}$ are given in Eq.~(\ref{Eq:F_2G_2}) and (\ref{Eq:TimeDependenceOfE}), respectively; the reason $F_{\rm s}$ does not appear in GR and Horndeski theorfies is that $F_{\rm s}=1$ in those theories. We assume that the bias parameters $\boldsymbol{\theta}_{\rm bias}$ take different values in all four BOSS samples. $f\sigma_8$, $\sigma_8$, and $F_{\rm s}\sigma_8$ have common values in NGC and SGC. $\xi_{f,{\rm s},{\rm t}}$ are common to all four BOSS samples. For the 2PCF analysis, we only consider $b_1\sigma_8$ and $f\sigma_8$. For example, if all four BOSS samples are used to constrain DHOST theories, the total number of parameters is $17$. Once again, note that the AP parameters are not varied in this analysis.

\subsection{\texorpdfstring{$M_1$ and $M_2$}{} factors}

\label{Sec:M1M2}

\begin{table}
\centering
\begin{tabular}{lccc}
\hline\hline
& $M_1$ & $M_2$ & $\#$ of params.\\
\hline
2PCF only ($z_{\rm eff}=0.38$) & $1.006$ & $1.013$ & $3$\\
2PCF only ($z_{\rm eff}=0.61$) & $1.006$ & $1.013$ & $3$\\
\hline
GR ($z_{\rm eff}=0.38$) & $1.049$ & $1.105$ & $8$\\
GR ($z_{\rm eff}=0.61$) & $1.049$ & $1.105$ & $8$\\
\hline
Horndeski ($z_{\rm eff}=0.38$) & $1.048$ & $1.104$ & $9$\\
Horndeski ($z_{\rm eff}=0.61$) & $1.048$ & $1.104$ & $9$\\
Horndeski ($z_{\rm eff}=0.38,\, 0.61$) & $1.044$ & $1.100$ & $16$\\
\hline
DHOST ($z_{\rm eff}=0.38$) & $1.048$ & $1.104$ & $10$\\
DHOST ($z_{\rm eff}=0.61$) & $1.048$ & $1.104$ & $10$\\
DHOST ($z_{\rm eff}=0.38,\, 0.61$) & $1.044$ & $1.100$ & $17$\\
\hline
\end{tabular}
\caption{
A summary of the $M_1$ (\ref{Eq:M1}) and $M_2$ (\ref{Eq:M2}) factor values used in our analysis. These values are calculated from the number of the Patchy mock simulations, $2048$ (Section~\ref{Sec:Data}), the number of data bins, $30$ for the 2PCF only and $202$ for the 2+3PCF (Section~\ref{Sec:NumberOfBins}), and the number of parameters summarised in the rightmost column (Section~\ref{Sec:ParameterSetting}).}
\label{Table:M1M2}
\end{table}

As mentioned in Section~\ref{Sec:CovarianceMatrix}, the number of the Patchy mock simulations used to calculate the covariance matrices for the 2PCF and 3PCF is finite, so the inverse of the covariance matrix must be multiplied by the Hartlap factor and the final parameter error by $M_1$. 

The $M_1$ factor~(\ref{Eq:M1}) is derived assuming that all parameters are constrained from a single data set. However, when constraining the common parameters $\xi_{f}$, $\xi_{\rm s}$, and $\xi_{\rm t}$ (\ref{Eq:TimeDependenceOfE}) from the four independent BOSS samples, as in the present analysis, the $M_1$ factor is expected to take a different form, but we do not know the correct correction factor corresponding to the $M_1$ factor in such as case. Therefore, when we use different galaxy samples simultaneously, we first count up all common and non-common parameters in the galaxy samples. Then, we calculate the $M_1$ factor using the number of data bins computed from a single galaxy sample and the number of the Patchy mocks, $2048$, corresponding to that galaxy sample, and multiply it by the final parameter error. Specifically, the multipole components of the 2PCF and 3PCF measured from a single galaxy sample are $\xi_0$, $\xi_2$, $\zeta_{000}$, $\zeta_{110}$, $\zeta_{202}$, and $\zeta_{112}$, for a total data bin number of $202$ (Section~\ref{Sec:NumberOfBins}). The number of parameters depends on the type of analysis; for example, we need $17$ parameters to test DHOST theories using all four galaxy samples (Section~\ref{Sec:ParameterSetting}).

Table~\ref{Table:M1M2} summarises the values of the $M_1$ and $M_2$ factors calculated in our analysis, leading to $M_2\sim1.1$ for all the 2+3PCF joint analyses. Thus, even without considering the Hartlap and $M_1$ factors in our analysis, the effect of finite mocks is at most $10\%$. In other words, since our analysis correctly considers these factors, the error due to the finite mock effect in the estimated parameter error is guaranteed to be $\lesssim10\%$.

\subsection{MCMC}

We apply the Metropolis-Hastings (MH) algorithm, an Markov Chain Monte Carlo (MCMC) method, implemented in the publicly available software package \textsc{Monte Python}~\citep{Audren:2012wb,Brinckmann:2018cvx} to estimate the posterior distribution of parameters in a multi-dimensional parameter space. In doing so, we set the super-update parameter to $20$, as recommended by the developers. In order to improve the convergence of the posterior distributions obtained by MCMC, we first perform an MCMC analysis with the number of steps set to $N_{\rm step}=200\,000$ and calculate the best-fit values and covariance matrix of the parameters. Then, we add the information of the best-fit values and covariance matrix and perform an MCMC analysis again with the same number of steps .

We ensure convergence of each MCMC chain, imposing $R-1 \lesssim {\cal O}(10^{-4})$ where $R$ is the standard Gelman-Rubin criteria~\citep{Gelman:1992StaSc...7..457G}. Furthermore, the convergence of the results is also checked through the following method. First, we create eight independent MCMC chains and compute the mean and standard deviation of the parameters from each chain. Then, from the eight means and standard deviations, we compute the standard deviation of the mean and the mean of the standard deviation and check that the ratio of them is less than about $20\%$ for all the results. 

\subsection{Mock tests}

We perform MCMC analyses on $100$ Patchy mock catalogues~\citep{Kitaura:2015uqa} using the same set of cosmological and nuisance parameters as in the actual BOSS galaxy data analysis. We then verify that our analysis can correctly return the values of the non-linear parameters predicted by GR for the Patchy mock catalogues designed under the assumption of a $\Lambda$CDM model. We also discuss the statistical scatter of the $100$ values of the parameters to be estimated.

\section{Fisher analysis}
\label{Sec:FisherAnalysis}

\begin{table*}

\centering
\begin{tabular}{lccccccc}
\hline\hline
& $(b_1\sigma_8)_{\rm fid}$ & $(f\sigma_8)_{\rm fid}$ & $(F_{\rm g}\sigma_8)_{\rm fid}$ & $(F_{\rm s}\sigma_8)_{\rm fid}$ & $(F_{\rm t}\sigma_8)_{\rm fid}$ & $(G_{\rm s}\sigma_8)_{\rm fid}$ & $(G_{\rm t}\sigma_8)_{\rm fid}$\\
\hline
& $1.362$ & $0.485$ & $0.552$ & $0.681$ & $0.194$ & $0.681$ & $0.386$ \\
\hline
& $\sigma_{\rm fisher}(b_1\sigma_8)$ & $\sigma_{\rm fisher}(f\sigma_8)$
& $\sigma_{\rm fisher}(F_{\rm g}\sigma_8)$ 
& $\sigma_{\rm fisher}(F_{\rm s}\sigma_8)$ & $\sigma_{\rm fisher}(F_{\rm t}\sigma_8)$ 
& $\sigma_{\rm fisher}(G_{\rm s}\sigma_8)$ & $\sigma_{\rm fisher}(G_{\rm t}\sigma_8)$\\
\hline
\multicolumn{8}{c}{GR} \\
\hline
Case 1 & $0.159$ & $0.093$ & $-$ & $-$ & $-$ & $-$ & $-$            \\
Case 2 & $0.154$ & $0.093$ & $0.418$ & $0.472$ & $0.312$ & $-$ & $-$\\
Case 3 & $0.159$ & $0.092$ & $0.802$ & $1.170$ & $2.387$ & $-$ & $-$\\
Case 4 & $0.155$ & $0.091$ & $0.420$ & $0.643$ & $0.426$ & $-$ & $-$\\
Case 5 & $0.151$ & $0.090$ & $0.330$ & $0.450$ & $0.291$ & $-$ & $-$\\
Case 6 & $0.155$ & $0.089$ & $0.396$ & $0.619$ & $0.411$ & $-$ & $-$\\
Case 7 & $0.151$ & $0.089$ & $0.315$ & $0.442$ & $0.283$ & $-$ & $-$\\
Case 8 & $0.153$ & $0.091$ & $0.363$ & $0.492$ & $0.324$ & $-$ & $-$\\
\hline
\multicolumn{8}{c}{Horndeski} \\
\hline
Case 2 & $0.154$ & $0.093$ & $2.361$ & $0.479$ & $3.156$ & $-$ & $29.25$\\
Case 3 & $0.159$ & $0.092$ & $0.816$ & $2.433$ & $2.633$ & $-$ & $1.312$ \\
Case 4 & $0.155$ & $0.092$ & $0.430$ & $0.726$ & $0.431$ & $-$ & $1.111$ \\
Case 5 & $0.151$ & $0.091$ & $0.331$ & $0.463$ & $0.295$ & $-$ & $1.044$ \\
Case 6 & $0.155$ & $0.091$ & $0.409$ & $0.715$ & $0.419$ & $-$ & $1.015$ \\
Case 7 & $0.151$ & $0.091$ & $0.316$ & $0.459$ & $0.285$ & $-$ & $0.946$ \\
Case 8 & $0.153$ & $0.091$ & $0.366$ & $0.498$ & $0.344$ & $-$ & $1.383$ \\
\hline
\multicolumn{8}{c}{DHOST} \\
\hline
Case 2 & $0.154$ & $0.093$ & $3.100$ & $3.840$ & $4.774$ & $35.22$ & $38.88$\\
Case 3 & $0.159$ & $0.093$ & $2.211$ & $3.652$ & $2.633$ & $2.976$ & $1.981$  \\
Case 4 & $0.157$ & $0.092$ & $0.726$ & $0.748$ & $0.446$ & $1.896$ & $1.112$  \\
Case 5 & $0.153$ & $0.092$ & $0.499$ & $0.476$ & $0.303$ & $1.636$ & $1.051$  \\
Case 6 & $0.156$ & $0.091$ & $0.710$ & $0.739$ & $0.431$ & $1.757$ & $1.017$  \\
Case 7 & $0.153$ & $0.091$ & $0.487$ & $0.468$ & $0.293$ & $1.519$ & $0.953$  \\
Case 8 & $0.154$ & $0.092$ & $0.595$ & $0.528$ & $0.344$ & $2.292$ & $1.580$  \\
\hline
\end{tabular}
\caption{
    The standard deviations of the parameters as predicted by the Fisher analysis, denoted as $\sigma_{\rm fisher}(\boldsymbol{\theta})$, are shown. These results are for the NGC at $z_{\rm eff}=0.38$. The parameter vectors of interest, $\boldsymbol{\theta}$, are different for each of the three gravity theories, GR, Horndeski, and DHOST theories (Eq.~\ref{Eq:param_vector}). The classification of the data vectors used is as shown in Eq.~(\ref{Eq:case}). The fiducial values of the parameters are calculated under the assumption of GR and are denoted as $(\boldsymbol{\theta})_{\rm fid}$. The scale range used for this Fisher analysis is $80\hMpc\leq r \leq 150\hMpc$.
}
\label{Table:Fisher}
\centering
\begin{tabular}{lccccccc}
\hline\hline
& $(E_f)_{\rm fid}$ & $(E_{\rm s})_{\rm fid}$ & $(E_{\rm t})_{\rm fid}$ & & $(\xi_{f})_{\rm fid}$ & $(\xi_{\rm s})_{\rm fid}$ & $(\xi_{\rm t})_{\rm fid}$\\
\hline
& $0.713$ & $1.000$ & $0.992$ & & $0.545$ & $0.000$ & $0.013$\\
\hline
& $\sigma_{\rm fisher}(E_f)$ & $\sigma_{\rm fisher}(E_{\rm s})$ & $\sigma_{\rm fisher}(E_{\rm t})$ &
& $\sigma_{\rm fisher}(\xi_{f})$ & $\sigma_{\rm fisher}(\xi_{\rm s})$ & $\sigma_{\rm fisher}(\xi_{\rm t})$\\
\hline
\multicolumn{8}{c}{Horndeski} \\
\hline
Case 2 & $0.502$ & $-$ & $75.28$ & & $1.146$ & $-$ & $124.3$\\
Case 3 & $2.558$ & $-$ & $6.018$ & & $5.844$ & $-$ & $9.934$   \\
Case 4 & $0.760$ & $-$ & $3.189$ & & $1.737$ & $-$ & $5.264$   \\
Case 5 & $0.488$ & $-$ & $2.767$ & & $1.114$ & $-$ & $4.567$   \\
Case 6 & $0.750$ & $-$ & $2.969$ & & $1.714$ & $-$ & $4.901$   \\
Case 7 & $0.483$ & $-$ & $2.522$ & & $1.103$ & $-$ & $4.163$   \\
Case 8 & $0.523$ & $-$ & $3.597$ & & $1.194$ & $-$ & $5.938$   \\
\hline
\multicolumn{8}{c}{DHOST} \\
\hline
Case 2 & $4.018$ & $57.31$ & $103.8$ & & $9.180$ & $93.85$ & $171.3$ \\
Case 3 & $3.824$ & $6.519$ & $9.837$ & & $8.737$ & $10.68$ & $16.24$     \\
Case 4 & $0.785$ & $2.835$ & $3.213$ & & $1.794$ & $4.642$ & $5.304$       \\
Case 5 & $0.503$ & $2.460$ & $2.771$ & & $1.150$ & $4.029$ & $4.574$       \\
Case 6 & $0.777$ & $2.633$ & $2.998$ & & $1.776$ & $4.311$ & $4.950$       \\
Case 7 & $0.495$ & $2.260$ & $2.527$ & & $1.130$ & $3.701$ & $4.172$       \\
Case 8 & $0.558$ & $3.542$ & $3.981$ & & $1.274$ & $5.801$ & $6.572$       \\
\hline
\end{tabular}
\caption{
    Same as Table~\ref{Table:Fisher}, but the standard deviations of the parameters defined in Eqs.~(\ref{Eq:Est}) and (\ref{Eq:TimeDependenceOfE}), calculated through variable transformations, are shown.
}
\label{Table:Fisher_xi}
\end{table*}

Before proceeding to the MCMC analysis using actual galaxies in Sections~\ref{Sec:GoodnessOfFit} and \ref{Sec:DataAnalysis}, in this section, we will understand how the 3PCF contains cosmological information through the Fisher analysis.

There are several reasons for performing the Fisher analysis before the MCMC analysis. First, calculating the Fisher matrix in Section~\ref{Sec:FisherMatrix} is less computationally intensive than performing the MCMC analysis, making it easier to compare the analysis results in various settings that take too much time in the MCMC analysis. Taking advantage of this, Section~\ref{Sec:InformationIn3PCFs} examines how the constraints on the parameters of interest change for various combinations of the multipole components of the 3PCF; in doing so, we focus only on the NGC sample at $z=0.38$ as a representative example. Section~\ref{Sec:Fisher_4samples} also discusses the relation among the values of the predicted parameter errors for the four BOSS samples, NGC and SGC at $z=0.38,\, 0.61$, and how the combination of the four BOSS samples affects the final results. Finally, in Section~\ref{Sec:ConsistencyCheckWithFisher}, we compare the results obtained from the above Fisher analysis with those obtained from the MCMC parameter estimation and check their consistency to confirm the validity of the final results in this paper.

In Section~\ref{Sec:SmallScales}, the Fisher analysis also allows us to estimate cosmological information at scales smaller than the scale range used in the MCMC analysis. The results are expected to motivate the construction of theoretical models applicable to smaller scales.

Finally, in Section~\ref{Sec:Priors}, we use the results of the Fisher analysis to determine the range of a flat prior used when performing the MCMC analysis.

\subsection{Fisher matrix}
\label{Sec:FisherMatrix}

From the likelihood function given in Eq.~(\ref{Eq:likelihood}), we calculate the Fisher matrix as
\begin{eqnarray}
    F_{ij} \hspace{-0.25cm} &=& \hspace{-0.25cm}
    -\left\langle \frac{\partial}{\partial \theta_i}\frac{\partial}{\partial \theta_j}\ln {\cal L} \right\rangle
    \nonumber \\
    \hspace{-0.25cm} &=& \hspace{-0.25cm}
    \frac{\partial\, \VEC{t}(\boldsymbol{\theta})}{\partial \theta_i} 
    \MAT{C}_{\rm Hartlap}^{-1}\frac{\partial\, \VEC{t}^{\rm T}(\boldsymbol{\theta})}{\partial \theta_j},
\end{eqnarray}
where we assumed that the covariance matrix $\MAT{C}$ is independent of the parameters. The indices $i$ and $j$ run over parameters of interest. In the limit of the Gaussian likelihood surface, the Cramer-Rao inequality shows that the Fisher matrix provides the minimum standard deviation on parameters, marginalized over all the other parameters: $\sigma(\theta_i)\geq\sigma_{\rm fisher}(\theta_i) = \left( F^{-1} \right)_{ii}^{1/2}$. We note that we adopt the inverse covariance matrix, $\MAT{C}_{\rm Hartlap}^{-1}$, that is non-Gaussian estimated from the Patchy mock simulations.

We consider three parameter vectors, depending on the gravity theory of interest:
\begin{eqnarray}
    \boldsymbol{\theta} = \{(b_1\sigma_8), (f\sigma_8)\}
    + \boldsymbol{\theta}_{\rm 3PCF},
\end{eqnarray}
with
\begin{eqnarray}
    &&\boldsymbol{\theta}_{\rm 3PCF} \nonumber \\=
    &&\hspace{-0.6cm}
    \begin{cases}
        \{(F_{\rm g}\sigma_8), (F_{\rm s}\sigma_8), (F_{\rm t}\sigma_8)\}, \hspace{2.99cm} \mbox{GR};\\
        \{(F_{\rm g}\sigma_8), (F_{\rm s}\sigma_8), (F_{\rm t}\sigma_8), (G_{\rm t}\sigma_8)\}, \hspace{0.7cm}
        \quad \mbox{Horndeski};\\
        \{(F_{\rm g}\sigma_8), (F_{\rm s}\sigma_8), (F_{\rm t}\sigma_8),
        (G_{\rm s}\sigma_8), (G_{\rm t}\sigma_8)\}, \hspace{0.2cm} \mbox{DHOST},\\
    \end{cases}
    \label{Eq:param_vector}
\end{eqnarray}
where $F_{\rm s}\sigma_8 = \sigma_8$ for GR and Horndeski theories. We obtain the results for $E_{f, \rm s, t}$ and $\xi_{f, \rm s, t}$ using the variable transformations in Eqs. (\ref{Eq:Est}) and (\ref{Eq:TimeDependenceOfE}). In particular, the results including $\xi_{f, \rm s, t}$ correspond to the parameter set (\ref{Eq:param_phys}) used in the MCMC analysis performed in Section~\ref{Sec:DataAnalysis}.

The fiducial values of the cosmological parameters needed to compute the Fisher matrix are the values in the $\Lambda$CDM model presented in Section~\ref{Sec:Introduction}. In doing so, we assume that the linear bias is $b_1=2$, and the values of the non-linear biases are zero: i.e., $b_2=b_{\rm s^2}=0$.

\subsection{Information contained in 3PCF multipoles}
\label{Sec:InformationIn3PCFs}

For the NGC sample at $z=0.38$, we perform Fisher analyses on the following eight data vectors consisting of combinations of the 2PCF and 3PCF multipole components, using the same settings as the MCMC analysis performed in Section~\ref{Sec:DataAnalysis} to investigate which components and how they affect parameter estimates.
\begin{eqnarray}
    && \mbox{Case 1} \quad \VEC{d} = \{\xi_0, \xi_2\}; \nonumber \\
    && \mbox{Case 2} \quad \VEC{d} = \{\xi_0, \xi_2, \zeta_{000}, \zeta_{110}\}; \nonumber \\
    && \mbox{Case 3} \quad \VEC{d} = \{\xi_0, \xi_2, \zeta_{202}, \zeta_{112}\}; \nonumber \\
    && \mbox{Case 4} \quad \VEC{d} = \{\xi_0, \xi_2, \zeta_{000}, \zeta_{202}\}; \nonumber \\
    && \mbox{Case 5} \quad \VEC{d} = \{\xi_0, \xi_2, \zeta_{000}, \zeta_{202}, \zeta_{110}\}; \nonumber \\
    && \mbox{Case 6} \quad \VEC{d} = \{\xi_0, \xi_2, \zeta_{000}, \zeta_{202}, \zeta_{112}\}; \nonumber \\
    && \mbox{Case 7} \quad \VEC{d} = \{\xi_0, \xi_2, \zeta_{000}, \zeta_{202}, \zeta_{110}, \zeta_{112}\}; \nonumber \\
    && \mbox{Case 8} \quad \VEC{d} = \{\xi_0, \xi_2, \zeta_{110}, \zeta_{112}\}.
    \label{Eq:case}
\end{eqnarray}
Case $1$ constrains $f\sigma_8$ using only the monopole and quadrupole 2PCFs. Cases $2$, $3$, and $4$ add to Case $1$ the two monopole 3PCFs ($\zeta_{000}$ and $\zeta_{110}$), the two quadrupole 3PCFs ($\zeta_{202}$ and $\zeta_{112}$), and the first terms of the monopole and quadrupole 3PCFs ($\zeta_{000}$ and $\zeta_{202}$), respectively. These three cases will highlight the importance of simultaneously considering both monopoles and quadrupoles in the 3PCF. Moreover, Cases $5$, $6$, and $7$ reveal the extent to which the final results can be improved by adding higher-order multipole components to Case $4$. Finally, Case $8$ only uses the higher-order multipoles, $\zeta_{110}$ and $\zeta_{112}$, for the monopole and quadrupole components.

We summarise the results of the Fisher analysis in Table~\ref{Table:Fisher}. In Horndeski and DHOST theories, the case $2$ results show that using only the monopole 3PCFs very weakly constrains the non-linear velocity parameters $G_{\rm s}\sigma_8$ and $G_{\rm t}\sigma_8$. On the other hand, in Case $3$, using only the quadrupole 3PCFs, we can mildly constrain the non-linear coefficients of both the density field and the velocity field. The reason is that the density and velocity fluctuations contribute to the quadrupole 3PCFs to the same extent (Figure~\ref{fig:decom_3PCF_quad}). Moreover, Cases $4$, $5$, $6$, $7$, and $8$, using both the monopole and quadrupole components, can constrain the non-linear coefficients more strongly than Cases $2$ and $3$. In particular, for the $G_{\rm s}\sigma_8$ and $G_{\rm t}\sigma_8$ constraints in DHOST theories, Case $7$ is $\sim20$ and $\sim40$ times better than Case $2$, respectively:
\begin{eqnarray}
    \sigma_{\rm fisher}(G_{\rm s}\sigma_8) \hspace{-0.25cm} &=& \hspace{-0.25cm}35.22 \quad \mbox{for Case 2}, \nonumber \\
    \sigma_{\rm fisher}(G_{\rm t}\sigma_8) \hspace{-0.25cm} &=& \hspace{-0.25cm}38.88 \quad \mbox{for Case 2}, \nonumber \\
    \sigma_{\rm fisher}(G_{\rm s}\sigma_8) \hspace{-0.25cm} &=& \hspace{-0.25cm}1.519 \quad \mbox{for Case 7},  \nonumber \\
    \sigma_{\rm fisher}(G_{\rm t}\sigma_8) \hspace{-0.25cm} &=& \hspace{-0.25cm}0.953 \quad \mbox{for Case 7}.
\end{eqnarray}
These results support the argument of this paper that we should use both monopole and quadrupole 3PCFs to study the non-linearity of the velocity field.

Case $7$, which uses all components of $\zeta_{000}$, $\zeta_{110}$, $\zeta_{202}$, and $\zeta_{112}$, provides the best constraints on $F_{\rm s}\sigma_8$, $G_{\rm s}\sigma_8$, and $G_{\rm t}\sigma_8$, as expected. Therefore, we can conclude that all these multipole components should be used in the MCMC analysis in Section~\ref{Sec:DataAnalysis}.

Case $7$ yields results that are about $10\%$ better than Case $5$, which uses $\zeta_{000}$, $\zeta_{110}$, and $\zeta_{202}$. This result indicates that while $\zeta_{202}$ contains the main cosmological information, $\zeta_{112}$ contains other information in addition to $\zeta_{202}$. Existing studies using \citet{Scoccimarro:1999ed}' decomposition method of the bispectrum tend to ignore the $M\neq0$ mode of the quadrupole component as not containing much cosmological information~\citep[e.g.,][]{Gagrani:2017MNRAS.467..928G,Rizzo:2022arXiv220413628R,D'Amico:2022arXiv220608327D}. However, our results show the importance of the $M\neq0$ modes because $\zeta_{202}$ contains only the $M=0$ mode, while $\zeta_{112}$ further contains the $M\neq0$ modes in addition to the $M=0$ mode (see also Section~\ref{Sec:3PCFs}).

By comparing the results of Case $4$, consisting of $\zeta_{000}$ and $\zeta_{202}$, with those of Case $8$, consisting of $\zeta_{110}$ and $\zeta_{112}$, we can find another viewpoint on the importance of higher-order multipole components. For example, the $(F_{\rm s}\sigma_8)$ constraint is better in Case $8$, and the $(G_{\rm s}\sigma_8)$ and $(G_{\rm t}\sigma_8)$ constraints are better in Case $4$. Also, the $(G_{\rm s,t}\sigma_8)$ result in Case $4$ is only about $30\%$ better than Case $8$. Thus, although $\zeta_{202}$ is more informative than $\zeta_{112}$, we interpret the information on both sides as overlapping to some extent.

We further calculate $\sigma_{\rm fisher}(\boldsymbol{\theta})$ for $\theta=E_f$, $E_{\rm s}$, $E_{\rm t}$, $\xi_f$, $\xi_{\rm s}$, and $\xi_{\rm t}$ through the variable transformations in Eqs. (\ref{Eq:Est}) and (\ref{Eq:TimeDependenceOfE}), summarising the results in Table~\ref{Table:Fisher_xi}. We find that both the monopole and quadrupole components of the 3PCF are needed to constrain $E_{\rm s}$, $E_{\rm t}$, $\xi_{\rm s}$, and $\xi_{\rm t}$ better. In Case $7$, the standard deviations of $E_{\rm s}$ and $E_{\rm t}$ are more than twice larger than the fiducial values of $E_{\rm s}$ and $E_{\rm t}$, i.e., $\sigma_{\rm fisher}(E_{\rm s,t})/(E_{\rm s,t})_{\rm fid}>2$, indicating that it is impossible to detect the $E_{\rm s}$ and $E_{\rm t}$ signals in the BOSS data. We can also confirm that for each of the $\xi_{\rm s}$ and $\xi_{\rm t}$ constraints in DHOST theories, the results of Case $7$ are $\sim25$ and $\sim40$ times stronger than those of Case $2$, respectively:
\begin{eqnarray}
    \sigma_{\rm fisher}(\xi_{\rm s}) \hspace{-0.25cm} &=& \hspace{-0.25cm} 93.85 \quad \mbox{for Case 2}, \nonumber \\
    \sigma_{\rm fisher}(\xi_{\rm t}) \hspace{-0.25cm} &=& \hspace{-0.25cm} 171.3 \quad \mbox{for Case 2}, \nonumber \\
    \sigma_{\rm fisher}(\xi_{\rm s}) \hspace{-0.25cm} &=& \hspace{-0.25cm} 3.701 \quad \mbox{for Case 7},  \nonumber \\
    \sigma_{\rm fisher}(\xi_{\rm t}) \hspace{-0.25cm} &=& \hspace{-0.25cm} 4.172 \quad \mbox{for Case 7}.
    \label{Eq:mono_vs_quad}
\end{eqnarray}

In GR, adding any multipole component of the 3PCF can only improve the $b_1\sigma_8$ and $f\sigma_8$ constraints by a few per cent. This result is consistent with the MCMC analysis of \citet{Sugiyama:2020uil} on the Patchy mock catalogues. Furthermore, the 3PCF-specific information, $\sigma_8$, is also uninformative compared to $f\sigma_8$. Specifically, $f\sigma_8$ can be determined with a precision of $\sim20\%$, while $\sigma_8$ can only reach a precision of $\sim60\%$. These results are for large scales ($r\geq 80\hMpc$); what happens when even smaller scales are used will be discussed in Section~\ref{Sec:SmallScales}.

\subsection{Fisher forecasts with all four BOSS samples}
\label{Sec:Fisher_4samples}

\begin{table*}

\centering
\begin{tabular}{lccccccc}
\hline\hline
\multicolumn{8}{c}{DHOST} \\
\hline
$\sigma_{\rm fisher}(\theta)/(\theta)_{\rm fid}$  
& $(b_1\sigma_8)$ & $(f\sigma_8)$ & $(F_{\rm g}\sigma_8)$ & $(F_{\rm s}\sigma_8)$ & $(F_{\rm t}\sigma_8)$ & $(G_{\rm s}\sigma_8)$ & $(G_{\rm t}\sigma_8)$\\
\hline
NGC at $z=0.38$ & $0.112$ & $0.188$ & $0.883$ & $0.687$ & $1.511$ & $2.231$ & $2.469$ \\
NGC at $z=0.61$ & $0.110$ & $0.181$ & $1.201$ & $0.880$ & $2.077$ & $2.789$ & $3.265$ \\
SGC at $z=0.38$ & $0.188$ & $0.311$ & $1.446$ & $1.142$ & $2.609$ & $3.642$ & $4.123$ \\
SGC at $z=0.61$ & $0.188$ & $0.304$ & $2.055$ & $1.444$ & $3.609$ & $4.721$ & $5.460$ \\
\hline
\end{tabular}
\caption{
The standard deviations computed by the Fisher analysis of Case $7$ in DHOST theories divided by the fiducial values of the parameters, $\sigma_{\rm fisher}(\boldsymbol{\theta})/(\boldsymbol{\theta})_{\rm fid}$, are shown, where $\boldsymbol{\theta}=\{(b_1\sigma_8), (f\sigma_8), (F_{\rm g}\sigma_8), (F_{\rm s}\sigma_8), (F_{\rm t}\sigma_8), (G_{\rm s}\sigma_8), (G_{\rm t}\sigma_8)\}$. These results are for the NGC and SGC at $z=0.38,\,0.61$.}
\label{Table:Fisher_4samples}
\centering
\begin{tabular}{lccccccc}
\hline\hline
\multicolumn{8}{c}{DHOST} \\
\hline
& $\sigma_{\rm fisher}(E_f)/(E_f)_{\rm fid}$ & $\sigma_{\rm fisher}(E_{\rm s})/(E_{\rm s})_{\rm fid}$ & $\sigma_{\rm fisher}(E_{\rm t})/(E_{\rm t})_{\rm fid}$ &
& $\sigma_{\rm fisher}(\xi_{f})$ & $\sigma_{\rm fisher}(\xi_{\rm s})$ & $\sigma_{\rm fisher}(\xi_{\rm t})$\\
\hline
NGC at $z=0.38$ & $0.626$ & $2.260$ & $2.541$ & & $1.130$ & $3.701$ & $4.172$ \\
NGC at $z=0.61$ & $0.890$ & $2.931$ & $3.384$ & & $2.084$ & $6.890$ & $7.955$ \\
SGC at $z=0.38$ & $1.035$ & $3.653$ & $4.234$ & & $1.867$ & $5.983$ & $6.950$ \\
SGC at $z=0.61$ & $1.469$ & $4.878$ & $5.677$ & & $3.439$ & $11.468$ & $13.345$ \\
\hline
\end{tabular}
\caption{
Same as Figure~\ref{Table:Fisher_4samples}, but $\sigma_{\rm fisher}(\boldsymbol{\theta})/(\boldsymbol{\theta})_{\rm fid}$ for $\boldsymbol{\theta}=\{E_{f}, E_{\rm s}, E_{\rm t}\}$ and $\sigma_{\rm fisher}(\boldsymbol{\theta})$ for  $\boldsymbol{\theta}=\{\xi_{f}, \xi_{\rm s}, \xi_{\rm t}\}$.
}
\label{Table:Fisher_xi_4samples}
\end{table*}

In this subsection, we repeat the analysis of Case $7$ in DHOST theories, performed in Section~\ref{Sec:InformationIn3PCFs}, for the other three BOSS samples, NGC at $z=0.61$ and SGC at $z=0.38,\,0.61$, and summarise the results in Tables~\ref{Table:Fisher_4samples} and~\ref{Table:Fisher_xi_4samples}.

Table~\ref{Table:Fisher_4samples} shows that the results for $(b_1\sigma_8)$ and $(f\sigma_8)$, which are mainly determined by the 2PCF, are slightly better for the sample at $z=0.61$ than for the sample at $z=0.38$ for both NGC and SGC. On the other hand, for the 3PCF-specific parameters, $(F_{\rm g,s,t}\sigma_8)$ and $(G_{\rm s,t}\sigma_8)$, the error is smaller for the $z=0.38$ sample than for the $z=0.61$ sample. This result reflects the different characteristics of the cumulative ${\rm S/N}$ between the 2PCF and the 3PCF, as discussed in Section~\ref{Sec:CumulativeSignalToNoiseRatio}. In other words, it suggests that higher number densities are more favourable than larger survey volumes for constraining the non-linear parameters, $(F_{\rm g,s,t}\sigma_8)$ and $(G_{\rm s,t}\sigma_8)$, using 3PCF measurements.

Table~\ref{Table:Fisher_xi_4samples} summarises the results of $E_{f,{\rm s},{\rm t}}$ and $\xi_{f,{\rm s},{\rm t}}$. As expected, the $z=0.38$ sample gives a smaller error than the $z=0.61$ sample for both $E_{f,{\rm s},{\rm t}}$ and $\xi_{f,{\rm s},{\rm t}}$. However, in the case of $\xi_{f,{\rm s},{\rm t}}$, the error at $z=0.38$ is almost twice as small as that at $z=0.61$, which is extremely favourable for the $z=0.38$ sample. For example, the $\xi_{s}$ results for the NGC samples are
\begin{eqnarray}
    \sigma_{\rm fisher}(\xi_{\rm s}) \hspace{-0.25cm} &=& \hspace{-0.25cm} 3.701\quad \mbox{for NGC at $z=0.38$},  \nonumber \\
    \sigma_{\rm fisher}(\xi_{\rm s}) \hspace{-0.25cm} &=& \hspace{-0.25cm} 6.890\quad \mbox{for NGC at $z=0.61$}.
\end{eqnarray}
This is because we parameterise the time evolution of $E_{f,{\rm s},{\rm t}}$ as $E_{f,{\rm s},{\rm t}} = \Omega_{\rm m}^{\xi_{f,{\rm s},{\rm t}}}$. That is, because $d\xi_{f,{\rm s},{\rm t}} = d \ln E_{f,{\rm s},{\rm t}} / (\ln \Omega_{\rm m})$, the errors in $\xi_{f,{\rm s},{\rm t}}$ are smaller for lower redshifts with smaller values of $\Omega_{\rm m}$. Specifically, in the LCDM model introduced in Section~\ref{Sec:Introduction}, $\Omega_{\rm m}(z=0.38)=0.54$ and $\Omega_{\rm m}(z=0.61)=0.65$, so $1/\ln \Omega_{\rm m}(z=0.38)=1.62$ and $1/\ln \Omega_{\rm m}(z=0.61)=2.32$. Even if $\sigma_{\rm fisher}(E_{f, \rm s,t})/(E_{f, \rm s,t})_{\rm fid}$ has the same value at the two redshifts of $z=0.38$ and $z=0.61$, the value of $\sigma_{\rm fisher}(\xi_{f, \rm s,t})$ at $z=0.38$ is $2.32/1.62=1.42$ times smaller than at $z=0.61$.

\subsection{Fisher forecasts using smaller scales}
\label{Sec:SmallScales}

So far, we have performed the Fisher analysis in the same setting as the MCMC analysis that will be performed in Section~\ref{Sec:DataAnalysis}. There, we have dealt with the behaviour of only large scales, $80\hMpc\leq r \leq 150\hMpc$. However, seeing how the parameter constraints improve when the minimum scale used, $r_{\rm min}$, is varied should be an excellent motivation for the future development of theoretical models. 

\begin{figure*}
    \scalebox{0.95}{\includegraphics[width=\textwidth]{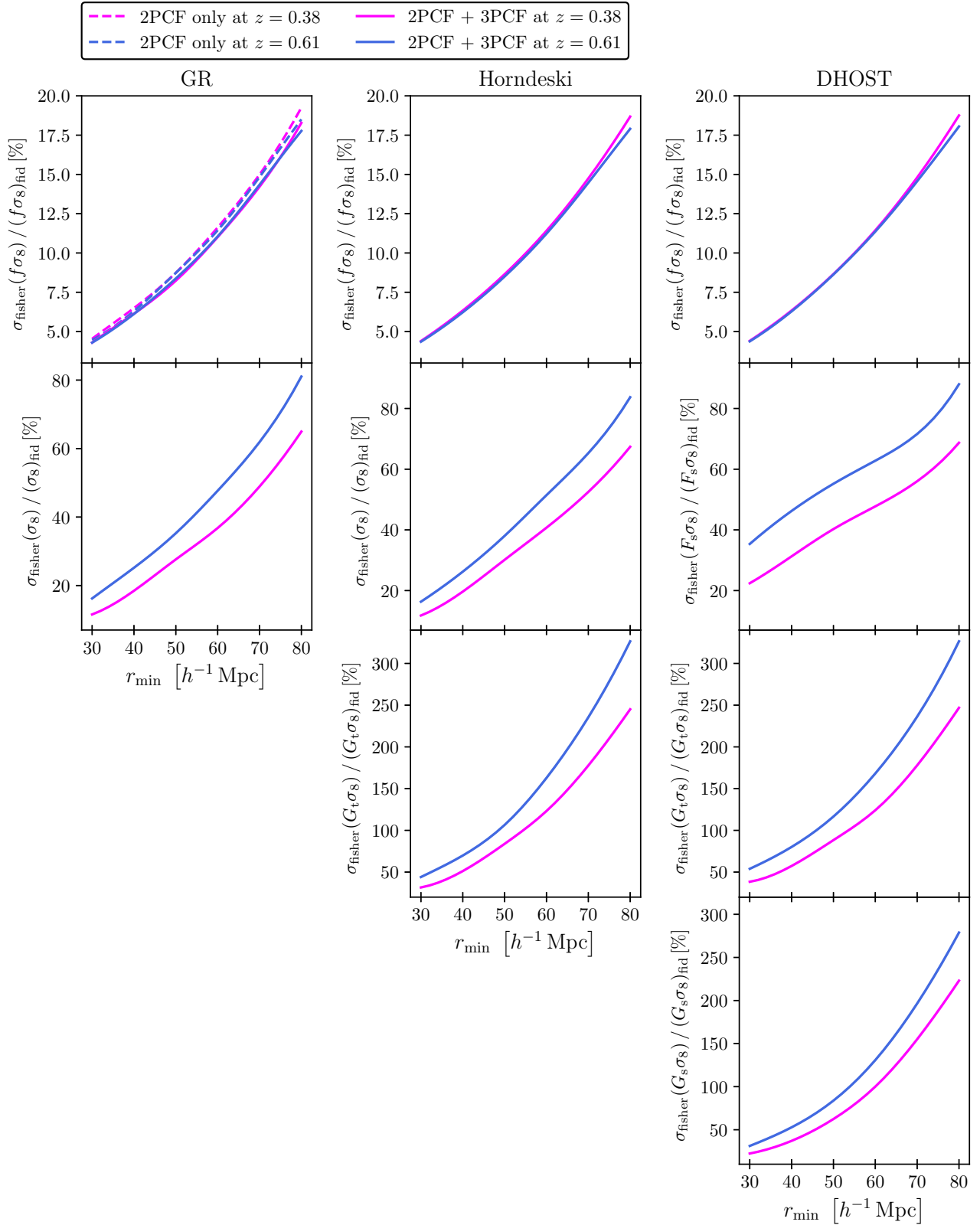}}
    \caption{The standard deviations computed by the Fisher analysis divided by the fiducial values of the parameter, $\sigma_{\rm fisher}(\theta)/(\theta)_{\rm fid}$, are shown as a function of the minimum scale used, $r_{\rm min}$. These results are for the NGC at $z_{\rm eff}=0.38,\,0.61$. The solid lines show the results using the multipole components of the 2PCF and 3PCF given in Case $7$ (\ref{Eq:case}), and the dashed lines are for the 2PCF-only analysis, Case $1$. The points at $r_{\rm min}=80\hMpc$ in the right panels correspond to the results in Table~\ref{Table:Fisher_4samples}.}
	\label{fig:fisher}
\end{figure*}

Figure~\ref{fig:fisher} plots $\sigma_{\rm fisher}(\theta)/(\theta)_{\rm fid}$ as a function of $r_{\rm min}$ for the three gravity theories, GR, Horndeski, and DHOST, at two redshifts of $z=0.38$ (magenta lines) and $z=0.61$ (blue lines). The multipole components of the 2PCF and 3PCF used here are Case $7$ (Eq.~(\ref{Eq:case})). First, even on the smaller scale, adding the 3PCF hardly improves the $f\sigma_8$ constraint compared to the case where only the 2PCF is used (compare solid and dashed lines in the top left panel of Figure~\ref{fig:fisher}.). On the other hand, at $r_{\rm min}=30\hMpc$, the $\sigma_8$ constraint reaches a precision of $\sim 10\%$, from which useful cosmological information may be extracted: e.g., $f=(f\sigma_8)/\sigma_8$ can be determined with a precision of $\sim10\%$. 

In addition, the non-linear velocity parameters, $G_{\rm s}\sigma_8$ and $G_{\rm t}\sigma_8$, can be determined with $30\mathchar`-50\%$ precision at $r_{\rm min}=30\hMpc$. Thus, future galaxy surveys with even larger volumes than the BOSS survey, such as DESI, Euclid, and PFS, may detect such non-linear coefficients of the velocity field.

Note that we obtained the Fisher analysis results using the IR-resumed tree-level solutions of the 2PCF and 3PCF given in Eqs.~(\ref{Eq:power_IR}) and (\ref{Eq:bispec_IR}). Although these models accurately describe the non-linear damping behaviour of BAO on large scales, they cannot predict the 2PCF and 3PCF on small scales with high accuracy. Therefore, to apply these models to smaller scales, it is necessary to account for non-linear effects, called loop correction terms. We leave to investigate how the results change when such a loop correction is added for future research.

\subsection{Flat priors}
\label{Sec:Priors}

\begin{table}

\centering
\begin{tabular}{lr}
\hline\hline
\multicolumn{2}{c}{Prior range} \\
\hline
$(b_1\sigma_8)_{\rm NGC,\, z=0.38}$ & $[0.60, 2.13]$ \\
$(b_1\sigma_8)_{\rm SGC,\, z=0.38}$ & $[0.08, 2.64]$ \\
$(b_1\sigma_8)_{\rm NGC,\, z=0.61}$ & $[0.54, 1.88]$ \\
$(b_1\sigma_8)_{\rm SGC,\, z=0.61}$ & $[0.07, 2.36]$ \\
\hline
$(f\sigma_8)_{\rm NGC,\, z=0.38}$ & $[0.03, 0.94]$  \\
$(f\sigma_8)_{\rm SGC,\, z=0.38}$ & $[0.00, 1.24]$ \\
$(f\sigma_8)_{\rm NGC,\, z=0.61}$ & $[0.05, 0.91]$  \\
$(f\sigma_8)_{\rm SGC,\, z=0.61}$ & $[0.00, 1.21]$ \\
\hline
$(F_{\rm g}\sigma_8)_{\rm NGC,\, z=0.38}$ & $[0.00, 2.99]$ \\
$(F_{\rm g}\sigma_8)_{\rm SGC,\, z=0.38}$ & $[0.00, 4.54]$ \\
$(F_{\rm g}\sigma_8)_{\rm NGC,\, z=0.61}$ & $[0.00, 3.44]$ \\
$(F_{\rm g}\sigma_8)_{\rm SGC,\, z=0.61}$ & $[0.00, 5.54]$ \\
\hline
$(F_{\rm s}\sigma_8)_{\rm NGC,\, z=0.38}$ & $[0.00, 3.02]$ \\
$(F_{\rm s}\sigma_8)_{\rm SGC,\, z=0.38}$ & $[0.00, 4.57]$ \\
$(F_{\rm s}\sigma_8)_{\rm NGC,\, z=0.61}$ & $[0.00, 3.27]$ \\
$(F_{\rm s}\sigma_8)_{\rm SGC,\, z=0.61}$ & $[0.00, 4.99]$ \\
\hline
$(F_{\rm t}\sigma_8)_{\rm NGC,\, z=0.38}$ & $[0.00, 1.66]$ \\
$(F_{\rm t}\sigma_8)_{\rm SGC,\, z=0.38}$ & $[0.00, 2.72]$ \\
$(F_{\rm t}\sigma_8)_{\rm NGC,\, z=0.61}$ & $[0.00, 1.97]$ \\
$(F_{\rm t}\sigma_8)_{\rm SGC,\, z=0.61}$ & $[0.00, 3.29]$ \\
\hline
$(\xi_{f})_{\rm NGC,\, z=0.38}$ & $[-5.11, 6.20]$   \\
$(\xi_{f})_{\rm SGC,\, z=0.38}$ & $[-8.79, 9.88]$   \\
$(\xi_{f})_{\rm NGC,\, z=0.61}$ & $[-9.88, 10.97]$  \\
$(\xi_{f})_{\rm SGC,\, z=0.61}$ & $[-16.65, 17.74]$ \\
\hline
$(\xi_{\rm s})_{\rm NGC,\, z=0.38}$ & $[-18.51, 11.10]$ \\
$(\xi_{\rm s})_{\rm SGC,\, z=0.38}$ & $[-29.91, 17.95]$ \\
$(\xi_{\rm s})_{\rm NGC,\, z=0.61}$ & $[-34.45, 20.67]$ \\
$(\xi_{\rm s})_{\rm SGC,\, z=0.61}$ & $[-57.34, 34.40]$ \\
\hline
$(\xi_{\rm t})_{\rm NGC,\, z=0.38}$ & $[-20.84, 12.53]$ \\
$(\xi_{\rm t})_{\rm SGC,\, z=0.38}$ & $[-34.74, 20.86]$ \\
$(\xi_{\rm t})_{\rm NGC,\, z=0.61}$ & $[-39.76, 23.88]$ \\
$(\xi_{\rm t})_{\rm SGC,\, z=0.61}$ & $[-66.71, 40.05]$ \\
\hline
\end{tabular}
\caption{The flat priors for the parameters that we employ in our MCMC analysis are shown. The results are calculated from Case $7$ in Eq.~(\ref{Eq:case}) assuming DHOST theories.
}
\label{Table:Prior}
\end{table}

As shown by the results of the Fisher analysis in Section~\ref{Sec:FisherAnalysis}, the constraints on the non-linear parameters $\xi_{f, \rm s,t}$ constrained by the 3PCF measured from BOSS are weak. Therefore, we need to set appropriate priors to efficiently perform the MCMC analysis.

We use the Fisher analysis results of Case $7$ in DHOST theories for the four BOSS galaxy samples, performed in Section~\ref{Sec:Fisher_4samples}. Then, we adopt a flat prior of $\theta_{\rm fid}\pm 5\sigma_{\rm fisher}(\theta)$ as the base setting for all parameters. If using several samples to constrain common parameters, we adopt a narrower range of priors for those samples. For example, at $z_{\rm eff}=0.38$, when constraining $f\sigma_8$ using both NGC and SGC samples, we adopt the prior computed in NGC. After this basic setting, we set a stronger prior based on further physical considerations below.

The linear bias $b_1$, the linear growth rate $f$, and $\sigma_8$ are always positive by definition: i.e., $b_1\sigma_8\geq0$ and $f\sigma_8\geq 0$.

In the case of GR, the non-linear parameters to be constrained are $F_{\rm g}\sigma_8$, $\sigma_8$, and $F_{\rm t}\sigma_8$. The non-linear local bias parameter $(1/2)(b_2/b_1)$ appearing in $F_{\rm g}$ is calculated to be $-0.02$ for $b_1=2.0$ using the fitting formula given by~\citet{Lazeyras:2016JCAP...02..018L}, which is sufficiently small compared to $17/21$. The tidal bias parameter $(b_{s^2}/b_1)$ appearing in $F_{\rm t}$ is also calculated to be $b_{s^2}/b_1=(-2/7)(1-1/b_1)=-0.14$ for the linear Lagrangian bias model~\citep[e.g.,][]{Desjacques:2016bnm}, and its value is also smaller than $2/7$. Therefore, even if the non-linear bias parameter is present, $F_{\rm g}\sigma_8$ and $F_{\rm t}\sigma_8$ are expected to be larger than zero: i.e., $F_{\rm g}\sigma_8\geq0$ and $F_{\rm t}\sigma_8\geq0$. We will discuss the validity of the analysis results when these conditions are imposed in Section~\ref{Sec:GR_bias} by comparing them with the results when $F_{\rm g}\sigma_8$ and $F_{\rm t}\sigma_8$ can take negative values.

In the cases of Horndeski and DHOST theories, the parameterisation we adopt describes the time evolution of the coefficients of the tidal and shift terms as powers of $\Omega_{\rm m}$ (Section~\ref{Sec:TimeDependencyOfParameters}), implicitly assuming that these coefficients are always positive: i.e., $F_{\rm s}\sigma_8\geq0$, $G_{\rm s}\sigma_8\geq0$, and $G_{\rm t}\sigma_8\geq0$. For $F_{\rm g}\sigma_8$ and $F_{\rm t}\sigma_8$, assuming that Horndeski and DHOST theories are not far from GR, we adopt $F_{\rm g}\sigma_8\geq0$ and $F_{\rm t}\sigma_8\geq0$, just like GR.

The Fisher analysis shows that the BOSS data cannot detect the $E_{{\rm s},{\rm t}}$ signals and only give them an upper limit (Section~\ref{Sec:FisherAnalysis}). This fact means that as $E_{{\rm s},{\rm t}}$ approach zero, the parameters $\xi_{{\rm s},{\rm t}}=\log_{\Omega_{\rm m}} E_{{\rm s},{\rm t}}$ can be as large as desired because of $\Omega_{\rm m} < 1$. Therefore, in this analysis, we set the upper limit of $\xi_{{\rm s},{\rm t}}$ to $(\xi_{\rm s,t})_{\rm fid}+3\sigma_{\rm fisher}(\xi_{\rm s,t})$, which is narrower than the basic setting. If $\xi_{{\rm s},{\rm t}}$ reach their upper bounds set here, we report only the lower bounds for those parameters as the final results.

We summarise the results of the above discussion in Table~\ref{Table:Prior}.

\section{Goodness of fit}
\label{Sec:GoodnessOfFit}

In this section, we examine the extent to which our analysis can give good fits to the 2PCF and 3PCF measurements from the BOSS data or Patchy mocks for a variety of cases, before presenting specific parameter constraint values in Section~\ref{Sec:DataAnalysis}. 

For this purpose, we calculate the minimum of $\chi^2 = -2 \ln {\cal L}$ (\ref{Eq:likelihood}), denoted $\chi^2_{\rm min}$, from the best-fit parameter values obtained from the joint analysis of the 2PCF and 3PCF. We use two multipole 2PCFs ($\xi_0$ and $\xi_2$), two monopole 3PCFs ($\zeta_{000}$ and $\zeta_{110}$), and two quadrupole 3PCFs ($\zeta_{202}$ and $\zeta_{112}$) in this analysis; the assumed gravity theories are GR, Horndeski, and DHOST theories. Tables~\ref{Table:results_chi2_Pvalue}-\ref{Table:results_chi2_Pvalue_rescaled} show the $\chi^2_{\rm min}$ divided by the degrees of freedom (DoF), i.e., the reduced $\chi^2_{\rm min}$, and the corresponding one-tailed $p$-values. At two redshift bins, $z=0.38$ and $z=0.61$, results are presented for NGC only, SGC only, and both NGC and SGC. In Horndeski and DHOST theories, we constrain the common parameters $\xi_{f,{\rm s},{\rm t}}$ among different redshift bins using the samples at both the redshift bins. Finally, we also include the results of the analysis using only 2PCF.

If the theoretical model fits the measurements well, the $p$-value should be close to $0.5$. A $p$-value close to $1$ does not mean that the theoretical model is correct, but that the theoretical model can explain the measurements within the error range, thanks to too large statistical errors in the measurements. On the other hand, a $p$-value close to $0$ indicates that the theoretical model cannot explain the measurements. In this paper, we decide that if $p<0.05$, attention should be paid to the consistency between the theoretical model and the measurements, and if $p<0.01$, there is an apparent discrepancy between them. We write in bold the $\chi^2$ and $p$ values shown in Tables~\ref{Table:results_chi2_Pvalue}-\ref{Table:results_chi2_Pvalue_rescaled} if $p<0.01$. Finally, we comment on the behaviour of $p$-values when combining different galaxy samples. For example, suppose that the reduced $\chi^2_{\rm min}$ is larger than $1$: i.e., $\chi^2_{\rm min}/{\rm DoF}>1$. In this case, if we increase the values of $\chi^2_{\rm min}$ and DoF by an equal factor while keeping the value of the reduced $\chi^2_{\rm min}$, the resulting $p$-value will be smaller than the original value, and conversely, if $\chi^2_{\rm min}/{\rm DoF}<1$, it will be larger than the original value. Since we treat the different galaxy samples as statistically independent, a similar situation occurs in analyses with multiple galaxy samples. Thus, if the $p$-value obtained from each galaxy sample is small, combining galaxy samples will yield a smaller $p$-value.

Section~\ref{Sec:GoFBOSS} reports an unexplained discrepancy between the 3PCF measured from the BOSS galaxy data at $z=0.38$ and our theoretical model on large scales, even considering DHOST theories, which is beyond GR. Section~\ref{Sec:GoFMonoQuad} shows that this discrepancy between the data and the theoretical model appears from the monopole 3PCF. Section~\ref{Sec:DHOSTnoprior} shows that the discrepancy still appears even when the parameter prior set introduced in Section~\ref{Sec:Priors} is removed. Section~\ref{Sec:GoFPatchy} confirms that the discrepancy does not appear in the analysis using the Patchy mock. Finally, as a temporary measure, we rescale the covariance matrix of the 3PCF at $z=0.38$ to generate acceptable $p$-values in Section~\ref{Sec:RescalingOfCovarianceMatrix}. Section~\ref{Sec:DataAnalysis} will report the parameter estimation results with and without rescaling the covariance matrix. 

\subsection{BOSS galaxies}
\label{Sec:GoFBOSS}

\begin{table*}
\centering
\begin{tabular}{lccc}
\hline\hline
\multicolumn{4}{c}{BOSS DR12} \\
\hline
\multicolumn{4}{c}{$\chi^2_{\rm min}/{\rm DoF}$ ($p$-value)} \\
\hline
& NGC + SGC & NGC & SGC \\
\hline
2PCF only ($z_{\rm eff}=0.38$) & $56.04/57$ $(0.511)$ & $32.18/28$ $(0.267)$ & $23.36/28$ $(0.715)$ \\
2PCF only ($z_{\rm eff}=0.61$) & $80.24/57$ $(0.023)$ & $42.08/28$ $(0.043)$ & $36.94/28$ $(0.120)$ \\
\hline
GR ($z_{\rm eff}=0.38$) & $\mathbf{488.38/396}$ $\mathbf{(0.001)}$ & $238.22/197$ $(0.024)$ & $\mathbf{248.72/197}$ $\mathbf{(0.007)}$ \\
GR ($z_{\rm eff}=0.61$) & $428.60/396$ $(0.125)$ & $218.48/197$ $(0.140)$ & $209.24/197$ $(0.262)$ \\
\hline
Horndeski ($z_{\rm eff}=0.38$) & $\mathbf{488.24/395}$ $\mathbf{(0.001)}$ & $236.38/196$ $(0.026)$ & $\mathbf{251.62/196}$ $\mathbf{(0.004)}$ \\
Horndeski ($z_{\rm eff}=0.61$) & $427.56/395$ $(0.125)$ & $218.06/196$ $(0.134)$ & $209.16/196$ $(0.247)$ \\
Horndeski ($z_{\rm eff}=0.38,\, 0.61$) & $\mathbf{918.76/792}$ $\mathbf{(0.001)}$ & $456.50/394$ $(0.016)$ & $458.94/394$ $(0.013)$ \\
\hline
DHOST ($z_{\rm eff}=0.38$) & $\mathbf{487.48/394}$ $\mathbf{(0.001)}$ & $235.80/195$ $(0.024)$ & $\mathbf{248.36/195}$ $\mathbf{(0.006)}$ \\
DHOST ($z_{\rm eff}=0.61$) & $427.62/394$ $(0.117)$ & $217.94/195$ $(0.125)$ & $209.06/195$ $(0.233)$ \\
DHOST ($z_{\rm eff}=0.38,\, 0.61$) & $\mathbf{918.04/791}$ $\mathbf{(0.001)}$ & $455.96/393$ $(0.015)$ & $458.88/393$ $(0.012)$ \\
\hline
\end{tabular}
\caption{The reduced $\chi^2$ and $p$-values (in round brackets) obtained from the joint analysis of the 2PCF and 3PCF are shown. These values are written in bold if $p<0.01$. The minimum $\chi^2$, denoted $\chi_{\rm min}^2$, is calculated from the best-fit parameters. The data used is the BOSS DR12 galaxy, split into two sky regions, NGC and SGC, and two redshift bins, $z=0.38$ and $z=0.61$. The joint analysis shows the results for three gravity theories, i.e., GR, Horndeski, and DHOST theories; for Horndeski and DHOST theories, also shown are the results using the two redshift bins to constrain the parameters $\xi_{f, {\rm s}, {\rm t}}$, which characterises the time evolution of the linear and non-linear effects of the velocity field. Furthermore, the results for the 2PCF-only analysis are shown. The combinations of the 2PCF and 3PCF multipole components used in this analysis correspond to Case $1$ and Case $7$ in Eq.~\ref{Eq:case}.
}
\label{Table:results_chi2_Pvalue}
\end{table*}

Table~\ref{Table:results_chi2_Pvalue} shows the results from the analysis method described in this paper. We have performed the MCMC analysis (Section~\ref{Sec:AnalysisSettings}) using $\xi_0$, $\xi_2$, $\zeta_{000}$, $\zeta_{110}$, $\zeta_{202}$, and $\zeta_{112}$ measured from the BOSS galaxy data (Section~\ref{Sec:Measurements}), the covariance matrix computed from the $2048$ Patchy mocks (Section~\ref{Sec:CovarianceMatrix}), and the flat prior of the parameter range (Section~\ref{Sec:Priors}).

First, we focus on the analysis case using only the 2PCF. For the NGC+SGC sample, the obtained $p$-values are $p=0.511$ at $z=0.38$ and $p=0.023$ at $z=0.61$. This $p=0.023$ indicates a small amount of a poor fit between the model and the measurements, but we consider it not problematic.

Next, turning to the joint analysis results of the 2PCF and 3PCF assuming GR, we find that the $p$-value at $z=0.38$ obtained for the NGC+SGC sample is extremely small, $0.001$. At $z=0.38$, the results for only NGC and only SGC are $p=0.024$ and $p=0.007$, indicating that the SGC sample is more problematic than the NGC. On the other hand, the $p$-value at $z=0.61$ for the NGC+SGC sample is $p=0.125$, indicating that our model explains the measured values without problems.

Finally, for Horndeski and DHOST theories, we find results similar to the GR case: the $p$-value is $0.001$ at $z=0.38$ and $p\sim0.1$ at $z=0.61$ for the NGC+SGC sample.

Thus, we conclude that there is an unexplained discrepancy between the 3PCF measurement from the BOSS sample at $z=0.38$ and the theoretical model we are using. Even Horndeski and DHOST theories, which are modified gravity theories beyond GR, cannot explain this discrepancy.

\subsection{Monopole- or Quadrupole-only 3PCF}
\label{Sec:GoFMonoQuad}

\begin{table*}
\centering
\begin{tabular}{lccc}
\hline\hline
\multicolumn{4}{c}{Joint analysis with monopole 3PCFs ($\zeta_{000}$ and $\zeta_{110}$) only} \\
\hline
\multicolumn{4}{c}{$\chi^2_{\rm min}/{\rm DoF}$ ($p$-value)} \\
\hline
& NGC + SGC & NGC & SGC \\
\hline
GR ($z_{\rm eff}=0.38$) & $\mathbf{252.08/196}$ $\mathbf{(0.004)}$ & $122.56/97$ $(0.041)$ & $127.32/97$ $(0.021)$ \\
GR ($z_{\rm eff}=0.61$) & $216.54/196$ $(0.150)$ & $109.00/97$ $(0.191)$ & $106.94/97$ $(0.230)$ \\
\hline
Horndeski ($z_{\rm eff}=0.38$) & $\mathbf{252.02/195}$ $\mathbf{(0.004)}$ & $122.54/96$ $(0.035)$ & $127.30/96$ $(0.018)$ \\
Horndeski ($z_{\rm eff}=0.61$) & $216.40/195$ $(0.140)$ & $109.02/96$ $(0.172)$ & $106.90/96$ $(0.210)$ \\
Horndeski ($z_{\rm eff}=0.38,\, 0.61$) & $\mathbf{469.58/392}$ $\mathbf{(0.004)}$ & $231.70/194$ $(0.033)$ & $234.68/194$ $(0.024)$ \\
\hline
DHOST ($z_{\rm eff}=0.38$) & $\mathbf{251.78/194}$ $\mathbf{(0.003)}$ & $122.48/95$ $(0.030)$ & $127.28/95$ $(0.015)$ \\
DHOST ($z_{\rm eff}=0.61$) & $216.40/194$ $(0.129)$ & $108.96/95$ $(0.155)$ & $106.90/95$ $(0.190)$ \\
DHOST ($z_{\rm eff}=0.38,\, 0.61$) & $\mathbf{469.86/391}$ $\mathbf{(0.004)}$ & $231.78/193$ $(0.029)$ & $234.94/193$ $(0.021)$ \\
\hline
\end{tabular}
\caption{
    Same as Table~\ref{Table:results_chi2_Pvalue}, except that only the monopole component of the 3PCF is used in the joint analysis of the 2PCF and 3PCF.
    The combination of the 2PCF and 3PCF multipole components used in this analysis corresponds to Case $2$ in Eq.~(\ref{Eq:case}).
}
\label{Table:results_chi2_Pvalue_mono}
\end{table*}

\begin{table*}
\centering
\begin{tabular}{lccc}
\hline\hline
\multicolumn{4}{c}{Joint analysis with quadrupole 3PCFs ($\zeta_{202}$ and $\zeta_{112}$) only} \\
\hline
\multicolumn{4}{c}{$\chi^2_{\rm min}/{\rm DoF}$ ($p$-value)} \\
\hline
& NGC + SGC & NGC & SGC \\
\hline
GR ($z_{\rm eff}=0.38$) & $280.18/252$ $(0.107)$ & $146.56/125$ $(0.091)$ & $132.68/125$ $(0.302)$ \\
GR ($z_{\rm eff}=0.61$) & $281.58/252$ $(0.097)$ & $148.04/125$ $(0.078)$ & $132.48/125$ $(0.306)$ \\
\hline
Horndeski ($z_{\rm eff}=0.38$) & $279.66/251$ $(0.103)$ & $144.68/124$ $(0.099)$ & $132.38/124$ $(0.287)$ \\
Horndeski ($z_{\rm eff}=0.61$) & $281.68/251$ $(0.089)$ & $148.14/124$ $(0.069)$ & $132.48/124$ $(0.285)$ \\
Horndeski ($z_{\rm eff}=0.38,\, 0.61$) & $563.66/504$ $(0.034)$ & $294.48/250$ $(0.028)$ & $265.08/250$ $(0.245)$ \\
\hline
DHOST ($z_{\rm eff}=0.38$) & $279.18/250$ $(0.099)$ & $144.52/123$ $(0.090)$ & $132.28/123$ $(0.268)$ \\
DHOST ($z_{\rm eff}=0.61$) & $281.68/250$ $(0.082)$ & $148.04/123$ $(0.062)$ & $132.50/123$ $(0.263)$ \\
DHOST ($z_{\rm eff}=0.38,\, 0.61$) & $563.26/503$ $(0.032)$ & $294.66/249$ $(0.025)$ & $265.24/249$ $(0.229)$ \\
\hline
\end{tabular}
\caption{
Same as Table~\ref{Table:results_chi2_Pvalue}, except that only the quadrupole component of the 3PCF is used in the joint analysis of the 2PCF and 3PCF.
The combination of the 2PCF and 3PCF multipole components used in this analysis corresponds to Case $3$ in Eq.~(\ref{Eq:case}).
}
\label{Table:results_chi2_Pvalue_quad}
\end{table*}

We investigate whether the discrepancy between the 3PCF measurement from the galaxy sample at $z=0.38$ and the theoretical model, shown in Table~\ref{Table:results_chi2_Pvalue}, originates from the monopole or quadrupole component.

For this purpose, Tables~\ref{Table:results_chi2_Pvalue_mono} and~\ref{Table:results_chi2_Pvalue_quad} show the joint analysis results using only monopole 3PCFs or only quadrupole 3PCFs in addition to the monopole and quadrupole 2PCFs. For a fair comparison with Table~\ref{Table:results_chi2_Pvalue}, the prior distributions of the parameters used here are those given in Table~\ref{Table:Prior}. For the NGC+SGC at $z=0.38$, the $p$-value obtained using the monopole 3PCFs is less than $0.01$ in all three gravity theories, whereas the $p$-value obtained using the quadrupole 3PCFs is $p\sim0.1$. Therefore, we can conclude that the monopole component of the 3PCF measurement at $z=0.38$ is inconsistent with the theoretical model.

\subsection{No prior in DHOST theories}
\label{Sec:DHOSTnoprior}

\begin{table*}
\centering
\begin{tabular}{lccc}
\hline\hline
\multicolumn{4}{c}{No prior in DHOST} \\
\hline
\multicolumn{4}{c}{$\chi^2_{\rm min}/{\rm DoF}$ ($p$-value)} \\
\hline
& NGC + SGC & NGC & SGC \\
\hline
DHOST ($z_{\rm eff}=0.38$) & $\mathbf{486.14/394}$ $\mathbf{(0.001)}$ & $235.80/195$ $(0.024)$ & $\mathbf{246.28/195}$ $\mathbf{(0.008)}$ \\
DHOST ($z_{\rm eff}=0.61$) & $428.16/394$ $(0.114)$ & $218.58/195$ $(0.119)$ & $209.58/195$ $(0.225)$ \\
DHOST ($z_{\rm eff}=0.38,\, 0.61$) & $\mathbf{916.04/791}$ $\mathbf{(0.001)}$ & $458.02/393$ $(0.013)$ & $456.78/393$ $(0.014)$ \\
\hline
\end{tabular}
\caption{
The analysis is repeated as in Table~\ref{Table:results_chi2_Pvalue}, except that the prior is removed for all parameters related to non-linear effects in DHOST theories, $F_{\rm g}\sigma_8$, $F_{\rm s}\sigma_8$, $F_{\rm t}\sigma_8$, $\xi_{f}$, $\xi_{\rm s}$, and $\xi_{\rm t}$, allowing them to vary from $-\infty$ to $+\infty$.
}
\label{Table:results_chi2_Pvalue_DHOSTnoprior}
\end{table*}

As an attempt to explain the discrepancy between the 3PCF measurement from the galaxy sample at $z=0.38$ and the theoretical model, we remove all flat prior for the non-linear parameters, $F_{\rm g}\sigma_8$, $F_{\rm s}\sigma_8$, $F_{\rm t}\sigma_8$, $\xi_{f}$, $\xi_{\rm s}$, and $\xi_{\rm t}$, set in Table~\ref{Table:Prior} and perform parameter fitting without imposing any prior. In particular, we investigate the possibility that imposing the conditions $F_{\rm g}\geq0$ and $F_{\rm t}\geq0$ on the parameters with the non-linear bias may have caused some problems fitting the monopole 3PCF. This subsection focuses on DHOST theories because they have the largest number of parameters to be varied.

Table~\ref{Table:results_chi2_Pvalue_DHOSTnoprior} summarises the results of the calculations and confirms that the $p$-value obtained from the NGC+SGC sample at $z=0.38$ is $0.001$, even if we assume no prior for the non-linear parameters. Therefore, we can conclude that the discrepancy between the galaxy data and the theoretical model at $z=0.38$ is not due to the prior imposed in Table~\ref{Table:Prior}.

\subsection{Patchy mocks}
\label{Sec:GoFPatchy}

\begin{table*}
\centering
\begin{tabular}{lccc}
\hline\hline
\multicolumn{4}{c}{MultiDark-Patchy mocks} \\
\hline
\multicolumn{4}{c}{$\chi^2_{\rm min}/{\rm DoF}$ ($p$-value)} \\
\hline
& NGC + SGC & NGC & SGC \\
\hline\vspace{0.07cm}
2PCF only ($z_{\rm eff}=0.38$) & $(57.76 \pm 12.25)/57$ $(0.447^{+0.416}_{-0.331})$ & $(28.84 \pm 7.92)/28$ $(0.421^{+0.408}_{-0.297})$ & $(27.75 \pm 8.15)/28$ $(0.478^{+0.401}_{-0.332})$ \vspace{0.07cm}\\
2PCF only ($z_{\rm eff}=0.61$) & $(56.50 \pm 9.54)/57$ $(0.494^{+0.332}_{-0.301})$ & $(27.99 \pm 6.79)/28$ $(0.465^{+0.352}_{-0.289})$ & $(27.52 \pm 6.76)/28$ $(0.490^{+0.345}_{-0.298})$ \\
\hline \vspace{0.07cm}
GR ($z_{\rm eff}=0.38$)                & $(364.08 \pm 29.31)/396$ $(0.873^{+0.115}_{-0.346})$ & $(181.04 \pm 20.14)/197$ $(0.786^{+0.186}_{-0.382})$ & $(180.91 \pm 21.55)/197$ $(0.788^{+0.189}_{-0.408})$ \vspace{0.07cm}\\ 
GR ($z_{\rm eff}=0.61$)                & $(362.41 \pm 24.97)/396$ $(0.886^{+0.099}_{-0.274})$ & $(179.10 \pm 18.41)/197$ $(0.815^{+0.158}_{-0.339})$ & $(181.45 \pm 18.02)/197$ $(0.780^{+0.181}_{-0.342})$ \\ 
\hline \vspace{0.07cm}
Horndeski ($z_{\rm eff}=0.38$)         & $(363.90 \pm 29.45)/395$ $(0.867^{+0.121}_{-0.353})$ & $(180.68 \pm 20.20)/196$ $(0.777^{+0.193}_{-0.386})$ & $(180.62 \pm 21.48)/196$ $(0.778^{+0.197}_{-0.410})$ \vspace{0.07cm}\\
Horndeski ($z_{\rm eff}=0.61$)         & $(361.99 \pm 25.00)/395$ $(0.882^{+0.102}_{-0.278})$ & $(178.48 \pm 18.49)/196$ $(0.810^{+0.162}_{-0.343})$ & $(181.21 \pm 17.92)/196$ $(0.768^{+0.189}_{-0.344})$ \vspace{0.07cm}\\
Horndeski ($z_{\rm eff}=0.38,\, 0.61$) & $(728.47 \pm 36.96)/792$ $(0.948^{+0.048}_{-0.203})$ & $(360.88 \pm 25.46)/394$ $(0.883^{+0.102}_{-0.284})$ & $(363.77 \pm 26.65)/394$ $(0.860^{+0.122}_{-0.319})$ \\
\hline \vspace{0.07cm}
DHOST ($z_{\rm eff}=0.38$)             & $(363.26 \pm 29.37)/394$ $(0.865^{+0.123}_{-0.355})$ & $(180.14 \pm 20.18)/195$ $(0.770^{+0.199}_{-0.388})$ & $(180.39 \pm 21.43)/195$ $(0.766^{+0.207}_{-0.412})$ \vspace{0.07cm}\\
DHOST ($z_{\rm eff}=0.61$)             & $(361.46 \pm 25.03)/394$ $(0.879^{+0.105}_{-0.282})$ & $(178.00 \pm 18.40)/195$ $(0.803^{+0.167}_{-0.345})$ & $(180.95 \pm 17.98)/195$ $(0.757^{+0.197}_{-0.348})$ \vspace{0.07cm}\\
DHOST ($z_{\rm eff}=0.38,\, 0.61$)     & $(728.23 \pm 37.03)/791$ $(0.946^{+0.050}_{-0.208})$ & $(360.45 \pm 25.33)/393$ $(0.879^{+0.105}_{-0.286})$ & $(363.47 \pm 26.59)/393$ $(0.855^{+0.127}_{-0.322})$ \\
\hline
\end{tabular}
\caption{Same as Table~\ref{Table:results_chi2_Pvalue}, except that the $\chi_{\rm min}^2$ is calculated from each of $100$ Patchy mocks, showing their mean values and standard deviations, and the means and $1\sigma$ errors of the corresponding $p$-values.}
\label{Table:results_chi2_Pvalue_mock}
\end{table*}

Table~\ref{Table:results_chi2_Pvalue_mock} shows the means and standard deviations of the $\chi_{\rm min}^2$ and the corresponding means and $1\sigma$ errors of the $p$-values obtained from the $100$ Patchy mock catalogues. The setup for the data analysis is the same as that performed in Table~\ref{Table:results_chi2_Pvalue}.

In all $30$ cases shown in Table~\ref{Table:results_chi2_Pvalue_mock}, the mean $p$-values obtained are almost always $p\gtrsim0.5$, both in the analysis using only the 2PCF and in the joint analysis with the 3PCF. This result means that our 2PCF and 3PCF theoretical templates fit well with the Patchy mock simulation data, indicating that the small $p$-values found in Table~\ref{Table:results_chi2_Pvalue} are a peculiar property of the BOSS galaxies.

As two representative examples, the rest of this subsection focuses on the DHOST theory analyses using only the SGC sample at $z = 0.38$ and all four galaxy samples (NGC+SGC at $z=0.38,0.61$). The reasons are as follows: (1) our primary goal is to test DHOST theories; (2) the analysis of the SGC at $z = 0.38$ in the BOSS data gives a p-value of $0.006$, which is the most significant discrepancy from the theoretical model among the four galaxy samples; (3) the analysis using all four BOSS galaxy samples gives our final results in Section~\ref{Sec:DataAnalysis}.

For the SGC sample at $z = 0.38$, the $\chi_{\rm min}^2$ values for the BOSS samples and the Patchy mocks are
\begin{eqnarray}
    \chi_{\rm min}^2\left( {\rm BOSS} \right) \hspace{-0.25cm} &=& \hspace{-0.25cm} 248.36, \nonumber \\
    \chi_{\rm min}^2\left( {\rm Patchy\, mocks} \right) \hspace{-0.25cm} &=& \hspace{-0.25cm} 180.39\pm21.43,
    \label{Eq:SS}
\end{eqnarray}
where ${\rm DoF}=195$. The above result means that assuming that the $\chi_{\rm min}^2$ follows a Gaussian distribution, the BOSS galaxy sample deviates from the Patchy mocks at the $3.2\sigma$ significance level.

For the NGC+SGC sample at $z=0.38,0.61$, we have
\begin{eqnarray}
    \chi_{\rm min}^2\left( {\rm BOSS} \right) \hspace{-0.25cm} &=& \hspace{-0.25cm} 918.04, \nonumber \\
    \chi_{\rm min}^2\left( {\rm Patchy\, mocks}\right) \hspace{-0.25cm}&=& \hspace{-0.25cm} 728.23\pm37.03
    \label{Eq:FF}
\end{eqnarray}
where ${\rm DoF}=791$. This result implies a discrepancy between the BOSS galaxy sample and the Patchy mocks at the $5.1\sigma$ level. Thus, we conclude that the discrepancy with the theoretical model in the BOSS galaxies cannot be explained by the statistical scatter of the Patchy mocks.

Although Table~\ref{Table:results_chi2_Pvalue_mock} has shown the results obtained from $100$ Patchy mocks, for a more detailed exploration, we perform MCMC analysis on all $2048$ publicly available Patchy mocks for the two examples above to see if it is possible to find realizations that return the similar $p$-values to the BOSS galaxy sample. For the SGC sample of $z=0.38$, only one Patchy mock catalogue gives $p=0.005$, close to the BOSS result. In this case, the Patchy mocks have a probability of $100\times(1/2048)=0.0488\%$ to reproduce the BOSS galaxy results. On the other hand, using all four galaxy samples, not a single catalogue among the $2048$ Patchy mocks reproduced the BOSS results. This result means that the BOSS result has less than a $0.0488\%$ probability of appearing in the Patchy mocks. These results are consistent with the $3.2\sigma$ and $5.1\sigma$ discrepancies between the BOSS and Patchy mock data presented in Eqs.~(\ref{Eq:SS}) and (\ref{Eq:FF}).

\begin{figure*}
    \scalebox{0.95}{\includegraphics[width=\textwidth]{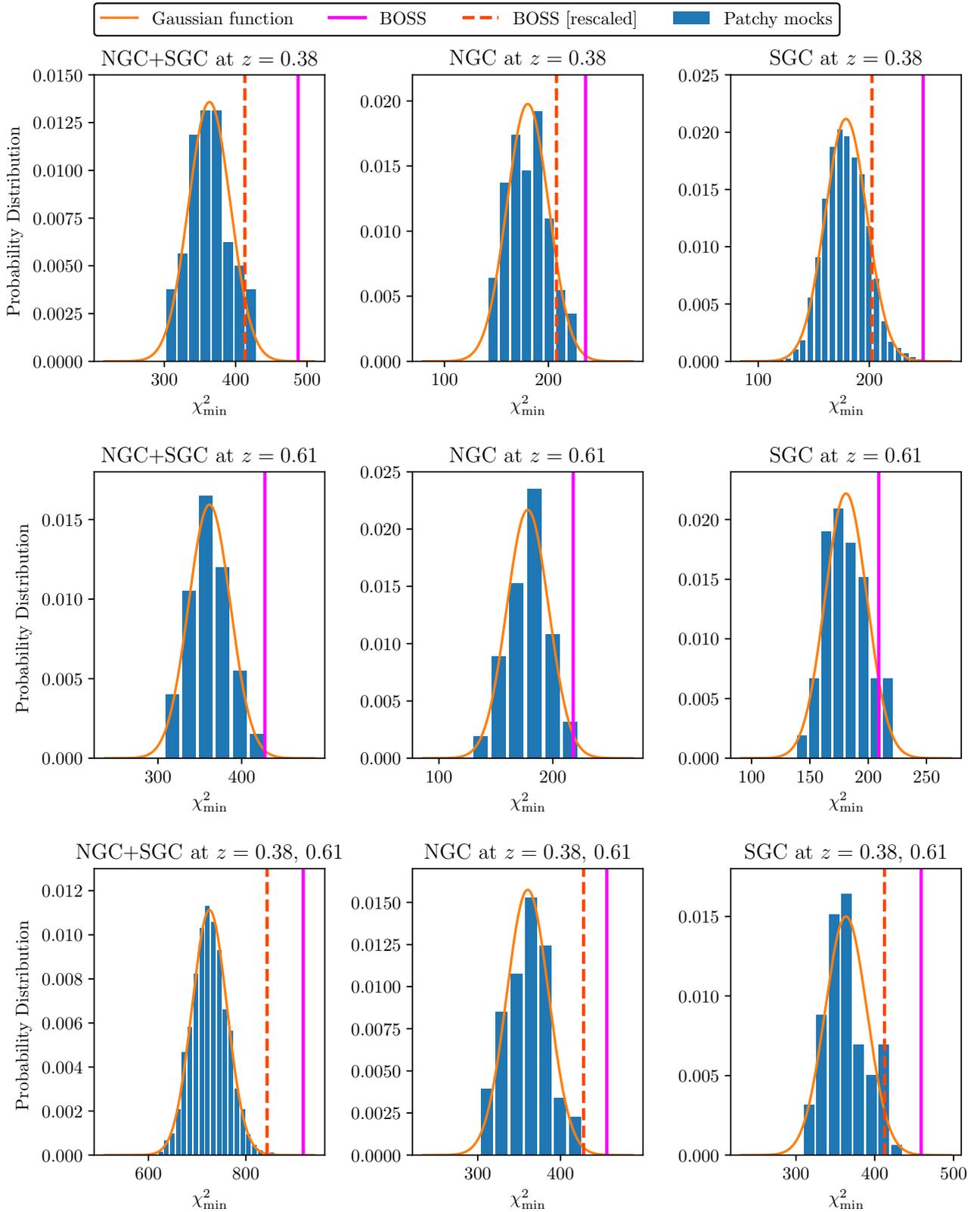}}
    \caption{
        Visualizations of the DHOST theory analysis results in Tables~\ref{Table:results_chi2_Pvalue}, \ref{Table:results_chi2_Pvalue_mock} and \ref{Table:results_chi2_Pvalue_rescaled}. The histograms of $\chi_{\rm min}^2$ computed from the Patchy mocks are shown. In the cases of SGC at $z=0.38$ (top right panel) and NGC+SGC at $z=0.38,0.61$ (bottom left panel), the histograms are computed from $2048$ Patchy mocks; otherwise, they are computed from $100$ Patchy mocks. Also, Gaussian functions (orange lines) with input values for the mean and standard deviation of $\chi_{\rm min}^2$ computed from the Patchy mocks are shown. The $\chi_{\rm min}^2$ values obtained from the BOSS data are plotted in magenta. Also shown are the results for the BOSS data using the rescaled 3PCF covariance matrix at $z=0.38$ (Section~\ref{Sec:RescalingOfCovarianceMatrix}) in dashed red lines.
}
	\label{fig:chi2}
\end{figure*}

Figure~\ref{fig:chi2} visualizes the results for the DHOST theory analysis in Tables~\ref{Table:results_chi2_Pvalue} and \ref{Table:results_chi2_Pvalue_mock}. As expected, the histogram of $\chi_{\rm min}^2$ computed from the Patchy mocks (blue bars) can be well approximated by a Gaussian function (orange line) with input values for the mean and standard deviation of $\chi_{\rm min}^2$ computed from the Patchy mocks. In the cases of SGC at $z=0.38$ (top right panel) and NGC+SGC at $z=0.38,0.61$ (bottom left panel), we compute the histograms from the $2048$ Patchy mocks; otherwise, we compute them from $100$ Patchy mocks. Also, we plot the $\chi_{\rm min}^2$ values obtained from the BOSS data in magenta.

\subsection{Rescaling of the covariance matrix}
\label{Sec:RescalingOfCovarianceMatrix}

\begin{table*}
\centering
\begin{tabular}{lccc}
\hline\hline
\multicolumn{4}{c}{$\chi^2_{\rm min}/{\rm DoF}$ ($p$-value)} \\
\hline
& NGC + SGC & NGC & SGC \\
\hline
GR ($z_{\rm eff}=0.38\, [{\rm rescaled}]$)                & $413.86/396$ $(0.258)$ & $209.98/197$ $(0.250)$ & $202.56/197$ $(0.378)$ \\
\hline
Horndeski ($z_{\rm eff}=0.38\, [{\rm rescaled}]$)         & $413.66/395$ $(0.249)$ & $208.24/196$ $(0.261)$ & $202.36/196$ $(0.363)$ \\
Horndeski ($z_{\rm eff}=0.38\, [{\rm rescaled}],\, 0.61$) & $844.18/792$ $(0.097)$ & $428.14/394$ $(0.114)$ & $412.46/394$ $(0.251)$ \\
\hline
DHOST ($z_{\rm eff}=0.38\, [{\rm rescaled}]$)             & $412.88/394$ $(0.246)$ & $207.82/195$ $(0.252)$ & $202.34/195$ $(0.344)$ \\
DHOST ($z_{\rm eff}=0.38\, [{\rm rescaled}],\, 0.61$)     & $843.66/791$ $(0.095)$ & $428.00/393$ $(0.108)$ & $412.36/393$ $(0.241)$ \\
\hline
\end{tabular}
\caption{
Same as Table~\ref{Table:results_chi2_Pvalue}, except that the covariance matrix of the 3PCF at $z=0.38$ is rescaled as in Eq.~(\ref{Eq:rescaling}).
}
\label{Table:results_chi2_Pvalue_rescaled}
\end{table*}

We have discussed the discrepancy between the 3PCF measured from the BOSS data at $z=0.38$ and the corresponding theoretical model. Unfortunately, this paper cannot provide a definitive answer to this question. 

There are three possible reasons for this discrepancy. The first concern is about the calculation of the covariance matrix. There may be physical effects that the Patchy mock used to calculate the covariance matrix needs to account for fully. For example, it is necessary to verify to what extent non-linear galaxy bias effects~\citep{Desjacques:2016bnm} and super-sample covariance effects~\citep{Takada:2013PhRvD..87l3504T} are correctly included in the Patchy mock. The second concern is about the theoretical model. For example, the theoretical model may have new physical effects dominating large scales at low redshifts. If so, we also need to account for that effect in the covariance matrix simultaneously. Finally, we are concerned with the observed galaxy data. There may be unknown observational effects that the weight function in Eq.~(\ref{Eq:w_c}) cannot explain. In any case, the findings in this section indicate the importance of discussing the validity of cosmological analyses that consider the 2PCF and 3PCF simultaneously.

This paper assumes that the discrepancy between the BOSS galaxy sample and the theoretical model is due to an improper covariance matrix for the 3PCF calculated with the Patchy mock. Therefore, as a temporary measure, we decided to rescale the 3PCF covariance matrix at $z=0.38$ to increase the obtained $p$-value to an acceptable value. Specifically, we rescale the 3PCF covariance matrix at $z=0.38$ as follows:
\begin{eqnarray}
    {\rm Cov}[{\rm 3PCF}]_{\rm rescaled} = A\, {\rm Cov}[{\rm 3PCF}].
    \label{Eq:rescaling}
\end{eqnarray}
where the rescaling factor $A$ is $A=1.15$ and $A=1.25$ for NGC and SGC, respectively. The values of $A$ are determined so that the resulting $p$-values at $z=0.38$ become similar to those at $z=0.61$.

Table~\ref{Table:results_chi2_Pvalue_rescaled} summarises the results of repeating the same analysis as Table~\ref{Table:results_chi2_Pvalue} using the rescaled covariance matrix; Figure~\ref{fig:chi2} visualizes the results for the DHOST theory analysis in Tables~\ref{Table:results_chi2_Pvalue_rescaled}. As expected, $p\gtrsim0.1$ for the NGC+SGC sample at $z=0.38$. Thus, if the discrepancy between the galaxy data and the theoretical model in the 3PCF measurement is due to the covariance matrix computed by the Patchy mocks, we find that we can solve this problem by increasing the resulting covariance matrix by $15-25\%$. We will give the results using this rescaled covariance matrix as the final result of this paper when we perform parameter estimation in Section~\ref{Sec:DataAnalysis} using the galaxy data at $z=0.38$.

\section{Results}
\label{Sec:DataAnalysis}

This section calculates the mean, standard deviation, $\pm1\sigma$ errors, and $95\%$ upper and lower bounds for the parameters computed from the likelihoods, where we perform parameter estimation for each BOSS DR12 galaxy and Patchy mock data. When using the Patchy mock data, we compute the mean, standard deviation, $\pm1\sigma$ errors, and $95\%$ limits for the parameters from each of the $100$ Patchy mocks; then, we calculate the means and standard deviations of them. All results here take into account both the NGC and SGC samples. We have already given the $\chi_{\rm min}^2$ and $p$-values calculated from the best-fit values of the parameters in the NGC+SGC columns of Tables~\ref{Table:results_chi2_Pvalue}, \ref{Table:results_chi2_Pvalue_mock}, and \ref{Table:results_chi2_Pvalue_rescaled}.

The main results of this paper are Eqs.~(\ref{Eq:Result_XT_1sigma})-(\ref{Eq:Result_XS_95}), which provide constraints in $\xi_{\rm t}$ and $\xi_{\rm s}$. In Figure~\ref{fig:2Dcontour_DHOST}, we plot the one- and two-dimensional likelihood distributions corresponding to these results. Finally, we summarise the measurement results for the 3PCF multipole components ($\zeta_{000}$, $\zeta_{110}$, $\zeta_{202}$, and $\zeta_{112}$) from the BOSS galaxies used in this analysis in Figures~\ref{fig:3PCFs_North_zbin1_mono}-\ref{fig:3PCFs_South_zbin3_quad}.

The combination of the 2PCF and 3PCF multipoles used in the joint analysis performed in this section corresponds to Case $7$ in Eq.~(\ref{Eq:case}); the analysis using only the 2PCF corresponds to Case $1$. In Section~\ref{Sec:Monopole3PCF}, the results of the joint analysis with only the monopole 3PCF, which corresponds to Case $2$, are also presented and compared with the final results obtained from Case $7$.

\subsection{Measurements}
\label{Sec:3PCFmeasurements}

\begin{figure*}
    \scalebox{0.95}{\includegraphics[width=\textwidth]{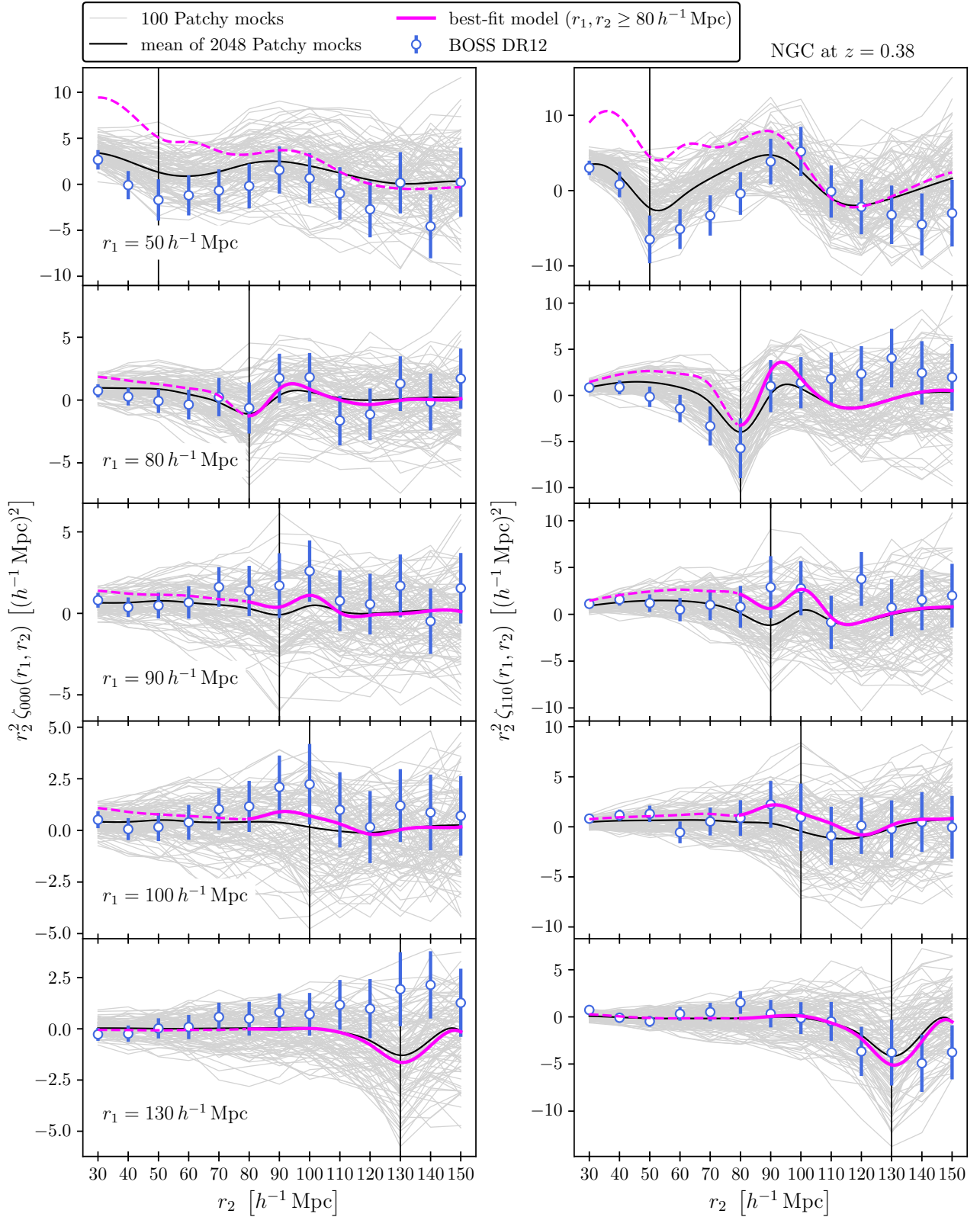}}
    \caption{
        Monopole 3PCFs ($\zeta_{000}$ and $\zeta_{110}$) measured from the NGC sample at $z=0.38$ (blue points). These plots are shown as a function of $r_2$, with $r_1$ fixed from the top to $50\hMpc$, $80\hMpc$, $90\hMpc$, $100\hMpc$, and $130\hMpc$. The error bars are the standard deviation of the 3PCF measurements computed from $2048$ Patchy mocks. Also plotted are the 3PCF measurements from $100$ Patchy mocks (grey) and the mean from the 3PCF measurements from $2048$ Patchy mocks (black). Finally, the results of the theoretical model calculated from the best-fit parameter values obtained from the DHOST theory analysis using all four BOSS samples (Sections~\ref{Sec:XF}-\ref{Sec:XS}) are shown by the magenta lines; they are shown as solid lines on the scales $r_1,r_2\geq 80\hMpc$ used in the analysis and as dashed lines on smaller scales.
    }
	\label{fig:3PCFs_North_zbin1_mono}
\end{figure*}

\begin{figure*}
    \scalebox{0.95}{\includegraphics[width=\textwidth]{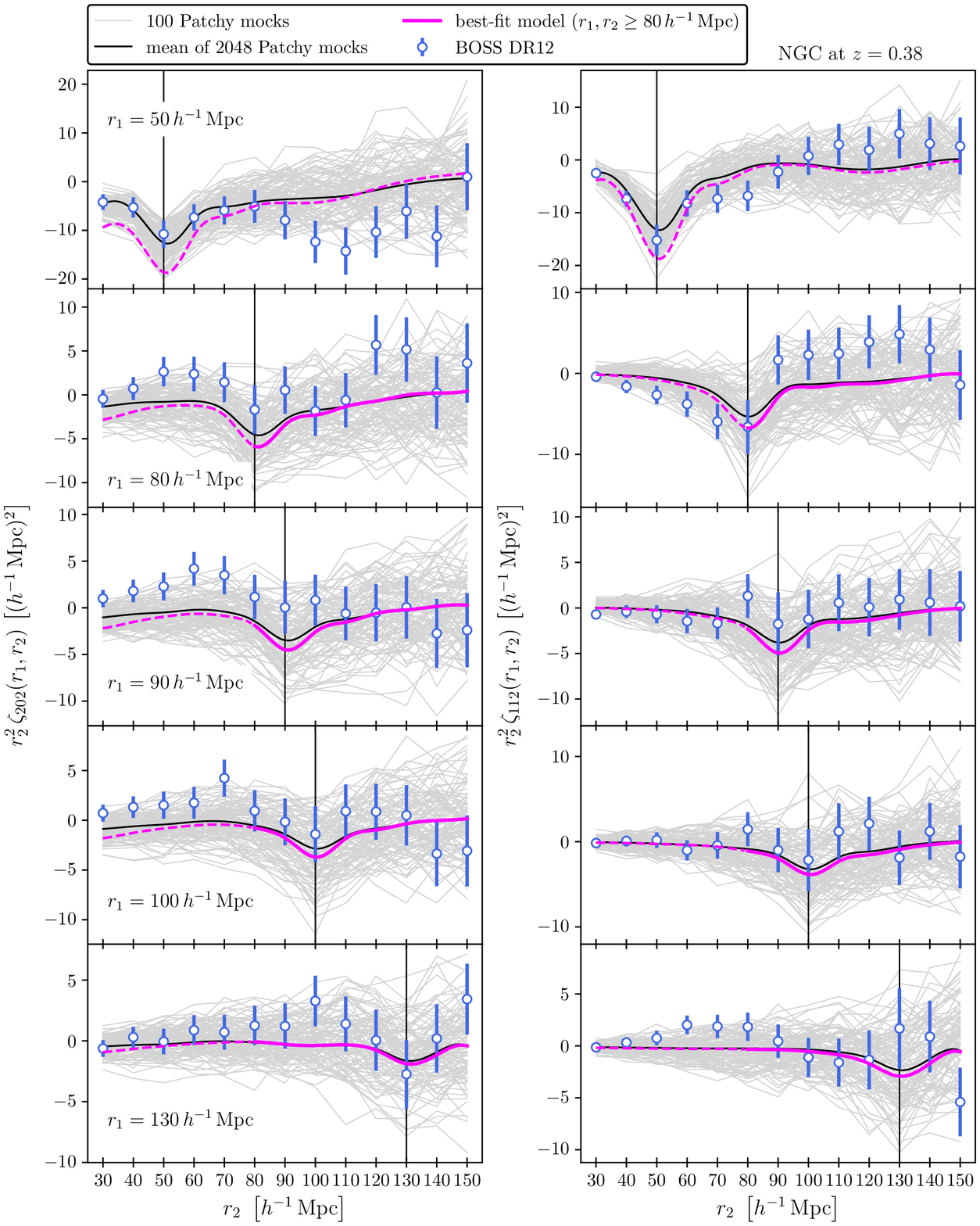}}
    \caption{Same as Figure~\ref{fig:3PCFs_North_zbin1_mono}, except that the quadrupole 3PCF results ($\zeta_{202}$ and $\zeta_{112}$) measured from the NGC sample at $z=0.38$ are shown.
    }
	\label{fig:3PCFs_North_zbin1_quad}
\end{figure*}

\begin{figure*}
    \scalebox{0.95}{\includegraphics[width=\textwidth]{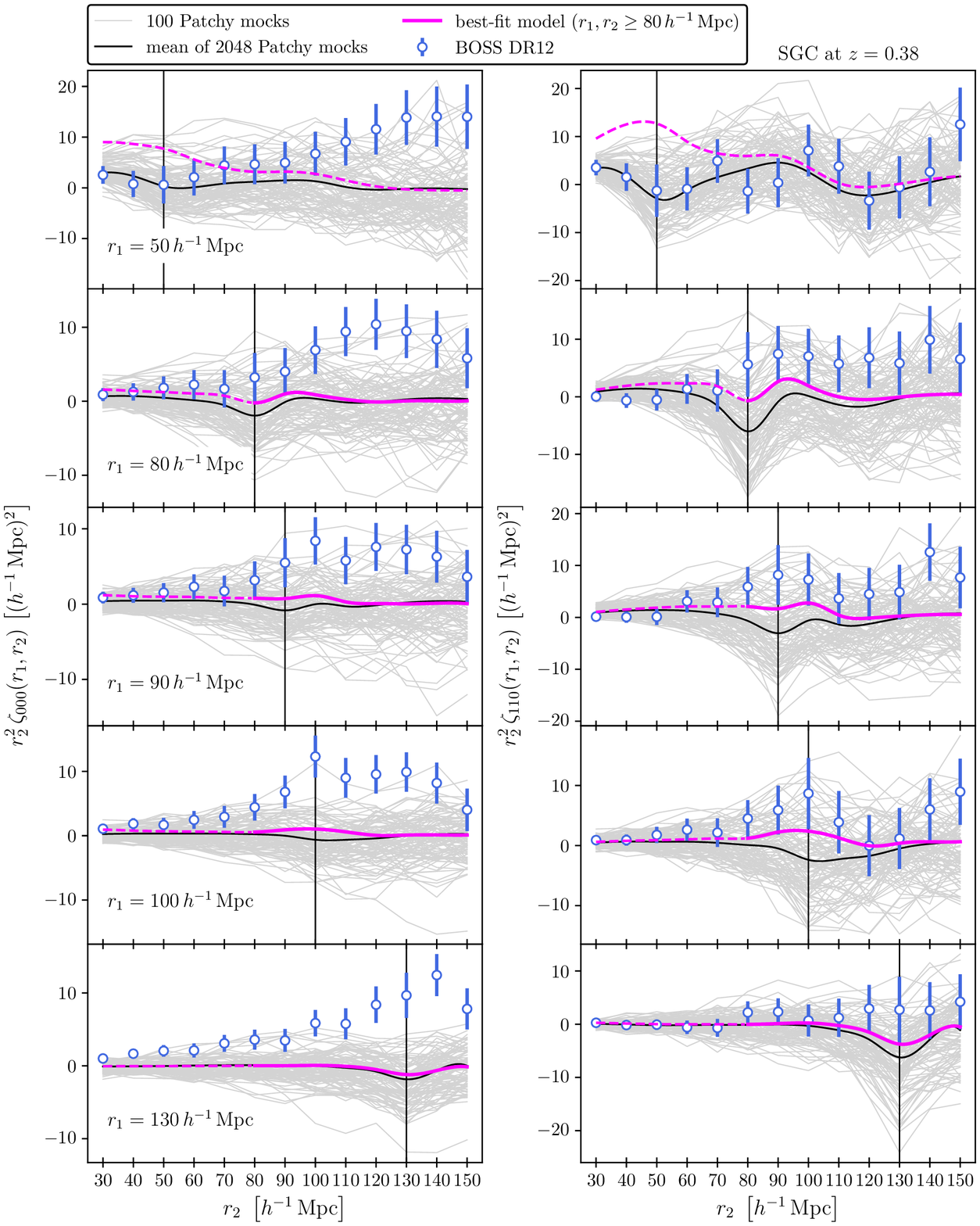}}
    \caption{Same as Figure~\ref{fig:3PCFs_North_zbin1_mono}, except that the monopole 3PCF results ($\zeta_{000}$ and $\zeta_{110}$) measured from the SGC sample at $z=0.38$ are shown.}
	\label{fig:3PCFs_South_zbin1_mono}
\end{figure*}

\begin{figure*}
    \scalebox{0.95}{\includegraphics[width=\textwidth]{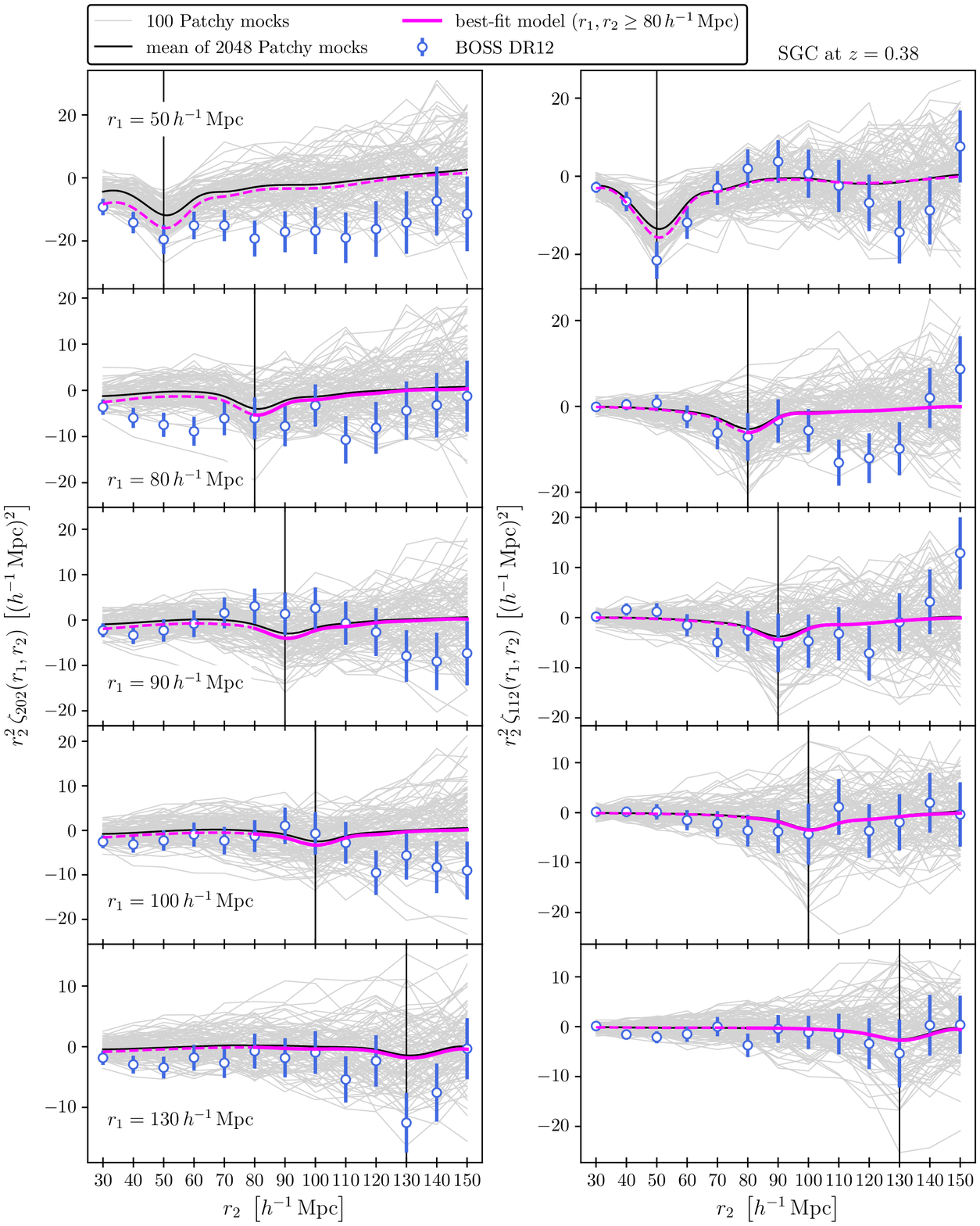}}
    \caption{Same as Figure~\ref{fig:3PCFs_North_zbin1_mono}, except that the quadrupole 3PCF results ($\zeta_{202}$ and $\zeta_{112}$) measured from the SGC sample at $z=0.38$ are shown.}
	\label{fig:3PCFs_South_zbin1_quad}
\end{figure*}

\begin{figure*}
    \scalebox{0.95}{\includegraphics[width=\textwidth]{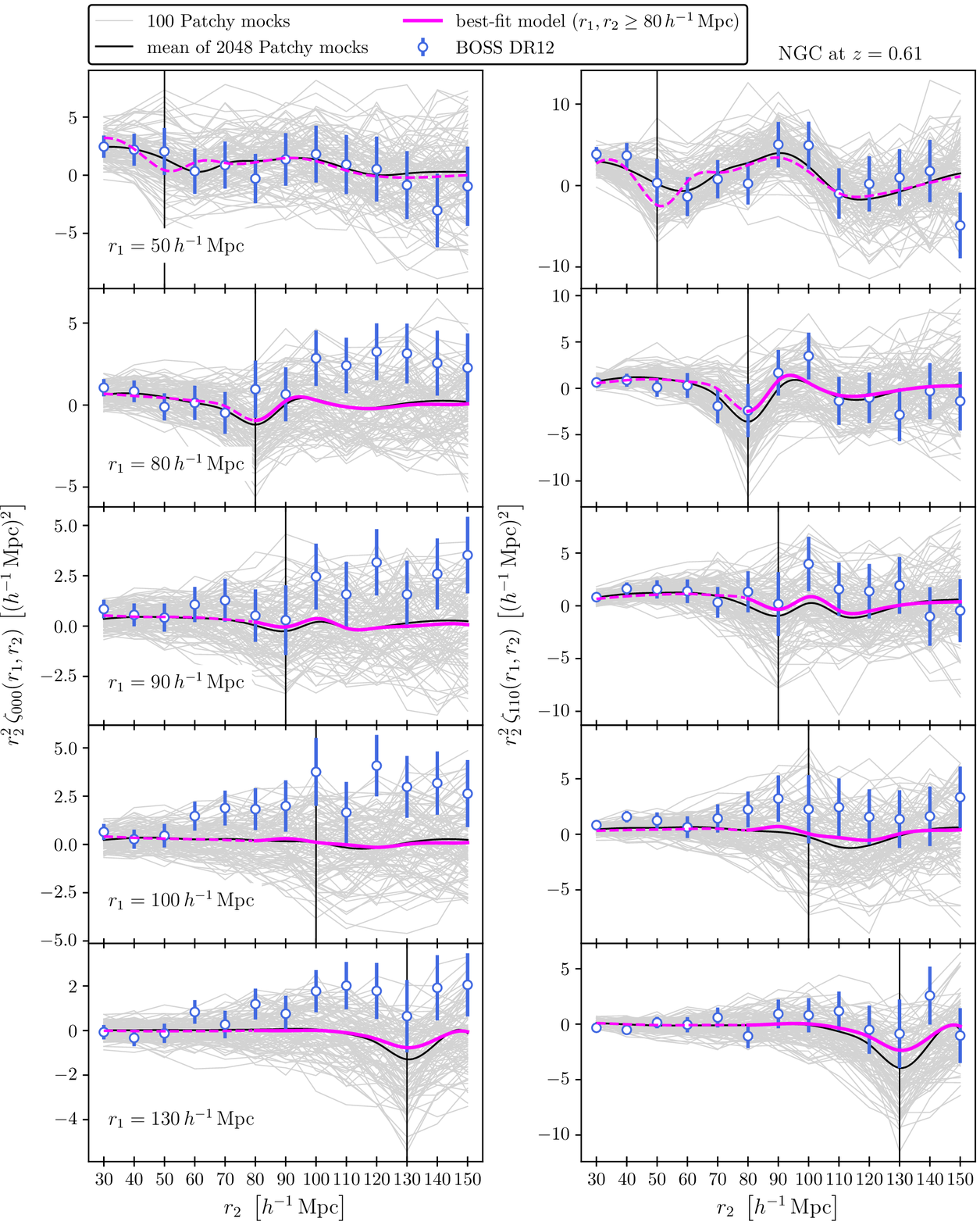}}
    \caption{Same as Figure~\ref{fig:3PCFs_North_zbin1_mono}, except that the monopole 3PCF results ($\zeta_{000}$ and $\zeta_{110}$) measured from the NGC sample at $z=0.61$ are shown.}
	\label{fig:3PCFs_North_zbin3_mono}
\end{figure*}

\begin{figure*}
    \scalebox{0.95}{\includegraphics[width=\textwidth]{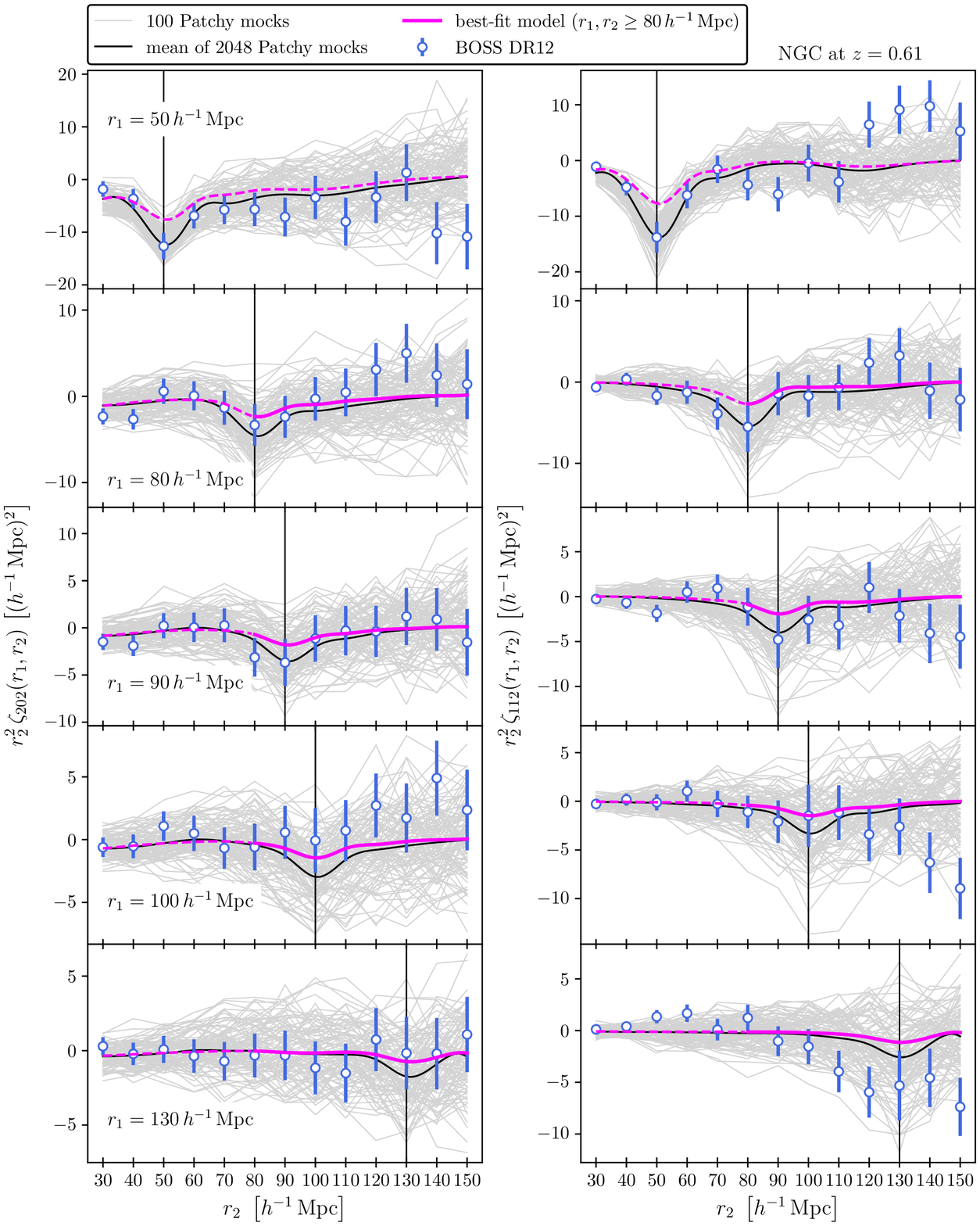}}
    \caption{Same as Figure~\ref{fig:3PCFs_North_zbin1_mono}, except that the quadrupole 3PCF results ($\zeta_{202}$ and $\zeta_{112}$) measured from the NGC sample at $z=0.61$ are shown.}
	\label{fig:3PCFs_North_zbin3_quad}
\end{figure*}

\begin{figure*}
    \scalebox{0.95}{\includegraphics[width=\textwidth]{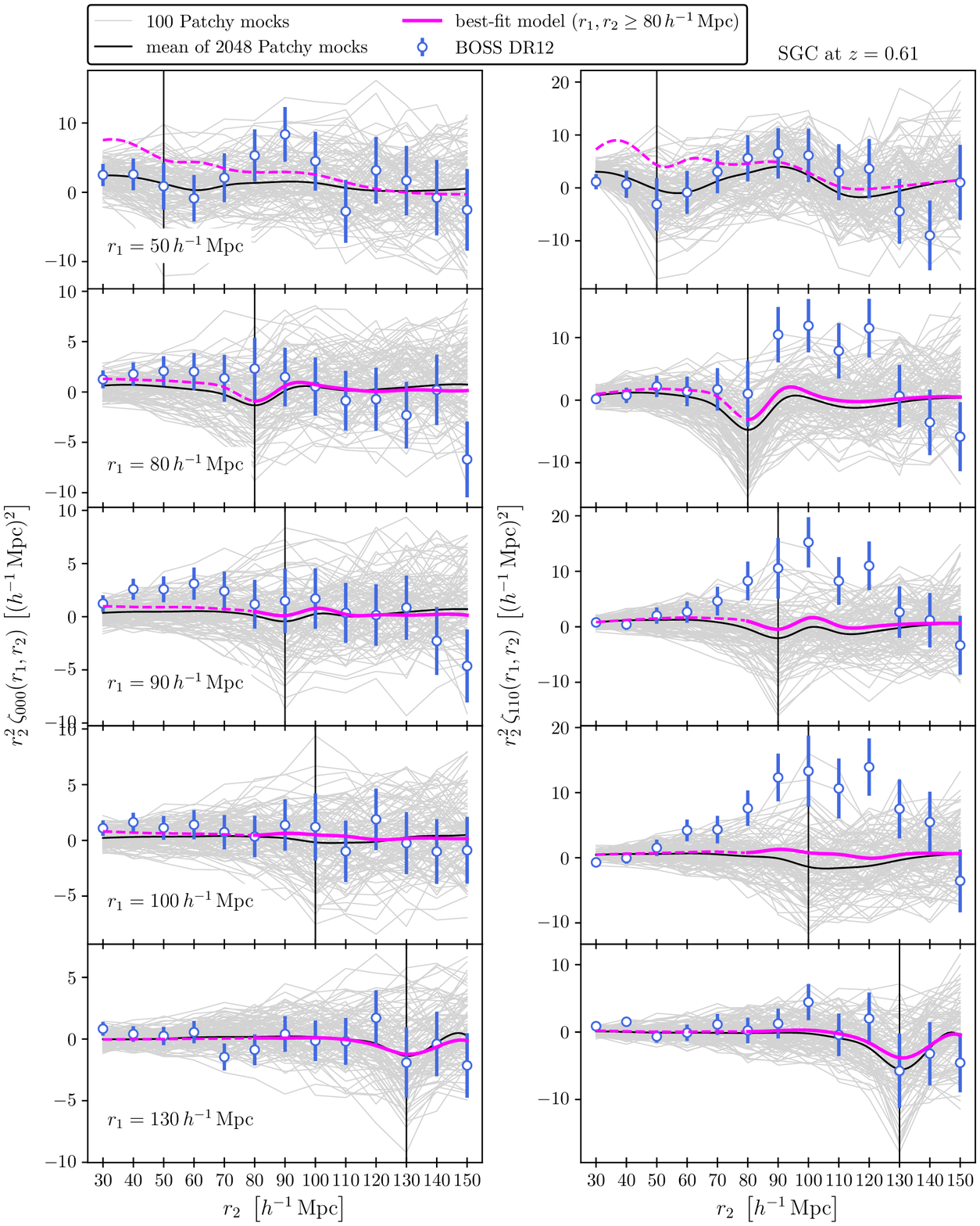}}
    \caption{Same as Figure~\ref{fig:3PCFs_North_zbin1_mono}, except that the monopole 3PCF results ($\zeta_{000}$ and $\zeta_{110}$) measured from the SGC sample at $z=0.61$ are shown.}
	\label{fig:3PCFs_South_zbin3_mono}
\end{figure*}

\begin{figure*}
    \scalebox{0.95}{\includegraphics[width=\textwidth]{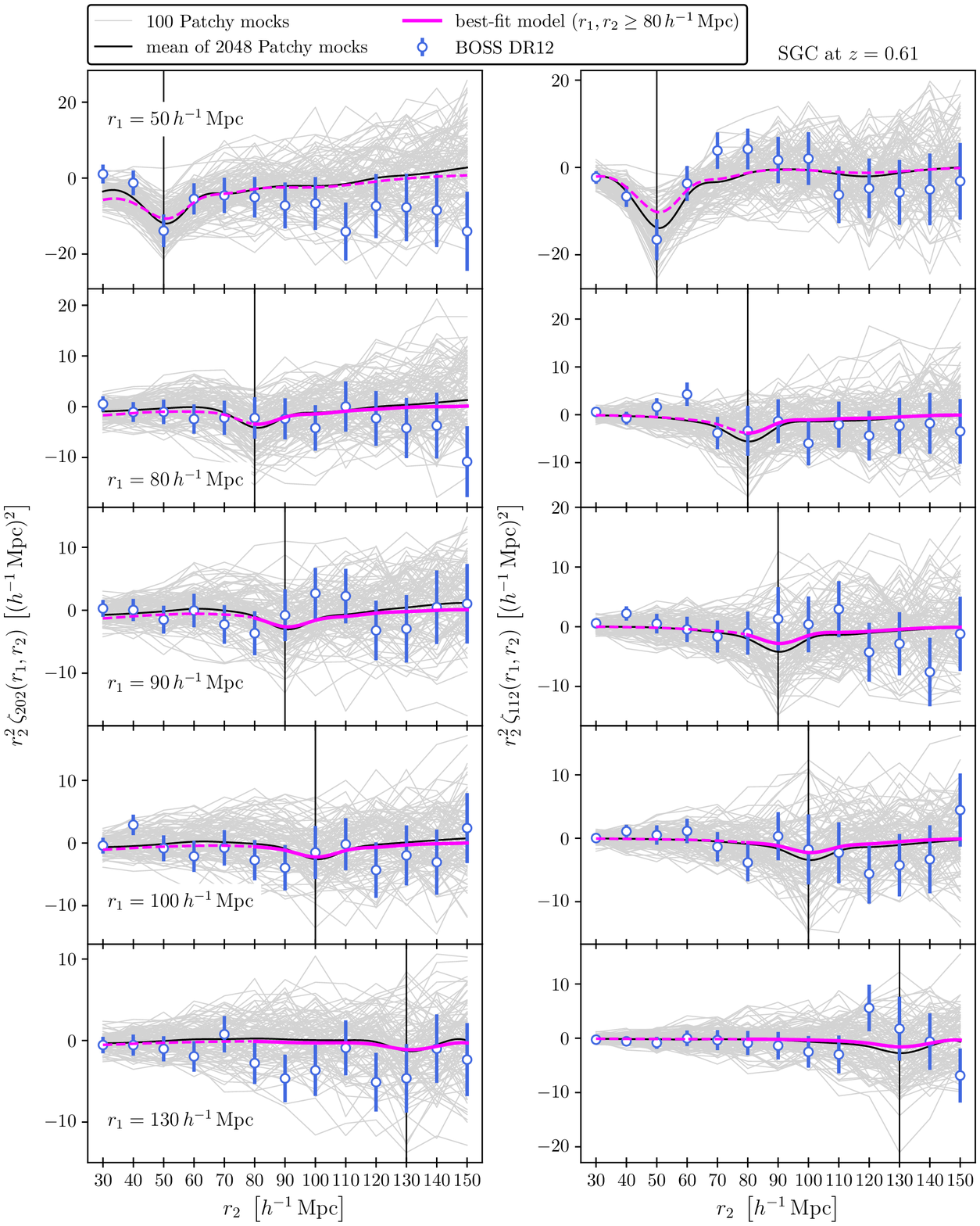}}
    \caption{Same as Figure~\ref{fig:3PCFs_North_zbin1_mono}, except that the quadrupole 3PCF results ($\zeta_{202}$ and $\zeta_{112}$) measured from the SGC sample at $z=0.61$ are shown.}
	\label{fig:3PCFs_South_zbin3_quad}
\end{figure*}

Figures~\ref{fig:3PCFs_North_zbin1_mono}-\ref{fig:3PCFs_South_zbin3_quad} plot the measurement results of the monopole 3PCFs ($\zeta_{000}$ and $\zeta_{110}$) and the quadrupole 3PCFs ($\zeta_{202}$ and $\zeta_{112}$) from the BOSS galaxies as a function of $r_2$ with $r_1$ fixed at $50\hMpc$, $80\hMpc$, $90\hMpc$, $100\hMpc$, and $130\hMpc$ from top to bottom; they are shown by blue circled points with $1\sigma$ error bars. Also plotted are the 3PCF measurements from $100$ Patchy mocks (grey) and the mean from the 3PCF measurements from $2048$ Patchy mocks (black). Finally, the theoretical models computed from the best-fit parameter values obtained from the DHOST theory analysis using all four BOSS samples are plotted with magenta lines; they are shown as solid lines on the scales $r_1,r_2\geq 80\hMpc$ used in the MCMC analysis and as dashed lines on smaller scales. Note that the theoretical model shown by the magenta dashed line does not need to explain the measurements from the galaxy data.

As can be seen from the lower left of Figure~\ref{fig:3PCFs_South_zbin1_mono}, the $\zeta_{000}$ measured from the SGC sample at $z=0.38$ shows a significant discrepancy with the theoretical model on large scales, which is to be expected from the results presented in Section~\ref{Sec:GoFMonoQuad}.

Theoretical predictions from Figures~\ref{fig:decom_3PCF_mono} and \ref{fig:decom_3PCF_quad} indicate that the monopole and quadrupole 3PCFs have trough-shaped signals at $r_1=r_2$. For example, this characteristic trough signal is seen in the blue data points for $\zeta_{112}$ measured in the NGC sample at $z=0.38$, shown in the first and second panels from the top in the right panel of Figure~\ref{fig:3PCFs_North_zbin1_quad}. However, due to the significant statistical scattering in the galaxy data, the trough signal is not necessarily found in the blue points of all panels in Figures~\ref{fig:3PCFs_North_zbin1_mono}-\ref{fig:3PCFs_South_zbin3_quad}.

In particular, for the monopole 3PCF, the BAO peak appears at $r_1\simeq r_2\simeq 100\hMpc$. Therefore, it is expected to cancel out the trough signal, resulting in a smooth line with no irregularities when plotting the 3PCF as a function of $r_2$ after fixing $r_1=100\hMpc$. For example, as seen from the second panel from the bottom in the right panel of Figure~\ref{fig:3PCFs_North_zbin1_mono}, the $\zeta_{110}$ measured from the NGC sample at $z=0.38$ shows that the trough-shaped signal disappears from the data points. Conversely, this is evidence of a BAO signal in the monopole 3PCF. Although plotting the 3PCF as a function of $r_1=r_2=r$ makes it easier to see the BAO signal from the galaxy data points (e.g., see Figure~\ref{fig:3PCF_std} and Figure $11$ in \citealt{Sugiyama:2020uil}), we do not plot such a figure because the subject of this paper is not the BAO signal.

\subsection{\texorpdfstring{$f\sigma_8$}{} constraints from the Patchy mocks in GR}
\label{Sec:fsigma8_mock}

\begin{table*}

\centering
\begin{tabular}{lcccccc}
\hline\hline
\multicolumn{7}{c}{MultiDark-Patchy mocks} \\
\hline
& $\langle f\sigma_8\rangle_{\rm mean}$ & $\langle f\sigma_8\rangle_{\rm std}$ & $\langle f\sigma_8\rangle_{-1\sigma}$ & $\langle f\sigma_8\rangle_{+1\sigma}$ & $\langle f\sigma_8\rangle_{>95\%}$ & $\langle f\sigma_8\rangle_{<95\%}$\\
\hline
2PCF only ($z_{\rm eff}=0.38$) & $0.445\, (0.491) \pm 0.078$ & $0.094 \pm 0.010$ & $-0.106 \pm 0.013$ & $0.091 \pm 0.009$ & $0.256 \pm 0.074$ & $0.635 \pm 0.090$ \\
2PCF only ($z_{\rm eff}=0.61$) & $0.457\, (0.485) \pm 0.068$ & $0.083 \pm 0.009$ & $-0.091 \pm 0.011$ & $0.082 \pm 0.008$ & $0.291 \pm 0.066$ & $0.624 \pm 0.077$ \\
\hline
GR ($z_{\rm eff}=0.38$) & $0.498 (0.491) \pm 0.085$ & $0.107 \pm 0.008$ & $-0.118 \pm 0.009$ & $0.105 \pm 0.012$ & $0.283 \pm 0.082$ & $0.718 \pm 0.093$ \\
GR ($z_{\rm eff}=0.61$) & $0.504 (0.485) \pm 0.070$ & $0.095 \pm 0.010$ & $-0.104 \pm 0.010$ & $0.092 \pm 0.011$ & $0.314 \pm 0.065$ & $0.698 \pm 0.082$ \\
\hline
\end{tabular}
\caption{
    Constraint results for $f\sigma_8$ obtained from the $100$ Patchy mocks. One hundred means, standard deviations, $\pm1\sigma$ errors, and $95\%$ upper and lower bounds are computed from the $100$ Patchy mocks; then, the means and standard deviations of them are shown. Values in parentheses are the input values for the Patchy mocks. Results are shown for two redshift bins at $z=0.38$ and $0.61$ in combination with the NGC and SGC samples. Also shown are the results for the 2PCF only analysis and the joint analysis with the 3PCF assuming GR. The $\chi_{\rm min}^2$ and $p$ values corresponding to this table are shown in the NGC+SGC column of Table~\ref{Table:results_chi2_Pvalue_mock}.
}
\label{Table:results_fsigma8_mock}

\vspace{0.5cm}

\centering
\begin{tabular}{lcccccc}
\hline\hline
\multicolumn{7}{c}{BOSS DR12} \\
\hline
& $(f\sigma_8)_{\rm mean}$ & $(f\sigma_8)_{\rm std}$ & $(f\sigma_8)_{-1\sigma}$ & $(f\sigma_8)_{+1\sigma}$ & $(f\sigma_8)_{>95\%}$ & $(f\sigma_8)_{<95\%}$\\
\hline
2PCF only ($z_{\rm eff}=0.38$) & $0.446$ & $0.086$ & $-0.096$ & $0.084$ & $0.273$ & $0.620$ \\
2PCF only ($z_{\rm eff}=0.61$) & $0.408$ & $0.086$ & $-0.095$ & $0.084$ & $0.236$ & $0.580$ \\
\hline
GR ($z_{\rm eff}=0.38$) & $0.561$ & $0.108$ & $-0.122$ & $0.098$ & $0.348$ & $0.785$ \\
GR ($z_{\rm eff}=0.61$) & $0.394$ & $0.091$ & $-0.099$ & $0.088$ & $0.208$ & $0.580$ \\
\hline
GR ($z_{\rm eff}=0.38\,[{\rm rescaled}]$) & $0.549$ & $0.108$ & $-0.122$ & $0.097$ & $0.337$ & $0.776$ \\
\hline
\end{tabular}
\caption{
Means, standard deviations, $\pm1\sigma$ errors, and $95\%$ upper and lower bounds for $f\sigma_8$ obtained in the 2PCF-only analysis and the joint analysis with the 3PCF using the BOSS DR12 galaxies, assuming GR for the joint analysis with the 3PCF. Results are shown for two redshifts at $z=0.38$ and $0.61$ using the NGC and SGC samples. Also shown are the results at $z=0.38$ for the analysis using the rescaled covariance matrix (\ref{Eq:rescaling}) to give an acceptable $p$-value. The $\chi_{\rm min}^2$ and $p$ values corresponding to this table are shown in the NGC+SGC column of Tables~\ref{Table:results_chi2_Pvalue} and \ref{Table:results_chi2_Pvalue_rescaled}.
}
\label{Table:results_fsigma8_galaxy}
\end{table*}

Table~\ref{Table:results_fsigma8_mock} shows the $f\sigma_8$ results obtained from the analysis of $100$ Patchy mocks, assuming GR.

The standard deviations of $f\sigma_8$ from the 2PCF-only analysis are almost identical to those obtained from the joint analysis with the 3PCF. This result is consistent with the results of the Fisher analysis in Section~\ref{Sec:FisherAnalysis}. Therefore, we can conclude that neither monopole 3PCF nor quadrupole 3PCF contributes to reducing the $f\sigma_8$ error. Nevertheless, note that we can constrain the growth rate function $f$ using the joint analysis with the 3PCF by combining the $\sigma_8$ constraint in Section~\ref{Sec:sigma8BOSS}. Furthermore, in the context of modified gravity theories, $f$ is extended to $E_{\rm f}$, and Section~\ref{Sec:XF} will constrain the parameter $\xi_f$ characterizing its time evolution.

Looking at the mean of $f\sigma_8$, the results obtained in the joint analysis with the 3PCF ($0.498$ at $z=0.38$ and $0.504$ at $z=0.61$) are slightly closer to the values input to the Patchy mock ($0.491$ and $0.485$) than those obtained with the 2PCF alone ($0.445$ and $0.457$). Thus, the 3PCF information helps reduce the bias in the $f\sigma_8$ mean values.

\subsection{\texorpdfstring{$f\sigma_8$}{} constraints from the BOSS DR12 galaxies in GR}
\label{Sec:fsigma8BOSS}

Table~\ref{Table:results_fsigma8_galaxy} summarises the results of the $f\sigma_8$ constraints obtained from the BOSS galaxy under the assumption of GR. "GR ($z=0.38$) [rescaled]" means the results using the rescaled covariance matrix (Section~\ref{Sec:RescalingOfCovarianceMatrix}). 

Note that the standard deviation result of $f\sigma_8$ does not change with and without rescaling the 3PCF covariance matrix at $z=0.38$: i.e., $(f\sigma_8)_{\rm std}=0.108$ in both cases. Thus, the $15-25\%$ difference in the 3PCF covariance matrix due to rescaling (Section~\ref{Sec:RescalingOfCovarianceMatrix}) does not propagate significantly to the final $f\sigma_8$ error. The reason for this may be mainly due to the effect of parameter degeneracy and other factors. However, due to the decisively different $p$-values obtained (see Tables~\ref{Table:results_chi2_Pvalue} and \ref{Table:results_chi2_Pvalue_rescaled}), we adopt the rescaled result at $z=0.38$ as the final result.

Comparing the results of the joint analysis with the 3PCF and the 2PCF-only analysis, the former has a larger $f\sigma_8$ error: i.e., $(f\sigma_8)_{\rm std}=0.108,\,0.091$ at $z=0.38,\, 0.61$ for the joint analysis with the 3PCF, and  $(f\sigma_8)_{\rm std}=0.086,\,0.086$ at $z=0.38,\, 0.61$ for the 2PCF-only analysis. Therefore, one may think that adding the 3PCF information has weakened the constraint on $f\sigma_8$. However, since Table~\ref{Table:results_fsigma8_mock} shows that the statistical uncertainty of $(f\sigma_8)_{\rm std}$ is $\sim 0.01$, the both results are statistically consistent at the $\lesssim 2\sigma$ level.

Our final results for the $f\sigma_8$ constraints are as follows. The 2PCF-only analysis gives, at the $1\sigma$ level, 
\begin{eqnarray}
    f\sigma_8 \hspace{-0.25cm}&=&\hspace{-0.25cm} 0.446^{+0.084}_{-0.096} \quad {\rm at}\, z=0.38 \nonumber \\
    f\sigma_8 \hspace{-0.25cm}&=&\hspace{-0.25cm} 0.408^{+0.084}_{-0.095} \quad {\rm at}\, z=0.61,
\end{eqnarray}
and the joint analysis with the 3PCF presents
\begin{eqnarray}
    f\sigma_8 \hspace{-0.25cm}&=&\hspace{-0.25cm} 0.549^{+0.097}_{-0.122} \quad {\rm at}\, z=0.38 \nonumber \\
    f\sigma_8 \hspace{-0.25cm}&=&\hspace{-0.25cm} 0.394^{+0.088}_{-0.099} \quad {\rm at}\, z=0.61.
    \label{Eq:result_fsigma8}
\end{eqnarray}
These $f\sigma_8$ constraints are consistent with the $f\sigma_8$ values ($f\sigma_8=0.485,\, 0.479$ at $z=0.38,\,0.61$) calculated from the cosmological parameters in a flat $\Lambda$CDM model (Section~\ref{Sec:Introduction}) given by Planck 2018~\citep{Aghanim:2018eyx}. However, the $f\sigma_8$ result in this analysis, which constrains $f\sigma_8$ with a $\sim20\%$ precision, is not as competitive as existing constraints~\citep[e.g.,][]{Alam:2016hwk,Ivanov:2020JCAP...05..042I,Lange:2022MNRAS.509.1779L,Kobayashi:2022PhRvD.105h3517K} because we only use large-scale information ($r\geq80\hMpc$).

\subsection{\texorpdfstring{$\sigma_8$}{} constraints from the Patchy mocks in GR}
\label{Sec:sigma8mock}

\begin{table*}

\centering
\begin{tabular}{lcccccc}
\hline\hline
\multicolumn{7}{c}{MultiDark-Patchy mocks} \\
\hline
& $\langle \sigma_8\rangle_{\rm mean}$ & $\langle \sigma_8\rangle_{\rm std}$ & $\langle \sigma_8\rangle_{-1\sigma}$ & $\langle \sigma_8\rangle_{+1\sigma}$ & $\langle \sigma_8\rangle_{>95\%}$ & $\langle \sigma_8\rangle_{<95\%}$\\
\hline
GR ($z_{\rm eff}=0.38$) & $0.741\, (0.691) \pm 0.347$ & $0.476 \pm 0.142$ & $-0.626 \pm 0.184$ & $0.235 \pm 0.191$ & $0.024 \pm 0.097$ & $1.668 \pm 0.617$ \\
GR ($z_{\rm eff}=0.61$) & $0.612\, (0.615) \pm 0.319$ & $0.415 \pm 0.165$ & $-0.550 \pm 0.220$ & $0.176 \pm 0.156$ & $0.003 \pm 0.023$ & $1.410 \pm 0.636$ \\
\hline
\end{tabular}
\caption{
Constraint results for $\sigma_8$ obtained from the $100$ Patchy mocks. One hundred means, standard deviations, $\pm1\sigma$ errors, and $95\%$ upper and lower bounds are computed from the $100$ Patchy mocks; then, the means and standard deviations of them are shown. Values in parentheses are the input values for the Patchy mocks. Results are shown for two redshift bins at $z=0.38$ and $0.61$ in combination with the NGC and SGC samples. Also shown are the results for the joint analysis with the 3PCF assuming GR. The $\chi_{\rm min}^2$ and $p$ values corresponding to this table are shown in the NGC+SGC column of Table~\ref{Table:results_chi2_Pvalue_mock}.
}
\label{Table:results_sigma8_mock}

\vspace{0.5cm}

\centering
\begin{tabular}{lcccccc}
\hline\hline
\multicolumn{7}{c}{MultiDark-Patchy mocks} \\
\hline
\multicolumn{7}{c}{Negative $F_{\rm g}$ and $F_{\rm t}$ allowed} \\
\hline
& $\langle \sigma_8\rangle_{\rm mean}$ & $\langle \sigma_8\rangle_{\rm std}$ & $\langle \sigma_8\rangle_{-1\sigma}$ & $\langle \sigma_8\rangle_{+1\sigma}$ & $\langle \sigma_8\rangle_{>95\%}$ & $\langle \sigma_8\rangle_{<95\%}$\\
\hline
GR ($z_{\rm eff}=0.38$) & $1.204\, (0.691) \pm 0.429$ & $0.628 \pm 0.109$ & $-0.771 \pm 0.184$ & $0.499 \pm 0.263$ & $0.151 \pm 0.269$ & $2.409 \pm 0.568$ \\
GR ($z_{\rm eff}=0.61$) & $1.004\, (0.615) \pm 0.441$ & $0.584 \pm 0.156$ & $-0.750 \pm 0.198$ & $0.374 \pm 0.260$ & $0.072 \pm 0.176$ & $2.140 \pm 0.719$ \\
\hline
\end{tabular}
\caption{
Same as Table~\ref{Table:results_sigma8_mock}, except that a prior with negative $F_{\rm g}$ and $F_{\rm t}$ allowed is adopted.
}
\label{Table:results_sigma8_mock_bias}

\vspace{0.5cm}

\centering
\begin{tabular}{lcccccc}
\hline\hline
\multicolumn{7}{c}{BOSS DR12} \\
\hline
& $(\sigma_8)_{\rm mean}$ & $(\sigma_8)_{\rm std}$ & $(\sigma_8)_{-1\sigma}$ & $(\sigma_8)_{+1\sigma}$ & $(\sigma_8)_{>95\%}$ & $(\sigma_8)_{<95\%}$\\
\hline
GR ($z_{\rm eff}=0.38$) & $0.702$ & $0.451$ & $-0.576$ & $0.221$ & $0.000$ & $1.563$ \\
GR ($z_{\rm eff}=0.61$) & $0.568$ & $0.404$ & $-0.547$ & $0.144$ & $0.000$ & $1.323$ \\
\hline
GR ($z_{\rm eff}=0.38\,[{\rm rescaled}]$) & $0.692$ & $0.459$ & $-0.591$ & $0.209$ & $0.000$ & $1.568$ \\
\hline
\end{tabular}
\caption{
    Means, standard deviations, $\pm1\sigma$ errors, and $95\%$ upper and lower bounds for $\sigma_8$ obtained in the joint analysis of the 2PCF and 3PCF using the BOSS DR12 galaxies, assuming GR. Results are shown for two redshifts at $z=0.38$ and $0.61$ using the NGC and SGC samples. Also shown are the results at $z=0.38$ for the analysis using the rescaled covariance matrix (\ref{Eq:rescaling}) to give an acceptable $p$-value. The $\chi_{\rm min}^2$ and $p$ values corresponding to this table are shown in the NGC+SGC column of Tables~\ref{Table:results_chi2_Pvalue} and \ref{Table:results_chi2_Pvalue_rescaled}.
}
\label{Table:results_sigma8_galaxy}
\end{table*}

Table~\ref{Table:results_sigma8_mock} summarises the results for $\sigma_8$ from $100$ Patchy mocks.

The mean values for $\sigma_8$ are $0.741$ and $0.612$ for $z=0.38$ and $z=0.61$, respectively, in good agreement with the mock input values ($0.691$ and $0.615$). Specifically, they agree to an accuracy of $7\%$ and $0.5\%$, respectively. Since $\sigma_8$ is the only physical parameter unique to the 3PCF in GR, the fact that we can estimate the $\sigma_8$ value with high accuracy guarantees the validity of our analysis. 

On the other hand, the $95\%$ lower limit of $\sigma_8$ is consistent with zero, so we cannot detect a statistically significant signal for $\sigma_8$ in our analysis.

\subsection{\texorpdfstring{$\sigma_8$}{} constraints from the Patchy mocks in GR with negative \texorpdfstring{$F_{\rm g}\sigma_8$}{} and \texorpdfstring{$F_{\rm t}\sigma_8$}{} allowed}
\label{Sec:GR_bias}

This subsection discusses the validity of the priors set in Section~\ref{Sec:Priors} for the parameters $F_{\rm g}\sigma_8$ and $F_{\rm t}\sigma_8$, which include non-linear bias parameters. We impose the assumption that $F_{\rm g}$ and $F_{\rm t}$ are positive, but there is no theoretical requirement that this assumption is correct since the values of the non-linear biases are uncertain. Therefore, as a test, we perform parameter estimation for $\sigma_8$ using a prior with negative $F_{\rm g}$ and $F_{\rm t}$ allowed to check if it returns the input values of the Patchy mocks. Specifically, the upper bounds of $F_{\rm g}\sigma_8$ and $F_{\rm t}\sigma_8$ given in Table~\ref{Table:Prior} are multiplied by $(-1)$ to set the lower bounds of $F_{\rm g}\sigma_8$ and $F_{\rm t}\sigma_8$. For example, we set $-2.99 \leq (F_{\rm g}\sigma_8)_{\rm NGC,\, z=0.38} \leq 2.99$. 

We summarise the results of this analysis in Table~\ref{Table:results_sigma8_mock_bias}. This table shows that the mean values for $\sigma_8$ are $1.204$ at $z=0.38$ and $1.004$ at $z=0.61$, which are about $1.5$ times larger than the input values, $0.691$ at $z=0.38$ and $0.615$ at $z=0.61$, in the Patchy mocks. Thus, if we allow negative values of $F_{\rm g}$ and $F_{\rm t}$, we cannot estimate the correct value of $\sigma_8$. We have no theoretical basis for explaining this fact, but as a result of numerical experiments, we conclude that it is reasonable to impose the conditions $F_{\rm g}\geq0$ and $F_{\rm t}\geq0$ in our analysis.

\subsection{\texorpdfstring{$\sigma_8$}{} constraints from the BOSS DR12 galaxies in GR}
\label{Sec:sigma8BOSS}

\begin{figure*}
    \scalebox{1.00}{\includegraphics[width=\textwidth]{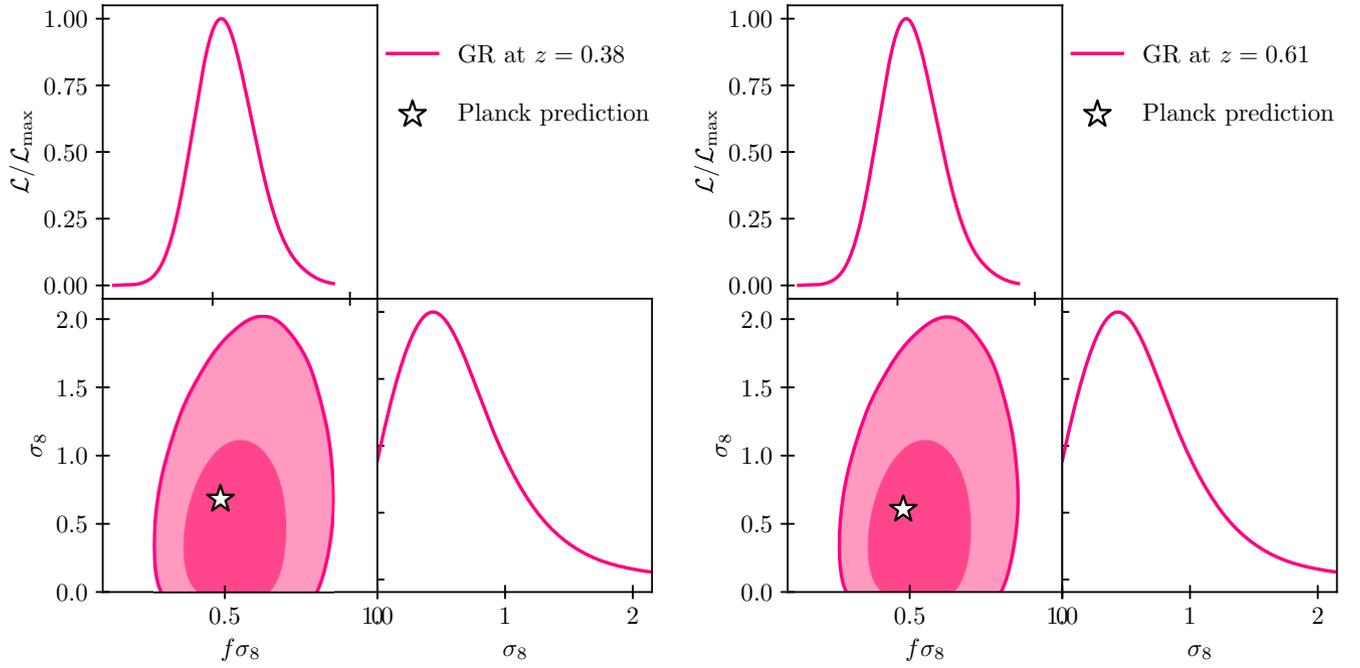}}
    \caption{
Marginalized two- and one-dimensional posteriors of $f\sigma_8$ and $\sigma_8$ for BOSS DR12. The contours indicate $68.27\%$ and $95.45\%$ confidence levels. Asterisks indicate predictions by Planck. The left and right panels show the cases at $z=0.38$ and $z=0.61$. The NGC and SGC samples are always combined to obtain this result. The rescaled 3PCF covariance matrix (Section~\ref{Sec:RescalingOfCovarianceMatrix}) at $z=0.38$ is used.
}
	\label{fig:2Dcontour_GR}
\end{figure*}

Table~\ref{Table:results_sigma8_galaxy} summarises the results for the $\sigma_8$ constraints obtained from the BOSS galaxies under the GR assumption. The ``GR ($z=0.38$) [rescaled]'' refers to the results obtained using the rescaled covariance matrix (Section~\ref{Sec:RescalingOfCovarianceMatrix}). Figure~\ref{fig:2Dcontour_GR} plots the marginalized one- and two-dimensional posteriors of $f\sigma_8$ and $\sigma_8$.

Similar to the results for the $f\sigma_8$ constraint in Section~\ref{Sec:fsigma8BOSS}, the results for the $\sigma_8$ constraint remain almost the same whether the covariance matrix is rescaled or not. Adopting the result using the rescaled covariance matrix as the final result, the $\sigma_8$ constraints at the $1\sigma$ level are
\begin{eqnarray}
    \sigma_8 \hspace{-0.25cm}&=&\hspace{-0.25cm} 0.692^{+0.209}_{-0.591} \quad {\rm at}\, z=0.38,\nonumber \\
    \sigma_8 \hspace{-0.25cm}&=&\hspace{-0.25cm} 0.568^{+0.144}_{-0.547} \quad {\rm at}\, z=0.61,
\end{eqnarray}
Also, as expected from the results of the Patchy mocks, the $95\%$ lower bounds for $\sigma_8$ reach $0$, so at the $95\%$ level, we get only the upper bounds:
\begin{eqnarray}
    \sigma_8 \hspace{-0.25cm}&<&\hspace{-0.25cm} 1.568\, (95\%\, {\rm CL}) \quad {\rm at}\, z=0.38, \nonumber \\
    \sigma_8 \hspace{-0.25cm}&<&\hspace{-0.25cm} 1.323\, (95\%\, {\rm CL}) \quad {\rm at}\, z=0.61. 
\end{eqnarray}
These results are consistent with the $\sigma_8$ values, ($\sigma_8=0.681,\,\, 0.606$ at $z=0.38,\,0.61$), calculated from the cosmological parameters in a flat $\Lambda$CDM model given by Planck 2018 (Section~\ref{Sec:Introduction}).

The ratio of the standard deviation to the mean for $\sigma_8$ is $(\sigma_8)_{\rm std}/(\sigma_8)_{\rm mean}=0.66$ at $z=0.38$ and $0.71$ at $z=0.61$, indicating that the galaxy sample at $z=0.38$ provides a better constraint on $\sigma_8$. This result is consistent with the Fisher analysis in Section~\ref{Sec:Fisher_4samples}.

\subsection{\texorpdfstring{$\xi_f$}{} constraints from the BOSS DR12 galaxies in Horndeski and DHOST theories}
\label{Sec:XF}

\begin{table*}
\centering
\begin{tabular}{lcccccc}
\hline\hline
\multicolumn{7}{c}{BOSS DR12} \\
\hline
& $(\xi_f)_{\rm mean}$ & $(\xi_f)_{\rm std}$ & $(\xi_f)_{-1\sigma}$ & $(\xi_f)_{+1\sigma}$ & $(\xi_f)_{>95\%}$ & $(\xi_f)_{<95\%}$\\
\hline
Horndeski ($z_{\rm eff}=0.38$) & $0.206$ & $1.016$ & $-0.777$ & $1.176$ & $-1.906$ & $2.201$ \\
Horndeski ($z_{\rm eff}=0.61$) & $1.142$ & $1.671$ & $-1.431$ & $1.862$ & $-2.302$ & $4.480$ \\
Horndeski ($z_{\rm eff}=0.38,\, 0.61$) & $0.562$ & $0.818$ & $-0.703$ & $0.913$ & $-1.079$ & $2.226$ \\
\hline
Horndeski ($z_{\rm eff}=0.38\,[{\rm rescaled}]$) & $0.202$ & $1.043$ & $-0.833$ & $1.201$ & $-1.921$ & $2.287$ \\
Horndeski ($z_{\rm eff}=0.38\,[{\rm rescaled}],\, 0.61$) & $0.485$ & $0.839$ & $-0.708$ & $0.967$ & $-1.216$ & $2.175$ \\
\hline
DHOST ($z_{\rm eff}=0.38$) & $0.458$ & $1.013$ & $-0.790$ & $1.188$ & $-1.564$ & $2.474$ \\
DHOST ($z_{\rm eff}=0.61$) & $1.248$ & $1.722$ & $-1.372$ & $1.981$ & $-2.318$ & $4.630$ \\
DHOST ($z_{\rm eff}=0.38,\, 0.61$) & $0.834$ & $0.829$ & $-0.686$ & $0.963$ & $-0.814$ & $2.484$ \\
\hline
DHOST ($z_{\rm eff}=0.38\,[{\rm rescaled}]$) & $0.129$ & $1.131$ & $-0.895$ & $1.078$ & $-2.096$ & $2.473$ \\
DHOST ($z_{\rm eff}=0.38\,[{\rm rescaled}],\, 0.61$) & $0.791$ & $0.830$ & $-0.691$ & $0.963$ & $-0.907$ & $2.447$ \\
\hline
\end{tabular}
\caption{
Means, standard deviations, $\pm1\sigma$ errors, and $95\%$ upper and lower bounds for $\xi_f$ obtained in the joint analysis of the 2PCF and 3PCF using the BOSS DR12 galaxies, assuming Horndeski or DHOST theories. The results for the two redshifts, $z=0.38$ and $0.61$, and their combined case are shown. Both NGC and SGC samples are used for all cases. Also shown are the results at $z = 0.38$ for the analysis using the rescaled covariance matrix (\ref{Eq:rescaling}) to give acceptable $p$-values. The $\chi_{\rm min}^2$ and $p$ values corresponding to this table are shown in the NGC+SGC column of Tables~\ref{Table:results_chi2_Pvalue} and \ref{Table:results_chi2_Pvalue_rescaled}.
}
\label{Table:results_XF_galaxy}

\vspace{1.5cm}

\centering
\begin{tabular}{lcccccc}
\hline\hline
\multicolumn{7}{c}{BOSS DR12} \\
\hline
& $(\xi_{\rm t})_{\rm mean}$ & $(\xi_{\rm t})_{\rm std}$ & $(\xi_{\rm t})_{-1\sigma}$ & $(\xi_{\rm t})_{+1\sigma}$ & $(\xi_{\rm t})_{>95\%}$ & $(\xi_{\rm t})_{<95\%}$\\
\hline
Horndeski ($z_{\rm eff}=0.38$) & $4.221$ & $4.693$ & $-5.982$ & $5.710$ & $-3.380$ & - \\
Horndeski ($z_{\rm eff}=0.61$) & $11.118$ & $7.354$ & $-7.204$ & $9.639$ & $-1.256$ & - \\
Horndeski ($z_{\rm eff}=0.38,\, 0.61$) & $5.298$ & $4.257$ & $-4.023$ & $6.092$ & $-1.865$ & - \\
\hline
Horndeski ($z_{\rm eff}=0.38\,[{\rm rescaled}]$) & $4.129$ & $4.704$ & $-5.968$ & $5.268$ & $-3.485$ & - \\
Horndeski ($z_{\rm eff}=0.38\,[{\rm rescaled}],\, 0.61$) & $5.151$ & $4.300$ & $-4.016$ & $6.112$ & $-2.098$ & - \\
\hline
DHOST ($z_{\rm eff}=0.38$) & $4.288$ & $4.589$ & $-6.002$ & $5.348$ & $-3.103$ & - \\
DHOST ($z_{\rm eff}=0.61$) & $11.361$ & $7.387$ & $-6.785$ & $10.112$ & $-1.183$ & - \\
DHOST ($z_{\rm eff}=0.38,\, 0.61$) & $5.349$ & $4.217$ & $-3.980$ & $5.915$ & $-1.688$ & - \\
\hline
DHOST ($z_{\rm eff}=0.38\,[{\rm rescaled}]$) & $3.745$ & $4.732$ & $-6.165$ & $5.262$ & $-3.921$ & - \\
DHOST ($z_{\rm eff}=0.38\,[{\rm rescaled}],\, 0.61$) & $5.414$ & $4.211$ & $-3.734$ & $6.007$ & $-1.655$ & - \\
\hline
\end{tabular}
\caption{Same as Table~\ref{Table:results_XF_galaxy}, except that the results for $\xi_{\rm t}$ are shown.}
\label{Table:results_XT_galaxy}

\vspace{1.5cm}

\centering
\begin{tabular}{lcccccc}
\hline\hline
\multicolumn{7}{c}{BOSS DR12} \\
\hline
& $(\xi_{\rm s})_{\rm mean}$ & $(\xi_{\rm s})_{\rm std}$ & $(\xi_{\rm s})_{-1\sigma}$ & $(\xi_{\rm s})_{+1\sigma}$ & $(\xi_{\rm s})_{>95\%}$ & $(\xi_{\rm s})_{<95\%}$\\
\hline
DHOST ($z_{\rm eff}=0.38$) & $5.232$ & $3.495$ & $-2.948$ & $4.904$ & $-0.744$ & - \\
DHOST ($z_{\rm eff}=0.61$) & $8.791$ & $6.841$ & $-6.984$ & $8.889$ & $-2.653$ & - \\
DHOST ($z_{\rm eff}=0.38,\, 0.61$) & $5.407$ & $3.360$ & $-2.988$ & $4.664$ & $-0.275$ & - \\
\hline
DHOST ($z_{\rm eff}=0.38\,[{\rm rescaled}]$) & $5.307$ & $3.519$ & $-2.854$ & $5.063$ & $-0.740$ & - \\
DHOST ($z_{\rm eff}=0.38\,[{\rm rescaled}],\, 0.61$) & $5.378$ & $3.438$ & $-2.777$ & $4.993$ & $-0.504$ & - \\
\hline
\end{tabular}
\caption{
Same as Table~\ref{Table:results_XF_galaxy}, except the results for $\xi_{\rm s}$ that varies in DHOST theories are shown.
}
\label{Table:results_XS_galaxy}
\end{table*}

Table~\ref{Table:results_XF_galaxy} summarises the constraint results for the parameter $\xi_{f}$, defined as $\xi_f = \ln_{\Omega_{\rm m}}\left( E_{f} \right) = \ln_{\Omega_{\rm m}}\left( f/\kappa_{\delta} \right)$ (\ref{Eq:TimeDependenceOfE}), characterising the time evolution of the amplitude of the linear velocity field. In GR and Horndeski theories, $\xi_f$ corresponds to the well-known parameter $\gamma$ since $\kappa_{\delta}=1$; in GR, $\xi_f=\gamma=6/11$~(\ref{Eq:Xi_GT}).

Using all the four galaxy samples, at the $1\sigma$ level, we obtain
\begin{eqnarray}
    \gamma \hspace{-0.25cm}&=&\hspace{-0.25cm} 0.485^{+0.967}_{-0.708} \quad \mbox{in Horndeski}, \nonumber \\
    \xi_f \hspace{-0.25cm}&=&\hspace{-0.25cm} 0.791^{+0.963}_{-0.691} \quad \mbox{in DHOST}, 
    \label{Eq:Result_XF_1sigma}
\end{eqnarray}
and at the $95\%$ confidence level, we have
\begin{eqnarray}
    & -1.216 < \gamma < 2.175\, (95\% {\rm CL}) & \quad \mbox{in Horndeski}, \nonumber \\
    & -0.907 < \xi_f < 2.447\, (95\% {\rm CL}) & \quad \mbox{in DHOST}. 
    \label{Eq:Result_XF_95}
\end{eqnarray}

All results in Table~\ref{Table:results_XF_galaxy} are consistent with GR within the $1\sigma$ level.

Note that the $\gamma$ constraints in Horndeski theories obtained here are not directly comparable to those obtained from existing studies by, e.g., ~\citet{Gil-Marin:2016wya}. The reason is that we simultaneously vary the $\xi_{\rm t}$ parameter characterising the tidal term in the non-linear velocity field in Horndeski theories, while \citet{Gil-Marin:2016wya} use the bispectrum model assuming GR.

\subsection{\texorpdfstring{$\xi_{\rm t}$}{} constraints from the BOSS DR12 galaxies in Horndeski and DHOST theories}
\label{Sec:XT}

Table~\ref{Table:results_XT_galaxy} summarises the constraint results for the $\xi_{\rm t}$ parameter, defined as $\xi_{\rm t} = \ln_{\Omega_{\rm m}}\left(E_{\rm t} \right) = \ln_{\Omega_{\rm m}}\left( \lambda_{\theta}/\kappa_{\delta} \right)$ (\ref{Eq:TimeDependenceOfE}), characterising the time evolution of the tidal term in the second-order velocity field. In GR, $\xi_{\rm t}=15/1144$~(\ref{Eq:Xi_GT}), and if $\xi_{\rm t}$ deviates from the GR value, it is evidence for Horndeski or DHOST theories.

Using all the four galaxy samples, at the $1\sigma$ level, we obtain
\begin{eqnarray}
    \xi_{\rm t}\hspace{-0.25cm}&=&\hspace{-0.25cm}  5.151^{+6.112}_{-4.016} \quad \mbox{in Horndeski}, \nonumber \\
    \xi_{\rm t} \hspace{-0.25cm}&=&\hspace{-0.25cm} 5.414^{+6.007}_{-3.734} \quad \mbox{in DHOST}, 
    \label{Eq:Result_XT_1sigma}
\end{eqnarray}
and at the $95\%$ confidence level, we have
\begin{eqnarray}
    &  \hspace{-0.25cm}
      -2.098 < \xi_{\rm t}\, (95\% {\rm CL}) & \quad \mbox{in Horndeski}, \nonumber \\
    & -1.655 < \xi_{\rm t}\, (95\% {\rm CL}) & \quad \mbox{in DHOST}. 
    \label{Eq:Result_XT_95}
\end{eqnarray}
Eqs.~(\ref{Eq:Result_XT_1sigma}) and (\ref{Eq:Result_XT_95}) are one of the main results in this paper. Since the $95\%$ upper bounds of $\xi_{\rm t}$ obtained in this analysis reach the upper bounds set by the flat prior distribution (Section~\ref{Sec:Priors}), we present only the $95\%$ lower bounds here. 

All results in Table~\ref{Table:results_XT_galaxy} are consistent with GR within the $95\%$ level.

\subsection{\texorpdfstring{$\xi_{\rm s}$}{} constraints from the BOSS DR12 galaxies in DHOST theories}
\label{Sec:XS}

Table~\ref{Table:results_XS_galaxy} summarises the constraint results for $\xi_{\rm s}$, defined as $\xi_{\rm s} = \ln_{\Omega_{\rm m}}\left( E_{\rm s} \right) = \ln_{\Omega_{\rm m}}\left( \kappa_{\theta}/\kappa_{\delta} \right)$ (\ref{Eq:TimeDependenceOfE}), characterising the time evolution of the shift term in the second-order velocity field. In GR or Horndeski theories, $\xi_{\rm s}=0$~(\ref{Eq:Xi_GT}) because $\kappa_{\delta}=\kappa_{\theta}=1$. If $\xi_{\rm s}\neq0$, then it is the specific signal appearing in DHOST theories. Note that $\xi_{\rm s}\neq0$ is a sufficient condition for detecting DHOST theories because there can be DHOST theories satisfying $\kappa_{\delta}=\kappa_{\theta}$ (see Section~\ref{Sec:TimeDependencyOfParameters}).

Using all the four galaxy samples, at the $1\sigma$ level, we obtain
\begin{eqnarray}
    \xi_{\rm s} \hspace{-0.25cm}&=&\hspace{-0.25cm} 5.378^{+4.993}_{-2.777} , 
    \label{Eq:Result_XS_1sigma}
\end{eqnarray}
and at the $95\%$ confidence level, we have
\begin{eqnarray}
    - 0.504 < \xi_{\rm s}\, (95\% {\rm CL}),
    \label{Eq:Result_XS_95}
\end{eqnarray}
where we show only the lower limit of $\xi_{\rm s}$ for the same reason as for $\xi_{\rm t}$. Eqs.~(\ref{Eq:Result_XS_1sigma}) and (\ref{Eq:Result_XS_95}) are the other main results of this paper in addition to Eqs.~(\ref{Eq:Result_XT_1sigma}) and (\ref{Eq:Result_XT_95}). 

All results in Table~\ref{Table:results_XS_galaxy} are consistent with GR within the $95\%$ level.

For all the results obtained from Tables~\ref{Table:results_XF_galaxy}, \ref{Table:results_XT_galaxy}, and \ref{Table:results_XS_galaxy}, the standard deviations of $\xi_{f, {\rm s}, {\rm t}}$ obtained by combining the samples $z=0.38$ and $z=0.61$ are smaller than those obtained at $z=0.38$ and $z=0.61$, respectively. Therefore, future galaxy surveys with more redshift bins should improve our $\xi_{f, {\rm s}, {\rm t}}$ constraints. 

Similar to the $f\sigma_8$ and $\sigma_8$ results in GR, we confirm that the constraints of $\xi_{f, {\rm s}, {\rm t}}$ are hardly affected by rescaling the covariance matrix by $15\%-25\%$ at $z=0.38$. This finding indicates that the results for $\xi_{f, {\rm s}, {\rm t}}$ presented here will not change significantly even if future re-analysis from a better mock simulation gives acceptable $p$-values.

\begin{figure*}
    \scalebox{0.60}{\includegraphics[width=\textwidth]{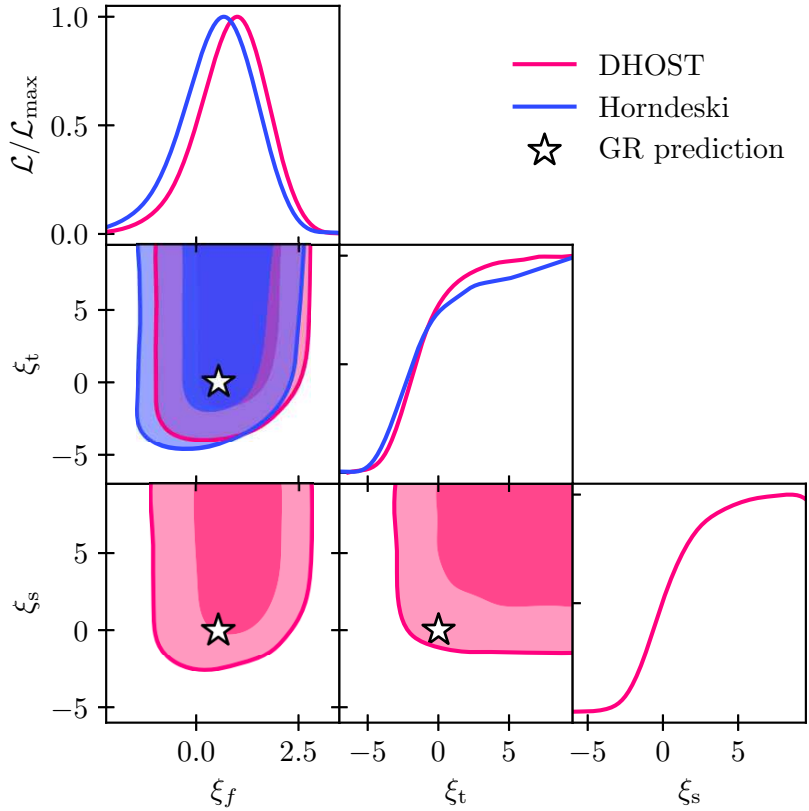}}
    \caption{Marginalized two- and one-dimensional posteriors of $\xi_{f}$, $\xi_{\rm t}$, and $\xi_{\rm s}$ for BOSS DR12. DHOST theories (red) vary these all three parameters, while Horndeski theories (blue) fix $\xi_{\rm s}$ to $\xi_{\rm s}=0$. The contours indicate $68.27\%$ and $95.45\%$ confidence levels. Asterisks indicate predictions by GR: $\xi_f=6/11$, $\xi_{\rm t}=15/1144$, and $\xi_{\rm s}=0$. The NGC and SGC samples at $z=0.38$ and $0.61$ are combined to obtain this result. The rescaled 3PCF covariance matrix (Section~\ref{Sec:RescalingOfCovarianceMatrix}) at $z=0.38$ is used.
    }
	\label{fig:2Dcontour_DHOST}
\end{figure*}

\subsection{Joint analysis with the monopole 3PCF only}
\label{Sec:Monopole3PCF}

This subsection presents the results of a joint analysis of the monopole and quadrupole 2PCFs ($\xi_0$ and $\xi_2$) with only the monopole 3PCFs ($\zeta_{000}$ and $\zeta_{110}$) and compares them with our main results, revealing the importance of the information in the quadrupole 3PCFs ($\zeta_{202}$ and $\zeta_{112}$). In other words, we compare the results corresponding to Case $2$ and Case $7$ in Eq.~\ref{Eq:case}. For simplicity, we focus here on the case where all four BOSS galaxy samples are used assuming DHOST theories and present a comparison of the results for $\xi_f$, $\xi_{\rm t}$, and $\xi_{\rm s}$. For Case $2$, as in Case $7$, we determine the parameter priors according to the method described in Section~\ref{Sec:Priors}, based on the results of the Fisher analysis.

The results of the joint analysis with the monopole 3PCFs are as follows:
\begin{eqnarray}
    (\xi_{\rm t})_{\rm mean} \pm (\xi_{\rm t})_{\rm std} \hspace{-0.25cm} &=& \hspace{-0.25cm} 257.451\pm145.68, \nonumber \\
    (\xi_{\rm s})_{\rm mean} \pm (\xi_{\rm s})_{\rm std} \hspace{-0.25cm} &=& \hspace{-0.25cm} 137.396\pm79.215.
    \label{Eq:mono_vs_quad_gal}
\end{eqnarray}
On the other hand, adding the quadrupole 3PCFs presents $(\xi_{\rm t})_{\rm std}= 4.211$ (Table~\ref{Table:results_XT_galaxy}) and $(\xi_{\rm s})_{\rm std}= 3.438$ (Table~\ref{Table:results_XS_galaxy}).

The addition of the quadrupole 3PCFs reduces the values of $(\xi_{\rm t})_{\rm std}$ and $(\xi_{\rm s})_{\rm std}$ by a factor of $\sim35$ and $\sim20$, respectively. This improvement is consistent with the Fisher analysis result in Section~\ref{Sec:FisherAnalysis} (see Eq.~(\ref{Eq:mono_vs_quad}) and Table~\ref{Table:Fisher_xi}). Therefore, we conclude that the quadrupole component of the 3PCF should always be used to constrain $\xi_{\rm t}$ and $\xi_{\rm s}$. Finally, the same should hold for testing other modified gravity theories through non-linear velocity fields.

\subsection{Consistency check with the Fisher analysis}
\label{Sec:ConsistencyCheckWithFisher}

\begin{table}
\centering
\begin{tabular}{lcc}
\hline\hline
\multicolumn{3}{c}{DHOST} \\
\hline
& $(\xi_f)_{\rm std}$ & $\sigma_{\rm fisher}(\xi_{f})$ \\
\hline
$z_{\rm eff}=0.38$ & $1.131$ & $0.967$\\
$z_{\rm eff}=0.61$ & $1.722$ & $1.782$\\
$z_{\rm eff}=0.38,\,0.61$ & $0.830$ & $0.850$\\
\hline
& $(\xi_{\rm t})_{\rm std}$ & $\sigma_{\rm fisher}(\xi_{\rm t})$ \\
\hline
$z_{\rm eff}=0.38$ & $4.732$ & $3.577$\\
$z_{\rm eff}=0.61$ & $7.387$ & $6.834$\\
$z_{\rm eff}=0.38,\,0.61$ & $4.211$ & $3.169$\\
\hline
& $(\xi_{\rm s})_{\rm std}$ & $\sigma_{\rm fisher}(\xi_{\rm s})$ \\
\hline
$z_{\rm eff}=0.38$ & $3.519$ & $3.147$\\
$z_{\rm eff}=0.61$ & $6.841$ & $5.906$\\
$z_{\rm eff}=0.38,\,0.61$ & $3.438$ & $2.778$\\
\hline
\end{tabular}
\caption{
Comparison of the standard deviations, $\sigma_{\rm fisher}(\theta)$ and $(\theta)_{\rm std}$, obtained from the Fisher analysis and MCMC for $\theta=\xi_f,\, \xi_{\rm t},\, \xi_{\rm s}$. The results are shown for each redshift of $z=0.38$ and $z=0.61$ and the combined case of the two redshifts. In all cases shown here, the NGC and SGC samples are used; the MCMC results at $z=0.38$ use the rescaled covariance matrix (Section~\ref{Sec:RescalingOfCovarianceMatrix}). All the values summarised here are those already given in Tables~\ref{Table:Fisher_xi_4samples}, \ref{Table:results_XF_galaxy}, \ref{Table:results_XT_galaxy}, and \ref{Table:results_XS_galaxy}. When combining the results for the different galaxy samples in Table~\ref{Table:Fisher_xi_4samples}, we use the standard error composition formula, assuming that each galaxy sample is independent.
}
\label{Table:ComparisonWithFisher}
\end{table}

This subsection discusses the consistency between the Fisher analysis results in Section~\ref{Sec:FisherAnalysis} and our final results from MCMC in this section. For this purpose, We compare the standard deviation of a parameter $\theta$ computed from the Fisher analysis, $\sigma_{\rm fisher}(\theta)$, with that estimated from MCMC, $(\theta)_{\rm std}$, where the parameters of interest are $\theta=\xi_f,\, \xi_{\rm t},\, \xi_{\rm t}$, which are the main target of this paper.

Table~\ref{Table:ComparisonWithFisher} summarises the cases for each redshift bin of $z=0.38$ and $z=0.61$ and for using both redshift bins, assuming DHOST theories. The values shown in this table are given from Tables~\ref{Table:Fisher_xi_4samples}, \ref{Table:results_XF_galaxy}, \ref{Table:results_XT_galaxy}, and \ref{Table:results_XS_galaxy}. When combining the results for the different galaxy samples in Table~\ref{Table:Fisher_xi_4samples}, we use the standard error combination formula, assuming that each galaxy sample is independent.

Table~\ref{Table:ComparisonWithFisher} shows that the MCMC results satisfy $(\theta)_{\rm std}\gtrsim\sigma_{\rm fisher}(\theta)$, indicating that the MCMC results are consistent with the Fisher analysis results, as expected. This result reinforces the validity of our main results shown in Tables~\ref{Table:results_XF_galaxy}, \ref{Table:results_XT_galaxy}, and \ref{Table:results_XS_galaxy}.

\subsection{Comments on bias effects on shift terms}
\label{Sec:CommentsOnShiftBias}

DHOST theories change the shift term of the non-linear density fluctuation from GR, which may introduce a new bias effect in the shift term, i.e., the shift bias parameter. Since $E_{f, \rm s, t}$ are the parameters that cancel the $\sigma_8$-dependence using the coefficients of the shift term of the density fluctuation, when the shift bias appears, $E_{f, \rm s, t}$ will also be contaminated by the bias effect. Furthermore, the shift bias may induce bias effects in linear and non-linear velocity fields. In such cases, we cannot use the parameterisation $E_{f,{\rm s},{\rm t}} = \Omega_{\rm m}^{\xi_{f,{\rm s},{\rm t}}}$ adopted in this paper to characterise the time dependence of $E_{f,{\rm s},{\rm t}}$ because the time dependence of the bias parameter is uncertain.

If we assume the presence of the shift bias effect, we propose simultaneously constraining all the six parameters $(F_{\rm g}\sigma_8)$, $(F_{\rm s}\sigma_8)$, $(F_{\rm t}\sigma_8)$, $(G_{\rm g}\sigma_8)$, $(G_{\rm s}\sigma_8)$ and $(G_{\rm t}\sigma_8)$~(\ref{Eq:F_2G_2}) that characterise the growth, shift, and tidal terms in the density and velocity fields in each galaxy sample as a more general test of modified gravity theories. In such an analysis, we should remove the relation $G_{\rm g}=G_{\rm s}-(2/3)G_{\rm t}$ imposed in DHOST theories. In particular, the $E_{\rm s}$ parameter, which represents the ratio of the coefficients of the shift terms of the non-linear density and velocity fields: $E_{\rm s}=(G_{\rm s}\sigma_8)/(F_{\rm s}\sigma_8)$, is always $E_{\rm s}=1$ in GR and Horndeski theories. Therefore, testing whether $E_{\rm s}=1$ in each galaxy sample verifies the theory of varying the shift term, such as DHOST-like theories. In other words, it should provide a means to test the LSS consistency relation, which DHOST-like theories violate (Section~\ref{Sec:Introduction}), using the galaxy 3PCF (or bispectrum).

\section{Conclusions}
\label{Sec:Conclusions}

This paper presents a joint analysis of the anisotropic two-point and three-point correlation functions measured from the publicly available BOSS DR12 galaxy data to test cosmological modified gravity theories. This paper has two important implications. First, it is the first work to extract cosmological information from actual galaxy data using the anisotropic component of the galaxy three-point correlation function induced by the RSD effect. Second, this analysis is the first attempt to constrain the non-linear effects of modified gravity theories from the galaxy three-point statistics.

We consider DHOST theories and their subclass, Horndeski theories, which are the candidates for modified gravity theories (see Section~\ref{Sec:DHOSTtheories}). They are quite general theoretical frameworks of scalar-tensor theories. Since the time evolution equation of the linear density fluctuations in these theories is scale-independent~(\ref{Eq:delta_evo}), the difference with GR appears only in the linear growth rate $f$ in the linear theory~\citep{Hirano:2019nkz}. On the other hand, the non-linear gravitational effect causes a difference in the scale-dependence of the density fluctuation, which allows us to examine the deviation from GR more clearly. Specifically, Horndeski theories change the tidal term of the second-order density fluctuation from GR, while DHOST theories change both the shift and tidal terms~(\ref{Eq:FG_m} and \ref{Eq:Param_FG})~\citep{Hirano:2018uar}. However, since non-linear bias parameters contaminate the density fluctuations, \citet{Yamauchi:2021arXiv210802382Y} have pointed out that one should investigate supposedly unbiased non-linear velocity fields induced by the RSD effect (see Section~\ref{Sec:TimeDependencyOfParameters} for a review). Specifically, they have suggested that one should constrain the parameters $\xi_{\rm t}$ and $\xi_{\rm s}$, which characterise the time evolution of the tidal and shift terms of the second-order velocity field: $\xi_{\rm t}=15/1144$ in GR and $\xi_{\rm s}=0$ in GR and Horndeski theories. Therefore, if $\xi_{\rm s}\neq 0$, then it is the signal specific to DHOST theories; they have also pointed out that in DHOST theories, the parameter $\gamma = \ln_{\Omega_{\rm m}}(f)$, which characterises the time dependence of the linear growth rate $f$, is extended to $\xi_{f} = \ln_{\Omega_{\rm m}}(f/\kappa)$ with $\kappa$ being the time-dependent function appearing in the shift term of the density fluctuation. To this end, we test DHOST and Horndeski theories by constraining these parameters $\xi_f$, $\xi_{\rm t}$, and $\xi_{\rm s}$ using the joint analysis method of the anisotropic 2PCF and 3PCF, established by~\citet{Sugiyama:2020uil}.

The following is a summary of the details of the analysis methodology and the findings obtained.

\begin{enumerate}

    \item Following~\citet{Sugiyama:2018yzo}, we apply the TripoSH decomposition method to the 3PCF to extract information about the anisotropic, i.e., \emph{quadrupole}, component of the 3PCF (see Sections~\ref{Sec:3PCFs} and~\ref{Sec:Estimator}). To simplify the data analysis, we then use only two monopole components ($\zeta_{000}$ and $\zeta_{110}$) and two quadrupole components ($\zeta_{202}$ and $\zeta_{112}$) from the decomposed 3PCF. For the 2PCF, we adopt the commonly used Legendre decomposition method and use the monopole and quadrupole components: i.e., $\xi_0$ and $\xi_2$. It is worth noting that $\zeta_{202}$ includes only the $M=0$ mode that appears in \citet{Scoccimarro:1999ed}'s decomposition formalism, while $\zeta_{112}$ includes $M\neq0$ modes in addition to the $M=0$ mode. Furthermore, the TripoSH-decomposed 3PCF allows a quantitative evaluation and detailed study of the survey window effect present in the measured 3PCFs (see Section~\ref{Sec:WindowCorrections}). Thus, this work is the first to extract information on the $M\neq 0$ modes from actual galaxy data, taking into account the window effect.

    \vspace{3mm}

    \item We only use data at large scales of $80\hMpc \leq r \leq 150\hMpc$, where higher-order non-linear corrections, called loop corrections, are not expected to contribute much to the 2PCF and 3PCF. In order to test modified gravity theories consistently using smaller scales, it is necessary to construct a model that includes the non-linear effects of modified gravity theories so that they are also included in the loop corrections. To our knowledge, only one such analysis has been performed so far for the case of the power spectrum in $f(R)$ gravity~\citep{Song:2015oza}. However, it is known that various uncertainties arise in the non-linear power spectrum in DHOST theories, such as IR cancellation breaking~\citep{Crisostomi:2019vhj,Lewandowski:2019txi} and UV divergence~\citep{Hirano:2020dom}. These theoretical uncertainties should also appear in the bispectrum. Therefore, focusing only on large scales is necessary to remove the theoretical uncertainties and safely constrain the non-linear effects of modified gravity theories. Our analysis is thus the second example of a consistent analysis incorporating the non-linear effects of modified gravity from spectroscopic galaxy surveys, and the first to use the galaxy three-point statistic.

    \vspace{3mm}

    \item As a theoretical model for the 3PCF, we use the IR-resummed model (\ref{Eq:bispec_IR}) proposed by~\citet{Sugiyama:2020uil} (see Section~\ref{Sec:3PCFsBAO}). This model can describe the BAO damping effect while keeping the shape of the 3PCF in the tree-level solution. For this model, we have investigated how the three decomposed non-linear effects, i.e., the growth, shift, and tidal terms, affect the 3PCF multipoles (see Figures~\ref{fig:decom_3PCF_mono} and~\ref{fig:decom_3PCF_quad} in Section~\ref{Sec:Parameters}). For example, in the quadrupole components ($\zeta_{202}$ and $\zeta_{112}$), the dominant term is the product of the linear density fluctuation and the linear velocity field that appears during the coordinate transformation from real space to redshift space; otherwise, the non-linear effects of the density and velocity fields contribute to the quadrupole component to the same extent. Figures~\ref{fig:3PCFs_North_zbin1_mono}-\ref{fig:3PCFs_South_zbin3_quad} in Section~\ref{Sec:3PCFmeasurements} show the $\zeta_{000}$, $\zeta_{110}$, $\zeta_{202}$, and $\zeta_{112}$ measured from the four BOSS galaxy samples and the corresponding theoretical models calculated using the best-fit parameters.

    \vspace{3mm}

    \item We have used the $2048$ publicly available Patchy mocks to compute the covariance matrices of the 2PCF and 3PCF in Section~\ref{Sec:CovarianceMatrix}. In our analysis, we ensure that the number of data bins in the 2PCF and 3PCF is sufficiently smaller than the number of the $2048$ mocks. In particular, the parameter $M_2$ (\ref{Eq:M2}), which represents the impact of a finite number of mocks on the final parameter error, is at most $M_2\sim1.1$ (see Section~\ref{Sec:M1M2}).

    \vspace{3mm}

    \item To understand the nature of the covariance matrix, we have calculated the cumulative ${\rm S/N}$ of the 2PCF and the 3PCF in Section~\ref{Sec:CumulativeSignalToNoiseRatio}. The results show that the cumulative ${\rm S/N}$ of the 3PCF has different characteristics from that of the 2PCF. In the case of the 2PCF, the galaxy sample at $z=0.61$ with a larger volume has a smaller covariance matrix than the sample at $z=0.38$, resulting in a larger ${\rm S/N}$ at $z=0.61$. On the other hand, for the 3PCF, the ${\rm S/N}$ at $z=0.38$ is comparable to the ${\rm S/N}$ at $z=0.61$. Therefore, the difference in survey volume cannot explain the relationship between the ${\rm S/N}$ of the 3PCF at $z=0.38$ and $0.61$. A possible explanation for this 3PCF ${\rm S/N}$ behaviour is that the covariance matrix of the 3PCF depends strongly on the number density of the galaxies~\citep[see][]{Sugiyama:2020MNRAS.497.1684S}: the BOSS sample at $z=0.38$ has a higher number density than the sample at $z=0.61$, even with a smaller survey volume (Table~\ref{Table:Grid}). We interpret this higher number density as why the ${\rm S/N}$ at $z=0.38$ is as high as that at $z=0.61$.

        \vspace{3mm}

    \item We have investigated the extent to which higher-order terms in the TripoSH decomposition of the 3PCF contain cosmological information by Fisher analysis (see Section~\ref{Sec:InformationIn3PCFs}). The results show that $\zeta_{202}$ is the main cosmological information in the quadrupole 3PCF, while other information is contained in the higher-order term $\zeta_{112}$ in addition to $\zeta_{202}$. Since $\zeta_{112}$ contains the $M\neq0$ modes in \citet{Scoccimarro:1999ed}'s decomposition formalism but not in $\zeta_{202}$, this result indicates the importance of the $M\neq0$ modes. 

        \vspace{3mm}

    \item In Section~\ref{Sec:GoodnessOfFit}, we have reported that at large scales ($\geq80\hMpc$), there can be statistically significant differences between the 3PCFs measured from the BOSS galaxies and the corresponding theoretical models, regardless of whether we assume GR, Horndeski or DHOST theories. For example, the $p$-value obtained from the SGC sample at $z=0.38$ is less than $0.01$, and the $p$-value obtained from the combined sample of the four BOSS samples is $0.001$ (see Section~\ref{Sec:GoFBOSS}). This result means that the discrepancies between the galaxy data and the theoretical models cannot be explained within the framework of scalar-tensor theory, even if they are due to unknown physical effects. Other results show that the discrepancy is mainly due to the monopole component of the 3PCF rather than the quadrupole component (see Section~\ref{Sec:GoFMonoQuad}) and that this discrepancy cannot be explained even if the prior distribution of the parameters is changed (see Section~\ref{Sec:DHOSTnoprior}). Finally, we have repeated the same analysis for the $100$ Patchy mocks as for the BOSS sample in Section~\ref{Sec:GoFPatchy}. The results show a statistically significant difference of more than $5\sigma$ between the $p$-values of the Patchy mocks and the BOSS galaxies. Therefore, the statistical variability of the Patchy mock galaxies cannot explain the low $p$-values ($p\sim0.001$) obtained from the BOSS galaxies.

        \vspace{3mm}

    \item In this paper, we assume that the discrepancy between the BOSS galaxy sample and the theoretical model is due to an inappropriate 3PCF covariance matrix computed from the Patch mocks. We then take a conservative approach by artificially rescaling the 3PCF covariance matrix at $z=0.38$ by $15\%$ for NGC and $25\%$ for SGC, resulting in acceptable $p$-values (see Section~\ref{Sec:RescalingOfCovarianceMatrix}). To confirm the validity of this method, we have presented in Section~\ref{Sec:DataAnalysis} the results of constraining the parameters of interest with and without rescaling the covariance matrix and have confirmed that there is no significant difference in the final results obtained in these two cases. We interpret this result as being due to a more significant degeneracy effect between the parameters than the $\sim 20\%$ difference in the 3PCF covariance matrix. Therefore, we do not expect that calculating the covariance matrix from simulation data that better reproduces the distribution of the BOSS galaxies will significantly change the results of the present paper.

        \vspace{3mm}

    \item We have constrained $f\sigma_8$ from the BOSS galaxies assuming GR in Section~\ref{Sec:fsigma8BOSS}. There, we have shown that adding isotropic and anisotropic 3PCF components ($\zeta_{000}$, $\zeta_{110}$, $\zeta_{202}$, and $\zeta_{112}$) does little to improve the results compared to the 2PCF-only analysis. Nevertheless, the analysis using the Patchy mocks shows that the 3PCF information does help to reduce the bias of the mean value of $f\sigma_8$ (see Section~\ref{Sec:fsigma8_mock}). Finally, we obtain $f\sigma_8=0.549^{+0.097}_{-0.122}$ at $z=0.38$ and $f\sigma_8=0.394^{+0.088}_{-0.099}$ at $z=0.61$ in the joint analysis of the anisotropic 2PCF and 3PCF assuming GR (\ref{Eq:result_fsigma8}). These $f\sigma_8$ results are not as competitive as existing constraints~\citep[e.g.,][]{Alam:2016hwk,Ivanov:2020JCAP...05..042I,Lange:2022MNRAS.509.1779L,Kobayashi:2022PhRvD.105h3517K} because we only use large-scale information ($r\geq80\hMpc$).

        One may think that adding the 3PCF information does not improve the $f\sigma_8$ results due to the focus on large scales only ($r\geq80\hMpc$). To test this concern, we have performed a Fisher analysis that includes small scales $(30\hMpc\leq r \leq 150\hMpc)$ and find that even if we extend the used scales to $30\hMpc$, there is no improvement in the $f\sigma_8$ results (see Section~\ref{Sec:SmallScales}). However, note that we use the IR-resummed tree-level model of the 3PCF in this Fisher analysis. Therefore, if we use a theoretical model with various loop corrections applicable down to small scales, parameter degeneracy may break, and it may still be possible to obtain improved $f\sigma_8$ constraints through a joint analysis of the 2PCF and 3PCF. 

        \vspace{3mm}

    \item We have constrained $\sigma_8$ from the BOSS galaxies assuming GR in Section~\ref{Sec:sigma8BOSS}. Thus, while the 3PCF information does not improve the $f\sigma_8$ constraints, it helps to break the degeneracy between parameters by providing information on $\sigma_8$: e.g., it allows us to constrain $f$. We have obtained $\sigma_8=0.692^{+0.209}_{-0.591}$ at $z=0.38$ and $\sigma_8=0.568^{+0.144}_{-0.547}$ at $z=0.61$ at the $1\sigma$ level. These results are consistent with $\sigma_8=0.681,\, 0.606$ at $z=0.38,\, 0.61$ calculated from the cosmological parameters in a flat $\Lambda$CDM model given by Planck 2018. The ratio of the standard deviation to the mean for $\sigma_8$ is $(\sigma_8)_{\rm std}/(\sigma_8)_{\rm mean}=0.66$ at $z=0.38$ and $0.71$ at $z=0.61$, indicating that the galaxy sample at $z=0.38$ provides a better constraint on $\sigma_8$. This result can be attributed to the higher number density of the sample at $z=0.38$ compared to that at $z=0.61$, similar to the argument of the cumulative ${\rm S/N}$ in Section~\ref{Sec:CumulativeSignalToNoiseRatio}.

       \vspace{3mm}

    \item Our main results, the constraints on the $\xi_{f}$, $\xi_{\rm t}$, and $\xi_{\rm s}$ parameters in DHOST theories, are summarised in Sections~\ref{Sec:XF}, \ref{Sec:XT}, and \ref{Sec:XS}. There, we obtain $\xi_f=0.791^{+0.963}_{-0.691}$ (\ref{Eq:Result_XF_1sigma}), $\xi_{\rm t}=5.414^{+6.007}_{-3.734}$ (\ref{Eq:Result_XT_1sigma}), and $\xi_{\rm s}=5.378^{+4.993}_{-2.777}$ (\ref{Eq:Result_XS_1sigma}) at the $1\sigma$ level; we also have $-0.907<\xi_f<2.447$ (\ref{Eq:Result_XF_95}), $-1.655<\xi_{\rm t}$ (\ref{Eq:Result_XT_95}), and $-0.504<\xi_{\rm s}$ (\ref{Eq:Result_XS_95}) at the $95\%$ confidence level. Since we cannot detect the signal of the tidal and shift terms in the second-order velocity field in the present analysis, we can only present the $95\%$ lower bounds of the $\xi_{\rm t}$ and $\xi_{\rm s}$ parameters. These results are consistent with the GR predictions $\xi_f=\gamma=6/11$, $\xi_{\rm t}=15/1144$, and $\xi_{\rm s}=0$ (see Figure~\ref{fig:2Dcontour_DHOST}). Moreover, we have checked the consistency of the estimated results from the BOSS galaxy sample with the Fisher analysis for the constraints on the $\xi_{f, \rm t,s}$ parameters in DHOST theories in Section~\ref{Sec:ConsistencyCheckWithFisher}.
        
        In Horndeski theories, we obtain $\xi_f=\gamma=0.485^{+0.967}_{-0.708}$ and $\xi_{\rm t}=5.151^{+6.112}_{-4.016}$ at the $1\sigma$ level, and $-1.216<\gamma<2.175$ and $-2.098<\xi_{\rm t}$ at the $95\%$ confidence level. The $\gamma$ constraint in Horndeski theories obtained here is not directly comparable to those obtained from existing studies by, e.g., ~\citet{Gil-Marin:2016wya} because we simultaneously vary the $\xi_{\rm t}$ parameter in Horndeski theories.

        \vspace{3mm}

    \item We have shown that the anisotropic component of the 3PCF contributes significantly to the constraints on the shape of the non-linear velocity field in Section~\ref{Sec:Monopole3PCF}. In particular, the constraints on the parameters $\xi_{\rm t}$ and $\xi_{\rm s}$ are $\sim35$ and $\sim20$ times better when the anisotropic component is added than when only the isotropic component is considered. This result strongly supports the main claim of this paper that the anisotropic three-point statistics should be considered to test the non-linearity of modified gravity theories.
 
\end{enumerate}

Below is a summary of some of the concerns and future enhancements to the results of this paper.

\begin{enumerate}

    \item In order to encourage the future development of the anisotropic 3PCF analysis, we comment on the situation beyond the assumptions used to derive the non-linear effects of DHOST theories that we focus on in this paper (see Section~\ref{Sec:LimitationsOfOurAssumptions}). First, our analysis can be applied to other modified gravity theories, such as $f(R)$ gravity models and brane-world models. In addition, it should also be possible to constrain effects such as the CDM-baryon relative velocity and massive neutrinos, which give rise to characteristic non-linear behaviour. The calculations of DHOST theories in this paper assume minimal coupling between the metric field and the scalar field, Gaussianity of the initial conditions, and the quasi-static limit, but we need additional correction terms if these assumptions are removed. In addition, since DHOST theories modify the shift term from GR, we cannot exclude the possibility of shift bias, which we do not consider in a $\Lambda$CDM model. In the presence of shift bias, we cannot use the $\xi_{\rm s}$ and $\xi_{\rm t}$ parameters to constrain DHOST theories, but we expect the $E_{\rm s}$ and $E_{\rm t}$ parameters constrained at each redshift to remain valid (Section~\ref{Sec:CommentsOnShiftBias}).

        \vspace{3mm}

    \item We also comment on some improvements in our analysis of the anisotropic 3PCF (see Section~\ref{Sec:LimitationsOf2PCFand3PCF}). First, as more mock catalogues are created in the future, increasing the number of multipoles in the 3PCF to be considered should improve the results of this work~\citep[e.g.,][]{Byun:2022arXiv220504579B}. Second, as shown in Figure~\ref{fig:fisher}, we can dramatically improve the current parameter constraints by using the theoretical model of the 3PCF, which is applicable to small scales (see Section~\ref{Sec:SmallScales}). Third, although we have used the shape of the linear power spectrum calculated by an $\Lambda$CDM model in a high-$z$ region in this work, it needs to be calculated in the framework of DHOST theories in the future~\citep[e.g.,][]{Hiramatsu:2020PhRvD.102h3525H}. Fourth, we have calculated the Gaussian function describing the damping effect of the BAO signal for a $\Lambda$CDM model, but we also need to constrain this function itself. Finally, we have neglected the Alcock-Paczy\'{n}ski (AP) effect in this work; the analysis method of the anisotropic 3PCF that includes the AP effect has been established by~\citet{Sugiyama:2020uil} using the Patchy mock and should be straightforward to apply to actual galaxy data. We hope that addressing these issues will further improve our results.

\end{enumerate}

Finally, in Appendix~\ref{Sec:HITOMI} we provide the software package that can reproduce all the results obtained in this paper, \textsc{HITOMI}. The aim of \textsc{HITOMI} is to make available all the programs we have used to complete the anisotropic 3PCF analysis, from downloading the SDSS DR12 galaxy data, measuring the 2PCFs and 3PCFs, computing the theoretical models, calculating the covariance matrices, the window function corrections, MCMC analysis, and producing figures and tables. This makes it easier for any user to see how partial improvements to \textsc{HITOMI}, e.g. improved 3PCF model calculations, feed through to the final parameter constraints. Furthermore, by replacing the BOSS galaxy data used in \textsc{HITOMI}, our analysis can be easily applied to future galaxy surveys such as DESI~\citep{Aghamousa:2016zmz}, Euclid~\citep{Laureijs:2011gra}, and PFS~\citep{Takada:2012rn}.

\section*{Acknowledgements}

NSS acknowledges financial support from JSPS KAKENHI Grant Number 19K14703. Numerical computations were carried out on Cray XC50 at Center for Computational Astrophysics, National Astronomical Observatory of Japan. The work of SH was supported by JSPS KAKENHI Grants No.~JP21H01080. The work of TK was supported by JSPS KAKENHI Grant No.~JP20K03936 and MEXT-JSPS Grant-in-Aid for Transformative Research Areas (A) ``Extreme Universe'', No.~JP21H05182 and No.~JP21H05189. The work of DY was supported in part by JSPS KAKENHI Grants No.~19H01891, No.~22K03627. SS acknowledges the support for this work from NSF-2219212. SS is supported in part by World Premier International Research Center Initiative (WPI Initiative), MEXT, Japan. H-JS is supported by the U.S. Department of Energy, Office of Science, Office of High Energy Physics under DE-SC0019091 and DE-SC0023241. This project has received funding from the European Research Council (ERC) under the European Union’s Horizon 2020 research and innovation program (grant agreement 853291). FB is a University Research Fellow.


\section*{Data Availability}

The data underlying this article are available at the SDSS data base (\url{https://www.sdss.org/dr12/}).



\bibliographystyle{mnras}
\bibliography{ms} 


\appendix

\section{HITOMI}
\label{Sec:HITOMI}

In order to improve the reproducibility of the results of this paper, we publish the complete set of program codes we used under the name \textsc{HITOMI}. The languages used in it are \textsc{c++} and \textsc{python}. Users can download the source files from the following link: \url{https://github.com/naonori/hitomi.git}. In particular, to reproduce the results of this paper, refer to the DEMO section of the linked page. There, it explains how to measure the 2PCF and 3PCF from the BOSS DR12 data, compute the theoretical model including the window function correction, compute the covariance matrix from the Patchy mocks, combine them to perform the Fisher and MCMC analyses, and finally summarise the obtained results in figures and tables. To illustrate these things, we recorded a video of us running the program and uploaded it to \textsf{YouTube}. The text editor used for this is \textsf{vim}.

\textsc{HITOMI} requires several external programs such as \textsc{montepython}~\citep{Brinckmann:2018cvx}, \textsc{CLASS}~\citep{Blas:2011rf}, \textsc{CUBA}~\citep{Hahn:2005CoPhC.168...78H}, \textsc{GSL}~\footnote{\url{http://www.gnu.org/software/gsl/}}, \textsc{FFTW}~\citep{FFTW05}, and \textsc{FFTLog}~\citep{Hamilton:2000MNRAS.312..257H}. We have written a script with the code to install the external programs needed to run \textsc{HITOMI} on the Cray XC50 at the Center for Computational Astrophysics of the National Astronomical Observatory of Japan. A video recording of the use of this script is available at the following link: \url{https://www.youtube.com/watch?v=vlP7XIXZsUM}. Of course, users of other PC clusters will have to install \textsc{HITOMI} according to their environment. Nevertheless, our installation instructions will be helpful to users as a demonstration.

\textsc{HITOMI} not only reproduces the results of this paper but also offers various options. For example, it can measure both the power spectrum and the bispectrum. \textsc{HITOMI} also provides the codes to simplify the 3PCF and bispectrum measurements for simulations with periodic boundary conditions with a global line-of-sight direction. It is also possible to measure the 2PCF and 3PCF (power spectrum and bispectrum) after the reconstruction of the galaxy distribution and compute the corresponding reconstructed models~\citep{Eisenstein:2006nk,Shirasaki:2021PhRvD.103b3506S}. Although not yet implemented, in the future, we plan to release a code to compute the bispectrum covariance matrix of galaxies based on perturbation theory, as was done by \citet{Sugiyama:2020MNRAS.497.1684S}. We also plan to release a code that performs an anisotropic BAO analysis using the anisotropic 3PCF, as in \citet{Sugiyama:2020uil}.

It is possible to modify parts of the \textsc{HITOMI} code, e.g., the theoretical calculation of the 3PCF, to investigate how the results propagate to the final parameter constraint results. It is also possible to replace the BOSS DR12 galaxy data with data from other galaxies or galaxy clusters, e.g. PFS~\citep{Takada:2012rn}, DESI~\citep{Aghamousa:2016zmz}, Euclid~\citep{Laureijs:2011gra}, SPHEREx~\citep{SPHEREx:2014arXiv1412.4872D}, CMB-S4~\citep{CMBS4:2019BAAS...51g.209C}, and eROSITA~\citep{eROSITA:2021A&A...647A...1P}, to perform data analysis of the 3PCF or bispectrum.

\bsp	
\label{lastpage}
\end{document}